\documentclass[aps,prb,twocolumn,longbibliography,notitlepage,superscriptaddress,nofootinbib]{revtex4-2}

\usepackage{comment}

\usepackage{ifthen}
\usepackage{sseq}
\usepackage{amsmath}\allowdisplaybreaks
\usepackage{amsfonts}
\usepackage{amssymb}
\usepackage{dsfont}
\usepackage{amsmath}
\usepackage{physics}
\usepackage{epsfig}
\usepackage{amsfonts}
\usepackage{savesym}
\usepackage{braket}
\usepackage{graphicx}
\usepackage{longtable}
\usepackage{afterpage}
\graphicspath{ {./figs/} }
\usepackage{epsfig}
\usepackage[scr=boondox]{mathalpha}
\usepackage{cancel}
\usepackage{xcolor}
\usepackage{tikz-cd}
\usepackage{tikz}
\def\sp{\text{}}
\usetikzlibrary{cd}
\usepackage{ stmaryrd }
\usepackage{mathtools}

\usepackage{mathtools}
\usepackage{bigdelim}
\usepackage{color}
\usepackage[linktocpage]{hyperref}
\hypersetup{colorlinks=true,citecolor=blue,linkcolor=blue, urlcolor=blue, breaklinks=true}

\usepackage{slashed}
\def\H{\mathcal{H}}

\def\Z{\mathbb{Z}}

\def\TT{\mathsf{T}}

\def\Aut{\mathrm{Aut}}
\def\Out{\mathrm{Out}}

\def\gb{\mathbf{g}}

\def\gbf{\mathbf{g}}
\def\hbf{\mathbf{h}}
\def\PH{\Xi}

\def\Oo{\mathrm{O}}
\def\QQ{\mathbb{Q}}
\def\Int{\text{eff}}

\newcommand{\ZZ}{{\mathbb Z}}

\newcommand{\RR}{{\mathbb R}}

\newcommand{\om}{{\omega_2}}
\def\Omc{\mathcal{O}}
\def\diag{\mathrm{diag}}
\newcommand{\ndiv}{\not\hspace{2.5pt}\mid}
\def\CS{\mathrm{CS}}

\def\ii{\mathrm{i}}

\newcommand{\eps}{\epsilon}

\def\Pmc{\mathcal{P}}

\def\Tt{\mathbf{T}}

\def\xx{\mathrm{x}}

\def\Sq{\mathrm{Sq}}

\def\ww{\mathrm{w}}
\def\d{\delta}
\def\r{\rho}

\def\n{\nu}

\def\W{\Omega}
\def\OO{\Omc}
\def\dd{\mathrm{d}}

\def\QU{\mathrm{QU}}

\def\Hom{\mathrm{Hom}}
\def\s{\sigma}
\def\t{\tau}

\def\SU{\mathrm{SU}}
\def\Sp{\mathrm{Sp}}
\def\PSp{\mathrm{PSp}}

\def\PP{\mathbb{P}}
\def\img{\mathrm{img}}
\def\RR{\mathbb{R}}

\def\q{\theta}

\def\ZZ{\mathbb{Z}}

\def\gcd{\mathrm{gcd}}
\def\s{\sigma}
\def\ww{\mathrm{w}}

\def\d{\delta}

\def\w{\omega}

\def\ufk{\mathfrak{u}}
\def\q{\theta}

\def\Bc{\mathrm{B}}

\def\n{\nu}
\def\U{\mathrm{U}}
\def\wfk{\mathfrak{w}}
\def\mod{\,\text{mod}\,}
\def\DD{\mathbb{D}}

\def\SO{\mathrm{SO}}
\def\Ufk{\mathfrak{U}}
\def\SU{\mathrm{SU}}
\def\PSO{\mathrm{PSO}}
\def\Spin{\mathrm{Spin}}
\def\Sp{\mathrm{Sp}}
\def\Oo{\mathrm{O}}
\def\Sq{\mathrm{Sq}}
\def\Pin{\mathrm{Pin}}

\def\Rr{\mathbf{R}}

\def\Hc{\mathrm{H}}
\def\ii{\mathrm{i}}
\def\PSU{\mathrm{PSU}}
\def\SU{\mathrm{SU}}
\def\SO{\mathrm{SO}}
\def\PO{\mathrm{PO}}
\def\hbf{\mathbf{h}}

\def\i{\iota}
\def\l{\lambda}

\def\efk{\mathfrak{e}}

\def\Tor{\mathrm{Tor}}
\begin{document}
\title{Non-perturbative constraints from symmetry and chirality on Majorana \\ zero modes and defect quantum numbers in (2+1)D
}

\author{Naren Manjunath}
\affiliation{Department of Physics, Condensed Matter Theory Center, and Joint Quantum Institute, University of Maryland, College Park, Maryland 20742, USA}

\author{Vladimir Calvera}
\affiliation{Department of Physics, Stanford University, Stanford, California 94305, USA}

\author{Maissam Barkeshli}
\affiliation{Department of Physics, Condensed Matter Theory Center, and Joint Quantum Institute, University of Maryland, College Park, Maryland 20742, USA}

\begin{abstract}
In (1+1)D topological phases, unpaired Majorana zero modes (MZMs) can arise only if the internal symmetry group $G_f$ of the ground state splits as $G_f = G_b \times \Z_2^f$, where $\Z_2^f$ is generated by fermion parity, $(-1)^F$. In contrast, (2+1)D topological superconductors (TSC) can host unpaired MZMs at defects even when $G_f$ is not of the form $G_b \times \Z_2^f$. In this paper we study how $G_f$ together with the chiral central charge $c_-$ strongly constrain the existence of unpaired MZMs and the quantum numbers of symmetry defects. Our results utilize a recent algebraic characterization of (2+1)D invertible fermionic topological states, which provides a non-perturbative approach based on topological quantum field theory, beyond free fermions. We study physically relevant groups such as $\U(1)^f\rtimes H,\SU(2)^f \times H, \U(2)^f\rtimes H $, generic Abelian groups, as well as more general compact Lie groups, antiunitary symmetries and crystalline symmetries. We present an algebraic formula for the fermionic crystalline equivalence principle, which gives an equivalence between states with crystalline and internal symmetries.  In light of our theory, we discuss several previously proposed realizations of unpaired MZMs in TSC materials such as Sr$_2$RuO$_4$, transition metal dichalcogenides and iron superconductors, in which crystalline symmetries are often important; in some cases we present additional predictions for the properties of these models.
 
\end{abstract}
\maketitle

\addtocontents{toc}{\protect\setcounter{tocdepth}{1}}

\tableofcontents

\section{Introduction}\label{sec:intro}

The possibility of topological superconductivity in electronic systems has generated intense interest in condensed matter physics \cite{read2000,kitaev2001,nayak2008,hasan2010,qi2010RMP,alicea2012review,Dassarma2015mzm,Sato2017tsc,sau2021}. One major reason for this interest is the possibility of realizing unpaired localized Majorana zero modes (MZMs) and their potential applications for topological quantum computation.\footnote{Here `unpaired' refers to having an odd number of MZMs, which necessarily leads to topological degeneracies.} In a mean-field treatment, in which phase fluctuations are ignored, topological superconductors (TSCs) can be modeled in terms of gapped many-body states of fermions. As such, they can be analyzed using theoretical techniques that have been developed to characterize and classify gapped topological states of matter in general, e.g. \cite{Kitaev2009periodic,ryu2010,Chen2013,kapustin2014SPTbeyond,kapustin2015fSPT,senthil2015,Chiu2016review,Freed:2016rqq,Wang2020fSPT,Barkeshli2019,barkeshli2021invertible,bulmashSymmFrac,aasen2021characterization}. 

Most topological superconductors of interest, such as the spinless $p$-wave superconductor in (1+1)D or the $p+ip$ superconductor in (2+1)D, are often modeled in terms of a free fermion Bogoliubov-de-Gennes (BdG) Hamiltonian. In a many-body context, these are examples of \it invertible \rm fermionic topological states of matter \cite{Freed:2016rqq,barkeshli2021invertible,aasen2021characterization}. Invertible topological states are defined by the property that they do not host deconfined anyon excitations and have a unique ground state on any spatial manifold with a fixed set of boundary conditions.\footnote{The term {\it invertible} refers to the fact that the ground state $|\Psi\rangle$ possesses an ``inverse" state $|\Psi^{-1}\rangle$, such that the $|\Psi\rangle \otimes |\Psi^{-1}\rangle$ can be adiabatically connected to a trivial product state.}  

In some cases topological superconductors host unpaired localized MZMs at their 0-dimensional defects. These defects may include boundaries of a (1+1)D system, fermion parity vortices of a (2+1)D system, or various other kinds of symmetry defects such as half-quantum vortices, lattice disclinations and dislocations, or corners of a system \cite{ivanov2001hqv,dassarma2006hqv,Asahi2012mzm,hughes2014mzm,Qian2014QSH,Hsu2020htsc,geier2018htsc,khalaf2018hoti,Benalcazar2014,wang2018htsc,wang2018maj,yan2018maj,liu2018maj,ueno2013sr}. An important theoretical question is to understand the fundamental constraints on realizing unpaired localized MZMs. For example, it is known that superconductivity is a crucial requirement, which translates into the statement that systems with $U(1)$ charge conservation symmetry preserved in the ground state are forbidden from hosting unpaired MZMs. However the constraints on unpaired MZMs are in general significantly stronger than simply requiring superconductivity.

Let $G_f$ be the internal (equivalently, on-site) fermionic symmetry group of the many-body ground state; here fermionic refers to the fact that fermion parity, $(-1)^F$, is included as a symmetry in $G_f$. $G_b = G_f/ \Z_2^f$ is then the symmetry group that acts on bosonic operators and $\Z_2^f$ is the group generated by fermion parity. In (1+1)D, unpaired MZMs can only exist if $G_f$ splits as a direct product: $G_f = \Z_2^f \times G_b$. This is a significantly stronger constraint than simply the requirement of superconductivity in the ground state.  The above constraint, for example, immediately rules out spin singlet superconductors, which have $\SU(2)^f$ internal symmetry. 

As we discuss later, one can interpret the case where $G_f$ is not of the form $\Z_2^f \times G_b$ as the case where the fermion carries fractional quantum numbers under $G_b$. Therefore the requirement that $G_f = \Z_2^f \times G_b$ amounts to the requirement that unpaired MZMs can only exist in (1+1)D if the fermion does not carry fractional quantum numbers under $G_b$. 

Remarkably, in (2+1)D the constraints are quite different and far richer. In particular, it is possible that unpaired MZMs can exist in systems where $G_f$ is not of the form $\Z_2^f \times G_b$. That is, unpaired MZMs in (2+1)D invertible topological phases are compatible with the fermion carrying fractional quantum numbers under $G_b$. 
(2+1)D topological states can also exhibit a non-trivial chiral central charge $c_-$, corresponding to the possibility of topologically protected chiral edge states. 
There is a rich set of constraints, involving $c_-$ and $G_f$, on when (2+1)D topological superconductors can host unpaired localized MZMs.

In addition to unpaired localized MZMs, symmetry defects in (2+1)D systems can also carry non-trivial quantum numbers of the $G_f$ symmetry group. These quantum numbers must also satisfy a rich set of constraints involving $c_-$, $G_f$, and whether unpaired MZMs exist. 

The main purpose of this paper is to study constraints from symmetry $G_f$ and chirality $c_-$ on unpaired MZMs and defect quantum numbers in (2+1)D invertible topological states. Our results utilize a non-perturbative approach that relies primarily on the properties of symmetry defects, and does not use explicit Hamiltonians. As such, this approach applies to generic interacting many-body systems of fermions, beyond the free fermion limit. This approach is based on a recent complete characterization and classification of invertible fermionic topological phases in (2+1)D using the framework of $G$-crossed braided tensor categories and Chern-Simons theory \cite{barkeshli2021invertible} (see also \cite{aasen2021characterization}). 

Our results emphasize that the possibility of unpaired MZMs in a TSC is constrained not just by the symmetry $G_f$ but also by the chiral central charge $c_-$ of the system, when $G_f$ is unitary. Such constraints are often straightforwardly encoded in the properties of symmetry defects; however, they may not always be apparent from a Hamiltonian perspective. This approach is therefore a useful complementary tool in fully understanding the physics of TSC systems. Symmetry groups $G_f$ with anti-unitary components can often impose constraints on when fermion parity vortices must carry a Kramers pair of localized MZMs. 

We note that Ref.~\cite{ning2021enforced} discusses constraints on (2+1)D invertible fermionic phases using a method similar to ours, but from the perspective of `enforced symmetry breaking,' i.e. that if $G_f$ and $c_-$ are specified, and $c_-$ is fixed, invertible phases can sometimes only be realized by breaking $G_f$ down to a subgroup. However, Ref.~\cite{ning2021enforced} does not discuss specific results on unpaired MZMs in topological superconductors. 

\begin{table*}[t]
    \centering
    \begin{tabular}{p{0.2\textwidth}|l|p{0.2\textwidth}|p{0.2\textwidth}|p{0.3\textwidth}}
    \hline
    \multicolumn{5}{c}{Results for unitary symmetries} \\
    \hline
        $G_f$ & $G_b$ & Allowed $c_-$ & Values of $c_-$ supporting unpaired MZMs & Physical model/other comments \\ \hline
        
        $G_b \times \Z_2^f$ & $G_b$ unitary  & $\frac{1}{2}\Z$ & $\frac{1}{2}\Z$ & Unpaired MZMs allowed at integer $c_-$ only if $G_b$ has a $\Z_{2n}$ or $\Z$ factor \\
        $\U(1)^f \times H$  & $\U(1)\times H$  & $\Z$ & - &Chern insulator \\
        $\Z_{4q}^f \times H$ & $\Z_{2q}\times H$  & $\Z$ & - &Charge $4q$ superconductor \\
        $\Z_{4q+2}^f \times H \cong \Z_2^f \times \Z_{2q+1} \times H$ & $\Z_{2q+1}\times H$  & $\frac{\Z}{2}$ & $\frac{1}{2}\Z$ &Charge $4q+2$ superconductor  \\
        $\SU(2)^f \times H$  & $\SO(3)\times H$ & $2\Z$ & - & Spin-singlet superconductor \\
        $\mathbb{D}_{8n}^f$& $\mathbb{D}_{4n} = \Z_{2n} \rtimes \Z_2$ &$\Z$ & $2\Z+1$ &  \\
         $\Z_2^f \times A$ & Abelian $A$ & $\tfrac{1}{2}\Z$ & $\tfrac{1}{2}\Z$ or $\Z + \frac{1}{2}$ & $\frac{1}{2}\Z$ if $A$ has a $\Z_{2n}$ or $\Z$ factor, and $\Z + \frac{1}{2}$ otherwise \\
        $U(1)^f \rtimes H$& $U(1) \rtimes H$& $\Z$ &$2\Z+1$ & 
${\bf g}$ symmetry defect has unpaired MZM iff ${\bf g}$ acts as charge conjugation and $c_-$ is odd
\\
        Fermion carries fractional quantum numbers under both $G_b^A$ and $G_b^B$& $G_b^A \times G_b^B$& $2\Z$ (odd $c_-$ only allowed if Eq.~\eqref{eq:GbaGbb} holds) &- &\\
       \hline
    \multicolumn{5}{c}{Unitary compact Lie groups} \\
    \hline
        $\Oo(n)^f$  & $\PO(n)$ & $\frac{\gcd(n,16)}{2}\ZZ$ & $\gcd(n,8)\ZZ + \frac{\gcd(n,16)}{2}$ &{$2c_-$ identical layers of a spinless $p+ip$ SC have $G_f = \Oo(2c_-)^f$; half-quantum vortex in any 2 layers hosts unpaired MZMs} \\
        $\SU(2n)^f$& $\SU(2n)^f/\ZZ_2^f$ & $2\gcd(4,n)\ZZ$ &- & \\
        $\U(n)^f$& $\U(n)^f/\ZZ_2^f$ & $\gcd(n,8)\ZZ$ & - & $n$ identical layers of Chern insulator   \\
        $\U(n)^f\rtimes \ZZ_2$& $(\U(n)^f/\ZZ_2^f)\rtimes \ZZ_2$ & $\gcd(n,8)\ZZ$ & $2\ZZ+1$ (only for odd $n$) &   \\
        $\Sp(n)^f$& $\PSp(n)$ & $\gcd(n,4)2\ZZ$ & - &  \\
       \hline 
    \end{tabular}
    \caption{Summary of mathematical results for internal unitary symmetry groups $G_f$. $G_b = G_f / \Z_2^f$ is the symmetry acting on bosonic operators and $H$ is an arbitrary group. The chiral central charge $c_-$ is in general either a half-integer or an integer. These results can be applied to spatial symmetries, if we modify the definition of $G_f$ in accordance with the fermionic crystalline equivalence principle stated in Section \ref{sec:cryst}.}
    \label{tab:summary} 
\end{table*}

\begin{table*}[t]
    \centering
    
     \begin{tabular}{l|l|p{0.25\textwidth}|p{0.45\textwidth}}
     \hline
    \multicolumn{4}{c}{Results for antiunitary internal symmetries $(c_-=0)$} \\
    \hline
        $G_f$ & $G_b$  &Choices of $n_1$ which imply unpaired MZMs   & Physical model/comments\\ \hline
         $\Z_n \times \Z_2^{\Tt} \times \Z_2^f$ &$\Z_n \times \Z_2^T$ & $n_1 = \ww_1$ (only if 8 divides $n$) & If $n=4$, unpaired MZMs are ($\mathcal{O}_4$) obstructed 
        \\ 
        $\Z_{2n}^f \times \Z_2^{\Tt}$ & $\Z_n\times \Z_2^{\Tt}$& - & \\
        $ A \times \ZZ_2^{\Tt} \times \ZZ_2^{f}$& $A \times \ZZ_2^{\Tt} $  & $n_1$ admits a lift to $\H^1(A,\ZZ_8)$.  & Class BDI + Abelian symmetries ($A$). 
               \\
        $\U(1)^f \times H^{\Tt}$ 
        &$\U(1) \times H^{\Tt}$ &- & Generalization of Class AIII TI \\ 
                $\ZZ_4^{\Tt f}\rtimes \ZZ_2$&$ \ZZ_2^{\Tt} \times \ZZ_2 $ & $n_1=s_1+\ww_1$ &  Fermions with $\ZZ_2$ eigenvalues $+1, -1$ form spinless $p+ip$ and $p-ip$ SC layers respectively \\
        
        $\ZZ_4^{\Tt f} \rtimes H$&$ \ZZ_2^{\Tt} \times H $  &  $n_1=s_1+\rho$ & $\rho(\gbf)\in\ZZ_2$ is defined by $\gbf\Tt \bar{\gbf}= \Tt^{1+2\rho(\gbf)} $
        \\ 
        \hline \hline
         $G_f$ & $G_b$  & Fermion parity flux carries Majorana Kramers pair & Physical model/comments\\ \hline
         $\Z_n \times \Z_4^{\Tt f}$ &$\Z_n \times \Z_2^{\Tt}$ & $n_1 = s_1$ & Class DIII TSC\\
        $\Z_{4n}^{\Tt f} $ &$\Z_{2n}^{\Tt}$ & $n_1 = s_1$ (only if $n$ is odd) & When $n$ is odd $\ZZ_{4n}^{\Tt f}\cong \ZZ_{4}^{\Tt f}\times \ZZ_{n}$\\
    \end{tabular}
    \caption{Constraints when the internal symmetry group $G_f$ contains antiunitary elements. $s_1: G_b \rightarrow \Z_2$ specifies which elements of $G_b$ are antiunitary. $n_1: G_b \rightarrow \Z_2$ is a homomorphism, further explained in Sec.~\ref{sec:BasicData}. If ${\bf g} \in G_b$ is a unitary operation, $n_1({\bf g}) = 1(0)$ indicates that a ${\bf g}$ symmetry defect will(will not) host unpaired MZMs (upper section of table). When ${\bf g}$ is time-reversal, $n_1({\bf g}) = 1$ instead implies that a fermion parity flux carries a degenerate Kramers pair of MZMs (lower section). If $\Z_n$ is unitary and $n$ is even, $\ww_1$ is the nontrivial homomorphism from $\Z_n \rightarrow \Z_2$. These results can also be applied to crystalline symmetries using the fCEP; there, the interpretation of $n_1$ as indicating unpaired MZMs/Majorana Kramers pairs may differ. 
    \label{tab:summaryAntiunitary} }
\end{table*}

\subsection{Summary of main results}

Our main results are summarized in Tables \ref{tab:summary}, \ref{tab:summaryAntiunitary} and \ref{tab:cryst}. In Table \ref{tab:summary} we consider various physically relevant unitary internal symmetry groups $G_f$, and list (i) the values of $c_-$ that permit an invertible phase, and (ii) the values of $c_-$ that additionally permit unpaired MZMs at symmetry defects. Note that $c_-$ can only be an integer or a half-integer for invertible fermionic phases. 
In Table \ref{tab:summaryAntiunitary} we present results for antiunitary internal symmetries. Here $c_-$ must be zero. Results for crystalline symmetries are given in Table~\ref{tab:cryst}. We briefly summarize these results below, emphasizing that they hold even in the presence of strong interactions. Applications to models of TSCs in the prior literature (mostly involving crystalline symmetries) are discussed in Sec.~\ref{sec:apps}.
Note we always consider (2+1)D systems, unless otherwise specified. 

\paragraph*{Notational remarks:} Below we often refer to the groups $\Oo(n)^f$, $\SU(n)^f$, $\Sp(n)^f$, $H^f$, etc. The superscript refers to the fact that a $\Z_2$ subgroup of the center of these groups is identified with fermion parity, $\Z_2^f$. We use the notation $k\Z$ for the subgroup of the rational numbers whose elements are $kn$ where $n$ is an integer and $k$ is a rational number, and $\Z + k$ for the set with elements $n+k$, where $n$ is an integer. In particular when $k = \frac{1}{2}$ we write $\frac{1}{2}\Z$ and $\Z + \frac{1}{2}$ respectively.\footnote{This is different from the symbol $\Z/2$, which is sometimes used for the group $\Z_2 \cong \Z/2\Z$.} The symbols $\Z_{2n}^{\bf T},\Z_{2n}^{\bf R},\Z_{2n}^{\bf{RT}}$ denote that the group $\Z_{2n}$ is generated by a time-reversal ${\bf T}$, a spatial reflection ${\bf R}$, or by the combination ${\bf RT}$ respectively. Similarly, $H^{\bf T}$ means that some element in $H$ is identified with ${\bf T}$ (as will be specified on a case-by-case basis). 
\begin{table*}[t]
    \centering
    
     \begin{tabular}{l|l|l|p{0.15\textwidth}|p{0.5\textwidth}}
     \hline
    \multicolumn{5}{c}{Results for crystalline symmetries}  \\
    \hline
        $G_f^{\sp}$ & $G_f^{\Int}$  &Allowed $c_-$ & $c_-$ which allows unpaired MZMs & Interpretation \\ \hline
       $C_{2k} \times \Z_2^f$ &$\Z_{4k}^{f}$ & $\Z$& - & No unpaired MZMs \\
       $C_{2k+1} \times \Z_2^f$ &$\Z_{2k+1} \times \Z_2^f$ &$\frac{1}{2}\Z$ & - & Unpaired MZMs only at fermion parity fluxes ($c_- \in \Z+\frac{1}{2}$)\\
       $C_{4k}^{f}$ &$\Z_{2k} \times \Z_2^f$ & $\frac{1}{2}\Z$& $\frac{1}{2}\Z$ & Unpaired MZMs at disclinations/corners with angle $\pi/k$ or at fermion parity fluxes (eg. spinless $p+ip$ SC) \\
       $\Z^2 \times \Z_2^f$ (translation) &$\Z^2\times \ZZ_2^f$ & $\frac{1}{2}\Z$& $\frac{1}{2}\Z$ & Unpaired MZMs at dislocations with Burgers vector along either $\hat{x}$ or $\hat{y}$\\ 
       $\Z^2 \times_{\pi} \Z_2^f$&$\Z^2 \times_{\pi} \Z_2^f$ & $\Z$& $\Z$& $G_f$ has $\pi$ flux per unit cell; unpaired MZMs allowed at dislocations with Burgers vector along $\hat{x}$ or $\hat{y}$ \\
       \hline \hline
      $\Z_2^{\bf R}\times \Z_2^f$  & $\Z_4^{{\bf T}f}$ & 0 & 0& Unpaired MZMs; Reflection axis can carry a Kitaev chain \\
      $\Z_4^{{\bf R}f}$  & $\Z_2^{\bf T}\times \Z_2^f$ & 0 & -&  \\
      $\Z_2^{\bf RT} \times \Z_2^f$ & $\Z_2\times\Z_2^f$ &$\frac{1}{2}\Z$ &$\frac{1}{2}\Z$ & For any $c_-$ unpaired MZMs can be found at endpoints of the reflection axis if it carries a generator of Class BDI TSCs \\
      $\Z_4^{{\bf RT}f}$  & $\Z_4^f$ & $\Z$ & -&  \\
    \end{tabular}
    \caption{Constraints when the symmetry group $G_f^{\sp}$ is spatial. $C_M$ denotes spatial rotations of order $M$; we use $C_{2M}^{f}$ when a $2\pi$ spatial rotation equals $(-1)^F$. $\Z_2^{\bf R}$ denotes the order-2 group generated by a unitary reflection symmetry while $\Z_2^{\bf R T}$ denotes the order-2 group generated by the anti-unitary reflection ${\bf R T}$. $G_f^{\Int}$ is the effective internal symmetry group, determined through the fermionic crystalline equivalence principle. $n_1: G_b \rightarrow \Z_2$ is discussed in Sec.~\ref{sec:BasicData}.} 
    \label{tab:cryst} 
\end{table*}

\subsubsection{Internal unitary symmetries}

We show that charge-conserving systems with $G_f = \U(1)^f \times H$ (where $H$ is arbitrary) must have integer $c_-$, but cannot host unpaired MZMs at symmetry defects (Section \ref{sec:charge}). Thus, for example, Chern insulators cannot host unpaired MZMs at symmetry defects. The only way for a charge-conserving system to have unpaired MZMs is if $H$ has charge-conjugating elements, so that the symmetry group instead becomes $U(1)^f \rtimes H$, where $\rtimes$ denotes the charge conjugation action. Here unpaired MZMs can exist, but only when $c_-$ is odd; moreover, in this case \textit{all} defects associated to charge-conjugating elements must host unpaired MZMs. If $c_-$ is even, topological insulators with the symmetry $\U(1)^f \times H$ or $\U(1)^f \rtimes H$ do exist, but do not host unpaired MZMs.

A simple example of such a charge conjugating symmetry is $G_f = \Oo(2)^f = U(1)^f \rtimes \Z_2$, where $H = \Z_2$. In a system of two identical layers of a spinless $p+ip$ superconductor, $G_f$ is the symmetry which permutes the fermions in the two layers while keeping the superconducting pairing term invariant. In this case, a defect of the $\Oo(2)^f$ reflections can host an unpaired MZM when $c_-$ is odd. Such defects correspond to `half-quantum vortices' in a spinful $p+ip$ TSC, which has previously been proposed as a mean-field model for the material $\text{Sr}_2\text{RuO}_4$.\footnote{The nature of the order parameter in $\text{Sr}_2\text{RuO}_4$ is however still not understood in light of recent experiments \cite{pustogow2019sr,Ishida2020sr,ghosh2021sr}.}

Interestingly, the example with $U(1)^f \rtimes H$ symmetry also captures the well-known possibility of creating unpaired MZMs by inducing a superconducting gap at an interface between two Chern insulators with equal and odd Chern numbers. Each endpoint of the superconducting interface can be viewed as a charge-conjugating symmetry defect if the system has a particle-hole symmetry in addition to $\U(1)^f$ (see Sec.~\ref{sec:charge} and Fig.~\ref{fig:chern_ins} therein for more details). The endpoint is a defect because, upon encircling it, a particle must cross the interface and hence transform into a hole.

Next, we consider spin rotation symmetry with $G_f = \SU(2)^f \times H$, where $H$ is any unitary symmetry (Section \ref{sec:spinrot}). We find that $c_-$ must be even, and that unpaired MZMs cannot exist for any $H$. This describes the situation in a 
spin-singlet superconductor. Thus, unless the spin rotation symmetry is broken or it interacts nontrivially with elements of $H$ (i.e. not as a direct product), unpaired MZMs are impossible to realize in (2+1)D  
spin-singlet superconductors. 

We show that if $G_f$ is abelian, unpaired MZMs can only exist if $G_f = \Z_2^f \times G_b$ for some abelian $G_b$ (Section \ref{sec:abelian}). On the other hand, if $G_f$ is abelian but not isomorphic to $\Z_2^f \times G_b$ (for example $\Z_{4}^f$), then the system cannot host unpaired MZMs, and can exist only at integer values of $c_-$. This extends previous results in Refs.~\cite{wang2016tsc,Wang2020fSPT}. Two remarkable applications of this result are to charge-$2q$ superconductors and to superconductors with an $M$-fold rotational point group symmetry, as we discuss in subsequent subsections. 

We obtain several results for compact Lie groups.  
In particular, we study the orthogonal families ($\Oo(n)^f, \SO(2n)^f$), unitary families ($\U(n)^f\rtimes \ZZ_2, \U(n)^f, \SU(2n)^f)$, and symplectic families $ \Sp(n)^f $.

We find a rich set of constraints for the family of groups $\Oo(n)^f$ (Section \ref{sec:ongeneral}). We show that in any invertible phase, $2c_-$ must be a multiple of $\gcd(n,16)$. Moreover, only a certain subset of these $c_-$ values is compatible with unpaired MZMs. For example, when $G_f = \Oo(2)^f$, invertible phases can exist for any integer $c_-$, but unpaired MZMs can only exist when $c_-$ is odd. When $G_f = \Oo(4)^f$, invertible phases can exist for any even $c_-$, but unpaired MZMs can exist only when $c_- = 2 \mod 4$. Note that these results also hold in the free fermion context, where an $\Oo(n)^f$ symmetric phase can be obtained by stacking $n$ identical layers of a spinless $p+ip$ SC. Our theory provides further information constraining a parameter $n_2$ which fixes the quantum numbers carried by the fermion parity fluxes (see Sec.~\ref{sec:ongeneral}); this is not apparent from free fermion constructions. 

In deriving the above constraints, we calculated the cohomology groups $\H^{*}(\PSO(2n),\ZZ)$ and $ \H^{*}(\PO(2n),\ZZ\oplus \ZZ^{\ww_1})$ (twisted coefficients) in degrees 6 and below  (App.~\ref{app:o2nf}). To our knowledge, these results have not appeared before in the literature. We also note that obtaining the full set of constraints requires us to calculate the ``$\mathcal{O}_4$ obstruction'' \cite{barkeshli2021invertible} of the invertible phase. The new technical results are primarily required for this calculation. Without accounting for the $\mathcal{O}_4$ obstruction, the constraints that can be put on $c_-$ and on unpaired MZMs will generally be weaker (see Sec.~\ref{sec:tHooftConstraint} for a discussion).

We discuss the symmetry group $\U(n)^f$, which is the symmetry of $n$ identical layers of a Chern insulator (Section \ref{sec:Unf}). We find that even if we allow interactions, $c_-$ must be a multiple of $\gcd(n,8)$, but unpaired MZMs are not allowed (similar to the case with $G_f = \U(1)^f$). However, if we consider an additional (unitary) charge conjugation symmetry, so that $G_f = \U(n)^f \rtimes \Z_2$, then unpaired MZMs are allowed only for odd $n$. 

The same constraints apply upon restriction to the subgroup $\SU(2n)^f$ (Sec. \ref{sec:SU2nf}). In particular, we find that when $G_f = \SU(8n)^f$, $c_-$ must be a multiple of 8, which recovers a result argued recently in Ref.~\cite{ning2021enforced}. Furthermore, the constraints even survive upon restriction to $\Sp(n)^f \subset \SU(2n)^f$ (Sec.~\ref{sec:symplectic}).

We study various finite subgroups of these compact Lie groups which illustrate the same phenomena and are particularly useful to study crystalline phases. For example, the constraints obtained for $G_f =\SO(2)^f, \Oo(2)^f, \Sp(1)^f$ and $\Oo(4)^f$ also apply to the finite subgroups $G_f =\ZZ_4^f, \DD_8^f, \QQ_8^f$ and $\frac{\QQ_8^f\times \QQ_8^f}{\langle (-1,-1) \rangle}\rtimes  \ZZ_2$, respectively. Here $\DD_8$ and $\QQ_8$ are the dihedral and quaternion groups with 8 elements. The $\ZZ_2$ in the last group acts by permuting the two $\QQ_8$ and the quotient identifies the $-1$ on each of the $\QQ_8$'s.

In many of the above examples, the constraints can be anticipated from a free fermion perspective. We will highlight those examples that are beyond free fermion constructions separately below.

\subsubsection{Internal antiunitary symmetries}

This work also contains several results for antiunitary internal symmetries, in which case $c_-$ must be zero. See Table \ref{tab:summaryAntiunitary} for a summary. Here, there are two physically distinct scenarios. The first is where we have an additional unitary symmetry whose symmetry defects can host unpaired MZMs. For example, we find that when $G_f = \Z_2^f \times \Z_2^{\bf T} \times \Z_n$, an elementary $\Z_n$ defect can only host unpaired MZMs when $n$ is a multiple of 8 (see App.~\ref{app:znz2T})\footnote{The case with $n=4$ was studied previously through cobordism theory (\textbf{Theorem 17} in Ref.~\cite{guo2020fermionic}), and the phase with $G_f=\ZZ_8\times\ZZ_2^{\Tt}\times\ZZ_2^{f}$ was briefly mentioned in the conclusion of Ref.~\cite{sullivan2020interacting}.}. We also study $G_f = \Z_4^{{\bf T}f} \rtimes \Z_2$, where the unitary $\ZZ_2$ symmetry anticommutes with time reversal on fermions. Here the existence of unpaired MZMs can be understood through a simple free fermion construction. 

On the other hand, if we consider systems such as the Class DIII TSC, there is a different situation in which the defining property of the TSC is not the existence of unpaired MZMs but instead the fact that fermion parity flux hosts a degenerate Kramers pair of MZMs. In Sec.~\ref{sec:int_AU} we provide a precise mathematical definition that captures this phenomenon in terms of the basic data in our theory. The concept of 'Majorana Kramers pair' also generalizes to systems with a $\Z_4^{{\bf T}f}$ subgroup. However, as we mention in Sec.~\ref{sec:int_AU}, we do not have a completely satisfactory interpretation of the phenomena associated to MZMs for general antiunitary symmetries.

\subsubsection{Crystalline symmetries}

The results in Tables \ref{tab:summary} and \ref{tab:summaryAntiunitary} apply to internal symmetries $G_f$. However, we can use the crystalline equivalence principle for fermions (fCEP), discussed in Section \ref{sec:cryst} and Appendix ~\ref{app:fCEPAnti}, to obtain analogous constraints for crystalline symmetries. The fCEP allows us to replace the true symmetry group $G_f$ with an effective internal symmetry group $G_f^{\Int}$ that is determined according to certain rules (for example, reflection is mapped to time-reversal, while `spinless' $n$-fold rotations with $\hat{C}_n^n = 1$ are mapped to `spinful' rotations with $\hat{C}_n^n = (-1)^F$, and vice versa). Upon doing so, our constraints for internal symmetries can all be applied to crystalline systems, including the rich class of higher-order TSCs. We summarize this in Table \ref{tab:cryst}.

The fCEP has been discussed formally in the literature in \cite{Thorngren2018,else2019crystalline,debray2021invertible}. In this paper we extend the previous results of \cite{Thorngren2018,else2019crystalline,debray2021invertible} by stating the fCEP in a computationally useful form as an explicit algebraic formula (Eq.~\eqref{eq:fCEP}), which was not presented in previous works. Furthermore, our result also applies to spatial symmetries that have antiunitary elements, which were not considered in real-space classification approaches that checked the fCEP in special cases \cite{song2020,zhang2020const,zhang2020realspace}.

An interesting prediction in Table \ref{tab:cryst} concerns systems with $G_f = C_{2k} \times \Z_2^f$ (in which the rotation operator acts on fermions as $\hat{C}_{2k}^{2k} = 1$). This applies for example to crystalline systems with point groups of order $2$, $4$, and $6$. States with such symmetry do not admit phases with $c_-=1/2 \mod 1$ and therefore do not admit unpaired MZMs at fermion parity vortices. They also cannot support unpaired MZMs at lattice disclinations. However if $G_f = C_{2k+1} \times \Z_2^f$, which applies to the case of point group symmetries of odd order, then phases with $c_-=1/2 \mod 1$ and unpaired MZMs at fermion parity vortices are allowed. When $G_f = C_{2k+1} \times \Z_2^f$, unpaired MZMs cannot exist at disclinations or corners unless also bound to fermion parity vortices in a system with $c_-=1/2 \mod 1$. 

Note that the familiar spinless $p+ip$ superconductors in the continuum have an $\SO(2)^f$ spatial rotational symmetry in the ground state, which arises from the order parameter breaking an underlying $\SO(2) \times \U(1)^f$ symmetry down to $\SO(2)^f$. On a lattice, this spatial symmetry is broken down to some discrete subgroup $C_{2M}^f$. Indeed, Table \ref{tab:cryst} states that this symmetry does allow $c_- = 1/2$ phases, such as the spinless $p+ip$ SC at weak pairing. Furthermore, only when $M$ is even does $G_f = C_{2M}^f$ allow unpaired MZMs at disclinations or corners without fermion parity vortices.

In Section \ref{sec:apps} we relate our formalism to previous proposals for higher-order TSCs in the literature. In several cases our theory suggests alternative interpretations of previous results. 

In Sec.~\ref{sec:mzmdisc}, we discuss examples predicting unpaired MZMs at the corners of a $C_2$ symmetric model (Sec.~\ref{sec:C2}), and Majorana Kramers pairs at the corners of a $C_4$ symmetric system (Sec.~\ref{sec:C4+TRS}). We argue that in these models, the above corner phenomena can also be seen at disclinations of angle $\pi$ and $\pi/2$ respectively. 

In Sec.~\ref{sec:mzmreflec} we discuss proposals for unpaired MZMs at the corners of models with reflection symmetry. In two examples involving a modified Dirac semimetal (Sec.~\ref{sec:MxMy}) and a $p+id$ SC (Sec.~\ref{sec:majkram}) the results do not directly follow from our theory, because the corners cannot be thought of as defects of the given symmetries. However, by identifying alternative reflection symmetries that were not commented upon in the original models, our theory gives a consistent explanation for the observed MZMs. 

In Sec.~\ref{sec:fesc} we discuss a model for unpaired MZMs at the corners of an iron-based heterostructure \cite{Zhang2019FeTSC}. Our considerations suggest that these corner modes do not necessarily imply a nontrivial bulk invariant unless additional rotational symmetries are identified. However the model may still be nontrivial in the bulk assuming only the stated translation symmetries, in which case it may have unpaired MZMs at lattice dislocations. 

Furthermore, in Sec.~\ref{sec:mzmdisloc} and ~\ref{sec:mzmdisc}, we study examples in which free fermion invariants are discussed, and we clarify their relationship to the corresponding invariants for interacting invertible phases. 

We also obtain some new results which build on previous work. We discuss proposals for unpaired MZMs at fermion parity fluxes in a superconducting version of the Hofstader model with magnetic translation symmetry studied in Ref.~\cite{shaffer2021hofsc} (Sec.~\ref{sec:magtrans}). Here our theory additionally suggests that unpaired MZMs can be found at lattice dislocations for models in the same symmetry class. Our theory also rules out unpaired MZMs at any symmetry defects if we generalize to charge-$4q$ superconductors with translation symmetry.

\subsubsection{Intrinsically interacting fermionic phases}
 
Although our results are all nonperturbative, they can in many cases be anticipated based on free fermion calculations. For example, the values of $c_-$ which permit unpaired MZMs in groups such as $\U(1)^f, \SU(2)^f$ and $\Oo(n)^f$ could perhaps be guessed from the properties of Chern insulators,  
spin-singlet superconductors, and layered $p$-wave superconductors respectively, together with knowledge of the $c_- = 8$ invertible bosonic $E_8$ phase based on $(E_8)_1$ Chern-Simons theory. Our results imply that these expectations are valid in the strongly interacting regime as well. Similarly, our results about unpaired MZMs in the tables are consistent with free fermion results obtained in several previous classifications (see e.g. \cite{Shiozaki2014TopologyCTSc,Benalcazar2014,geier2018htsc}). 

On the other hand, when the symmetry does not admit free fermion realizations of invertible phases, our theory becomes especially useful. For example, consider the case where $G_f$ is unitary and abelian, and does not split as $G_b \times \Z_2^f$, for some $G_b$. From Table \ref{tab:summary} we see that unpaired MZMs are not allowed for any $c_-$.

We can apply the aforementioned result to the symmetry group $G_f = \Z_{4q}^f$, which is the symmetry of charge-$4q$ superconductors. $q$ is an integer. Here the ground state is a condensate of bound states of $4q$ fermions; these phases do not admit any free fermion description, because the Hamiltonian necessarily involves interactions among $4q$ fermions. The above result then states that such a system cannot have unpaired MZMs. This is an example of a `negative' result provided by our theory, which goes beyond free fermion classification results.

By combining our interacting classification with known results on free fermions, it is also possible to identify ``intrinsically interacting" invertible phases that have no free fermion analogs. Thus our theory also contains `positive' results which predict new interacting phases. As an illustration, we study a symmetry group $G_f$ with $4^4 \times 2$ elements, which is the central product of 4 copies of $\mathbb{D}_8^f$, meaning that the order 2 rotations in each $\mathbb{D}_8$ are identified with the fermion parity (see Appendix \ref{sec:D8xL}). This group admits an invertible phase with $c_-=4$ that cannot be realized through free fermions, i.e. the phase is intrinsically interacting. It is also intrinsically fermionic because bosonic invertible phases must have $c_- = 0 \mod 8$.

\subsubsection{(1+1)D invertible fermionic phases}

We briefly discuss (1+1)D systems and show that the relevant constraints can be reproduced from the (2+1)D equations (Sec.~\ref{sec:1d}). The idea is to consider a stack of (1+1)D chains and apply the constraints for (2+1)D invertible phases to this stack.

\subsubsection{Organization of paper }

The rest of this paper is organized as follows. In Section \ref{sec:internal} we introduce the basic data and equations describing invertible fermionic phases in (2+1)D and physically interpret the resulting constraints for internal symmetries. In Section \ref{sec:cryst} we state the crystalline equivalence principle in a computationally useful form, which allows us to apply our theory to crystalline symmetries. In Section \ref{sec:examples} we work out various nontrivial mathematical consequences of the constraints. In Section \ref{sec:apps} we discuss a number of applications involving TSC materials and models that have been studied in the literature, with a focus on spatial symmetries. We then conclude and discuss future directions.

\section{Review of classification of invertible fermionic phases}\label{sec:internal}
 
\subsection{Overview}\label{sec:class_sum}
Let $G_f$ be the symmetry group of the invertible fermionic phase (assumed to be internal in this section). Define $G_b = G_f/\Z_2^f$ where $\Z_2^f$ denotes the fermion parity symmetry. Let $({\bf g},a)$ dentote an element of $G_f$, where ${\bf g} \in G_b, a \in \Z_2^f$. Then, the group law in $G_f$ can be written as 
\begin{equation}\label{eq:grouplaw}
    ({\bf g}_1, a_1)({\bf g}_2,a_2) = ({\bf g}_1{\bf g}_2, a_1+a_2+\om({\bf g}_1,{\bf g}_2))
\end{equation}
where $\om$ is a 2-cocycle representative of the group $\H^2(G_b,\Z_2)$. The homomorphism $s_1: G_b \rightarrow \Z_2$ specifies if ${\bf g} \in G_b$ is unitary ($s_1({\bf g}) = 0$) or antiunitary ($s_1({\bf g}) = 1$).

If the symmetry is of the form $G_f = G_b \times \Z_2^f$, we can set $\om= 0$. If $[\om]$ belongs to a nontrivial class, there is some combination of operations $\{{\bf g}_i\} \in G_b$ with $\prod_i {\bf g}_i = {\bf 0}$, such that
\begin{equation}
    \prod_i ({\bf g}_i,0) = ({\bf 0},1).
\end{equation}
Thus a sequence of symmetry operations which acts as the identity on bosonic operators instead acts as $(-1)^F$ on fermionic operators. The symmetry acts through a projective representation of $G_b$ (which is still however a linear representation of $G_f$). In this situation the fermion is said to have fractional $G_b$ quantum numbers. 

For example, consider a system with $G_b = SO(3)$ in which a $2\pi$ rotation within $G_b$ equals 1 or $(-1)^F$. Such a system has integer and half-integer spin respectively. As another example, consider a system with $G_b = \Z_2^{\bf T}$ time-reversal symmetry implemented by ${\bf T}$, satisfying ${\bf T}^2 = (-1)^F$; we can also express this as $\om({\bf T},{\bf T}) = 1 \mod 2$. In this case the relevant fractional $G_b$ quantum number is the Kramers degeneracy carried by a fermion.

The main result of Ref.~\cite{barkeshli2021invertible} is that each invertible fermionic phase with symmetry group $G_f$ is classified by the data $(c_-,n_1,n_2,\nu_3)$, where $c_-$ is either an integer or a half-integer, and $n_1: G_b \rightarrow \Z_2, n_2: G_b \times G_b \rightarrow \Z_2, \nu_3: G_b \times G_b \times G_b \rightarrow U(1)$ are functions (`cochains') in 1,2 and 3 variables respectively. They satisfy the following consistency conditions:
\begin{align}\label{eq:iFTPEqs}
    dn_1 &= 0 \\
    dn_2 &= \mathcal{O}_3[c_-,n_1] \label{eq:O3def}\\
    d\nu_3 &= \mathcal{O}_4[c_-,n_1,n_2] \label{eq:O4def}.
\end{align}
The operator $d$ is a derivative defined on cochains, see Appendix \ref{app:coho}. The quantities $\mathcal{O}_3, \mathcal{O}_4$ have complicated expressions and will be discussed further below (see Eqs.~\eqref{eq:O3full} and ~\eqref{eq:O4simple} respectively). Different sets of data $(c_-,n_1,n_2,\nu_3)$ can describe the same physical system, and are thus equivalent. The classification of invertible phases for a given $G_f$ is obtained by finding all possible solutions to the above equations, and then modding out by such equivalences.

The purpose of this paper is to solve and physically interpret these equations for several interesting choices of $G_f$. In the rest of this section, we will explain the general meaning of the above data and equations. While the mathematical content in this section is the same as in Ref.~\cite{barkeshli2021invertible}, we include several examples and additional physical arguments to aid readers who are new to the formalism.  We refer the reader to \cite{Barkeshli2019,Barkeshli2020Anomaly,barkeshli2021invertible} for background on the general formalism that we use here.

\subsection{Definition of \texorpdfstring{$(c_-,n_1,n_2,\nu_3)$}{(c-,n1,n2,nu3)}}\label{sec:BasicData}
\subsubsection{$c_-$}
Let us fix $G_f$ (and hence $\om$). The quantity $c_-$ denotes the chiral central charge of the invertible fermionic phase. It can be measured from the thermal Hall conductance $\kappa_{xy}$ at a temperature $T$ through the relation
\begin{equation}
    \kappa_{xy} = c_-\frac{\pi^2 k_B^2}{3h}T.
\end{equation}
$c_-$ is an integer or a half-integer; in a topological superconductor, $2c_-$ is the spectral Chern number of the associated BdG Hamiltonian. 

The basic data fully specify the quantized topological properties of symmetry defects in the system, such as their fusion rules, their quantum numbers under $G_f$, and so on. Symmetry defects are objects labelled by a group element in $G_f$. In a Hamiltonian picture, a ${\bf g}$ defect is a static modification of the Hamiltonian along some spatial cut, so that a particle which crosses the cut gets acted upon by ${\bf g}$. Topologically distinct defects corresponding to the same group element are denoted $a_{\bf g}, b_{\bf g},$ and so on.  

$c_-$ determines the nature of the fermion parity $({\bf 0},1)$ fluxes in the system. These are also sometimes referred to as `fermion parity defects' or `fermion parity vortices.' If $c_-$ is a half-integer, there is a single fermion parity flux, denoted $\sigma_{({\bf 0},1)}$, which is nonabelian and hosts an unpaired MZM. (A notational remark: for fermion parity fluxes, we will usually drop the subscript and instead just write $\sigma$.) It satisfies the fusion rules $\sigma \times \psi = \psi \times \sigma = \sigma; \sigma \times \sigma = 1 + \psi$, where $\psi$ denotes a fermion. This is the situation in the weak-pairing spinless $p+ip$ superconductor, where $c_-=1/2$.

If $c_-$ is an integer, there are two fermion parity fluxes, which do not carry an unpaired MZM. When $c_-$ is even we denote them by $m$ and $e = m \times \psi$. They satisfy $e\times e = m \times m = 1$. If $c_-$ is odd, we instead denote the fermion parity fluxes as $v, \bar{v} = v \times \psi$. In this case, $v\times v = \bar{v}\times \bar{v} = \psi$. When $c_-$ is an unspecified integer, by convention we will use $e,m$.

Note that if we gauge the fermion parity symmetry of an invertible fermion phase, the fermion parity fluxes are promoted to anyons with the same fusion rules as given above. The topological twist $\theta$ of these anyons satisfies $\theta = e^{i 2\pi \frac{c_-}{8}}$ and thus encodes $c_- \mod 8$. 

\subsubsection{$n_1$ and MZMs}
Next we have a parameter $n_1: G_b \rightarrow \Z_2$. The following interpretation of $n_1$ is valid for unitary as well as antiunitary symmetries. Suppose we gauge the fermion parity. Then, if $n_1({\bf g}) = 1$, any fermion parity flux (say we denote it as $m$) is converted to $m \times \psi$ when acted upon by ${\bf g}$. Thus the action of ${\bf g}$ changes the fermion parity of $m$. We write this as ${^{\bf g}}m = m \times \psi$.

When ${\bf g}$ is a unitary operation, there is a second equivalent interpretation which is useful: $c_-$ and $n_1$ together completely specify which symmetry defects carry unpaired MZMs. First suppose $c_- \in \Z$. Now, if $n_1({\bf g}) = 0$, then any $({\bf g},a)$ defect in the invertible fermionic phase is abelian, i.e. does not carry an unpaired MZM, while if $n_1({\bf g}) = 1$, every $({\bf g},a)$ defect carries an unpaired MZM.

When $c_-$ is a half-integer, we saw that the $({\bf 0},1)$ defect always hosts an unpaired MZM. If $n_1({\bf g})=0$, the situation is the same as when ${\bf g}={\bf 0}$: the $({\bf g},1)$ defect hosts an unpaired MZM while the $({\bf g},0)$ defect does not. On the other hand, $n_1({\bf g})=1$ implies that a $({\bf g},0)$ defect hosts an unpaired MZM, but a $({\bf g},1)$ defect does not. 

When ${\bf g}$ is the time-reversal operation, a ${\bf g}$ defect is not in general a well-defined concept in a Hamiltonian framework.\footnote{In a Euclidean space-time path integral, one can think of a time-reversal defect as an orientation-reversing wall in space-time. See e.g. \cite{barkeshli2019tr}.} However, there is still an interpretation of $n_1({\bf g})=1$: it means that a fermion parity flux carries two degenerate states with different fermion parities which are related by ${\bf g}$. These states can be identified with $m$ and $e = m \times \psi$. Table \ref{tab:summaryAntiunitary} contains several examples for which $n_1 = s_1$, i.e. $n_1$ is nonzero only for the subgroup generated by time-reversal. This latter interpretation applies to all those examples. We provide a much more detailed interpretation of the equation $n_1=s_1$ in Sec.~\ref{sec:int_AU} below.

By considering the fusion of a $({\bf g}_1,a)$-defect with a $({\bf g}_2,b)$-defect into a $({\bf g}_1,a)({\bf g}_2,b)$-defect, we can prove that $n_1$ should be a homomorphism:
\begin{align}
    n_1({\bf g}_1) + n_1({\bf g}_2) &= n_1({\bf g}_1{\bf g}_2) \mod 2 \nonumber \\
    \implies \quad dn_1 &= 0 \mod 2.
\end{align}

\subsubsection{$\om$ and fractional quantum numbers of $\psi$}

The content of Eq.~\eqref{eq:grouplaw} is summarized by the definition
\begin{equation}
    \eta_{\psi}({\bf g}_1,{\bf g}_2) := (-1)^{\om({\bf g}_1,{\bf g}_2)},
\end{equation}
where ${\bf g}_1, {\bf g}_2 \in G_b$. Physically, $\eta_{\psi}({\bf g}_1,{\bf g}_2)$ measures the phase difference between acting on the fermion $\psi$ by ${\bf g}_1$ and ${\bf g}_2$ separately as opposed to ${\bf g}_1{\bf g}_2$. A detailed discussion of the meaning of the $\eta$ symbols in general can be found in Refs.~\cite{Barkeshli2019,Barkeshli2020Anomaly,bulmashSymmFrac,barkeshli2021invertible}.

The symbols $\eta_{\psi}$ also determine the fractional $G_b$ quantum numbers carried by $\psi$. Given some $G_b$, the choices of $[\om]$ are classified by $\H^2(G_b,\Z_2) \cong (\Z_2)^{r}$ for some $r \ge 0$. Thus there are $r$ $\Z_2$ invariants which together specify $[\om]$. We denote these as $Q_{\psi}^i \mod 1$, $i = 1,2, \dots , r$. Each $Q_{\psi}^i \in \{0, 1/2\} \mod 1$ is a $G_b$ quantum number of $\psi$, and is defined by some `gauge-invariant' combination of $\eta_{\psi}$ symbols. The precise definition of $Q_{\psi}^i$ is symmetry-dependent. Whenever $Q^i_{\psi}=\frac{1}{2} \mod 1$ (i.e. $Q^i_{\psi}$ is a \textit{fractional} $G_b$ quantum number), there exists a sequence of group operations which act trivially on any bosonic operator but transform any fermionic operator by a minus sign.  

\paragraph*{Example.} For example, consider a system with $G_f = \U(1)^f$ charge conservation symmetry. Here $G_f$ and $G_b = U(1)^f/\Z_2^f$ are both isomorphic to $\U(1)$, but the operator generating a rotation by $\theta$ in $G_b$ only generates a rotation by $\theta/2$ in $G_f$. Note that two $\theta = \pi$ rotations in $G_b$ together act trivially on any bosonic operator, but transform a fermionic operator by a sign. Thus $\om(\pi,\pi) = 1$. We can also define
\begin{equation}
    \eta_{\psi}(\pi,\pi) = -1 =: e^{2\pi i Q_{\psi}}.
\end{equation}
This means that $\psi$ has $G_b$ charge $Q_{\psi} = \frac{1}{2}$. Since $\H^2(U(1),\Z_2) \cong \Z_2$, we only define one invariant $Q_{\psi}$.

\subsubsection{$n_2$ and fractional quantum numbers of fermion parity fluxes}

We can similarly define $\eta$ symbols for the fermion parity fluxes, say $e$ and $m$.
$\eta_a({\bf g}, {\bf h}) \in U(1)$ for $a = e,m,\psi$ essentially describes the phase difference between applying ${\bf g}$ and ${\bf h}$ sequentially as compared to ${\bf gh}$ (See \cite{Barkeshli2019} for a detailed definition). The $\eta$ symbols in general encode the fractional $G_b$ quantum numbers of excitations. 

Just as $\eta_{\psi}$ encodes $\om$, $\eta_e$ and $\eta_m$ together encode a function $n_2: G_b \times G_b \rightarrow \ZZ_2$. For example, when $c_-=0$, we can define
\begin{align}
\eta_m({\bf g},{\bf h}) &= (-1)^{n_2({\bf g},{\bf h})} \nonumber \\
\eta_e({\bf g},{\bf h}) &= (-1)^{(n_2+\om)({\bf g},{\bf h})}.
\end{align}
The $\eta$ symbols for general $c_-$ in terms of $n_2$ are given in Ref.~\cite{barkeshli2021invertible}. 

\paragraph*{Example.} 
Consider a crystalline topological insulator with charge conservation and translation symmetries, with $c_-=0$ and a fermion per unit cell (implying a filling $\nu = 1$). Here $G_f = \U(1)^f \times \ZZ^2$. 
Note that transporting an $e$ (or $m$) particle around a single unit cell results in an overall phase of -1, because the fermion within the unit cell is seen as a $\pi$-flux for the $e$ and $m$ particles. Thus the elementary translations do not commute on a state with a fermion parity flux ($T_{\bf x} T_{\bf y}=- T_{\bf y} T_{\bf x}$ in such a state). This is the `nontrivial $G_b$ quantum number' encoded by the $\eta_e, \eta_m$ symbols. In particular, we can choose a gauge where $\eta_e(\gbf,\hbf)=\eta_m(\gbf,\hbf)= (-1)^{X_1(\gbf)X_2(\hbf)}$ where $X_i$ is the projection to the $i$-th component of $\ZZ^2$. Then, on a state with an $e$ or $m$ particle, we have $T_{\bf x} T_{\bf y}(T_{\bf y} T_{\bf x})^{-1} = \eta_e(T_{\bf x},T_{\bf y})/\eta_e(T_{\bf y},T_{\bf x}) = -1$. 

Although we will not require them for this paper, we note for completeness that the theory also contains a set of $U$ symbols $U_{\bf g}(a,b; a \times b)$ for $a,b \in \{e,m,\psi\}$, which can all be set to 1 for unitary symmetries. Through appropriate `gauge-invariant' combinations of the $\eta$ and $U$ symbols, we can define fractional $G_b$ quantum numbers $Q_e^j, Q_m^j$ for the fermion parity fluxes, where $j = 1, \dots , k$ is an index and $k$ is the number of independent fractional quantum numbers. For a discussion of gauge transformations on the basic data, see Appendix A of Ref.~\cite{barkeshli2021invertible}. 

\subsubsection{$\mathcal{O}_3$ obstruction to defining $n_2$}

By requiring that the symmetry action on the fermions and the fermion parity fluxes respect their fusion rules, we can obtain various relations among the $\eta$ symbols. For example, when the symmetry is unitary, we can show that
\begin{equation}\label{eq:eta:U}
    \eta_a \eta_b = \eta_c
\end{equation}
whenever $c$ is a fusion product of $a$ and $b$, and $a,b,c$ correspond to $\psi$ or to a fermion parity flux. (For antiunitary symmetries, see Eq.~\eqref{eq:eta_AU}.) This places constraints on the fractional $G_b$ quantum numbers (or equivalently, on $n_2$). For example, if $G_f = U(1)^f$ as above and $c_-=0$, we can define $e^{2\pi i Q_a} := \eta_a(\pi,\pi)$ for $a = e,m,\psi$. Then the relation $e \times m = \psi$ along with Eq.~\eqref{eq:eta:U} implies that $Q_e + Q_m = Q_{\psi} \mod 1$. 

Another important relation (again written for unitary symmetries) takes the form 
\begin{equation}\label{eq:etacoc}
    \eta_a({\bf g},{\bf h})\eta_a({\bf gh},{\bf k}) = \eta_{^{\bf \bar{g}}a}({\bf h},{\bf k})\eta_a({\bf g},{\bf hk}).
\end{equation}
Here $^{\bf \bar{g}}a$ is the result of permuting $a$ by ${\bf \bar{g}} = {\bf g}^{-1}$. Note that if $n_1 \ne 0$, there is a group element which permutes $e$ and $m$. In that case the above condition may introduce further constraints on $Q_e$ and $Q_m$. These relations serve to constrain $n_2$ in terms of $n_1, c_-$ and $\om$. 

The constraint on $n_2$ is summarized by a cocycle $\mathcal{O}_3 \in Z^3(G_b,\Z_2)$. The general form of $\mathcal{O}_3$ is 
\begin{equation}\label{eq:O3full}
    \mathcal{O}_3 := n_1 \cup (\om + s_1 \cup n_1) + c_- \om \cup_1 \om \mod 2.
\end{equation}
For reference, the same equation is written after expanding the $\cup$-cup products in Appendix \ref{app:coho}. $\mathcal{O}_3$ is viewed as an obstruction, because if $\mathcal{O}_3$ is in a non-trivial cohomology class of $\mathcal{H}^3(G_b, \Z_2)$, then the equation $d n_2 = \mathcal{O}_3$ cannot be solved, as the lhs is a 3-coboundary and thus trivial in $\mathcal{H}^3(G_b, \Z_2)$. 
We will discuss the different terms in Eq. \ref{eq:O3full} more fully in Secs.~\ref{sec:int_U},~\ref{sec:int_AU} below. In particular, for a unitary symmetry, $\mathcal{O}_3$ encodes how $\om$ and $c_-$ fix whether or not symmetry defects can host unpaired MZMs. 

\subsubsection{$\mathcal{O}_4$ obstruction and $\nu_3$}

Above we discussed constraints at the level of $n_2$. Even if $n_2$ is well-defined, however, the invertible phase may not be well-defined if the data $(\om, c_-, n_1, n_2)$ have a nontrivial obstruction (often referred to as 't Hooft anomaly), given by a class $[\mathcal{O}_4] \in \H^4(G_b,U(1))$. The obstruction is trivial if there exists some $\nu_3: C^3(G_b,U(1))$ such that
\begin{equation}\label{eq:H4anom}
    d\nu_3 = \mathcal{O}_4[c_-,n_1,n_2].
\end{equation}
When $n_1 = 0$, the expression for $\mathcal{O}_4$ (both unitary and antiunitary symmetries) is
\begin{equation}\label{eq:O4simple}
    \mathcal{O}_4 = \frac{c_- \mathcal{P}(\omega_2)}{8} + \frac{1}{2}n_2 \cup (\omega_2 + n_2) \mod 1
\end{equation}
where $\mathcal{P}:\H^2(G_b,\ZZ_2)\rightarrow \H^4(G_b, \ZZ_4)$ is the Pontryagin square which is a cohomology operation that refines the square - in the sense that $2\mathcal{P}(\omega_2)=2\omega_2\cup \omega_2 \mod 4$ (see Appendix \ref{app:coho}). Expressions for $\mathcal{O}_4$ when $n_1 \ne 0$ can be found in Ref.~\cite{barkeshli2021invertible}.

If $[\mathcal{O}_4]$ vanishes, we can define a consistent set of fusion and braiding data for all the defects; this requires an additional parameter $\nu_3({\bf g},{\bf h},{\bf k})$, which satisfies Eq.~\eqref{eq:H4anom}. As described below, $\nu_3$ specifies the $G_b$ quantum numbers of $G_b$ symmetry defects. However, if $[\mathcal{O}_4]$ is nonvanishing, the system with data $(\om,c_-,n_1,n_2)$ can only be defined in conjunction with another system that cancels the obstruction. For example, such a system may exist on the boundary of a bosonic SPT in (3+1) dimensions.

\paragraph*{Examples.} 
First consider an integer quantum Hall state with $G_f = \U(1)^f$. There are no obstructions, so $\nu_3$ can be defined consistently. In this case Ref.~\cite{barkeshli2021invertible} showed that $\nu_3$ directly encodes the Hall conductance. Now just as the Hall conductance measures the $\U(1)$ charge bound to magnetic flux, in general $\nu_3$ encodes the $G_b$ quantum numbers associated to $G_b$ symmetry defects.

An example of a system with nontrivial $\mathcal{O}_4$ (and hence no solution for $\nu_3$) is a translationally symmetric system in (2+1)D with spin-1/2 per unit cell. Here $G_f = \ZZ^2 \times \SU(2)^f$ and $G_b = \ZZ^2 \times \SO(3)$. The statement that fermions carry spin-1/2 under $\SO(3)$ implies that $\om$ is nontrivial, while the statement that there is a fermion in each unit cell implies that $n_2$ is nontrivial. In this case, the data $(\om,c_- = 0, n_1 = 0, n_2)$ give a nontrivial $\mathcal{O}_4$. This recovers a version of the well-known Lieb-Shultz-Mattis theorem, which states that there cannot be an invertible state in 2+1 dimensions satisfying the above criteria.

While solving the obstruction equations, it may happen that certain choices of $n_2$ that satisfy Eq.~\eqref{eq:O3full} do not satisfy Eq.~\eqref{eq:H4anom}. In this sense the $\mathcal{O}_4$ obstruction is a more restrictive constraint on $c_-$ and $n_1$, so it is not enough to solve only the $\mathcal{O}_3$ equation and ignore $\mathcal{O}_4$. We mention several examples illustrating this point in Section \ref{sec:tHooftConstraint}. Note that the results in Tables~\ref{tab:summary},~\ref{tab:summaryAntiunitary},~\ref{tab:cryst} have been obtained after solving the equations for $\mathcal{O}_3$ as well as $\mathcal{O}_4$.
 
\textbf{Remark:} In this paper we identify $\U(1) \cong \RR/\ZZ$ and we will interchangeably use multiplicative and additive notation for $\nu_3$, which are related as follows:
\[\nu_3^{\text{multiplicative}} = \exp(2\pi i \nu_3^{\text{additive}})\]
where $\nu_3^{\text{additive}}$ is to be understood as a real number modulo 1. 

\subsection{\texorpdfstring{Equations for $c_-=\frac{1}{2} \mod 1$}{c-=1/2 mod 1}}\label{sec:int_chalf}

We start with $c_-=1/2$, which is mathematically the simplest case. Here there are two main results: (i) the symmetry must be unitary; (ii) $\om=0$. 

(i) is due to the fact that an antiunitary operation takes $c_- \rightarrow - c_-$. For (ii), note that systems with half-integer $c_-$ have a fermion parity flux $\sigma$ which can absorb a fermion:
\begin{equation}
    \sigma \times \psi = \sigma.
\end{equation}

We can define the symbols $\eta_{\psi}, \eta_{\sigma}$ as discussed above. From the fusion rule, we can show that
\begin{equation}
    \eta_{\psi} \eta_{\sigma} = \eta_{\psi \times \sigma} = \eta_{\sigma};
\end{equation}
this implies that $\eta_{\psi} = 1$. But since we also have $\eta_{\psi} = (-1)^{\om}$ by definition, we conclude that for a system with half-integer $c_-$,
\begin{equation}
    \om = 0 \mod 2.
\end{equation}
In particular, $\psi$ must carry integer $G_b$ quantum numbers. In this case the obstruction $\mathcal{O}_3$ vanishes; thus the constraint on $n_2$ is simply $dn_2 = 0$. The constraint on $\nu_3$ given by Eqs.~\eqref{eq:O4def},~\eqref{eq:O4simple}.

\subsection{Equations for unitary symmetry with integer
\texorpdfstring{$c_-$}{c-}}\label{sec:int_U}

For unitary symmetries, taking $s_1 = 0$ in Eq.~\eqref{eq:O3full} gives
\begin{equation}\label{eq:n2unitary}
    dn_2 = n_1 \cup \om + c_- \om \cup_1 \om \quad\mod 2.
\end{equation}
In practice, an equation such as \eqref{eq:n2unitary} is useful because it converts the complex problem of predicting unpaired MZMs in strongly interacting invertible phases to a formal mathematical calculation that can be tackled with tools from group cohomology. However Eq.~\eqref{eq:n2unitary} also contains some valuable physical intuition, which we will now discuss through examples.  

First take $c_-$ to be even, so that the second term vanishes, and consider the remaining term $(n_1 \cup \om)({\bf g},{\bf h},{\bf k}) = n_1({\bf g}) \om({\bf h},{\bf k})$. It is nonzero only if $n_1$ and $\om$ are both nonzero, i.e. the system carries unpaired MZMs and the fermion also has fractional $G_b$ quantum numbers. 

\paragraph*{Example.} As a first example, consider $G_b = \ZZ_2$ and $G_f = \ZZ_4^f$. If ${\bf g}_0$ is the generator of $G_b$, then ${\bf g}_0^2 = (-1)^F$. Defining $\eta_a({\bf g}_0,{\bf g}_0) = (-1)^{Q_a}$, we see that $Q_{\psi} = \frac{1}{2}$.
The fermion parity fluxes $e$ and $m$ satisfy
\begin{equation}
    e\times e = m\times m = 1, \quad e \times m = \psi.
\end{equation}
To be consistent with the above fusion rules, the quantum numbers should satisfy
\begin{align}
    2 Q_e = 2 Q_m &= 0 \mod 1 \nonumber \\
    Q_e + Q_m &= Q_{\psi} \mod 1.
\end{align}
This is the content of Eq.~\eqref{eq:eta:U}. If $n_1 = 0$, these are the only equations. They can always be solved consistently, and the choice of $Q_e, Q_m$ will implicitly fix $n_2$. However, if $n_1({\bf g}_0) = 1$, there is an additional constraint which can be seen by taking ${\bf g} = {\bf h} = {\bf k} = {\bf g}_0$ in Eq.~\eqref{eq:etacoc}: 
\begin{align}
    \eta_m({\bf g}_0,{\bf g}_0) &= \eta_{{^{{\bf g}_0}}m}({\bf g}_0,{\bf g}_0) \nonumber \\
\implies    Q_e &= Q_m \mod 1.
\end{align}
Taking $Q_{\psi} = \frac{1}{2}$, we see from the last two equations that $Q_e = \pm \frac{1}{4} \mod 1$. But this contradicts the first equation. Thus we cannot solve all the equations for $\eta$ consistently in this case. This is mathematically expressed by the fact that $n_1 \cup \om$ specifies a nontrivial obstruction $[\mathcal{O}_3] \in \H^3(G_b,\Z_2)$, and so Eq.~\eqref{eq:n2unitary} has no solution.

\paragraph*{Example.} As a second example, consider $G_f=\Oo(4)^f$, for which $G_b = [\SO(3)_L \times \SO(3)_R] \rtimes \Z_2$. Again assume $c_-$ is even. The quantum numbers $Q_a = (s_{a,L}, s_{a,R})$ are now given by the spin of $a = e,m,\psi$ under $\SO(3)_L$ and $\SO(3)_R$ respectively. In terms of $\eta$ symbols, we can define $e^{2\pi i s_{a,L(R)}} := \eta_a({\bf g},{\bf g})$, where ${\bf g}$ is a $\pi$ rotation in $\SO(3)_{L(R)}$ about any axis. From the definition of $G_f$, we have $Q_{\psi} = (\frac{1}{2},\frac{1}{2})$, i.e. the fermion carries spin-1/2 under both $\SO(3)$ subgroups. 

Now let the group $\Z_2$ which interchanges $L$ and $R$ be generated by ${\bf h}$, and assume that $n_1({\bf h}) = 1$. Since ${\bf h}$ permutes $e,m$ and also interchanges $L,R$, we must have 
\begin{equation}
    s_{e,L} = s_{m,R}; \quad s_{e,R} = s_{m,L}.
\end{equation}
(This can also be derived formally from Eq.~\eqref{eq:etacoc}.) Thus if $Q_e = (\frac{1}{2},0)$, this condition forces $Q_m = (0,\frac{1}{2})$. In contrast to the previous example, these choices do not conflict with $Q_{\psi} = (\frac{1}{2},\frac{1}{2})$. Note that such solutions exist because the term $n_1 \cup \om$ is a 3-coboundary. Thus we can solve Eq.~\eqref{eq:n2unitary} consistently for some $n_2$, which in turn sets the values of $Q_e, Q_m$.

The second term in Eq.~\eqref{eq:n2unitary} can be understood similarly. For convenience, let us set $n_1 = 0$ so that the first term is zero. Also let us consider some odd $c_-$. In this case, $\psi$ has a `square root': there are fermion parity fluxes $v,\bar{v} = v \times \psi$ which satisfy
\begin{equation}
    v^2 = \bar{v}^2 = \psi.
\end{equation}
Thus $v,\bar{v}$ must carry representations of $G_f$ which square to the one defined by $\om$; in other words, we must have $2 Q_v = 2Q_{\bar{v}} = Q_{\psi}$, for some appropriately defined quantum numbers. The equation $dn_2 = c_- \om \cup_1 \om$ can be solved for odd $c_-$ if and only if such representations exist. 

\paragraph*{Example.} To illustrate the point above, consider $G_b = SO(3)$. Suppose $\psi$ carries spin-1/2 under $G_b$. If $c_-$ is odd, this would mean that $v,\bar{v}$ must carry `spin $\pm 1/4$' representations, which do not exist. Hence we cannot have an invertible phase with $c_-$ odd and with spin-1/2 fermions. 

 In the most general case, the rhs of Eq.~\eqref{eq:n2unitary} gives an obstruction whenever the equations for the $\eta$ symbols lead to contradictory results for the quantum numbers of the fermion parity fluxes. Note that we can have the interesting situation in which the two terms in Eq.~\eqref{eq:n2unitary} are both nontrivial but mathematically cancel each other, giving $dn_2 = 0$. The physical interpretation here is that the odd chiral central charge and the symmetry together conspire to give a consistent set of $\eta$ symbols.

\subsection{Equations for antiunitary symmetries}\label{sec:int_AU}

For antiunitary internal symmetries $s_1 \neq 0$ and $c_- = 0$. In this case, Eq.~\eqref{eq:eta:U} is modified as follows \cite{barkeshli2021invertible}:
\begin{equation}\label{eq:eta_AU}
    \eta_e({\bf g},{\bf h}) \eta_m({\bf g},{\bf h}) = \eta_{\psi}({\bf g},{\bf h}) (-1)^{s_1({\bf g}) n_1({\bf h})}
\end{equation}
with $\eta_a^2 = 1$. Eq.~\eqref{eq:n2unitary} correspondingly gets modified \cite{Wang2020fSPT,barkeshli2021invertible}:
\begin{equation}\label{eq:n2AU}
    dn_2 = n_1 \cup (\om + s_1 \cup n_1).
\end{equation}

\def\Dd{\mathtt{D}}

In this section we discuss the meaning of $n_1({\bf T})=1$, and the meaning of Eq.~\eqref{eq:n2AU}. We understand the meaning of $n_1({\bf T})=1$ whenever $G_f$ has a $\Z_4^{{\bf T}f}$ subgroup. In these cases it implies that $e$ and $m$ have a `fermionic Kramers degeneracy'. Eq.~\eqref{eq:n2AU} simply enforces that these properties- a $\Z_4^{{\bf T}f}$ subgroup with nontrivial $n_1$, and a fermionic Kramers degeneracy- are equivalent.

\paragraph*{Example with time-reversal alone.} Consider a system with $G_b = \Z_2^{\bf T }$. If $[\om] = 0$, time-reversal acts as $\Tt^2 = 1$, $\psi$ does not carry any Kramers degeneracy, and $G_f = \Z_2^{\bf T}\times \Z_2^f$. If $[\om] \ne 0$, time-reversal acts as $\Tt^2 = (-1)^F$, $\psi$ carries a Kramers degeneracy, and $G_f = \Z_4^{{\bf T}f}$. We will consider both possibilities together. Note that in the latter case we can write $\om({\bf g},{\bf h}) = s_1({\bf g}) s_1({\bf h}) := (s_1 \cup s_1)({\bf g},{\bf h})$.\footnote{This is because the only nonzero value of $\om$ is at ${\bf g} = {\bf h} = {\bf T}$.}

It will be useful to define the following invariants. If ${\bf T}$ leaves $a \in \{e,m,\psi\}$ invariant after gauging fermion parity, we define $
    \eta_{a}({\bf T},{\bf T}) \in \{1,-1\}$. This measures the local ${\bf T}^2$ eigenvalue of $a$. When ${\bf T}$ takes $a \rightarrow a \times \psi$, as when $n_1({\bf T})=1$ and $a=e,m$, the correct invariant turns out to be
 \begin{equation}\label{eq:Qaperm}
\eta_a^{{\bf T}} := \eta_a({\bf T},{\bf T}) U_{\bf T}(a,\psi; a \times \psi).  
\end{equation}
One can show that $\eta_a^{{\bf T}}$ also corresponds to the local ${\bf T}^2$ eigenvalue of $a=e,m$, although the arguments are more involved \cite{metlitski2014,bulmashSymmFrac}. Here $U_{\bf T}$ is the $U$ symbol mentioned previously; unlike for unitary symmetries, it cannot be set to 1 identically. This invariant was stated previously in Ref.~\cite{bulmashSymmFrac,tata2021anomalies,aasen2021characterization}. 

Instead of looking at Eq.~\eqref{eq:n2AU} term by term, we can write a gauge-invariant equation that summarizes the constraint. Ref.~\cite{bulmashSymmFrac}  showed that 
\begin{equation}
    \left(\eta_a^{{\bf T}}\right)^2 = \eta_{\psi}({\bf T},{\bf T}) = (-1)^{\om({\bf T},{\bf T})}.
\end{equation}
Thus, if $\psi$ does not have a Kramers degeneracy, $e$ and $m$ have ${\bf T}^2$ eigenvalue $\pm 1$. On the other hand, if $\psi$ has a Kramers degeneracy, $e$ and $m$ must have local ${\bf T}^2$ eigenvalue $\pm i$. This is referred to as a `fermionic Kramers degeneracy', see also Ref.~\cite{metlitski2014}. This is the content of Eq.~\eqref{eq:n2AU} when $G_b = \Z_2^{\bf T}$. Note that a fermionic Kramers degeneracy requires both $n_1$ and $\om$ to be nontrivial: if either $n_1=0$ or $\om = 0$, the ${\bf T}^2$ eigenvalue of $m$ must be $\pm 1$.

\subsubsection{Interpretation of $n_1 = s_1$}\label{sec:interpn1s1}
  
There is familiar interpretation of fermionic Kramers degeneracy in the context of TSCs. Note that the solution with $G_b = \Z_2^{\bf T}$ and $n_1 \ne 0$ describes the nontrivial Class DIII TSC. We know (see e.g. \cite{Qi2009TRS-TSC,Zhang2013MKP}) that in this case, inserting a $\pi$ flux induces a Kramers pair of MZMs, corresponding to $e$ and $m$. This property is equivalent to having a fermionic Kramers degeneracy, as was argued in Refs.~\cite{metlitski2014,bulmashSymmFrac}.\footnote{In particular, if we consider the generator of (1+1)D invertible phases in Class DIII, there is a Kramers pair of MZMs at the ends of the (1+1)D system. The symmetry acts locally at each end with eigenvalue $\pm i$ under ${\bf T}^2$ \cite{metlitski2014}. A dimensional reduction argument reviewed in Ref.~\cite{bulmashSymmFrac} shows that this eigenvalue is indeed equal to the quantity $\eta_m^{\bf T}$ defined for the Class DIII TSC.} We conclude that in the context of (2+1)D invertible phases, the following statements are equivalent when $G_f\cong \Z_4^{{\bf T}f}$: i) $n_1({\bf T})=1$; ii) $e,m$ are fermionic Kramers; iii) a fermion parity flux hosts a Kramers pair of MZMs. 

\paragraph*{Generalization.} We can generalize this result as follows. First consider any internal symmetry group $G_f$ that has a $\Z_4^{{\bf T}f}$ subgroup, and $n_1, \om$ restrict to $s_1$ and $s_1 \cup s_1$ respectively on this subgroup.\footnote{For example we can take $G_b = \Z_2^T \times H$, and $G_f = \Z_4^{{\bf T}f}\rtimes_{\rho} H$ where $H$ is unitary. Here we can set $n_1 = s_1 + \rho, \om = s_1^2 + s_1 \rho$.} Then, as above, this choice of $n_1$ and $\om$ implies that a fermion parity flux hosts a Majorana Kramers pair, and also that $e,m$ are fermionic Kramers.

Now suppose $G_b$ has a $\Z_{2n}^{\bf T}$ subgroup instead of just $\Z_{2}^{\bf T}$. When $n$ is odd, we can just write $G_b = \Z_2^{\bf T} \times \Z_n$, so the example is already covered. For $G_b = \Z_{2n}^{\bf T}$ with $n$ even, however, we find that $n_1 = s_1$ is possible only if $\om=0$. Here, the fermion is non-Kramers, and it is not even clear what nontrivial phenomenon is encoded by the data $n_1 = s_1$.

Thus, when $G_f$ does not have a $\Z_4^{{\bf T}f}$ subgroup, we do not have a good understanding of how to interpret $n_1=s_1$. We leave this as a subject for future study.

\subsubsection{A sufficient condition for a solution:\texorpdfstring{ $\om = s_1 \cup u_1$}{ w2 = s1u1}}\label{sec:SuffConditionus}

Suppose we are given some $\om$ of the form
\begin{equation}\label{eq:AUsuff}
    \om = s_1 \cup u_1 \mod 2,
\end{equation}
where $u_1 \in Z^1(G_b, \Z_2)$ is a 1-cocycle. 
Now, Eq.~\eqref{eq:n2AU} can always be mathematically solved if we take $n_1=u_1$. Indeed, we can always choose $n_2 = 0$. Then the $\mathcal{O}_4$ obstruction also vanishes, and we can always set $\nu_3 = 1$. Thus we obtain a useful sufficient condition for a topological phase with $n_1 \ne 0$ when $G_f$ is antiunitary.

Examples in which $n_1 = s_1 + \rho$ and $\om = s_1^2 + s_1 \rho$ all fall into this category. Here we assume that $\rho$ is nonzero only in the unitary subgroup of $G_b$, and the full symmetry group can be written as $G_f = \Z_4^{{\bf T}f} \rtimes_{\rho} H$ for some unitary $H$. As we discussed above, a fermion parity flux hosts a Majorana Kramers pair in these examples.

Eq.~\eqref{eq:AUsuff} is not a necessary condition, however. A simple counterexample involves $G_b = \Z^2$. Since $\H^3(\Z^2, \Z_2)$ is trivial, Eq.~\eqref{eq:n2AU} can be solved for any choice of $n_1, s_1, \om$. Moreover, since $\H^4(G_b,U(1))$ is also trivial, there is no $\mathcal{O}_4$ obstruction in this case, and we can obtain a well-defined invertible phase which does not satisfy $\om = s_1 \cup u_1$. Indeed, the symmetry $G_b = \Z^2$ and $s_1=x+y$ corresponds to the simplest nontrivial magnetic space group and is the symmetry that describes the square lattice Neel antiferromagnet state.

\section{Extending the constraints to crystalline symmetries}\label{sec:cryst}

The formalism above is strictly valid only when $G_f$ is an internal symmetry. However, with the aid of the fermionic crystalline equivalence principle (fCEP) \cite{Thorngren2018,Else2019,zhang2020realspace,debray2021invertible}, it can be applied in situations where $G_f$  
acts on space. This is of practical value, since a large number of material candidates for TSCs exploit crystalline symmetries in various ways. 

We note that the fCEP has only been discussed for completely general symmetries in \cite{Thorngren2018}; furthermore \cite{Thorngren2018} used the more abstract language of vector bundles. Below we present a concrete formula for the fCEP that is more useful in our context.

\subsection{Fermionic crystalline equivalence principle}

We consider a bosonic symmetry group $G_b$ that may include spatial symmetries. We assume that the clean 2d system without any defects is defined on the infinite plane, so that the spatial elements in $G_b$ are specified by a map $(\vec{R},\rho_s):G_b\to \RR^2 \rtimes \Oo(2)$ where $\RR^2 \rtimes \Oo(2)$ is the group of continuous translations, rotations and reflections in 2 dimensions. Here $\vec{R}: G_b \rightarrow \RR^2$ and $\rho_s: G_b \rightarrow \Oo(2)$.
When $(\vec{R},\rho_s)$ is non-trivial, we include it as part of the symmetry data of the system. Note that previous works checking the CEP for bosonic and fermionic systems \cite{song2020, zhang2020realspace,zhang2020const} considered $G_b=H \times G_{\text{space}}$ where $H$ is internal and $G_{\text{space}}$ is a point-group or wallpaper group symmetry. But our result below is expected to hold for arbitrary $G_b$.

The fCEP states that the classification of invertible fermionic topological phases with \textit{spatial} symmetry $G_f^{\sp}$ defined by the data $(G_b,s_1^{\sp},\om^{\sp},(\vec{R},\rho_s))$ is in 1-1 correspondence with the classification of invertible fermionic topological phases with an effective internal symmetry $G_f^{\Int}$ that has data $(G_b,s_1^{\Int}, \om^{\Int})$ where $G_b$ acts trivially on space. $s_1^{\Int}, \om^{\Int}$ are completely determined by $s_1^{\sp},\om^{\sp},\rho_s$.

We conjecture that the internal symmetry data are given by
\begin{equation}\label{eq:fCEP}
    \begin{split}
s_1^{\Int}&= s_1^{\sp}+w_1    \\
 \omega_2^{\Int} &= \omega_2^{\sp} + w_2 + w_1(s_1^{\sp}+w_1)
    \end{split}
\end{equation}
where $w_1 = \rho_s^* \ww_{1,r},w_2 = \rho_s^* \ww_2$ are obtained by pulling back the Stiefel-Whitney classes $\ww_{1,r},\ww_2$ that generate $ \H^1(\Oo(2),\Z_2),\H^2(\SO(2),\Z_2)$ respectively. If there are no reflections in $G_f$, $w_1 = 0$; if there are no rotations, $w_2=0$. In App.~\ref{app:fCEPAnti} we explain the notation further, check each term in the above formula with examples and give some heuristic justification.

Note that this formula treats unitary translations exactly as if they were internal symmetries: they do not appear in Eq.~\eqref{eq:fCEP}. The formula can also be straightforwardly generalized to $d$ spatial dimensions, as we discuss in App.~\ref{app:fCEPAnti}.

Each term above can be understood as follows. The term $w_1$ in the equation for $s_1^{\Int}$ implies that spatial orientation-reversing symmetries are mapped to internal antiunitary symmetries. 

The term $w_2$ in the equation for $\om^{\Int}$ implies that if a $2\pi$ spatial rotation acts as the identity in $G_f^{\sp}$, it should act as $(-1)^F$ in $G_f^{\Int}$, and vice versa. The term $w_1^2$ implies that if two successive reflections in $G_b$ act as the identity in $G_f^{\sp}$, they should act as $(-1)^F$ in $G_f^{\Int}$, and vice versa. The term $w_1 s_1^{\sp}$ implies that if $G_b$ has time-reversal  and reflection symmetries generated by ${\bf T},{\bf R}$ respectively that act on fermions as ${\bf RT} = {\bf TR}$, the fCEP maps them to a pair of anti-unitary internal symmetries ${\bf T},{\bf R}'$ that satisfy ${\bf R}'{\bf T} = (-1)^F{\bf TR}'$.

This principle has not been proven in complete generality for all invertible phases, but its predictions have been tested in several examples using different techniques. For example, Ref.~\cite{zhang2020realspace} provides evidence through a real space construction that if $G_b = H \times G_{\text{space}}$ with $s_1^{\sp}=0$, where $H$ is an internal symmetry and $G_{\text{space}}$ is a 2d wallpaper group, then $s_1^{\Int}$ and $\omega_2^{\Int}$ should be defined as in Eq.~\ref{eq:fCEP}. 

As a nontrivial check on Eq.~\ref{eq:fCEP} when $s_1^{\sp}\neq 0$ and $w_1 \neq 0$, we work out an example with $G_f =\ZZ_2^{{\bf RT}} \times \ZZ_4^{{\bf T}f}$ (see App.~\ref{app:fCEPAnti}). This means ${\bf R}^2 = {\bf T}^2 = (-1)^F$ and $({\bf RT})^2 = +1$. According to an edge argument \cite{YaoTScwithRT} using this $G_f$ gives a $\ZZ_8$ classification; we confirm that our theory combined with Eq.~\ref{eq:fCEP} gives the same classification.

Our general strategy to study crystalline invertible phases is thus to identify $s_1^{\sp},\om^{\sp}$, say from the system Hamiltonian, and then compute $s_1^{\Int}, \om^{\Int}$. Thereafter we study solutions to the $\mathcal{O}_3$ and $\mathcal{O}_4$ obstructions using $s_1^{\Int}, \om^{\Int}$ instead of the original spatial symmetry data. 

\subsection{Meaning of $n_1$ for crystalline symmetries}

\begin{figure}
    \centering
    \includegraphics[width=0.45\textwidth]{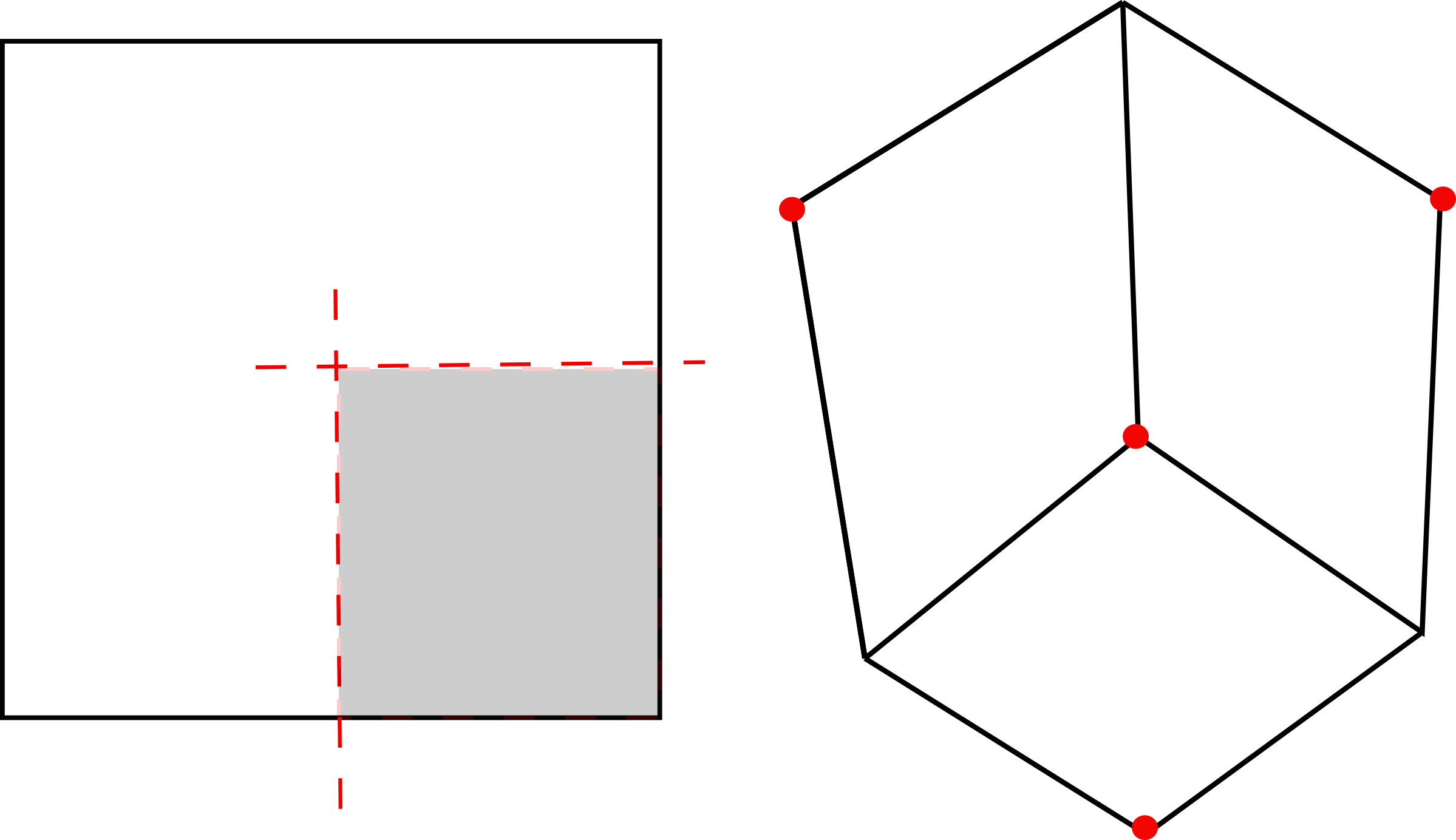}
    \caption{A $\pi/2$ disclination in a square lattice is constructed by removing a quadrant of angle $\pi/2$ (grey region) and reconnecting the severed edges. If the center of the disclination hosts an unpaired MZM (red circle), each corner must also carry an unpaired MZM so that the system on the defect lattice has a well-defined fermion parity.}
    \label{fig:disc_corner}
\end{figure}

If ${\bf g} \in G_b$ is a spatial symmetry operation, we have the following interpretations for $n_1({\bf g})=1$. When ${\bf g}$ is a lattice translation, this means that there is an unpaired MZM at a dislocation defect with Burgers vector equal to the translation defined by ${\bf g}$. When ${\bf g}$ is a discrete rotation by the angle $2\pi/n$, $n_1({\bf g})=1$ means that a $2\pi/n$ disclination defect hosts an unpaired MZM. It also means that corners of angle $2\pi/n$ host unpaired MZMs; the relationship between disclinations and corners is illustrated for $C_4$ symmetry in Fig.~\ref{fig:disc_corner}.\footnote{Note that if we consider a different symmetry, e.g. $C_2$, and construct a disclination of angle $\pi$, we might obtain one unpaired MZM at the disclination and another at the boundary of the system, but in that case the precise location of the boundary MZM is not fixed by symmetry alone.}

Finally, when ${\bf g}$ is a (unitary) reflection, we expect that $n_1({\bf g})=1$ means that the reflection line invariant under ${\bf g}$ hosts a (1+1)D invertible phase with unpaired MZMs at its ends. This expectation is based on results from real-space constructions of crystalline topological phases \cite{song2017,song2019topcryst,zhang2020const,zhang2020realspace}.\footnote{The assumption here is that the physical predictions of the real-space constructions hold more generally beyond their idealized limits.}  
For (2+1)D invertible fermionic phases in the presence of a reflection line, the real-space construction tells us to deposit a (1+1)D invertible fermionic phase on the reflection axis while treating the reflection as an onsite symmetry. In particular, if the symmetry of the (2+1)D system is $\Z_2^f \times \Z_2^{\bf R}$, the corresponding generator of (1+1)D invertible phases has unpaired MZMs at its ends. 

For general reflections, one needs to check whether the chosen (1+1)D decoration becomes trivialized or leads to spurious gapless degrees of freedom. In our language, the two latter conditions mean that we need to take additional equivalences into consideration and make sure that $\Omc_3$ and $\Omc_4$ are trivial.

\subsection{Remark on terminology}

TSCs protected by crystalline symmetries are often called `higher-order' TSCs, where the order refers to the spatial dimension to which unpaired MZMs are bound. In our formalism, an invariant is higher-order if it is trivialized upon breaking the crystalline symmetries. There is also a notion of `weak TSCs' in the literature. These are described by data that become trivial if we break translation symmetries.

\section{Examples of nontrivial constraints}\label{sec:examples}

\subsection{Constraints for (1+1)D invertible phases}\label{sec:1d}

(1+1)D invertible fermionic phases are not directly captured by the formalism of Ref.~\cite{barkeshli2021invertible}. However, some of the most attractive proposals for experimentally realizing unpaired MZMs use (1+1)D physics, therefore we discuss them briefly for completeness. 
 
The classification of (1+1)D invertible fermionic phases, including their complete stacking rules, has been presented in Refs.~\cite{turzillo2019fmps,bourne2021,aksoy2022}. Let $G_f, G_b$ and $\om$ be defined as usual. Then the classification consists of three parameters $n_0^{1d} \in \Z_2, n_1^{1d} \in \H^1(G_b,\Z_2)$ and $\nu_2^{1d} \in C^2(G_b,U(1))$. $n_0^{1d}$ specifies the number of unpaired MZMs modulo 2 at each end of a finite spatial segment. For each ${\bf g} \in G_b$, $n_1^{1d}({\bf g}) = 1$ specifies that the boundaries of the system have a degeneracy, and that the local ${\bf g}$ action at the boundary changes the local fermion parity. $\nu_2^{1d}({\bf g},{\bf h})$, along with the other data, specifies the projective representation of $G_f$ at each end of the system. 

In this case the basic constraint on MZMs is that $n_0^{1d}$ must be zero if $[\om] \ne 0$. This is proven in Ref.~\cite{turzillo2019fmps} using properties of matrix product states. We can intuitively understand this as follows: when $n_0^{1d} = 1$ and $[\om] \ne 0$, the fermion carries fractional $G_b$ charge, which can be absorbed by the unpaired MZM. Therefore the total $G_b$ charge in the system would change by a fractional amount in this case, leading to a contradiction. Interestingly, this argument closely resembles the one which enforces $\om = 0$ in (2+1)-dimensional invertible phases with half-integer $c_-$. 

From this result, we can immediately conclude that systems with the following symmetries cannot support MZMs in (1+1) dimensions, since they have nontrivial $[\om]$: $G_f = U(1)^f$ charge conservation, $G_f = SU(2)^f$ spin rotation symmetry with spin-1/2 fermions; systems in Class DIII with $G_f = \Z_4^{{\bf T}f}$; and so on.

The parameters $n_1^{1d},\nu_2^{1d}$ satisfy the equations \cite{turzillo2019fmps,Wang2020fSPT}
\begin{align}
    \dd n_1^{1d} &= 0 \\
    \dd\nu_2^{1d} &= (-1)^{n_1^{1d} \cup \om}.
\end{align}

We can verify these equations using the (2+1)D classification, in the following manner. Suppose the (2+1)D symmetry group is $G_f^{2d} = G_f^{1d}\times\ZZ$, with $n_0^{1d},n_1^{1d},\nu_2^{1d}$ defined using $G_f^{1d}$. Let $x$ be the generator of $\H^1(\Z,\Z_2)$. Then there is a $(c_-=0)$ (2+1)D invertible phase obtained by stacking $G_f^{1d}$ phases along the direction generated by $\Z$. Let this phase be described by the data
\begin{align}
    n_1^{2d} &=n_0^{1d}\cup x \\
    n_2^{2d} &= n_1^{1d}\cup x \\
    \nu_3^{2d} &= \nu_2^{1d}\cup x.
\end{align}
By applying the conditions on $n_1^{2d}, n_2^{2d}, \nu_3^{2d}$ written in Ref.~\cite{barkeshli2021invertible}, we observe that we can recover the above equations for (1+1)D phases. The equation for $n_1^{1d}$ that we obtain is
\begin{equation}
    dn_1^{1d} = n_0^{1d} \cup \om,
\end{equation}
but in order to solve this the rhs must be zero as a cohomology class. Thus we must either have $n_0^{1d}=0$ or $[\om] = 0$, as we claimed earlier. 

The equation for $\nu_2^{1d}$ is derived as follows. When $c_-=0$ and $[\om]=0$ or $n_1=0$, the $\Omc_4$ obstruction reduces to 
\begin{align}
    [\Omc_4] &= \frac{1}{2}n_2(\om+n_2) \\
    &= \frac{1}{2}n_1^{1d}x(n_1^{1d}x +\om ).
\end{align}
Now using $[x^2]=0$, we can rewrite $[\Omc_4] = \frac{1}{2}n_1^{1d}\om \cup x $. 

Note that this is only a useful consistency check, and not a complete derivation of the (1+1)D equations. Moreover, it is not a prescription for dimensionally reducing a (2+1)D phase into a (1+1)D phase. This is an interesting problem that we hope to study in the future.

\subsection{$G_f = \U(1)^f \times H$ and $G_f = \U(1)^f \rtimes H$}\label{sec:charge}
Suppose $G_f$ has a subgroup $U(1)^f$ corresponding to charge conservation. The notation $U(1)^f$ implies that this subgroup contains the fermion parity operation as its order 2 element, so $[\om] \ne 0$. We define the `bosonic' charge conservation symmetry as $U(1)_b := U(1)^f/\ZZ_2^f$. The charge of the fermion under $U(1)_b$ is $Q_{\psi} = \frac{1}{2}$.

First consider the case where $G_b = U(1) \times H$ for some symmetry group $H$, which may be unitary or antiunitary. In this case, we prove that $n_1=0$, i.e. there cannot be unpaired MZMs at symmetry defects. Formally, we argue using Eqs.~\eqref{eq:n2unitary} or ~\eqref{eq:n2AU}. Note that $U(1)$ is continuous, so $n_1$ must vanish within $U(1)_b$. Now, if $n_1$ is nonzero within $H$, we can show that $\mathcal{O}_3$ is a nontrivial 3-cocycle for each $c_- \in \Z$, so there is no solution for $n_2$. 

For a more physical derivation, we define 
\begin{equation}
    e^{2\pi i Q_a} := \eta_a(\pi,\pi)
\end{equation}
where $\pi$ is the order 2 element in $\U(1)_b$. Note that $Q_{\psi} = \frac{1}{2} \mod 1$. From Eq.~\eqref{eq:eta:U} we have
\begin{align}
    2 Q_e = 2 Q_m &= 0 \mod 1 \nonumber \\
    Q_e + Q_m &= Q_{\psi} \mod 1.
\end{align}
When $n_1 \ne 0$, we can use Eq.~\eqref{eq:etacoc} to obtain an additional relation
\begin{equation}
    Q_e = Q_m.
\end{equation}
The derivation is the same as the one given in Appendix \ref{app:abelian} for abelian groups $\U(1)^f \times H$. The last two equations force $Q_e = \frac{1}{4}$, which contradicts the first equation. Thus, if $G_f = \U(1)^f \times H$, the system cannot support unpaired MZMs at symmetry defects. 

\begin{figure}
    \centering
    \includegraphics[width=0.45\textwidth]{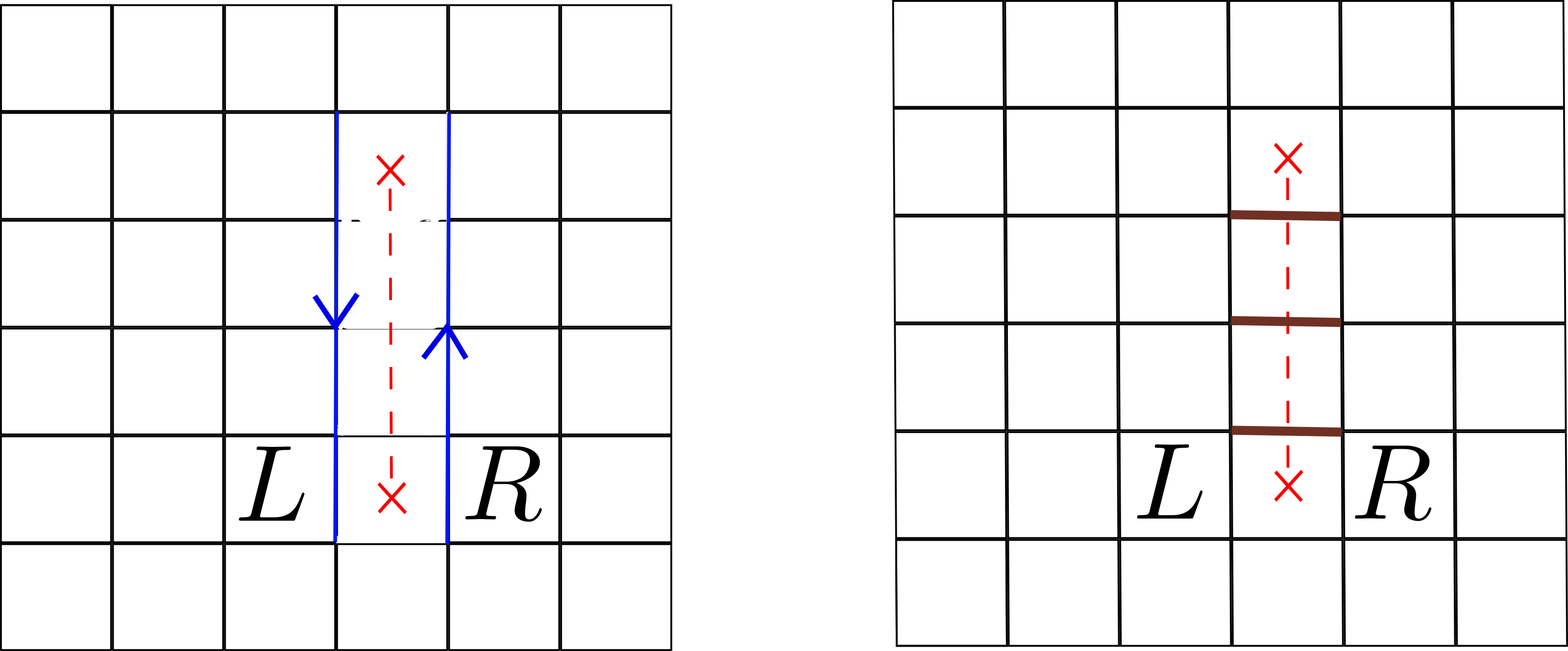}
    \caption{Construction of a defect of particle-hole symmetry in a Chern insulator. A line of hopping terms is removed along a cut, leading to gapless chiral edge states (blue lines). Upon reconnecting the edges with pairing terms (brown links), the edge states are gapped out. When $c_-$ is odd, an unpaired MZM is trapped at each defect, marked with a red 'X'.}
    \label{fig:chern_ins}
\end{figure}
The situation may be different if $G_f$ does not split as $\U(1)^f \times H$. In particular, consider $G_f = \Oo(2)^f$, for which $G_b = \Oo(2) = S\Oo(2) \rtimes \ZZ_2$. $G_b$ has a normal subgroup $S\Oo(2) \cong \U(1)_b$ corresponding to charge conservation; now there is an additional $\ZZ_2$ subgroup which corresponds to a unitary particle-hole symmetry. Now even though $[\om]$ is nonzero, unpaired MZMs are allowed at defects of the $\ZZ_2$ particle-hole symmetry, if $c_-$ is odd. The mathematical derivation is given in Appendix \ref{app:insulator}. 

Physically, we can see this by considering a free fermion system with Chern number 1 and with particle-hole symmetry. Now we can cut out a line of hopping terms $t_{ij} c_{Li}^{\dagger} c_{Rj} + h.c.$ from the Hamiltonian (see Fig.~\ref{fig:chern_ins}) and reglue them with superconducting pairing terms $\Delta c_{Li} c_{Rj} + h.c. $ instead. Each end of the cut corresponds to a $\ZZ_2$ symmetry defect, because a particle that goes around such a defect in the reglued system gets transformed to a hole, and vice versa. Prior to the regluing, the Chern number 1 system has a complex fermion mode propagating along the edge of the cut as shown in Fig. \ref{fig:chern_ins}. The superconducting pairing terms serve to gap out this edge, resulting in unpaired MZMs at each end of the cut. Thus we conclude that the particle-hole symmetry defects host unpaired MZMs.

It is interesting to note that we would have unpaired MZMs even in the absence of exact particle-hole symmetry, because the cutting and regluing procedure only depends on having odd Chern number. If there is no particle-hole symmetry, the ends of the cut cannot be thought of as symmetry defects; however, they will still host unpaired MZMs.

\subsection{$G_f = \SU(2)^f\times H$}\label{sec:spinrot}

Next we consider $G_f=\SU(2)^f\times H$, where $\SU(2)^f$ corresponds to spin conservation. In this case, we have the following results: (i) $c_-$ must be even; (ii) unpaired MZMs are not allowed at $H$ symmetry defects; (iii) the fermion parity fluxes must transform linearly under $H$. The last result is beyond Eq.~\eqref{eq:n2unitary}; it comes from the $\mathcal{O}_4$ obstruction.

We can get the first constraint by restricting to $G_f=\SU(2)^f$. In this case, $G_b=\SO(3)$ and $\om=\ww_2$ is the second SW class of the vector irrep of $\SO(3)$ (indicating that the fermion has spin-1/2). As $\om\neq 0$, we must have $c_-\in \ZZ$. Next, we note that \[
\om\cup_1\om = \Sq^1(\ww_2) = \ww_3
\]
is a mathematical equality\footnote{This is true by, for example, a Theorem by Wu that can be found as Theorem 4.5 in Ref.~\cite{thomas1960cohomology}.}. Here $\ww_3$ is the third SW class and is a generator for $\H^3(\SO(3),\ZZ_2)$. $\H^1(\SO(3),\ZZ_2)$ is trivial so $n_1$ must be trivial. Then the $\mathcal{O}_3$ obstruction reduces to
\[
\dd{n_2} = c_- \ww_3\mod 2
\]
but as $\ww_3$ is a non-trivial class, we need $c_-= 0 \mod 2$. In terms of $G_b$ quantum numbers, the above obstruction can be understood as follows: if $c_-$ were odd, the fermion parity fluxes must carry projective representations that square to the spin-1/2 representation carried by $\psi$ (but such representations do not exist). Therefore $c_-$ must be even.

Next, we argue that $n_1=0$. $n_1$ must vanish within $\SO(3)$ because it is connected. Moreover, it must vanish within $H$ because $\mathcal{O}_3 = n_1 \cup \om$ is nontrivial whenever $n_1$ is nontrivial in $H$. \footnote{When $n_1\neq 0$ there exists $\hbf \in H$ such that $n_1(\hbf)=1$. Then one can check that $\iota_{X}\iota_{Z}\iota_{\hbf}\Omc_3=1 \mod 2$ is a nontrivial invariant for $\mathcal{O}_3$. Here $\iota_{\gbf}$ is the slant product defined in App.~\ref{app:slant} and $X(Z)$ is the $\pi$ rotation in $\SO(3)$ around the $X(Z)$ axis. }

The $\mathcal{O}_4$ obstruction gives a further constraint that $n_2$ must be trivial. In particular, the fermion parity fluxes $e,m$ cannot carry any fractional $H$ quantum numbers. We prove this using Eq.~\eqref{eq:O4simple} in Appendix \ref{app:tHooftObstruction}. 

We can also see that $n_2=0$ by arguing that a system with spin-1/2 fermions and with some nontrivial $n_2 \in \H^2(H,\Z_2)$ has the same obstruction as the boundary of a nontrivial (3+1)D bosonic SPT with symmetry $H \times \SO(3)$. Such SPTs are classified by $\H^4(H \times \SO(3),U(1))$. Using the Kunneth formula we see that this has a subgroup $\H^2(H,\H^2(\SO(3),U(1))) \cong \H^2(H,\Z_2)$. Phases classified by this subgroup can be constructed by decorating a codimension 2 junction of $H$ defects (which can be formally thought of as a vortex for $H$) with a (1+1)D bosonic SPT with $\SO(3)$ symmetry. On the boundary of the (3+1)D phase, this codimension 2 junction appears to carry spin-1/2, which is the boundary signature of the SPT. But in our (2+1)D invertible phase, a nontrivial $n_2$ precisely implies that a codimension 2 junction of $H$ defects carries a fermion with spin-1/2. Therefore it is to be expected that the anomaly in the invertible fermion phase is cancelled by the (3+1)D bosonic SPT.

\subsection{Constraints when \texorpdfstring{ $G_f$}{Gf} is abelian}\label{sec:abelian}

If $G_f$ is a unitary abelian symmetry group, we obtain the following general restriction on unpaired MZMs, using Eq.~\eqref{eq:n2unitary}: either $n_1=0$ or $\om=0$. Note that $G_f$ must be a product of factors, each of the form $\Z_n, \Z$ or $\U(1)$. If $c_-$ is a half-integer, we must have $\om = 0$, so the claim is automatically satisfied. Therefore we focus on integer $c_-$. Here we will show that if $[\om] \ne 0$, we must have $n_1 = 0$. 

There are two ways in which $[\om]$ can be nontrivial for some abelian $G_f$: either
\begin{equation}
    G_f = \Z^f_{2N_0} \times A
\end{equation}
where $A$ is an abelian group, and $N_0$ is even; or
\begin{equation}
    G_f = \U(1)^f \times A.
\end{equation} 
In either case we find that $n_1$ must be zero. A corollary of this result is that systems with nontrivial $n_1$ and $[\om]$ must have $G_f$ non-abelian, if it is unitary.

Let us explain how this can be derived in the language of $G_b$ quantum numbers. First assume $c_-$ is even, and denote the fermion parity fluxes by $e,m$. In either of the above cases, the subgroup with nontrivial $[\om]$ has an order 2 element ${\bf h}$, and we can define $e^{2\pi i Q_a} := \eta_a({\bf h},{\bf h})$ for $a = \psi, e, m$\footnote{Note that $Q_e, Q_m$ are not fully gauge-invariant quantities if $n_1({\bf h})=1$. But the consistency relations involving $Q_e, Q_m$ still hold.}. Note that $Q_{\psi} = \frac{1}{2}$. If $n_1 = 0$, Eq.~\eqref{eq:eta:U} implies that  
\begin{equation}
    2 Q_e = 2 Q_m = Q_e + Q_m - Q_{\psi} = 0 \mod 1.
\end{equation}
Now if $n_1({\bf k}) = 1$ for any ${\bf k} \in G_b$, we obtain the additional constraint
\begin{equation}
    Q_e = Q_m.
\end{equation}
These equations are all derived in Appendix \ref{app:abelian}. They do not admit any consistent solution; therefore we must set $n_1 = 0$, proving our claim for even $c_-$. For odd $c_-$, the fermion parity fluxes are denoted as $v, \bar{v} = v \times \psi$; in this case we have a slightly different system of equations (see Appendix \ref{app:abelian}),
\begin{align}
    2 Q_v = 2 Q_{\bar{v}} &= Q_{\psi} \mod 1 \nonumber \\
    Q_v &= - Q_{\bar{v}} \mod 1
\end{align}
and if we have $n_1 \ne 0$, we obtain another constraint
\begin{equation}
Q_v = Q_{\bar{v}}    
\end{equation}
which is again inconsistent with the previous equations when $Q_{\psi} = \frac{1}{2}$. Mathematically, these constraints all originate from the fact that the term $n_1 \cup \om$ in Eq.~\eqref{eq:n2unitary} is a nontrivial 3-cocycle, and thus cannot be written as $dn_2$ for any choice of $n_2$.
Note that if we had started with $\om = 0$, i.e. with $Q_{\psi} = 0$, we could have solved the equations for each $c_-$, irrespective of $n_1$. We conclude that when $G_f$ is unitary and $[\om]$ is nontrivial, unpaired MZMs can only exist if $G_f$ is nonabelian.

The result is different when $G_f$ is antiunitary. When $G_f = U(1)^f \times A$, with $A$ antiunitary, we must still have $n_1 = 0$, and the arguments above still hold. However, if $G_f = \ZZ_{2N_0}^f \times A$, there is a simple counterexample, namely the class DIII time-reversal invariant TSC which has $G_f = \Z_4^{{\bf T}f}$. Here $n_1, \om$ can both be nontrivial. However, as we pointed out in Sec.~\ref{sec:int_AU}, this system displays Majorana Kramers pairs at fermion parity fluxes rather than unpaired MZMs.

\subsection{Orthogonal groups $G_f = \Oo(n)^f$}\label{sec:ongeneral}

When $G_f= \Oo(n)^f$, we find the following:
\begin{enumerate}
    \item $2c_-$ must be a multiple of $\gcd(n,16)$.
    \item When $n=2k$ is even, there are unpaired MZMs iff $c_- = \gcd(k,8) \mod (2\gcd(k,8)) $. In other words, $c_-$ is an odd multiple of the smallest allowed $c_-$ for the specific group.
    \item We obtain a description of the quantum numbers of the fermion parity vortices (see below for each case). 
    \item We find the full classification of invertible fermionic phases with $\Oo(n)^f$ symmetry. 
\end{enumerate}

The bosonic symmetry groups are the 'projective orthogonal groups' $\PO(n) := \Oo(n)/\ZZ_2$. We calculated their group cohomology up to degree 6 in App.~\ref{app:o2nf} \footnote{For this we use a spectral sequence calculation. At some point, we need to use the existence of the free fermion root phase (see below) to determine the value of a differential.}. We summarize the relevant cohomology results and then proceed to the derivation. Let $q: \Oo(n)\to \PO(n)$ be the quotient map and $q^*:\H(\PO(n),\ZZ_2)\to \H(\Oo(n),\ZZ_2)$ be the pull-back via $q$. Note that $\H(\Oo(n),\ZZ_2)$ is a polynomial ring generated by the first $n$ Stiefel-Whitney classes $\ww_i$. Here $\ww_i$ has degree $i$, and $\ww_j$ is set to $0$ whenever $j>n$ (See \textbf{Theorem 1.2} of Ref.~\cite{brown1982cohomology}). 

First, $\H^1(\PO(2n),\ZZ_2)\cong \ZZ_2$ is generated by a class $\xx_1$ that satisfies $q^*\xx_1=\ww_1$. An explicit formula is $\xx_1(\gbf) = \frac{1+(-1)^{\det(\tilde{\gbf})}}{2}$ where $\tilde{\gbf}$ is any lift of $\gbf \in \PO(2n)$ to $\Oo(2n)$. $\xx_1(\gbf)$ measures whether $\gbf$  acts non-trivially on an odd number of directions. 

Secondly, $\H^2(\PO(2n),\ZZ_2) = \ZZ_2\times\ZZ_2$. It is generated by $\xx_2$ and $\xx_1^2$. $\xx_2$ is defined as the class corresponding to the extension of $\PO(2n)$ by $\ZZ_2$ isomorphic to $\Oo(2n)$. In other words, for $G_f=\Oo(2n)^f$ we have $G_b=\PO(2n)$ and $\om=\xx_2$. This means that a $\pi$ rotation in $\PO(2n)$ squares to fermion parity. $\xx_2\cup_1\xx_2=\Sq^1(\xx_2)$ depends on the parity of $n$. For $n$ odd, $\Sq^1(\xx_2)=\xx_1\xx_2$. While for $n$ even, $\Sq^1(\xx_2)=\xx_3\neq0 $ and  $ \xx_1 \xx_2 =0 $. 

Finally, $\H^3(\PO(4n),\U(1)) = \ZZ \times \ZZ_2$ is generated by a Chern-Simons term for $\PO(4n)$ and $(-1)^{\ww_1^3}$. For $n>1$, $\H^3(\PO(4n+2),\U(1)) = \ZZ \times \ZZ_2^2$ is generated by a Chern-Simons term, $(-1)^{\xx_1^3}$ and $(-1)^{\xx_3}$. Here $\xx_3\in \H^3(\PO(4n+2),\ZZ_2)$ is such that $q^*\xx_3=\ww_3$, the third SW class of $\Oo(4n+2)$. For $\PO(2)$, $\xx_3$ is trivial because $\ww_3$ is trivial in $\Oo(2)$.

We define a root phase for $G_f=\Oo(2n)^f$ by stacking $2n$ identical layers of a spinless $p+ip$ SC. These phases have $n_1=\ww_1$ and $c_-=n$\footnote{Ref.~\cite{Chen2019freeinteracting} argued that for a free fermion phase in 2+1d transforming in a real representation ($\rho$) of $G_f=\ZZ_2^f\times G_b$, the chiral central charge is $1/2$ the dimension of the representation and $n_1$ is given by $(-1)^{n_1(\gbf)}=\det(\rho(\gbf)) $ ($n_1=\gamma$ in their notation). We can use this result for $G_f=\Oo(n)^f$ by restriction to the $\ZZ_2^f\times\ZZ_2$ subgroup where $\ZZ_2$ is a reflection along a single axis. 
 }.

\subsubsection{$G_f = \Oo(2)^f$}\label{sec:o2f}

Consider $G_f = \Oo(2)^f$, for which $G_b = \Oo(2)$ so we will write $\ww_i$ instead of $\xx_i$. Here $\om=\ww_2$ is nontrivial; this corresponds to the fermion having charge $\frac{1}{2}$ with respect to the $\SO(2)$ subgroup of $G_b$. 

If $n_1 \ne 0$, we need $c_-$ to be odd. The mathematical argument is as follows. As $\PO(2)\cong\Oo(2)$, we have the relation $\ww_2\cup_1\ww_2=\Sq^1(\ww_2)=\ww_1\ww_2$. Then, $\mathcal{O}_3=(n_1+c_-\ww_1)\ww_2$, which is a trivial cocycle iff $n_1+c_-\ww_1=0$.

Here is a more physical argument. First, for even $c_-$, we can define the $\SO(2)$ charge of the fermion parity fluxes as $Q_e, Q_m$. The symmetry defines $Q_{\psi} = \frac{1}{2}$. Using the constraints on the $\eta$ symbols, we find that
\begin{align}
    2Q_e = 2Q_m &= 0 \nonumber \\
    Q_e + Q_m &= \frac{1}{2} \nonumber \\
    Q_e &= Q_m.
\end{align}
These equations (all taken modulo 1) are clearly inconsistent. However, if $c_-$ is odd, we find a different set of equations,
\begin{align}
    2Q_v = 2Q_{\bar{v}} &= \frac{1}{2} \nonumber \\
    Q_v &= -Q_{\bar{v}},
\end{align}
which does permit a consistent solution where $v,\bar{v}$ have charge $\pm 1/4$. 

A simple construction of a system with $c_-=1$ is that of a spinful $p+ip$ superconductor defined in terms of two flavors of fermions $c_1,c_2$. At weak pairing the superconducting pairing term in such a system is of the form $\sum_{\bf k}\Delta({\bf k}) (c_{1,{\bf k}} c_{1,-{\bf k}} + c_{2,{\bf k}} c_{2,-{\bf k}}) + h.c.$. There is an $\Oo(2)^f$ symmetry which transforms $c_{1,{\bf k}}$ and $c_{2,{\bf k}}$ into each other for each ${\bf k}$, but leaves the pairing term invariant. The reflection symmetry defect here corresponds exactly to the `half-quantum vortex' \cite{dassarma2006hqv} which is expected to carry a MZM (see Sec.~\ref{sec:hqv} for more discussion). A similar example was analyzed in Ref. \cite{barkeshli2021invertible}, in which the $\Oo(2)^f$ symmetry is broken down to a discrete subgroup $\mathbb{D}_8^f$ but the mathematical constraints are equivalent.

\subsubsection{$G_f = \Oo(4)^f$}\label{sec:o4f}
For $G_f = \Oo(4)^f$, the result is as follows. $c_-$ must be an even integer. If $c_- = 2 \mod 4$, $n_1$ must be nontrivial, and the fermion parity fluxes must transform projectively under $\SO(4)^f \subset \Oo(4)^f$. If $c_- = 0 \mod 4$, $n_1$ must be zero, and fermion parity fluxes must transform as a linear representation of $\SO(4)^f$. Regardless of the value of $c_-$, the fermion parity fluxes can transform projectively under orientation reversing elements ('odd reflections') of $\Oo(4)^f$.

To get some intuition, note that there is a simple construction of a free fermion system with $c_-=2$ which supports unpaired MZMs. Consider a stack of 4 identical layers of a spinless $p+ip$ superconductor at weak pairing; this has the required internal symmetry $\Oo(4)^f$ which permutes the fermions in each layer, but keeps the pairing term $\sum_{{\bf k},\alpha}  \Delta({\bf k}) c_{\alpha,{\bf k}} c_{\alpha_-{\bf k}} + h.c.$ invariant ($1\le \alpha \le 4$ is the layer index). An unpaired MZM can be introduced by inserting an HQV into just two of the layers, ignoring the other two. This {shows that there is a solution} with $c_- = 2 \mod 4$ and $n_1 \ne 0$. 

In this case $G_b = \PO(4)= (\SO(3)_{L}\times\SO(3)_{R})\rtimes\ZZ_2$ where the reflection acts by permuting the ``left" (L) and ``right" (R) $\SO(3)$ symmetries. The fermion has spin $(s_{L,\psi},s_{R,\psi}) = (\frac{1}{2},\frac{1}{2})$ under the two $\SO(3)$ subgroups. If we consider some odd $c_-$, the spin of a fermion parity flux is constrained by the relations $2s_{v,L(R)} = s_{\psi,L(R)} = \frac{1}{2} \mod 1$, which have no solution (this is similar to the result for a single $\SO(3)$ symmetry). Therefore $c_-$ must be even.

Eq.~\eqref{eq:n2AU} does not impose any nontrivial constraints when $n_1 = 0$. Now consider the unique nontrivial choice $n_1 = \xx_1$, which is nonzero on the $\ZZ_2$ reflection which also interchanges $L \leftrightarrow R.$ For this $n_1$, there is in fact a solution for $n_2$. This can be seen by writing down the equations constraining the spin of the fermion parity fluxes:
\begin{align}
    2s_{e,L(R)} = 2s_{m,L(R)} &= 0 \mod 1 \\
    s_{e,L(R)} + s_{m,L(R)} &= \frac{1}{2} \\
    s_{e,L(R)} &= s_{m,R(L)}.
\end{align}
The first two relations arise in order to have compatibility with the fusion rules. The third is due to the permutation which simultaneously takes $e \leftrightarrow m, L \leftrightarrow R$. The only solutions are $(s_{e,L},s_{e,R}) = (\frac{1}{2},0)$ and $(s_{m,L},s_{m,R}) = (0,\frac{1}{2})$, or vice versa. This means that $n_2$ must be nontrivial, and is fixed by the choice of $n_1$. We can also argue this formally using group cohomology (see Appendix \ref{app:Orthogonal(4n)}).

At the level of the $\mathcal{O}_3$ obstruction we can have $n_1$ non-trivial for any allowed $c_-$. Nevertheless, $\mathcal{O}_4$ imposes the additional constraint $n_1 = \frac{c_-}{2} \xx_1$. This has the perhaps surprising consequence that when $c_- = 2 \mod 4$, there is no solution with $n_1 = 0$, i.e. the system is forced to have unpaired MZMs. See App.~\ref{app:o2nf} for the mathematical details, and App.~\ref{app:WreathProducts} for related examples generated by so-called `wreath products'. 

Note that there is freedom to adding $\xx_1^2 $ to $n_2$ regardless of the value of $c_-$. We can restrict to  $\ZZ_2^f\times\ZZ_2\subset \Oo(4)^f$  where the first is generated by $-1$ and the second by the  diagonal matrix with diagonal $[-1,1,1,1]$. $n_2=\xx_1^2$ means that $\ZZ_2$ acts projectively on the fermion parity flux.

\subsubsection{$G_f=\Oo(2n+1)^f$}

In this case, $\PO(2n+1)=\SO(2n+1)$ and $G_f = \SO(2n+1)\times \Z_2^f$. The extension of $G_b$ by $\Z_2^f$ is trivial, {i.e.} $\om = 0$. Thus we can have any $c_- \in \tfrac{1}{2}\ZZ$ and there are unpaired MZMs at fermion parity fluxes if and only if $2c_-$ is an odd integer. 

For $n=0$, the classification of invertible phases with $G_f$ symmetry is simply $\tfrac{1}{2}\ZZ$. On the other hand, when $n>0$, it is $\tfrac{1}{2}\ZZ \times \ZZ$, with the second factor generated by a fermionic phase with $c_-=0$, $n_1=0$, $n_2 =\ww_2$.

\subsubsection{$G_f=\Oo(4n+2)^f$}

For $G_b=\PO(4n+2)$ and $n>0$, there is no longer a simple expression for $G_b$ as for $n=0$. Nevertheless, we can gain a lot of information by restricting to the diagonal $\Oo(2)$ subgroup. Let $r:\PO(2) \hookrightarrow \PO(4n+2)$ be the diagonal inclusion map \footnote{$r$ is defined by first taking the diagonal inclusion of $\Oo(2) \subset \Oo(4n+2)$ (take an $\Oo(2)$ matrix and send it to the direct sum of $2n+1$ copies of the matrix) and then taking the quotient by $-1$. } and let $r^*:\H(\PO(2),\ZZ_2) \to \H(\PO(4n+2),\ZZ_2)$ be the pull-back via $r$. We find that $c_-\in \ZZ$ and $n_1 = c_- \xx_1$.

Suppose we have a $G_f=\Oo(4n+2)^f$ phase with $n_1=k\xx_1$, where  $(k\in \{0,1\})$ and $c_-\in \ZZ$. We have $r^*\xx_1 = \ww_1 \mod 2$. Thus $r^*n_1 = kr^*\ww_1=k\ww_1$ while $c_-$ does not change. From the constraints on $\Oo(2)^f$, we must have $k=c_- \mod 2$, and $c_-$ is forced to be an integer. 

The root phase has an odd chiral central charge $c^*_- = 2n+1$. As $\gcd(c^*_-,8)=1$, we can always find integers $m,l$ such $c^*_- m+ 8l=1$. Then if we stack $m$ copies of the root phase with $l$ copies of the $E_8$ state \cite{Kitaev2006} we obtain a phase with $c_-=1$. In this way, we can construct phases with any $c_-\in \ZZ$.

The classification of invertible phases with symmetry $G_f = \Oo(2)^f$ is $\ZZ \times \ZZ \times \ZZ_2$ where the factors correspond to $c_-$, $\frac{\sigma_H-c_-}{8}$ and a bosonic SPT with $\nu_3 =(-1)^{\xx_1^3}$, respectively. For $G_f=\Oo(4n+2)^f $ and $ (n>0)$, the classification is $\ZZ \times \ZZ \times \ZZ_2\times\ZZ_2$ where the extra $\Z_2$ corresponds to a bosonic SPT with $\nu_3= (-1)^{\xx_3}$ where $q^*\xx_3 = \ww_3$. 

\subsubsection{$G_f=\Oo(4n)^f$}

Depending on the divisibility of $n$ by powers of two we obtain different constraints, which are summarized in Table \ref{tab:summary}. 

If $n$ is odd, we consider the diagonal $\Oo(4)^f$ subgroup\footnote{By diagonal subgroup of $\Oo(4)$ we mean the image of $\Oo(4)$ under the map that sends $R\in \Oo(4)$ to the direct sum of $n$ copies of $R$.} to show that $c_-$ is even and $n_1=(c_-/2)\ww_1 $. This is the strongest possible constraint because the root phase has $c_-^*=2n$ and $n_1=n\ww_1 \mod 2$. Furthermore as $\gcd(2n,8)=2$, by stacking the appropriate number of root phases and $E_8$ states we can get any $c_-\in 2\ZZ$.

When $n$ is even, restricting to the diagonal $\Oo(4)^f$ subgroup takes $\ww_1$ to zero, so we get no constraint for $n_1$. To make progress, we need to consider the larger subgroups $\Oo(8)^f$ and $\Oo(16)^f$. In App.~\ref{app:ConstrainsO(8)O(16)}, we argue that for $\Oo(8)^f$ we must have $c_- =0 \mod 4$ and $n_1=(c_-/4)\ww_1$; and for $\Oo(16)^f$ we must have $c_-=0 \mod 8 $, with no constraint on $n_1$.

We can repeat the arguments used for $\Oo(4n+2)^f$ and $\Oo(8n+4)^f$ to show the following. For $\Oo(16n+8)^f$, we must have $c_-\in 4\ZZ$ and $n_1= (c_-/4)\ww_1$ and these constraints are saturated by a suitable stack of the root phase and $E_8$ states. For $\Oo(16n)$, we must have $c_-\in 8\ZZ$ and $n_1$ can be non-trivial for any $c_-$. The constraints are sharp again.

The classification of invertible phases with symmetry $G_f = \Oo(4n)^f$ is $\ZZ \times \ZZ \times \ZZ_4$ corresponding to (up to normalization) $c_-$, $\frac{\sigma_{H}-c_-}{8}$, and a third factor generated by a (fermionic) phase with $c_-=0$, $n_1=0$,  $n_2= \ww_1^2 $. 
\subsection{Unitary groups }
\subsubsection{$G_f = \SU(2n)^f$}\label{sec:SU2nf}

We now consider special unitary groups. As these groups are simply connected, $n_1$ must be trivial. We can derive constraints on $c_-$ by considering appropriate subgroups. We define the root phase for this symmetry to be a stack of $2n$ layers of a Chern insulator (although these have a larger symmetry $\U(2n)^f$, they also belong to the classification for $\SU(2n)^f$). These states have $c_-=2n$. We consider the following cases \footnote{The constraint for $G_f=\SU(2)^f$ can be found in Sec.~\ref{sec:spinrot}. The constraint for $G_f=\SU(4)^f\cong \Spin(6)^f$ can be defined by looking at the $\Spin(5)^f$ subgroup and use the results of Sec.~\ref{sec:tHooftConstraint}. Finally, the constraint for $\SU(8)^f$ is derived in App.~\ref{app:HSUNZ2} using a spectral sequence calculation.
}

\paragraph{$n$ is odd:} We consider the diagonal $\SU(2)^f$ subgroup to show that $c_- =0  \mod 2$. \footnote{In $\SU(nm)$, we define the diagonal $\SU(n)$ subgroup to be the image of the map that sends ${\bf g}\in\SU(n)$ to the direct sum of $m$ copies of ${\bf g}$.}

\paragraph{$n/2$ is odd:} We consider the diagonal $\SU(4)^f$ subgroup to show that $c_- = 0 \mod 4$. 

\paragraph{$n/2$ is even:} We consider the diagonal $\SU(8)^f$ subgroup to show that $c_- = 0 \mod 8$. 

In all cases, some combination of the root phase and $E_8$ have the minimal allowed $c_-$. These constraints depend on computing $\mathcal{O}_4$ and cannot be inferred just from $\mathcal{O}_3$; see Sec.~\ref{sec:tHooftConstraint} for more examples of this type.

\subsubsection{$G_f = \U(n)^f$ and $\U(n)^f \rtimes \Z_2$}\label{sec:Unf}
Next we discuss the family of unitary groups $\U(n)^f$. This is the symmetry of $n$ identical layers of a Chern insulator. All the results in this case can be obtained by restricting to the (diagonal) $\U(1)^f$ and $\SU(n)^f$ subgroups of $\U(n)^f$. 

In particular, from the diagonal $\U(1)^f$ subgroup we see that $c_-$ must be an integer, and that there can be no unpaired MZMs since in that case there would be an obstruction due to the $n_1 \cup \omega_2$ term in Eq.~\eqref{eq:n2unitary}. By restricting to $\SU(n)^f$, we use the results of the previous section to see that $c_-$ must be a multiple of $\gcd(n,8)$.

One situation where the system can have unpaired MZMs in the presence of $\U(n)^f$ symmetry is if there is an additional charge conjugation symmetry. Suppose the full symmetry is $\U(n)^f \rtimes \Z_2$, where $\rtimes$ denotes charge conjugation \footnote{$\ZZ_2$ acts as the unique non-trivial outer automorphism of $\U(n)$.}. By restricting to the diagonal $\U(1)^f \rtimes \Z_2$ subgroup, we see that unpaired MZMs can exist at $\Z_2$ defects when $c_-$ is odd (see the discussion of $G_f = \Oo(2)^f$ in Section \ref{sec:o2f}). By restricting to $\SU(n)^f$, we see that $n$ must be odd (otherwise $c_-$ would have to be even, leading to a contradiction). 

We can also understand this based on our discussion of $\Oo(2)^f$ symmetry defects in Section \ref{sec:charge}. There we discussed a procedure to create an unpaired MZM in a $c_-=1$ system with $\U(1)^f \rtimes \Z_2$ symmetry by introducing suitable SC pairing terms. A similar procedure allows us to create $n$ MZMs in an $n$-layer system with $\U(n)^f$ symmetry. But in order to have a residual unpaired MZM, $n$ must be odd. 

\subsection{Symplectic groups $G_f=\Sp(n)^f$}\label{sec:symplectic}

We consider the groups $G_f=\Sp(n)^f$. The group cohomology of $G_b=\PSp(n)=\Sp(n)/\ZZ_2$ is calculated in App.~\ref{app:SymplecticGroup}. As $G_b$ is connected, we must have $n_1=0$. So we only look for constraints on $c_-$.
We first find the constraint $c_-=0\mod 2$ by restriction to the diagonal $\Sp(1)^f\cong\SU(2)^f$ subgroup \footnote{The diagonal subgroup is the image of the map that sends $R\in \Sp(1)$ to the direct sum of $n$ copies of $R$.}. There are only two (equivalent) choices for $n_2$: $n_2=0,\ww_2$. The $\mathcal{O}_4$ obstruction is 
\begin{equation}
\Omc_4 =\frac{c_-}{2} \frac{\Pmc(\ww_2)}{4} \mod 1
\end{equation}
which is non-trivial only when $c_-= 0 \mod 2\gcd(n,4)$. The full classification is $\ZZ^2$ and is generated by $\frac{c_-}{2\gcd(4,n)}$ and a generalization of the spin-Hall conductivity. 

\subsection{Constraints when $G_b$ is a direct product}\label{sec:dirprod}

In this section we prove two results. First, if the symmetry group is of the form $G_b=G_b^A \times G_b^B$ and $\om = \om^{\!\! A}+\om^{\!\! B}$ with both $\om^{A},\om^{B}$ non-trivial, there can be no unpaired MZMs. Physically, this means that if the fermion carries fractional quantum numbers under both $G_b^A,G_b^B$, then there cannot be unpaired MZMs at any symmetry defects, because either the $G_b^A$ or $G_b^B$ quantum numbers at these defects will not be consistently defined. Furthermore, in this case $c_-$ can be odd only if $\Sq^1(\om^{\!\!i})=0$ for $i=A,B$. 

Consider some $n_1 \in \H^1(G_b,\Z_2)$. This has the form $n_1 = n_1^A + n_1^B$, where $n_1^i \in \H^1(G_b^i,\Z_2)$ for $i=A,B$. For this $n_1$, Eq.~\eqref{eq:n2unitary} reads
\begin{align}
    dn_2 &= n_1^A \om^A + n_1^B \om^B + n_1^B\om^A + n_1^A \om^B \nonumber \\
    &+ c_-(\om^A \cup_1 \om^A + \om^B \cup_1 \om^B + \dd(\om^A \cup_1 \om^B)).
\end{align}
The possible obstructions in $\H^3(G_b,\Z_2)$ can be classified by the K\"{u}nneth formula (see e.g. \cite{Sato1999}):
\begin{align}\label{eq:kun1}
    \H^3(G_b,\Z_2) &= \H^3(G_b^A,\Z_2) \times \H^3(G_b^B,\Z_2) \nonumber \\
    & \times \H^1(G_b^A,\H^2(G_b^B,\Z_2)) \times \H^1(G_b^B,\H^2(G_b^A,\Z_2)). 
\end{align}
Suppose $n_1^A \ne 0$. Now since $[\om^B]$ is nontrivial, $n_1^A \om^B$ is also nontrivial as a 3-cocycle, and is classified by the third factor in Eq.~\eqref{eq:kun1} above. None of the other terms can cancel this obstruction, since they all belong to other factors in the K\"{u}nneth decomposition. Therefore we cannot solve for $n_2$ unless we set $n_1^A = 0$. Similarly we argue that $n_1^B = 0$, and hence $n_1 = 0$. We conclude that for this choice of symmetry group, unpaired MZMs are not possible. 

Setting $n_1 = 0$, the equation for $n_2$ reduces to 
\begin{equation}\label{eq:GbaGbb}
    dn_2 = c_-(\om^A \cup_1 \om^A + \om^B \cup_1 \om^B).
\end{equation}
Again using the K\"{u}nneth decomposition, we can see that an odd $c_-$ is allowed only if both $\om^A \cup_1 \om^A, \om^B \cup_1 \om^B$ are trivial in $\H^3(G_b,\Z_2)$. Actually, this condition can be strengthened by looking at $\mathcal{O}_4$: under the Bockstein map (see App.~\ref{app:coho}) $\beta: \H^2(G_b,\Z_2) \rightarrow \H^3(G_b,\Z)$, the image of both $\om^A, \om^B$ must be trivial. The argument is given in Appendix \ref{app:dirprod}.

\def\Xx{\mathbf{X}}
\def\t{\tau}
\subsection{ 
$G_f = \Z_4^{{\bf T}f} \rtimes \Z_2$}\label{sec:Z2XZ4T}

The usual symmetry which defines the class DIII TSC is $G_f = \Z_4^{\Tt f}$. Let's add to Class DIII a $\ZZ_2$ unitary symmetry, generated by $\Xx$, that anticommutes with time-reversal on the fermions, {i.e.} $\Tt \Xx = (-1)^F \Xx \Tt$ and $\Xx^2=1$. In this case $G_b=\ZZ_2^\Tt \times \ZZ_2$. Let $\t$ and $x$ be the projections to the $\ZZ_2^\Tt$ and $\ZZ_2$ subgroups of $\H^1(G_b,\Z_2)$. Then $s_1=\t$ and $\om=\t\cup(\t+x)$. Therefore, by the condition stated in Eq.~\eqref{eq:AUsuff} there is a solution with $n_1= \t+x$ and $n_2=0$, $\nu_3=1$. 

Note that under restriction to $G_f'=\ZZ_2^f\times \ZZ_2$, $n_1$ restricts to $x$, which is a generator of the $\ZZ_8$ classification of invertible phases with symmetry $G_f'$. These phases can be constructed from free fermions by stacking $\nu$ layers of a spinless $p+ip$ SC with fermions carrying $\Xx$ eigenvalue $-1$ along with their $\Tt$ partners, which in this case corresponds to $p-ip$ SC layers with fermions carrying $\Xx$ eigenvalue +1. \cite{Gu2014interaction}.

\subsection{Additional constraints from $\mathcal{O}_4$ obstruction}\label{sec:tHooftConstraint}

 Although we considered both $\mathcal{O}_3$ and $\mathcal{O}_4$ in the preceding sections, we emphasize that the $\mathcal{O}_4$ obstruction can give stronger constraints on $c_-, n_1, n_2$ that cannot be deduced from $\mathcal{O}_3$ alone. In this section we give some examples of this. The mathematical calculations involving $\mathcal{O}_4$ are organized in Appendix \ref{app:tHooftObstruction}. Here we will only state the results:
\begin{enumerate}
    \item We show in App.~\ref{app:SON} that when $G_b = \SO(N)$ and $G_f = \Spin(N)$ (i.e. the nontrivial extension of $G_b$ by $\Z_2^f$), and $N \ge 5$, the choice $c_-=2 \mod 4$ is not obstructed by $\mathcal{O}_3$ but has a nontrivial $\mathcal{O}_4$ obstruction. Thus $c_-$ needs to be a multiple of 4 in order to have a well-defined invertible phase.
    \item An example of a finite group carrying the same obstruction as $G_f=\Spin(5)^f$ is $G_f= \DD_8^f \circ \QQ_8^f $ which is the central product of $\DD_8$ and $\QQ_8$ and has $G_b=\ZZ_2^4$. The calculation can be found in Appendix~\ref{app:finite}. 
    Note that $G_f$ in this example is isomorphic to the group of order 32 in Table II of Ref.~\cite{ning2021enforced} \footnote{To show this one can use the characterization of extra-special 2-groups - the extension is characterized in terms of its quadratic form, in particular the dimension of its cokernel and its Arf invariant.  }.
\item Conversely, when $c_- = 0 \mod 4$ and $G_b$ is a unitary compact Lie group, we show that the data $n_1 = 0, n_2 = 0$ are non-anomalous for very general choices of $\om$, as long as a certain technical condition is satisfied.\footnote{
The constraint is that $\mathrm{Tor}^{\ZZ}_1[\H^5(G_b,\ZZ),\ZZ_4]$ is of the form $\ZZ_2^L$ for some integer $L$.
} See App.~\ref{app:SomeGfAdmitc4} for details. 

In this situation, the only nontrivial data is $[\om]$. We can understand two stacked copies of the non-anomalous phase as equivalent to a stack of 3 systems: (i) a trivial fermionic phase, (ii) the $E_8$ phase, which is a bosonic phase of matter with $c_-=8$ that exists for any symmetry group, and (iii) an additional bSPT with cocycle $\nu_3$ satisfying $\beta(\nu_3) = \om \cup \om \mod 2$. $\beta$ is a Bockstein map defined in App.~\ref{app:coho}.

One application of this result is to the group with $G_b = (\Z_2 \times \Z_2)^L$, such that $G_f$ is a central product of $L$ copies of $\mathbb{D}_8$ (i.e. the order 2 rotation in each $\mathbb{D}_8$ is identified with $(-1)^F$). This symmetry group satisfies the condition to admit a $c_-=4$ solution. Interestingly, a free fermion construction can only give solutions with $c_- = 2^{L-1}$. Thus for $L>3$, the $c_-=4$ solution must correspond to an intrinsically interacting phase. See Appendix \ref{sec:D8xL} for the details.

In Ref.~\cite{ning2021enforced}, it was argued that there cannot exist any invertible phases with $c_-=4$ for the group $\SU(8)^f$. We have checked that in this case, the technical condition required to ensure a $c_-=4$ solution actually does not hold, and moreover we explicitly identify $[\mathcal{O}_4]$ as the order 2 element of $\H^4(\SU(8)/\ZZ_2,\U(1)) \cong \Z_4$ (See App.~\ref{app:HSUNZ2} for details). Therefore our result agrees with the one obtained in Ref.~\cite{ning2021enforced} through different arguments.

\item We show that when $G_f = \Z_{n} \times \Z_2^{\Tt} \times\ZZ_2^f$, the constraint from $\mathcal{O}_3$ only forbids unpaired MZMs when $n$ is even. On the other hand, the $\mathcal{O}_4$ obstruction also forbids unpaired MZMs when $n$ is a multiple of 4. Thus if the system is to have unpaired MZMs, $n$ must be a multiple of 8. See Appendix \ref{app:znz2T}.

If we break time reversal symmetry, we get a phase with $G_f'=\ZZ_{n}\times\ZZ_2^f$ and $n_1=\ww_1$. This phase generates a $\ZZ_2$ factor in the classification of invertible phases of fermions with $G_f'=\ZZ_{n}\times\ZZ_2^f$ that is known to be generated by free fermions: see Ref.~\cite{grigoletto2021spin}. Table 2 in this paper shows that one of the phases with $n_1 = \ww_1$ is given by a stack of a spinless $p+ip$ SC of fermions with charge $n/2$ under $\Z_n$, and a spinless $p-ip$ SC of fermions with charge $0$.

\end{enumerate}

\section{Applications}\label{sec:apps}

We now discuss several examples showing how our theory relates to different proposals for detecting unpaired MZMs at symmetry defects. In many of the examples discussed below, our theory provides additional predictions beyond those discussed in the preceding literature. 

\subsection{Unpaired MZMs at half-quantum vortices in a spinful $p + ip$ SC 
}\label{sec:hqv}

Here we discuss a spinful $p+ip$ superconductor, modeled as a stack of two identical layers of spinless $p+ip$ superconductors, corresponding to two flavors of fermions $c_{\alpha,i}$, with $\alpha = 1,2$ the layer index, and $i$ a position index. This system 
has the symmetry $G_f = \Oo(2)^f$ discussed in Sec. \ref{sec:charge}. The spinful $p+ip$ superconductor has been proposed as a mean-field model for strontium ruthenate, $\text{Sr}_2\text{RuO}_4$, although this has called into question by recent experiments \cite{pustogow2019sr,Ishida2020sr,ghosh2021sr}.

For each $i$, the global $\Oo(2)^f$ symmetry acts on the fermions $c_{1,i}, c_{2,i}$ through the standard $2\times 2$ matrix representation $R$. Denote a group element in $\Oo(2)^f$ as $(\theta, r)$ where $\theta \in [0,2\pi)$ is a rotation and $r \in \{0,1\}$ is a reflection. Then, we have
\begin{align}
    U((\theta,r)) \binom{c_{1,i}}{ c_{2,i}} U((\theta,r))^{\dagger} &= R((\theta,r)) \binom{c_{1,i}}{c_{2,i}}, \\
    R((\theta,r)) &= \begin{pmatrix}(-1)^r \cos \theta & - \sin \theta \\ (-1)^r \sin \theta & \cos \theta \end{pmatrix}.
\end{align}
Note that $G_b$ is also isomorphic to the group $\Oo(2)$, but a $2\pi$ rotation in $G_b$ must act on fermions as $(-1)^F$. The symmetry operators $U_b$ of $G_b$ act as follows:
\begin{equation}
    U_b((\theta,r)) \binom{c_{1,i}}{ c_{2,i}} U_b((\theta,r))^{\dagger} = R((\theta/2,r)) \binom{c_{1,i}}{ c_{2,i}}.
\end{equation}
A group element ${\bf g} \in G_b$ lifted to $G_f$ can thus be written as ${\bf g} = (\theta/2,r)$, where $\theta \in [0,2\pi)$. Note that the symmetry $\Oo(2)^f$ is present in any fermionic system with two identical layers, including any spinful system where the two spin flavors are decoupled and have identical Hamiltonians. However, an arbitrary 2-layer system may have additional symmetries. 

We now specialize to the case where each layer is a spinless $p+ip$ SC at weak pairing, so that the full symmetry is indeed $\Oo(2)^f$, and $c_- = 1$ for the whole system. We claim that $\det R({\bf g}) = (-1)^{n_1({\bf g})}$. That is, if ${\bf g}$ contains a reflection, so that $\det R({\bf g}) = -1$, then a ${\bf g}$-defect hosts an unpaired MZM. In particular, consider a defect of the group element ${\bf h}$ with $R({\bf h}) = \begin{pmatrix}0 & 1 \\ 1 & 0\end{pmatrix}$, which interchanges the fermions. An ${\bf h}$ defect can be created by deforming the Hamiltonian along a cut such that an operator $c_{1,i}$ is permuted into $c_{2,i}$ upon crossing the cut, and vice versa. But each endpoint of such a cut is precisely a half-quantum vortex. As shown for example in \cite{ivanov2001hqv,dassarma2006hqv}, by making a suitable basis transformation on the fermion flavors, we can map the above system with the half-quantum vortex onto a new double layer system in which the first layer feels a $\pi$ flux (which hosts an unpaired MZM) while the second layer feels zero flux. Any other reflection can be related to ${\bf h}$ by a rotation, and the corresponding defect will also host an unpaired MZM.  

This physics can be completely explained by our theory. When $G_f = \Oo(2)^f$, $[\om]$ is nontrivial, and corresponds to the fermion having isospin-1/2 under the $S\Oo(2)$ subgroup of $G_b$. From Section \ref{sec:charge}, we see that any invertible phase with $G_f = \Oo(2)^f$ and $c_- = 1$ must have $n_1 \ne 0$; in particular, $n_1({\bf g}) = 1$ whenever ${\bf g}$ contains a reflection. The same analysis holds if $\Oo(2)^f$ is replaced by its finite subgroup $\DD_{8n}^f$, for integer $n$.

\subsection{Unpaired MZMs at lattice dislocations}\label{sec:mzmdisloc}

In this section we review different proposals to realize unpaired MZMs at lattice dislocations, which are translation symmetry defects.

\subsubsection{\texorpdfstring{Layered $p+ip$ SCs}{ Layered p+ip SCs}}

Ref. \cite{Asahi2012mzm} discussed a method to realize unpaired MZMs at dislocations in a layered $p+ip$ superconductor. We reproduce their prediction and then comment on an additional invariant in our theory.

The Hamiltonian is $H = \sum_{\bf k}\Psi^{\dagger}_{\bf k}H({\bf k}) \Psi_{\bf k}$, where $\Psi^{\dagger}_{\bf k} = (c^{\dagger}_{\bf k},c_{-\bf k})$, and
\begin{align}
    &H({\bf k}) \nonumber \\ &= \begin{pmatrix} 2 t_x \cos k_x + 2 t_y \cos k_y - \mu& d_x \sin k_x - i d_y \sin k_y \\
    d_x \sin k_x + i d_y \sin k_y& \mu - 2t_x \cos k_x - 2t_y \cos k_y\end{pmatrix}.
\end{align}
The full phase diagram as a function of $\mu, t_x, t_y$ is characterized in Ref.~\cite{Asahi2012mzm} by two $\Z_2$ invariants $\nu_x, \nu_y$, and a $\Z$ invariant $\nu$, which describes the Chern number of the above BdG Hamiltonian. $\pi\nu_x$ is the Berry phase of the ground state wavefunction between the time-reversal invariant momenta $(\pi,0)$ and $(\pi,\pi)$. $\nu_x$ is defined mod 2 because the Berry phase is defined mod $2\pi$. $\nu_y$ is the Berry phase of the wavefunction between $(0,\pi)$ and $(\pi,\pi)$.
In particular, $\nu_x =1$ when $|t_x + \frac{\mu}{2}| < |t_y|$, and $\nu_y =1$ when $|t_y + \frac{\mu}{2}| < |t_x|$. 

Ref.~\cite{Asahi2012mzm} claims that if the only invariant is, say, $\nu_y=1$, the system can be decomposed into a set of weakly coupled Kitaev chains along $x$, stacked in the $y$ direction. If we assume this picture, such a system should also have an unpaired MZM at dislocations with Burgers vector $\hat{y}$. This is because we essentially terminate one of the Kitaev chains at the dislocation, meaning that it will host an unpaired MZM. 

Indeed, Ref.~\cite{Asahi2012mzm} predicted that a dislocation on a square lattice with Burgers vector $\vec{b}$ hosts an unpaired MZM when 
\begin{equation}
    \vec{b} \cdot (\nu_x \hat{x} + \nu_y \hat{y}) = 1 \mod 2
\end{equation}
where $\nu_x,\nu_y$ are as defined above. Thus $\nu_x b_x + \nu_y b_y \mod 2$ is the parity of MZMs bound to the dislocation. This agrees with the above argument, where we took $\nu_y = b_y = 1$.

Let us now understand these results using our formalism. Mathematically, the simplest group which contains translation symmetries is $G_f = \Z_2^f \times \Z^2$, which also describes the above model. Since $[\om] = 0$ we can choose $c_-$ to be an integer or a half integer. Whether a $G_b$ defect can host unpaired MZMs or not is determined by the parameter $n_1$. The choices of $n_1$ are classified by $\H^1(\Z^2,\Z_2) \cong \Z_2^2$. This tells us that there are 4 possible situations in which unpaired MZMs can be realised at dislocations along the $x$ and/or $y$ directions independently. Thus, if we only consider the invariants $c_-$ and $n_1({\bf r}) = \nu_x r_x + \nu_y r_y$, we recover the invariants stated in Ref. \cite{Asahi2012mzm}.

On the other hand, the classification of Ref.~\cite{Asahi2012mzm} ignores the freedom in choosing $[n_2] \in \H^2(\Z^2,\Z_2) \cong \Z_2$. In this example, states with trivial and nontrivial $[n_2]$ class differ by a fermion per unit cell. In our classification, states with a fermion per unit cell are topologically non-trivial, as they give rise to nontrivial $G_b$ quantum numbers for the fermion parity fluxes. In particular, translations in the $x$ and $y$ directions anticommute on the fermion parity fluxes, although they commute on fermions. Thus in our theory the final classification is $\Z\times \Z_2^3$.\footnote{In Ref.~\cite{zhang2020realspace}, a real space classification theory was used to obtain the classification of fermion SPT phases with $G_f = \Z^2 \times \Z_2^f$ and $c_-=0$ to be $\Z_2 \times \Z_4$ as opposed to our result $\Z_2^3$. The former result is inconsistent with the stacking rules for invertible phases proposed in \cite{barkeshli2021invertible}.}  

Ref.~\cite{rao2021theory} finds the same classification ($\ZZ\times\ZZ_2^3)$ for free fermion BdG Hamiltonians with only translation symmetry and argues that the classification is robust against interactions and disorder. Our result shows that this classification is sharp: every invertible fermionic phase protected by $G_f=\ZZ^2\times \ZZ_2^f$ has a free fermion representative. 

\subsubsection{MZMs at dislocations in a model for $\text{Sr}_2\text{RuO}_4$}

This example is taken from Refs.~\cite{raghu2010tsc-sr,hughes2014mzm}. Ref.~\cite{raghu2010tsc-sr} proposed a pairing mechanism for $\text{Sr}_2\text{RuO}_4$ which involves $d$ orbitals. It
argued that this mechanism results in a higher-order TSC that hosts unpaired MZMs at dislocations, and that this is possible even for $c_-=0$.

Ref.~\cite{hughes2014mzm} argued that in a suitable parameter regime, the band structure of the three-dimensional material $\text{Sr}_2\text{RuO}_4$ breaks up into a quasi-2d band (meaning that the leading contribution to its dispersion is proprtional to $(\cos(k_x) + \cos(k_y))$) and two quasi-1d bands (i.e. their dispersions have leading contributions proportional to $\cos(k_x)$ and $\cos(k_y)$ respectively). The quasi-1d bands are modelled by the `RKK' Hamiltonian proposed in Ref.~\cite{raghu2010tsc-sr}: 

\begin{align}
    H({\bf k}) &= \begin{pmatrix}
        \epsilon_{xz}({\bf k}) & \Gamma({\bf k}) \\ \Gamma({\bf k}) & \epsilon_{yz}({\bf k})
    \end{pmatrix} \otimes \mathds{1}_{\text{spin}}
\end{align}
with the pairing terms for the two orbitals given by
\begin{align}
    \Delta_j &= i {\bf d}_j({\bf k}) \cdot \vec{\sigma} \sigma^y, \quad j = 1,2 \nonumber \\
    {\bf d}_1 &= \hat{z} \Delta_0 \sin(k_x) \cos(k_y) \nonumber \\
    {\bf d}_2 &= i\hat{z} \Delta_0 \sin(k_y) \cos(k_x). 
\end{align}
The claim is that the entire model can be expressed as a stack of decoupled (2+1)D systems along $\hat{z}$. The quasi-2d bands form a topologically nontrivial $p+ip$ superconductor in the $x-y$ plane. And the pairing for the quasi-1d bands realizes a system that hosts unpaired MZMs at dislocations in the $\hat{x}$ and $\hat{y}$ directions.

Let us discuss the conclusion of Ref.~\cite{hughes2014mzm} using our theory. The full symmetry group of the (3+1)D system is $G_f = \Z^3 \times \Z_2^f$ but we can understand it using purely (2+1)D physics. The system restricted to the $x-y$ plane has the symmetry $G_f = \Z^2 \times \Z_2^f$. The quasi-2D bands give a system with $c_-=1/2$, while the quasi-1D bands give a system with $n_1({\bf x}) = n_1({\bf y}) = 1$, where ${\bf x},{\bf y}$ are the elementary translations in $\Z^2$. Thus the nontrivial phase realized by the RKK model corresponds to the data $(c_-=1/2, n_1({\bf x}) = n_1({\bf y}) = 1)$. Our theory is consistent with the results of Ref.~\cite{hughes2014mzm}, if there is only translational symmetry.

\subsection{Unpaired MZMs at lattice disclinations and corners}\label{sec:mzmdisc}

\subsubsection{Inversion symmetry and monolayer WTe$_2$}\label{sec:C2}

Tungsten ditelluride, $\text{WTe}_2$, is a 3d material with nonsymmorphic symmetries corresponding to the space group Pnm21 (number 31). Monolayer $\text{WTe}_2$, on the other hand, can undergo a lattice distortion into a phase denoted as 1T', which has a rectangular lattice with only translation and twofold rotation symmetries. The presence of spin-orbit coupling implies that $\SU(2)^f$ spin rotation symmetry is broken. This phase is known to support quantum spin Hall insulators in the normal state of $\text{WTe}_2$ \cite{Qian2014QSH}. 

Ref. \cite{Hsu2020htsc} proposed that monolayer $\text{WTe}_2$ stabilizes a higher-order TSC phase in which MZMs are found at opposite corners of the inversion symmetric system (see Refs.~\cite{geier2018htsc,khalaf2018hoti} for other discussions of this phase, and Ref.~\cite{Benalcazar2014} for a related classification of TSCs with unpaired MZMs at lattice disclinations). 

To model this system, we take $G_b = \Z_2$ (ignoring the translation symmetry, which does not contribute anything nontrivial in this example) and assume that a $2\pi$ rotation acts as $(-1)^F$. This implies that the spatial symmetry is $G_f = \Z_4^f$.  The fCEP maps this system onto one with an effective internal symmetry $G_f^{\Int} = \Z_2 \times \Z_2^f$. From Table \ref{tab:cryst}, we see that our theory also has a solution with $n_1 \ne 0$, corresponding to MZMs at $\ZZ_2$ symmetry defects. 

Our usual understanding is that an inversion symmetry defect is a disclination of angle $\pi$, constructed using a cut-and-glue procedure; this was not studied in Ref.~\cite{Hsu2020htsc}. In their model, the boundary MZMs are localized at a pair of opposite corners, where the $p$-wave pairing vanishes. Since we cannot prove that having an unpaired MZM at a $\pi$ disclination implies having unpaired MZMs at opposite corners, our theory cannot directly verify this prediction. We note that the choice of corners which host unpaired MZMs is fixed by the nodes of the pairing term; but in a disclination, this detail does not matter. In the higher-order TSC phase our theory predicts that there will always be an unpaired MZM at the disclination.

The full classification of invertible phases with $G_f = \Z_2 \times \Z_2^f$ is $\Z_8 \times \frac{1}{2} \Z$ where the second factor is due to $c_-$. Ref. \cite{Hsu2020htsc} presented a $\ZZ_4$ valued band invariant (`symmetry indicator') $\kappa$ for higher order TSCs with inversion symmetry (this invariant is discussed in more detail in Ref. \cite{Skurativska2020}). It is interesting that $\kappa = 1$ corresponds to the spinless $p+ip$ TSC with MZMs at fermion parity vortices and where inversion symmetry does not play any role, while $\kappa = 2$ corresponds to the inversion symmetric higher-order TSC with odd $c_-$, and with MZMs at corners. 
Note that if $\kappa$ is really treated as a topological invariant, it would not respect stacking: two identical $p+ip$ states with $\kappa = 1$ do not give a state with $\kappa = 2$ upon stacking. We suggest that $\kappa$ is really a $\Z_2 \times \Z_2$ invariant in disguise. If we define $k=0,1$ when $n_1 = 0$ and $n_1 \ne 0$ respectively, then $\kappa = [2c_-]_2 + 2k \mod 4$, where $2c_- \mod 2, k \mod 2$ are the two $\Z_2$ invariants.

\subsubsection{Example with $G_b = C_4 \times \Z_2^{\Tt}$}\label{sec:C4+TRS}
Ref. \cite{wang2018maj} discussed a different proposal for `Majorana corner states' (see also Refs. \cite{yan2018maj,liu2018maj}). Ref. \cite{wang2018maj} argues that a second-order time-reversal invariant TSC can be realized by proximitizing a $C_4$ symmetric
2d TI with an $s$-wave superconductor. It claims that in this phase, there is a Majorana Kramers pair at each corner. We verify this and argue that the system appears to be described by a nontrivial $n_2$. 

The superconducting system has a $C_4$ symmetry whose generator satisfies ${\bf h}^4 = (-1)^F$. There is also a time reversal symmetry with $\Tt^2=(-1)^F$ and ${\bf h T} = {\bf Th}$. To model this we consider $G_b = \Z_4\times\Z_2^{\bf T}$, and define $\om^{\sp} = w_2 + s_1^2$ as described in the model. By the fCEP, we should consider an equivalent internal symmetry with $\om^{\Int} = \om^{\sp} + w_2 = s_1^2$. This implies that $G_f^{\Int} = \Z_4 \times \Z_4^{{\bf T}f}$ (we ignore translations throughout, as they do not affect the discussion). It turns out that there is a solution with with $n_1=0$ and $n_2 = w_1 s_1$, where $w_1 \in \H^1(\Z_4,\Z_2)$. 

In this example, it is easiest to interpret $n_2$ by looking at the remaining parameter $\nu_3$, which is derived from $n_2$ using Eq.~\eqref{eq:O4def} and determines the $G_b$ quantum numbers of symmetry defects. In particular, with $n_2$ as above, we obtain a parameter $\nu_3$ whose functional form indicates that a $\Z_4$ defect has local ${\bf T}^2$ eigenvalue $\pm i$\footnote{Specifically, $\nu_3 = \pm\frac{1}{4} w_1 s_1^2 \mod 1$. Moreover, there is a bosonic SPT with $G_b$ symmetry corresponding to $2\nu_3$. From the Kunneth formula, we can check that this SPT generates the subgroup $\H^2(\Z_2^{\bf T},\H^1(\Z_4,U(1)))$. The interpretation here is that a $\Z_4$ defect has ${\bf T}^2$ eigenvalue -1. To arrive at the above interpretation for $\nu_3$ we simply take the square root of this eigenvalue.}.

In Sec.~\ref{sec:int_AU} we noted that ${\bf T}^2 = \pm i$ indicates the presence of a Majorana Kramers pair, or a `fermionic Kramers degeneracy'. In that context this degeneracy was associated to a fermion parity flux and described by $n_1 = s_1$. Here, since we found above that $n_2 = w_1 s_1$ implies that a $\Z_4$ defect carries a local ${\bf T}^2 = \pm i$ eigenvalue, we conclude that in this model a $C_4$ defect (i.e. a disclination or corner with angle $\pi/2$) carries a Majorana Kramers pair. This precisely reproduces the conclusion of Ref.~\cite{wang2018htsc}, and further predicts that a $\pi/2$ disclination will also host a Kramers pair of MZMs.

Note that there is also a solution for this symmetry with $n_1 = s_1$, but it does not involve the $C_4$ symmetry, and so does not describe the above example.

\subsection{MZMs in reflection symmetric systems}\label{sec:mzmreflec}
\subsubsection{TSCs with mirror symmetry}\label{sec:MxMy}
\begin{figure}
    \centering
    \includegraphics[width=0.25\textwidth]{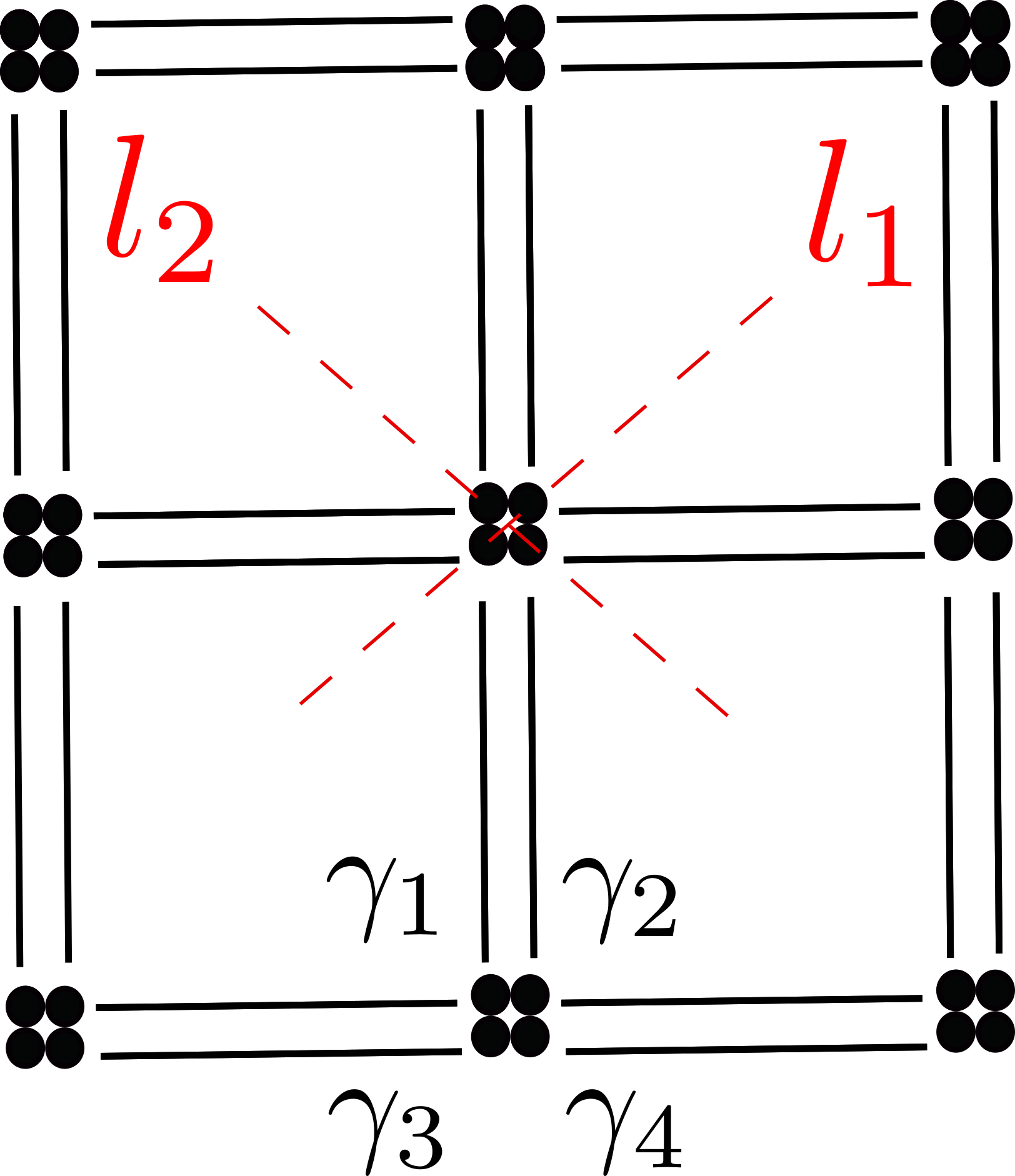}
    \caption{Schematic for a TSC with mirror symmetry. The symmetries of the model given in Ref.~\cite{wang2018htsc} are mirror reflections along the usual $x$ and $y$ directions. However, we find unpaired MZMs at the corners of the sample only if we consider a different set of reflections along the diagonals $l_1,l_2$.}
    \label{fig:mxmy}
\end{figure}

This example is taken from Ref. \cite{wang2018htsc} which takes a model for a Dirac semimetal and gaps out the Dirac nodes along with a $p+ip$ superconducting pairing term. We show that although our theory supports their conclusion about unpaired MZMs at the corners of a reflection-symmetric system, the symmetries originally identified in Ref. \cite{wang2018htsc} are not sufficient to establish this claim. This example also gives us intuition for the meaning of $n_1 \ne 0$ for reflection symmetries.

The system is modelled by four Majorana fermions $\gamma_{i,\bf r}, 1 \le i \le 4$ per site (see Fig.~\ref{fig:mxmy}), with the Hamiltonian
\begin{align}
    H &= -2it \sum\limits_{\bf r}  (\gamma_{2,\bf r} \gamma_{{1,\bf r}+\hat{x}} + \gamma_{4,\bf r} \gamma_{3,{\bf r}+\hat{x}} \nonumber \\
    & \quad - \gamma_{2,\bf r} \gamma_{4,{\bf r}+\hat{y}} + \gamma_{1,\bf r} \gamma_{3,{\bf r}+\hat{y}}) + h.c.
\end{align}
Ref.~\cite{wang2018htsc} states that $G_b$ is generated by two mirror symmetries $\mathcal{M}_x, \mathcal{M}_y$ which satisfy
\begin{equation}\label{eq:mirr1}
    \mathcal{M}_x^2 = \mathcal{M}_y^2 = (\mathcal{M}_x\mathcal{M}_y)^2 = (-1)^F
\end{equation}
(see Appendix \ref{app:symmetryAnalysisC4T} for a derivation.) Upon adding a $p$-wave superconducting term, the paper observes that the system hosts MZMs at its corners.

While $\mathcal{M}_x,\mathcal{M}_y$ are indeed symmetries of the model, we believe that they are not sufficient to explain the presence of unpaired MZMs at corners. Eq.~\eqref{eq:mirr1} implies that the spatial symmetry is the quaternion group $\mathbb{Q}_8$. Using the fCEP, the equivalent internal symmetry is $\Z_2 \times \Z_2^{\bf T} \times \Z_2^f$, i.e. $\om^{\Int} = 0$ and $s_1\neq 0$. But in this case there is no solution with $n_1 \ne 0$. Therefore our theory predicts that if these were the full symmetries, unpaired MZMs would not be allowed at the corresponding symmetry defects. Another issue is that in this example, the corners do not correspond to defects of $\mathcal{M}_x$ or $\mathcal{M}_y$.

One possible resolution is the following. The system also has two reflection symmetries $\mathcal{M}_1, \mathcal{M}_2$ about the lines $l_1, l_2$ shown in Fig.~\ref{fig:mxmy}. We find that $\mathcal{M}_1$ takes $(x,y) \rightarrow (y,x)$ and $\vec{\gamma} = (\gamma_1,\gamma_2,\gamma_3,\gamma_4) \rightarrow \vec{\gamma}' = (-\gamma_4,-\gamma_2,\gamma_3,-\gamma_1)$ (here we have suppressd the position index). In the Majorana basis, an operator satisfying $\mathcal{M}_1 \vec{\gamma}^T \mathcal{M}_1^{\dagger} = \vec{\gamma}'^T$ is
\begin{equation}
    \mathcal{M}_1 = \frac{1+\gamma_1 \gamma_4}{\sqrt{2}}\gamma_3.
\end{equation}
From this we see that $\mathcal{M}_1^2$ leaves all positions invariant and also satisfies $\mathcal{M}_1^2 = -\gamma_1 \gamma_4$ within each site. This implies that $\mathcal{M}_1^4 = -\mathds{1}_{4\times 4}$ acts trivially on the Majorana fermions, as $\mathcal{M}_1^4 \gamma_i (\mathcal{M}_1^{\dagger})^4 = \gamma_i$. Thus $\mathcal{M}_1$ generates a orientation reversing $\Z_4$ subgroup of $G_f$. Similarly, we can show that 
\begin{equation}
    \mathcal{M}_2 = \frac{1+\gamma_2 \gamma_3}{\sqrt{2}}\gamma_{1}
\end{equation}
 generates an orientation reversing $\Z_4$ subgroup of $G_f$. 
The full symmetry $G_f$ includes these reflections and the lattice translations, as well as an onsite symmetry that permutes the Majorana fermions. 
Note that the reflections mentioned in Ref.~\cite{wang2018htsc} can be expressed in terms of $\mathcal{M}_1,\mathcal{M}_2$ and the internal symmetry operations. 

Suppose we consider just the reflection about $l_1$ (or $l_2$). The spatial symmetry is $G_f = \Z_4^{\bf R} \times \Z_2^f$. Now we can use the fCEP. The effective internal symmetry is $G_f^{\Int} = \Z_4^{\Tt} \times \Z_2^f$. From Table \ref{tab:summaryAntiunitary}, there is a solution with $n_1=s_1$. 

Now if $n_1({\bf g})=1$ where ${\bf g}$ is originally a reflection, we expect the system to carry a Kitaev chain along the reflection axis, with unpaired MZMs at its ends. This would be a generator of (1+1)D invertible phases with $\Z_4 \times \Z_2^f$ onsite symmetry. In this model, the corners of the system lie along either $l_1$ or $l_2$, and it is clear that they carry unpaired MZMs. Therefore our theory now supports the conclusion of Ref.~\cite{wang2018htsc}.

\subsubsection{MZMs in a mirror symmetric model for ${\mathrm{Sr}}_{2}{\mathrm{RuO}}_{4}$}
Ref.~\cite{ueno2013sr} studied a model for ${\mathrm{Sr}}_{2}{\mathrm{RuO}}_{4}$ with a mirror symmetry $\mathcal{M}$ that takes $(x,y,z) \rightarrow (x,y,-z)$. Let $H = \sum_{{\bf k},\alpha} \Psi^{\dagger}_{{\bf k} \alpha} \mathscr{H}({\bf k}) \Psi_{{\bf k}\alpha}$, where $\Psi_{{\bf k}\alpha} = (c_{{\bf k}\alpha}, c^{\dagger}_{{\bf k}\alpha})^T$ and $\alpha$ indexes spin and orbital degrees of freedom. 
If we fix $z=0$ in this model, $\mathcal{M}$ takes $c_{(k_x,k_y)\alpha} \rightarrow \pm c_{(k_x,k_y)\alpha}$ where spin up and spin down fermions have eigenvalues $+1,-1$ respectively.\footnote{Ref.~\cite{ueno2013sr} defines a reflection operator $\tilde{M}$ under which fermions in the $z=0$ layer have eigenvalue $\pm i$, which seems to imply that the symmetry is $\Z_4^{{\bf R}f}$. But $\tilde{M}$ takes $c \rightarrow \pm i c$ and $c^{\dagger} \rightarrow \pm i c^{\dagger}$. This transformation is not consistent, because if $c \rightarrow e^{i \theta} c$, we must necessarily have $c^{\dagger} \rightarrow e^{-i \theta} c^{\dagger}$. 
However, by defining $\mathcal{M} = i\tilde{M}$, we do find a valid symmetry operator under which $c \rightarrow \pm c$.} 
In particular $\mathcal{M}^2$ acts trivially on fermions. So if we consider only the reflection symmetry, $G_f = \Z_2^{\bf R}\times \Z_2^f$.

The paper argues that there is a superconducting phase in which the system decouples into 2d layers. And that the $z=0$ layer can be viewed as a stack of a spinless $p+ip$ SC and a spinless $p-ip$ SC formed by fermions which are even and odd under $\mathcal{M}$ respectively. 
We can explain these results by modelling the mirror symmetry as an internal, unitary $G_f^{\Int} = \Z_2 \times \Z_2^f$ symmetry acting within the $z=0$ layer. Now the TSC state proposed in Ref.~\cite{ueno2013sr} is the same as the generator of invertible phases with internal symmetry $\Z_2\times\Z_2^f$, as proposed by Gu and Levin \cite{Gu2014interaction}. 
In our theory, this state is given by the data $c_-=0$ and $n_1 \ne 0$. 

Ref.~\cite{ueno2013sr} did not give a proposal to isolate an unpaired MZM at a symmetry defect in this model. It may be possible to implement a reflection symmetry defect in the (3+1)D system as a cross-cap; its action on the $z= 0$ layer would then act as an on-site $\Z_2$ symmetry defect whose endpoints would host unpaired MZMs. Such a defect would not arise in a real material, although perhaps it could be engineered in a synthetic quantum many-body system. 
 
\subsubsection{\texorpdfstring{$p+id$ superconductor}{p+id sc}}\label{sec:majkram}

This proposal is from Ref. \cite{wang2018htsc}. The system under study is a spinful $p+id$ superconductor and the symmetry reported is an antiunitary $C_4$ symmetry that we write as $G_b = \Z_4^{\Tt}$ (in the paper it is written as $C_4\mathcal{T}$). The fermionic symmetry group is $G_f = \Z_8^{{\bf T}f}$. We have $s_1 \ne 0$. The model predicts unpaired MZMs at corners; our theory cannot directly confirm this from the given symmetry, but it can do so if we consider an additional reflection symmetry that is also present in the Hamiltonian that was not commented upon in \cite{wang2018htsc}.

The model is of spinful fermions in which the Hamiltonian has a kinetic term, a standard $p$-wave pairing term, and an additional $d$-wave pairing term:
\begin{widetext}
\begin{align}
    H({\bf k}) = \int d{\bf k} \left[ c^{\dagger}_{\bf k} \left(\frac{{\bf k}^2}{2m} - \mu\right) c_{\bf k} + (\Delta_p c^T_{\bf k} ({\bf k} \cdot \vec{\sigma})  i \sigma^y c_{-{\bf k}} + i \Delta_d c^T_{\bf k} (k_x^2 - k_y^2)  i \sigma^y c_{-{\bf k}}+ \text{h.c.})\right) ]
\end{align}
\end{widetext}
The $d$-wave pairing term separately breaks rotation and time-reversal symmetries, but is symmetric under $\Z_8^{{\bf T}f}$.  In the absence of this term, the system is a time-reversal invariant TSC in Class DIII, with gapless Majorana edge states. Ref. \cite{wang2018htsc} shows that the $d$-wave pairing term changes sign under a rotation. It gaps out the boundaries, but in such a way that the mass terms have opposite signs on adjacent boundaries that meet at a corner. Since each corner is a node of the mass term, it must host an unpaired MZM. 

Now we describe the same system using our theory. 
We ignore the translations as in the previous example. By the fCEP, we should apply our theory with an effective internal symmetry under which the fermion has trivial $G_b$ quantum numbers, i.e. $G_f^{\Int} = \Z_4^{{\bf T}} \times \Z_2^f$. For this symmetry, Table \ref{tab:summaryAntiunitary} shows that there is a solution with $n_1 = s_1$. But we do not know how to properly define defects of $\Z_4^{{\bf T}}$, and so it is not clear that $n_1 = s_1$ implies unpaired MZMs at the corners.

\def\ua{\uparrow}
\def\da{\downarrow}
There is however a resolution involving reflection symmetries. Note that the transformation ${\bf R}: c_{(k_x,k_y)} \rightarrow i \sigma^x c_{(k_x,-k_y)}$ is a unitary reflection. Combined with the generators of $\Z_4^{\bf T}$, which take $c_{(k_x,k_y)} \rightarrow i \sigma^y c_{(k_y,-k_x)}$ (along with a complex conjugation operation), we get a pair of antiunitary reflections ${\bf R}_1, {\bf R}_2$ that take $c_{(k_x,k_y)} \rightarrow ic_{(\pm k_y, k_x)}$ and leave opposite corners invariant. In particular, ${\bf R}_1^2 = {\bf R}_2^2 = 1$. 

Now we can consider each reflection separately with $G_f = \Z_2^{{\bf R}_{1(2)}}\times \Z_2^f$. Using the fCEP, we find that $G_f^{\Int} =\Z_2 \times \Z_2^f$, which does admit a solution with nontrivial $n_1$\footnote{Here $\om^{\sp}=0$ and $\om^{\Int} = \ww_1^2 + \ww_1 s_1 = 0$ where $\ww_1 = s_1$. This uses the most general formulation of the fCEP with both reflection and time-reversal symmetries. Therefore this example can be thought of as another nontrivial check on the fCEP formula, Eq.~\eqref{eq:fCEP}.}. This solution means that each reflection axis hosts a 1d TSC with unpaired MZMs at its ends, explaining the observation in Ref.~\cite{wang2018htsc}. 

The argument is as follows. Since ${\bf R}_{1(2)}^2 = 1$, if we restrict to either invariant reflection invariant line, the symmetry acts as time-reversal with ${\bf T}^2=+1$. Therefore, the effective 1d system is in Class BDI, for which there is indeed a TSC with unpaired MZMs at the ends \cite{fidkowski2011}.  
Applying this argument to both reflection axes leads to a solution with unpaired MZMs at each corner of the original model.

\subsection{\texorpdfstring{Magnetic translation symmetry}{Charge-q SCs}}\label{sec:magtrans}

Ref.~\cite{shaffer2021hofsc} studies `Hofstadter superconductors', which are invertible phases that result from adding $p$-wave pairing terms to the Hofstadter model. It studies symmetry conditions under which a $c_-=1/2$ state can be realized. We express their results using our theory, and furthermore study the possiblility of unpaired MZMs at dislocations in this model. We then comment on the properties of charge-$2Q$ Hofstadter SCs for general $Q$.

In the regular Hofstadter model the symmetry group $G_f$ consists of magnetic translations with flux $\phi = 2\pi\frac{p}{q}$ per unit cell, with $p,q$ coprime integers. For concreteness, we write the pure Hofstadter Hamiltonian in Landau gauge as
\begin{equation}
    H_0 = -t \sum\limits_{\bf r} c^{\dagger}_{\bf r} c_{{\bf r} + \hat{x}} +e^{-2\pi i \frac{p}{q} r_x} c^{\dagger}_{\bf r} c_{{\bf r} + \hat{y}} + h.c.
\end{equation}
$H_0$ has a $U(1)^f$ symmetry defined by the operators $U_{\theta}, \theta \in [0,2\pi)$, which satisfy $U_{\theta} c_{\bf r} U_{\theta}^{\dagger} := e^{i \theta}c_{\bf r}$ for each ${\bf r}$. The generators $T_{\bf x}, T_{\bf y}$ of the magnetic translations have the following action on fermionic operators:
\begin{align}
    T_{\bf x} c_{\bf r} T_{\bf x}^{\dagger} &:= e^{-2\pi i \frac{p}{q} r_y} c_{{\bf r} + \hat{x}} \\
    T_{\bf y} c_{\bf r} T_{\bf y}^{\dagger} &:= c_{{\bf r} + \hat{y}}.
\end{align}
This results in the following commutation relation on single fermionic operators:
\begin{equation}
    T_{\bf x} T_{\bf y} = e^{2\pi i \frac{p}{q}} T_{\bf y} T_{\bf x},
\end{equation}
Importantly, when $q$ is even, we have the relation
\begin{equation}\label{eq:HofSC}
    T_{\bf x} T_{\bf y}^{q/2} = (-1)^F T_{\bf y}^{q/2} T_{\bf x}.
\end{equation}
Thus the commutator of $T_{\bf x}$ and $T_{\bf y}^{q/2}$ acts trivially on bosonic operators but by a minus sign on fermionic operators. This relation is not true when $q$ is odd. 

Suppose we now add a pairing term of the form
\begin{equation}
    H_{\Delta} := \Delta \sum\limits_{\bf r} (c_{\bf r}c_{{\bf r} + \hat{x}} + c_{\bf r}c_{{\bf r} + \hat{y}}) + h.c.
\end{equation}
Adding the pairing term breaks the $U(1)^f$ symmetry down to $\Z_2^f$ fermion parity. The magnetic translation symmetry is also broken: in the above model, $T_{\bf y}$ remains a symmetry, while $T_{\bf x}$ is broken. Now for odd $q$, $T_{\bf x}^q$ is the smallest translation along $\hat{x}$ that is still a symmetry of the paired Hamiltonian. Note that $T_{\bf x}^q$ commutes with $T_{\bf y}$. The new symmetry group is $G_f = \Z^2 \times \Z_2^f$, and has $[\om] = 0$. 

However, for even $q$, the smallest translation along $\hat{x}$ is $T_{\bf x}^{q/2}$, which \textit{anticommutes} with $T_{\bf y}$. Thus the new unit cell in this case encloses $\pi$ flux rather than zero flux. This symmetry group has $[\om] \ne 0$.

From this follows several consequences that were also derived in Ref.~\cite{shaffer2021hofsc}, using the representation theory of the pairing terms in momentum space. Paired states with even $q$ must have $c_- \in \Z$ in order to support $\pi$ flux per unit cell, while states with odd $q$ can have half-integer $c_-$. Thus the minimal nonzero value of $c_-$ is $q/2 \mod 1$. This implies that fermion parity vortices can only host unpaired MZMs if $q$ is odd.

Using our theory we can additionally discuss the possibility of unpaired MZMs at dislocation defects: this was not commented upon in Ref.~\cite{shaffer2021hofsc}. Let us restrict to the case with integer $c_-$. For $G_b = \Z^2$, all obstructions at the level of $n_2$ or $\nu_3$ vanish. Thus we can choose an arbitrary nonzero $n_1$, irrespective of the value of $q$. This means hat unpaired MZMs can be found at $\hat{x}$ or $\hat{y}$ dislocation defects independently, although the Hamiltonian may need to be modified to realize these phases. 

Another extension of this idea is to consider charge $2Qe$ superconductors instead of charge-$2e$ superconductors, for different even $Q$. The crucial difference here is that the $U(1)$ bosonic charge conservation symmetry is broken down to a discrete $\Z_Q$ subgroup, corresponding to the fact that $Q$ bosons can be created or annihilated simultaneously through suitable interactions. For odd $Q$, the charge conservation symmetry acting on the fermions takes the form $\Z_2^f \times \Z_Q$ (this is isomorphic to $\Z_{2Q}^f$ for odd $Q$). On the other hand, for even $Q$, the charge conservation symmetry is of the form $\Z_{2Q}^f$, with $[\om] \ne 0$. For even $Q$, $c_-$ is therefore forced to be an integer, irrespective of any other symmetries (see Table \ref{tab:summary}). 

Suppose we additionally consider a magnetic translation symmetry, and study the properties of dislocation defects in this case. For even $Q$, the $\Z_{2Q}^f$ charge conservation symmetry implies that unpaired MZMs cannot exist even at dislocation defects, if the symmetry does not include charge conjugating elements. The arguments are identical to those given in Sec.~\ref{sec:charge}: essentially, the fermion has charge $1/2$ under the bosonic $\Z_Q$ symmetry, but when $n_1 \ne 0$ we cannot consistently extend this to define a $\Z_Q$ charge for the fermion parity fluxes.

\subsection{Unpaired MZMs in iron superconductors}\label{sec:fesc}

In this example we consider a proposal involving a 2d heterostructure of monolayer $\text{FeTe}_{1-x}\text{Se}_x$ (FTS) with monolayer FeTe \cite{Zhang2019FeTSC}. This work predicts unpaired MZMs at the corners of the system (to be discussed below). We study the proposal and argue that the observations do not correspond to a well-defined 2d bulk invariant in our theory, unless additional unidentified symmetries are involved. However, the symmetries do admit possible bulk invariants that, if non-zero, would imply unpaired MZMs at dislocations. 

Monolayer FTS is a superconductor with a normal-state band structure that likely corresponds to a quantum spin Hall insulator. Monolayer FeTe has ferromagnetic (FM) and antiferromagnetic (AFM) edges along, say, the $\hat{y}$ and $\hat{x}$ directions respectively. The symmetry generator along $\hat{x}$ in the heterostructure is a half unit cell translation combined with a time reversal operation, which generates a group we denote as $\Z^{\bf T}$. The full spatial symmetry is $\Z \times \Z^{\bf T}$.

The proposal is that if the FTS layer is made superconducting, the heterostructure realizes a higher-order TSC phase with unpaired MZMs at the corners An exchange coupling with the FM edge in FeTe gaps out the edges of the heterostructure that run along $\hat{y}$. This edge is argued to be in the same phase as a 1d spinless $p$-wave superconductor, if the FM interaction is sufficiently strong. In the normal state, there is a gapless state running along each $\hat{x}$ edge, which is also gapped by superconductivity. Since the AFM edge is topologically trivial, \cite{Zhang2019FeTSC} argues that each corner lies at the junction of a spinless 1d $p$-wave SC and a trivial phase, and so there should be an unpaired MZM at each corner. 

Importantly, corner MZMs are not necessarily a bulk topological property unless protected by a rotational symmetry. This is because we can always add ancilla degrees of freedom and tune the edge Hamiltonian so as to remove the corner MZMs. 

While our theory does not predict corner MZMs, it has solutions with $n_1 \ne 0$ when $G_b = \Z\times \Z^{\bf T}$. In particular, if $n_1(\hat{y})=1$, a dislocation with Burgers vector $\hat{y}$ will bind an unpaired MZM; this is a bulk invariant that would be interesting to check in the above model. There are also solutions with $n_1(\hat{x})=1$ and $n_1(\hat{x}+\hat{y})=1$, but these translations are antiunitary so we cannot treat dislocations with Burgers vector $\hat{x}$ or $\hat{x}+\hat{y})$ as symmetry defects in this case; it would be interesting to understand what observable consequences they imply.

\section{Discussion}

We have shown that the chiral central charge $c_-$, together with the symmetry of the ground state $G_f$, imposes nonperturbative constraints on whether an invertible phase can host unpaired MZMs or not. There are two obstructions, captured by the quantities $\mathcal{O}_3$ and $\mathcal{O}_4$, and the $\mathcal{O}_4$ obstruction typically imposes stronger constraints than $\mathcal{O}_3$. We also demonstrated how to apply our results when there are crystalline symmetries, using the fermionic crystalline equivalence principle. We stated a concrete algebraic formula for the fCEP (Eq.~\eqref{eq:fCEP}) that was not presented in previous work, and illustrate various checks on its correctness in App.~\ref{app:fCEPAnti}.

The richness of these constraints can be illustrated just by looking at order 2 symmetries. We recall the following results from Tables \ref{tab:summary},\ref{tab:summaryAntiunitary},\ref{tab:cryst}. When ${\bf h}$ generates a unitary internal $G_b = \Z_2$ symmetry, unpaired MZMs are allowed if ${\bf h}^2 = 1$ and are forbidden if ${\bf h}^2 = (-1)^F$. The latter condition automatically means that a charge-4 superconductor cannot have unpaired MZMs, and must have integer $c_-$. This is a result that cannot be derived from free-fermion techniques. The analogous statement for a charge $4n$ SC is also true, if we replace $G_b = \Z_2$ by $\Z_{2n}$. 

The situation is reversed when ${\bf h}$ is a unitary $C_2$ rotation: unpaired MZMs are allowed at disclinations or system boundaries if ${\bf h}^2 = (-1)^F$, while if ${\bf h}^2 = 1$, we can only have integer $c_-$, and unpaired MZMs are not allowed. The former result is the basis for several proposals of inversion-symmetric higher-order TSCs.

If ${\bf h}$ is a reflection (unitary or antiunitary), and ${\bf h}^2=1$, unpaired MZMs can be realized by placing a Kitaev chain on the reflection axis. This idea has also been utilized in several higher-order TSC proposals. Unpaired MZMs cannot be realized if ${\bf h}^2 = (-1)^F$.

Finally, when ${\bf h}={\bf T}$ is an antiunitary internal symmetry, and ${\bf T}^2 = (-1)^F$, we can have a phase with Majorana Kramers pairs bound to fermion parity flux (the Class DIII TSC). It can be mathematically characterized in terms of a local ${\bf T}^2$ eigenvalue $\pm i$ associated to a fermion parity flux. 

Our theory can mathematically describe further variations on these symmetries, such as antiunitary rotations and reflections, although the physical interpretation of the data is not always clear in this case, as we mention below.

If the system has translation symmetry, unpaired MZMs are allowed either at fermion parity fluxes (when $c_-$ is a half-integer) or at lattice dislocations (for any value of $c_-$). This is also true if the system has $\pi$ flux per unit cell. An open problem is to extend these results to understand all the constraints on $c_-$ and on unpaired MZMs for systems with 2d wallpaper group symmetries.

In obtaining constraints on several compact Lie groups, we derived results about the group cohomology of their $\ZZ_2$ quotient, which to our knowledge includes some new mathematical results. For general Lie groups we can derive many of the constraints by considering the right subgroups. For example, the constraints for $G_f=\U(n)^f$ can be easily obtained by studying its $\U(1)^f$ and $\SU(n)^f$ subgroups. 

The solution for $G_f=\Oo(n)^f$ is particularly useful because it can be used to find solutions to $\mathcal{O}_3, \mathcal{O}_4$ for arbitrary groups $G_f'$ as follows. First, for any group $G_f'$, we consider real representations defined by a map $\rho:G_f' \to \Oo(n_{\rho})$ that satisfies $\rho((-1)^{F}) = - 1$. Next, define a map on the quotient $\tilde{\rho}: G_f'/\ZZ_2^f \to \PO(n_{\rho}) $ by $\tilde{\rho}(\gbf)=\rho(\tilde{\gbf})$. Here $\tilde{\gbf}$ is a lift of $\gbf$ to $G_f'$.\footnote{Note that the map is well defined because the other lift differ by $(-1)^F$ so that the image would be $\rho(\tilde{\gbf}(-1)^F)=\rho(\tilde{\gbf})\rho((-1)^F)$ but $\rho((-1)^F)=-1$ is equivalent to the identity in $\PO(n_{\rho})$.} We then construct solutions of Eqs.~\ref{eq:iFTPEqs} in $G_b' =G_f'/\ZZ_2^f$ by pull-back of the solutions in $\PO(n_{\rho})$ using $\tilde{\rho}$. \footnote{This may be viewed as a free to interacting map, similar to \cite{Chen2019freeinteracting}, in the more general case of $G_f\neq \ZZ_2^{f}\times G_b$. Another way to view this map is as a generalization of the spin cobordism version of the Chern character defined in Ref.~\cite{grigoletto2021spin}.} 

There is an important unanswered question in this work. Unpaired MZMs at ${\bf g}$ defects for systems with unitary symmetries are described by a parameter $n_1({\bf g})=1$. For antiunitary symmetries with ${\bf T}^2 = (-1)^F$, $n_1({\bf T})=1$ indicates a Majorana Kramers pair bound to a fermion parity flux. However, we do not systematically understand the meaning of $n_1({\bf g})=1$ for general antiunitary symmetries, because we do not have a well-defined construction for antiunitary defects. For example, if $G_f = \Z_4^{\bf T} \times \Z_2^f$ (internal) or $G_f = C_{8}^{{\bf T}f}$ (antiunitary $\pi/2$ rotation), the theory has a solution with $n_1 \ne 0$ for the generator of $G_b$, but we do not understand what this means physically. Fully understanding the mathematical data for antiunitary symmetries is still an open question.  

There are also several broader theoretical questions. We gave an example of an intrinsically interacting fermionic phase in Sec.~\ref{sec:tHooftConstraint}. We would like to fully understand the map between free and interacting fermion systems, particularly in the context of crystalline symmetries, and systematically identify intrinsically interacting phases. Other open questions are to develop a full dimensional reduction procedure to relate the data between (1+1)D and (2+1)D invertible phases (see Ref~\cite{Tantiv2017} for a partial answer in the case of `supercohomology' fermionic SPT phases), and in the other direction, to understand the (3+1)D classification using methods similar to ours. In the latter case the situation is much more complicated. Finally, it would be interesting to study constraints on $c_-$ in the setting of non-invertible fermionic topological states that have intrinsic topological order, which are also referred to as fermionic symmetry-enriched topological (SET) phases, and for which a theory was recently developed in Refs.~\cite{bulmashSymmFrac,bulmashAnomaly,aasen2021characterization}.

\section{Acknowledgements}

We thank Jay Deep Sau, Sankar Das Sarma and Ryohei Kobayashi for discussions, and Yu-An Chen for participating in the initial stages of the project. This work is supported by NSF CAREER DMR- 1753240 (MB and NM). VC was supported by the U.S. Department of Energy, Office of Basic Energy Sciences, Division of Materials Sciences and Engineering, under Contract No. DE-AC02-76SF00515. 

\appendix
\section{Results from group cohomology}\label{app:coho}
\subsection{Cup product definitions used in the main text}

Here we state some basic definitions in group cohomology. A more detailed discussion of the quantities defined here can be found in Appendix B of Ref.~\cite{barkeshli2021invertible}. Let $C^n(G_b,M)$ be the set of functions in $n$ variables from $G_b$ to the abelian group $M$. Each element of $C^n(G_b,M)$ is called an $n$-cochain. We commonly take $M = \Z, \Z_2, U(1)$: for example, in defining $n_1, n_2, \mathcal{O}_3$ we take $M = \Z_2$, and in defining $\nu_3, \mathcal{O}_4$ we take $M = U(1)$. 

The differential $\dd$ takes an element in $C^m(G_b,M)$ to an element in $C^{m+1}(G_b,M)$, and satisfies $\dd \circ \dd = 0$. Examples are shown below. 

When $M$ is a ring, there is a binary operation on cochains called the cup product, denoted $\cup: C^n(G_b,M) \times C^m(G_b,M) \rightarrow C^{n+m}(G_b,M)$, defined as 
\begin{equation}
    (f_n \cup f_m)({\bf g}_1, \dots, {\bf g}_{n+m}) := f_n({\bf g}_1, \dots, {\bf g}_{n}) f_m({\bf g}_1, \dots, {\bf g}_{m}).
\end{equation}
Similarly we can define cup-i products, denoted $\cup_i$, for $i \ge 0$. These are binary operations $\cup_i: C^n(G_b,M) \times C^m(G_b,M) \rightarrow C^{n+m-i}(G_b,M)$. 

Here we give the expanded form of some of the equations used regularly in the main text that used cup products or other cohomology operations. If ${\bf g},{\bf h},{\bf k},{\bf l} \in G_b$, and $f_i \in C^i(G_b,M)$,  
\begin{widetext}
\begin{align}
    \dd f_1({\bf g},{\bf h}) &:= f_1({\bf g}) + {^{\bf g}} f_1({\bf h}) - f_1({\bf gh}) \\
    \dd f_2({\bf g},{\bf h},{\bf k}) &:= -(f_2({\bf g},{\bf h}) + f_2({\bf gh},{\bf k})) + (f_2({\bf g},{\bf hk}) + {^{\bf g}
    }f_2({\bf h},{\bf k})) \\ 
    f_1 \cup f_2 ({\bf g},{\bf h},{\bf k}) &:= f_1({\bf g}) f_2({\bf h},{\bf k}) \\
    f_2 \cup_1 f_2({\bf g},{\bf h},{\bf k}) &:= f_2({\bf g},{\bf hk}) f_2({\bf gh},{\bf k}) + f_2({\bf g},{\bf hk})f_2({\bf h},{\bf k}) \mod 2\\
    \dd f_3({\bf g},{\bf h},{\bf k},{\bf l}) &:= f_3({\bf g},{\bf h},{\bf k})- f_3({\bf gh},{\bf k},{\bf l}) + f_3({\bf g},{\bf hk},{\bf l}) \nonumber 
    \\ & \quad -f_3({\bf g},{\bf h},{\bf kl}) + {^{\bf g}}f_3({\bf h},{\bf k},{\bf l}).
\end{align}
The superscript ${^{\bf g}}f_n$ refers to the action of $G_b$ on $M$. In Eq.~\eqref{eq:O3def}, which defines $\Omc_3$, the action is trivial. In Eq.~\eqref{eq:O4def}, while defining $d\nu_3$, we assume that antiunitary operations act on the $\U(1)$ coefficients by conjugation. Therefore ${^{\bf g}}\nu_3({\bf h},{\bf k},{\bf l}) = (1-2 s_1({\bf g}))\nu_3({\bf h},{\bf k},{\bf l})$. We will interchangeably use multiplicative and additive notation for $\U(1)$ by identifying $x \mod 1$ (in additive notation) with $e^{2\pi \ii x}$ (in multiplicative notation).

\subsection{Slant products}\label{app:slant}

In calculations we often use the so-called slant product. Let's start by reviewing the results of Ref.~\cite{Tantiv2017}. The set of $G$-chains of degree $d$ is denoted by $C_d(G,\ZZ)$ and corresponds to the abelian group generated by elements of the $d$-fold product of $G$ ($G^{\times d}$). We write $c_m=[\gbf_1|\dots|\gbf_d]$ for the generator corresponding to $(\gbf_1,\dots,\gbf_d)\in G\times \cdots \times G$. There is a differential $\partial: C_{m}(G,M)\to C_{m-1}(G,M)$ defined on the basis elements $c_m=[\gbf_1|\dots,|\gbf_m]$ as 
\[\partial c_m = [\gbf_2|\dots|\gbf_m]+\sum_{j=1}^{m-1}(-1)^j[\gbf_1|\dots|{\gbf}_{j}{\gbf}_{j+1}|
\dots|\gbf_m]+(-1)^{m}[\gbf_1|\dots|\gbf_{m-1}].\]
The definition on more general chains follows by $\ZZ$-linearity.

The slant product 
\[\i: \,C^{n+m}(G,M) \times C_{m}(G,\ZZ)  \longrightarrow C^{n}(G,M)\]
is defined as follows. First, consider a $G$-cochain  $\w_{n+m}\in C^{n+m}(G,M)$ and an integer $G$-chain, $c_m=[\gbf_1|\dots|\gbf_m]\in C_{m}(G,\ZZ)$. The slant product between them is defined by the formula 
\begin{equation}
    \begin{split}
     (\i_{c_m}\w_{n+m})(a_{1},\dots,a_{n})&:= \sum_{I}(-1)^{\s(I) }   \w_{n+m}(\dots,a_{i_j-j},\gbf_{j},a_{i_{j}-j+1},\dots) \\
     I &= (i_1 < \dots < i_{m})\\
     \s(I) &= \sum_{j=1}^{m} (n+j+i_{j}) \mod 2
    \end{split}
\end{equation}
where the sum is over all ordered subsets of $\{1,2,\dots,m+n\}$ with $m$ elements (denoted by $I$). The $I$-th summand has a sign $(-1)^{\sigma(I)}$ and a $\omega_{m+n}(\cdots)$ part. The argument for the latter is determined by the following procedure: Fill the $j$-th slot of $\omega_{m+n}$ with $\gbf_j$ ($j\in \{1,2,\dots, m\}$). Then fill the remaining $n$ slots with $a_1,a_2,\dots,a_n$ from left to right. The intuition for $\s(I)$ is that it counts how many times we moved the $\gbf_k$'s from the ``reference order": $a_1,\dots,a_n, \gbf_1,\dots,\gbf_m$ for which $i_j= n+j$. The definition of the slant product on general chains is then defined by $\ZZ$-linearity. 

As reviewed in Ref.~\cite{Tantiv2017}, for $G=A$ abelian,
\begin{equation}
    \dd (\iota_{c_m}\w_{n+m})=\iota_{c_m}\dd{\w_{n+m}} + (-1)^{m+n} \iota_{\partial c_m} \w_{n+m}.
\end{equation}
If we have $\partial c_m =0$, $\iota_{c_m}$ maps cocycles/coboundaries to cocycles/coboundaries in lower degree. This implies that we get a well defined map in cohomology $\iota_{c_m}:\H^{m+n}(A,M)\rightarrow \H^{n}(A,M)$. 

The most common case is to take $c_1^\gbf:=[\gbf]$ so that $\partial c_1^\gbf=[\gbf]-[\gbf]=0$. Another useful example is when there is a subgroup $\ZZ_m \subset A$ that is generated by $\gbf$. Then we can define the quantity $c_3^{\bf g}$: 
\begin{equation}
    c_{3}^\gbf := \sum_{j=0}^{m-1}[\gbf|\gbf^j|\gbf]\Rightarrow 
\partial c_{3}^\gbf = \sum_{j=0}^{m-1}\left(
[\gbf^j|\gbf] - 
[\gbf\gbf^j|\gbf] +
[\gbf|\gbf^j\gbf]-
[\gbf|\gbf^j]\right)=0,
\end{equation}
where we shifted the summation $j\rightarrow j+1$ on the terms with minus signs and used that $\gbf^m=\gbf^0$. This generalizes to
\begin{equation}
c_{2n+1}^\gbf=\sum_{j_1,\dots,j_n=0}^{m-1} [\gbf| \gbf^{j_1}|\gbf|\gbf^{j_2}|\gbf|\dots|\gbf^{j_{n-1}}|\gbf|\gbf^{j_n}|\gbf] \Longrightarrow \partial c_{2n+1}^\gbf=0.
\end{equation}
The notation for the summand means that the odd slots are $\gbf$, while slot $2l$ is $\gbf^{j_l}$

Note that the same procedure almost works for the following chains
\begin{equation}
c_{2n}^\gbf=\sum_{j_1,\dots,j_n=0}^{m-1} [ \gbf^{j_1}|\gbf|\dots|\gbf|\gbf^{j_n}|\gbf] \Longrightarrow \partial c_{2n}^\gbf= m c_{2n-1}^\gbf.
\end{equation}
If we consider $M=\ZZ_{l}$ and $m = 0 \mod l $, we can use $c_{2n}^{\gbf}$ to calculate invariants because $\partial c_{2n}^{\gbf}$ is zero modulo $l$. We denote
\begin{equation}
    \nabla^\gbf_{j}=\iota_{c_{j}^\gbf}.
\end{equation}

\end{widetext}

\subsection{Pontryagin square}

The Pontryagin square is a cohomology operation that refines the square coming from the cup product. In particular, consider $\w\in \H^{2d}(G,\ZZ_{2L})$, then $\Pmc(\w)\in \H^{4d}(G,\ZZ_{4L})$ such that $2\Pmc(\w)=2\w\cup \w \mod 4L$ in $\H^{4d}(G,\ZZ_{4L})$. An explicit expression for this operation is as follows. Take $W$ an integral lift of $\w$, {i.e.} $W\in C^{2d}(G,\ZZ)$ such that $W = \w \mod 2L$. Then 
\[
\Pmc(\w) = W \cup W - W \cup_1 \dd{W} \mod 4L
\]
is independent of the lift $W$ as a cohomology class in $\H^{4d}(G,\ZZ_{4L})$.

\subsection{Spectral sequences}

In this appendix we discuss the Lyndon-Hochschild-Serre spectral sequence (LHSS) that we use for some calculations in the appendices. An introduction to the LHS which is oriented towards physicists is given in App.~K of Ref.~\cite{Manjunath2020fqh}, whose notation we also adopt here. This appendix assumes that background, and discusses how to use the graded ring structure of $\H(G,M)$ to simplify the calculation, when $M$ is a ring. For a more complete mathematical discussion see Refs.~\cite{ramos2017spectral, brown2012cohomology} and the references therein. 

Consider a group $G$ that is the extension of some group $Q$ by some other group $N$, {i.e.} there is a short exact sequence 
\begin{equation}
    1 \rightarrow N \xrightarrow{\iota} G \xrightarrow{\pi} Q \rightarrow 1.
\end{equation}
When $N$ is an abelian group, the extension is specified by two objects.  $\r:Q\rightarrow \Aut(N)$ specifies how $Q$ acts on $N$, and a 2-cocycle $\mu_2 \in Z^2_{\r}(Q,N)$ specifies $G$. The notation means we consider cochains with coefficients in $N$ that are closed under the twisted differential specified by $\rho$. When $N$ is a non-abelian group, there is a similiar, but more complicated, characterization of the extensions. The simplest case of the latter is when the group $G$ is a semi-direct product. In this case, the extension is simply specified by $\r: Q \rightarrow \Out(N)$.

The LHSS is a tool to calculate $\H(G,M)$ with $M$ any $G$-module (the differential can be twisted) assuming that the group cohomology of $Q$ and $N$ are known. Furthermore, when $M$ has a ring structure it also gives the graded-ring structure of $\H(G,M)$. 

In the calculations below, we take $M=\Z_n,\Z,R=\Z^2$. $\Z_n$ and $\Z$ are rings with the obvious multiplication and trivial $G$-action. We can give $\Z^2$ a ring structure by identifying it with $\Z[\eps]/(\eps^2-1)$, the polynomial ring in one variable $\eps$ modulo the relation $\eps^2=1$. We can let $G$ act on $R$ by sending $a+b\eps \to a+b(-1)^{\rho(\gbf)}\eps$. This allows us to calculate the cohomology with $\Z$-coefficients for trivial and twisted action at the same time, as well as uncover extra structure, using methods from Ref.~\cite{316315}.

The setup of the LHSS is the following. Imagine we have a notebook with enumerated pages. Starting on page 2 ($E_2$) we define a two dimensional lattice with non-trivial entries in the first quadrant, where the $(p,q)$ element is $E_2^{p,q} := \H^{p}(Q,\H^q(N,M))$ for $p,q\ge 0$ and $E_2^{p,q}=0$ otherwise. Here $Q$ acts on the coefficients in two ways: it acts on $\H^q(N,M)$ via a map induced by $\rho$, and also acts directly on $M$.

Then the basic result is that the LHSS converges to the desired cohomology $\H(G,M)$, as expressed by
\begin{equation}
    E_2^{p,q} = \H^p(Q,\H^q(N,M)) \Rightarrow \H^{p+q}(G,M).
\end{equation}
This means that, starting from the $E_2$ page, we can take successive differentials (as defined in App.~K of Ref.~\cite{Manjunath2020fqh}) to obtain higher pages $E_n, n>2$. The LHSS stabilizes at a page denoted $E_{\infty}$ (pronounced `$E$-infinity'), in which all differentials vanish. The group cohomology $\H(G,M)$ can then be obtained by carrying out a sequence of group extensions involving groups on the $E_{\infty}$ page.

Now if $M$ is a ring, i.e. it has a multiplication operation compatible with addition, and if the action of any group $K$ on $M$ is compatible with this multiplication, we can define a cup product in the group cohomology with $M$ coefficients:
\begin{equation}
    \cup:\H^{n}(K,M)\times \H^{m}(K,M)\to \H^{n+m}(K,M).
\end{equation}
In this case there is also extra structure to the LHSS. Consider the $r$th row of the $E_2$-page, $E_2^{*,r}$. The elements of $E_2^{*,r}$ can be obtained as cup products of elements in $E_2^{*,0}$ and a set of generators in $E_2^{*,r}$. Additionally, $E_2^{0,*} \cong \H^*(Q,M)$ has a ring structure that imposes relations between the generators of different rows $E_2^{*,r}$. This is useful because in the examples we consider, the whole $E_2$-page is specified by a small number of generators. This statement also applies to higher pages, including $E_{\infty}$.

The cup product structure is also useful in determining differentials. For example, the differential $\dd_2$ is given by a map $\dd_2^{p,q}: E_2^{p,q} \rightarrow E_2^{p+2,q-1}$ for each $p,q>0$ \cite{Manjunath2020fqh}. A useful fact is that this differential satisfies a Leibniz rule of the form  
\begin{equation}
    \dd_2(x\cup y)=(\dd_2 x) \cup y + (-1)^{\abs{x}}x\cup (\dd_2y),
\end{equation}
where $x$ and $y$ are cohomology classes and $\abs{x}$ is the degree of $x$. Thus, knowing how $\dd_2$ acts on the generators in the $E_2$-page allows us to fully determine $\dd_2$. A similar statement applies to the higher differentials $\dd_i, i>2$. 

Suppose we have determined all differentials and obtained the $E_{\infty}$ page of the LHSS. A further application of the above structure is to solve the extension problem which is necessary to obtain $\H(G,M)$ from the $E_{\infty}$ page. To determine $\H^{n}(G,M)$ we have to solve $(n-1)$ group-extensions. However, the fact that the correct solution to the extension problem must be consistent with the ring structure of $\H^{*}(G,M)$ imposes several constraints that simplify the calculations.

We will also make use of what are called `edge morphisms'. These are two maps
\begin{align}\label{eq:EdgeMorphism}
        \iota^*_{\circ}: &\H^q(G,M)\hookrightarrow E_{\infty}^{0,q}\\
        \pi^*_{\circ}: &E_{\infty}^{p,0} \hookrightarrow \H^p(G,M).
\end{align} 
Suppose the group extension is split, {i.e.} there is a group homomorphism  $\sigma: Q\rightarrow G$ such that $\pi\sigma \cong Id$. Then $\sigma ^* \pi^*$ must be an isomorphism. Thus $\pi^*$ is injective and $\sigma^*$ is surjective. From this it can be shown \cite{ramos2017spectral} that $\pi^*_{\circ}$ is also injective, i.e. that there is a copy of $\H(Q,M)$ inside $\H(G,M)$. What is more, $E_{2}^{*,0} =E_{\infty}^{*,0}$, so no non-trivial differential hits the $q=0$ row.

\subsection{Bockstein homomorphisms}

The previous section was about how we can the cohomology of a group that lies in an SES, if the SES contains other groups whose cohomology is known. In this section we will fix the group $G$ but get further information  about its group cohomology by connecting different coefficients. 

Consider a short exact sequence of $G$-modules
\begin{equation}
    M_1 \xrightarrow{\iota_1} M_2 \xrightarrow{\iota_2} M_3.
\end{equation}
This induces a long exact sequence (LES) in group cohomology
\begin{widetext}
\begin{equation}
    \cdots\rightarrow\H^p(G,M_1) \xrightarrow{\iota_1^*}\H^p(G, M_2) \xrightarrow{\iota_2^*} \H^p(G,M_3) \xrightarrow{b_{(p)}} \H^{p+1}(G,M_1) \rightarrow \cdots
\end{equation}
\end{widetext}
where $b_{(p)}:\H^{p}(G,M_3)\rightarrow \H^{p+1}(G,M_1)$ is a collection of maps that increases the degree by one and measures the obstruction to lifting classes in $\H^{p}(G,M_3)$ to $\H^{p}(G,M_2)$. We abuse notation and package each $b_{(p)}$ into a single object $b:\H(G,M_3)\rightarrow\H(G,M_1)$. This map is called the \textit{Bockstein homomorphism} or the \textit{Bockstein map}. There are explicit formulas for the Bockstein obtained by taking lifts, differentiating and then using $\iota_{i}$ to map the obtained class to the correct coefficients. 

In particular, consider the SES of trivial $G$-modules 
\begin{equation}
    1\rightarrow \ZZ \rightarrow \RR \rightarrow \U(1)\rightarrow 1.
\end{equation}
We denote the Bockstein map for this SES by $\beta$. At the cochain level, $\beta$ corresponds to taking an $\RR$-lift of a representative of $\zeta\in \H(G,\U(1))$ and applying $\dd$.

More generally, let $\s\in \H^1(G,\ZZ_2)$ specify the sign-changing action of $G$ on $\Z$ or $\U(1)$ coefficients. Then in Table~\ref{tab:SES_Bock} we summarize the Bocksteins we use in this work. When $\s$ is trivial, we omit it from the notation. In the notation of the table, $\d^{\s} \equiv \beta_2^{\s}$ (in writing $\beta_2^{\s}$, the subscript refers to the group $\Z_2$ on the right of the SES which defines $\beta_2^{\s}$, and the superscript to the action of $G$ on coefficients).

There are many relations between the different Bocksteins. These can be obtained from a commutative diagram that involves two SESs of coefficients. For example, the diagram 
\begin{equation}
    \begin{tikzcd}
  1 \arrow[r] & M_1 \arrow{r}{m_1}\arrow{d}{f_1} & M_2 \arrow{r}{m_2}\arrow{d}{f_2} & M_3 \arrow{r}\arrow{d}{f_3} & 1  \\
  1 \arrow{r} & N_1\arrow{r}{n_1} & N_2\arrow{r}{m_2} & N_3\arrow[r] & 1
\end{tikzcd}
\end{equation}
implies the commutativity of
\begin{equation}
    \begin{tikzcd}
 \cdots \arrow{r} & \H^{d}(G,M_3) \arrow{r}{b_M}\arrow{d}{f_3^*} & \H^{d+1}(G,M_1) \arrow{r}\arrow{d}{f_1^*} & \cdots \\
  \cdots\arrow{r} & \H^{d}(G,N_3)\arrow{r}{b_N} &\H^{d+1}(G,N_1)\arrow[r] & \cdots
\end{tikzcd}
\end{equation}
which implies 
\begin{equation}
    b_N \circ f_3^* =f_1^* \circ b_M.
\end{equation}

In particular, consider the following maps: the inclusion  $j_n:\ZZ_n \hookrightarrow \U(1)$ that sends $j_n(k)=k/n$, $\rho_n:\ZZ\rightarrow \ZZ_n$ that sends $\rho_n(k)=k\mod n$, and $\q_{n}(\mu_n)$ which are multiplication by $n(1/n)$. Again we denote the action of $G$ on each coefficient module by the superscript $\sigma$. These maps fit in the commutative diagrams 
\begin{equation}
    \begin{tikzcd}
  1 \arrow[r] & \ZZ^{\sigma} \arrow{r}{\q_n}\arrow{d}{\cong} & \ZZ^{\sigma} \arrow{r}{\r_n}\arrow{d}{\mu_{n}} & \ZZ_n^{\sigma} \arrow{r}\arrow{d}{j_n} & 1  \\
  1 \arrow{r} & \ZZ^{\sigma}\arrow{r}{\iota} & \RR^{\sigma}\arrow{r}{\pi} & \U(1)^{\sigma}\arrow[r] & 1
\end{tikzcd}
\end{equation}
\begin{equation}
    \begin{tikzcd}
  1 \arrow[r] & \ZZ^{\sigma} \arrow{r}{\q_n}\arrow{d}{\rho_m} & \ZZ^{\sigma} \arrow{r}{\r_{mn}}\arrow{d}{\mu_{n}} & \ZZ_n^{\sigma} \arrow{r}\arrow{d}{\cong} & 1  \\
  1 \arrow{r} & \ZZ_{m}^{\sigma}\arrow{r}{\iota} & \ZZ_{mn}^{\sigma}\arrow{r}{\pi} & \ZZ_{n}^{\sigma}\arrow[r] & 1
\end{tikzcd}
\end{equation}
They imply
\begin{equation}
    \begin{split}
        \beta^{\s} \circ j_n^* &= \beta_n^{\s}\\
    \beta_{m,n}^{\sigma} &= \r_m^*\circ \beta_{n}^{\sigma}\\
    \Rightarrow \beta_{m,n}^{\sigma} &= \r_m^*\circ \beta^{\sigma}\circ j_n^*.
    \end{split}
\end{equation}

Finally, it is also known that (see e.g. \textbf{Theorem 2.3} of Ref.~\cite{greenblatt2006homology})
\begin{equation}
    \rho_2^*\circ\delta^{\s}(\omega) = \Sq^1(\omega)+\s \cup \omega
\end{equation}
where $\omega\in \H(G,\ZZ_2)$.

\begin{table}[h]
    \centering
    \begin{tabular}{c|c}
    SES & Bockstein \\\hline
         $1\rightarrow \ZZ^{\sigma} \rightarrow \RR^{\sigma} \rightarrow \U(1)^{\sigma}\rightarrow 1$ & $\beta^{\sigma}$  \\
         $1\rightarrow \ZZ^{\s} \rightarrow \ZZ^{\s} \rightarrow \ZZ_{n}^{\s}\rightarrow 1$ & $\beta_{n}^{\sigma}$\\
         $1\rightarrow \ZZ^{\s}_{m} \rightarrow \ZZ_{mn}^{\s} \rightarrow \ZZ_{n}^{\s}\rightarrow 1$ & $\beta_{m,n}^{\sigma}$\\
         $1\rightarrow \ZZ^{\s} \rightarrow \ZZ^{\s} \rightarrow \ZZ_{2}\rightarrow 1$ & $\delta^{\sigma}$
    \end{tabular}
    \caption{Commonly used short exact sequences and the notation for their Bockstein homomorphisms.}
    \label{tab:SES_Bock}
\end{table}

\subsection{Comment about continuous groups}

When $G$ is not a discrete group, $\H^*(G,M)$ is taken to mean group cohomology with measurable cochains. ($M$ is an arbitrary $G$-module.) 

On the other hand, a useful cohomology related to a group $G$ is the singular cohomology of its classifying space ($\Bc G$) denoted by $\Hc^*(\Bc G, M)$. There are several results about these cohomology groups and they come with a geometrical intuition that is sometimes useful.

As reviewed in \textbf{Theorem J.1} of Ref.~\cite{Manjunath2020fqh}, if $G$ is finite-dimensional, locally compact, $\sigma$-compact group and $M$ is discrete,  there is an isomorphism
\begin{equation}\label{eq:EquivalenceOfCohomologies}
    \H^{n}(G,M) \cong \Hc^{n}(\Bc G, M).
\end{equation}
The above assumptions are in particular true for finite dimensional compact Lie groups and finite groups. 

We use the isomorphism in Eq.~\ref{eq:EquivalenceOfCohomologies} to identify $\H^{n}(G,M)$ and $\Hc^{n}(\Bc G, M)$ in the remaining appendices. In later appendices, the spectral sequence calculations involving Lie groups are done using the singular cohomology of the classifying space $\Hc^{n}(\Bc G, M)$.

\section{Algebraic formula for the fCEP }\label{app:fCEPAnti}
\subsection{Notation}

First we discuss the notation used for Eq.~\eqref{eq:fCEP} in detail. Consider the group 
$$G_E = (\RR^d \rtimes \Oo(d)) \times \Z_2^{\bf T}$$ 
where $d=D-1$ is the space dimension, which is the group of translations, rotations, reflections, and time reversal. The symmetry of the TQFT describing topological phases of matter contains $G_E$ as a subgroup. A symmetry which acts on both space and time is implemented via the map 
$$\bar{\rho}: G_b \rightarrow G_E.$$ 
We can write $\bar{\rho} = (\vec{R},\rho_s,\rho_t)$ where the three terms project to $\RR^d,\Oo(d),\Z_2^{\bf T}$ respectively. Note that the restriction $(\vec{R},\rho_s)$ of $\bar{\rho}$ fully specifies the action of $G_b$ on space.

We first define the following classes, which generate their respective groups:
\begin{align}
    [\ww_{1,r}] &\in \H^1(\Oo(d),\Z_2) \nonumber \\
    [\ww_{1,t}] &\in \H^1(\Z_2^{\bf T},\Z_2) \nonumber \\
    [\ww_2] &\in \H^2(\SO(d),\Z_2).
\end{align} 
Correspondingly we can define 
\begin{align}
    [w_1] &= [\rho_s^* \ww_{1,r}] \in \H^1(G_b,\Z_2) \nonumber \\
    [s_1^{\sp}] &= [\rho_t^* \ww_{1,t}] \in \H^1(G_b,\Z_2) \nonumber \\
    [w_2] &= [\rho_s^* \ww_2] \in \H^2(G_b,\Z_2).
\end{align}
Here the pullback $\rho^*$ is defined so that $\rho^* f({\bf g}_1,\dots ,{\bf g}_i) = f(\rho({\bf g}_1),\dots ,\rho({\bf g}_i))$ where ${\bf g}_1,\dots {\bf g}_i \in G_b$. Note that $s_1^{\sp}$ has the same information as $\rho_t$ but differs in that it takes an antiunitary element in $G_b$ to the number $1\in \Z_2$ while $\rho_t$ takes an antiunitary element to ${\bf T} \in \Z_2^{\bf T}$. This is only a formal distinction but will be useful in the next section.

The data for the spatial symmetry are thus given by  $(G_b,s_1^{\sp},\w_2^{\sp},(\vec{R},\rho_s))$. Note that the data for the effective internal symmetry are fully specified by $(G_b,s_1^{\Int},\om^{\Int})$ because $(\vec{R},\rho_s)$ are both trivial and can be dropped. The two sets of data are related by the fCEP formula, Eq.~\eqref{eq:fCEP}.

\subsection{Checks}

Below we give examples showing why each term in the above formula should be present.

\subsubsection{$w_2$ }
To see why the term $w_2=\rho_s^* \ww_2$ should be present, note that a single layer of a spinless $p+ip$ SC with $c_-=1/2$ has a unitary $C_2$ rotational symmetry that acts as $\hat{C}_2^2 = (-1)^F$ on fermions. Conversely, a single layer of a spinless $p+ip$ SC with $c_-=1/2$ can only have a $\Z_2$ internal unitary symmetry if the generator ${\bf h}$ satisfies ${\bf h}^2 = +1$. For these two states to be related by the fCEP, $\w_2^{\Int}$ and $\w_2^{\sp}$ should differ by a nontrivial 2-cocycle in $\H^2(G_b,\Z_2)$. This nontrivial cocycle can be constructed using $\rho_s$ as $ \rho_s^* \ww_2 = w_2$ because $\ww_2$ is the cocycle that corresponds to lifting $2\pi$ rotations to $(-1)^F$. Therefore
\begin{equation}
    \om^{\Int} = \om^{\sp} + w_2.
\end{equation}
Note that in this example, $s_1^{\sp}=0, w_1=0$, so the above result is consistent with the general formula. 

\subsubsection{$w_1^2$ }

Next, note that we can only have a nontrivial TSC with $\Z_2^{\bf R}$ reflection symmetry if it acts as ${\bf R}^2 = +1$ on fermions. In this case, we can place a 1d Kitaev chain on the reflection axis. We cannot do so if ${\bf R}^2 = (-1)^F$, for in that case the onsite symmetry on the reflection axis would be $\Z_4^f$, and there are no 1d invertible phases with unpaired MZMs in this case.

On the other hand, consider the equivalent internal symmetry, and assume that reflections map to time-reversal. Here we know that the only nontrivial TSC has ${\bf T}^2= (-1)^F$. This means that in this example, $\om^{\sp}$ and $\om^{\Int}$ should differ by a nontrivial 2-cocycle in $\H^2(G_b,\Z_2)$. This nontrivial cocycle can be constructed using $\rho_s$ as $\rho_s^* \ww_{1,r}^2 = w_1^2$ because $\ww_{1,r}^2$ is the cocycle that corresponds to lifting $\Rr^2$ to $(-1)^F$. Thus we should have
\begin{equation}
    \om^{\Int} = \om^{\sp} + w_1^2.
\end{equation} 
Moreover, note that in this example $w_2=0$ and $ s_1^{\sp}=0$, so the above is consistent with the general formula. 

\subsubsection{$s_1^{\sp} w_1$}

In the previous literature, the formula in Eq.~\ref{eq:fCEP} has appeared in the case where there are no anti-unitary symmetries. We now argue for the new term $s_1^{\sp}\cup w_1^{\sp}$ as follows. 

Consider a fermionic system in (2+1)D with a reflection $\Rr$ and time-reversal symmetry $\Tt$ such that 
\begin{equation}
    \Tt^2 =\Rr^2 = (-1)^F ; \quad (\Rr \Tt)^2 =1.
\end{equation}
This can be written as $G_f=\ZZ_2^{\Rr \Tt} \times \ZZ_{4 }^{\Tt f}$. This corresponds to $G_b= \ZZ_2^{\Rr \Tt} \times \ZZ_2^{\Tt} $. Let $\tau,\sigma \in \H^1(G_b,\Z_2)$ be defined as follows. If an element in $G_b$ is written as $[a,b]$ with $a \in \Z_2^{\bf RT}, b \in \Z_2^{\bf T}$, then $\tau([a,b])=a+b$ and $\sigma([a,b])=a$. The symmetry data is $s_1^{\sp}=\tau$, $\w_2^{\sp}=\tau^2 +\s^2$ and $\rho_s([a,b])= \diag((-1)^{a},1)$. 

Note that $w_1^{\sp}=\sigma$ and $w_2^{\sp}=0$. Applying the fCEP using Eq.~\eqref{eq:fCEP} gives $s_1^{\Int}= \tau+\sigma$ and $\w_2^{\Int}= \sigma (\sigma+\tau)$. This corresponds to $G_f^{\Int}= \Z_4^{\Rr' f}\rtimes \ZZ_2^{\Rr' \Tt}$ -  $\Rr'$ is an anti-unitary internal symmetry that squares to $(-1)^F$ and anticommutes with the unitary $\Z_2$ symmetry $\Rr' \Tt$. This is the symmetry of the example in Sec.~\ref{sec:Z2XZ4T} which has a $\Z_8$ classification and it is generated by free fermions. This is consistent with the result found in Ref.~\cite{YaoTScwithRT}. 

If we did not include the term $s_1^{\sp}\cup w_1$ in Eq.~\ref{eq:fCEP}, we would find that $\w_2^{\Int}=\s^2$ and $G_f^{\Int}= \Z_4^{\Rr' f }\times \ZZ_2^{\Rr' \Tt}$. Which is equivalent to $G_f = \Z_4^{\Tt f}\times\Z_2$ so that $\om = s_1^2$. Let $x$ be projection to the second factor. We find that $n_1\in\{0,s_1\}$ and $n_2=0$. There is only one non-trivial solution with $n_1=n_2=0$ for $\nu_3$ and it is equal to $\frac{1}{2}x^3$. Note that even if we do not fully apply the stacking rules, these results would imply a classification of $\Z_2\times \Z_2$ or $\Z_4$, which cannot be equal to the $\Z_8$ proposed in Ref.~\cite{YaoTScwithRT}.

\subsection{Heuristic argument}
Here we assume the abstract statement of the fCEP without further justification, i.e. that the classification of fermionic topological phases with the spatial symmetry data $(G_b,s_1^{\sp},\om^{\sp},(\vec{R},\rho_s))$ is isomorphic to that of fermionic topological phases with some internal symmetry given by the data $(G_b,s_1^{\Int},\om^{\Int})$. We now discuss some heuristic arguments to connect this abstract statement to the concrete formula in Eq.~\eqref{eq:fCEP}. We note that the abstract statement of the fCEP is argued for in Refs.~\cite{Thorngren2018,else2019crystalline,debray2021invertible}. One can directly derive the formula, with several assumptions, by using the mathematical statements given in these references, but we will not present this here. 

A result that we use below is that $\H^i(G_E,\ZZ_2)$ can be identified with $\H^i(\Oo(d)\times\Z_2^{\Tt},\ZZ_2)$, i.e. the $\RR^d$ translations do not contribute to the cohomology (see e.g. Theorem J7 of Ref.~\cite{Manjunath2020fqh}).

Now if we believe that the fCEP exists, then for fermionic phases it must relate the data $s_1^{\sp},\om^{\sp},\bar{\rho}$ to $s_1^{\Int},\om^{\Int}$. Clearly it should take the form
\begin{align}
    s_1^{\Int} &= s_1^{\sp} + f_1 \nonumber \\
    \om^{\Int} &= \om^{\sp} + f_2
\end{align}
where $f_i \in \H^i(G_b,\Z_2)$. To include $\bar{\rho}$ in the above formula, it is natural to construct $f_i$ by considering cohomology classes $\kappa_i \in \H^i(G_E,\Z_2)$ and then setting $f_i = \bar{\rho}^* \kappa_i$, i.e. $f_i({\bf g}) = \kappa_i(\bar{\rho}({\bf g}))$ for ${\bf g} \in G_b$; the pullback $\bar{\rho}^*$ was defined previously.

Now $\H^1(G_E,\Z_2) \cong \H^1(\Oo(d)\times \Z_2^{\bf T},\Z_2)$ is generated by the cocycles $\ww_{1,r},\ww_{1,t}$. Thus $f_1$ must be a linear combination of $\rho_s^* \ww_{1,r} = w_1$ and $\rho_t^*\ww_{1,t} = s_1^{\sp}$. We note that the space-time orientation-reversing elements are specified by $s_1^{\sp} + w_1$ for the spatial symmetry, and by $s_1^{\Int}$ for the internal symmetry. Since both the spatial symmetry and its effective internal symmetry should have the same orientation-reversing action on space-time, we should set
\begin{equation}
    s_1^{\Int} = s_1^{\sp} + w_1.
\end{equation}

Next we discuss $f_2$. Note that $\H^2(G_E,\Z_2)\cong \H^2(\Oo(d)\times \Z_2^{\bf T},\Z_2)$ is generated by the cocycles $\ww_{1,t}^2, \ww_{1,t} \ww_{1,r}, \ww_{1,r}^2, \ww_2$. Thus $f_2$ should be some linear combination of $\rho_t^* \ww_{1,t}^2 = (s_1^{\sp})^2, \rho_t^* \ww_{1,t} \rho_s^* \ww_{1,r} = s_1^{\sp} w_1, \rho_s^* \ww_{1,r}^2 = w_1^2, \rho_s^* \ww_2 = w_2$. The previously discussed examples show that the last three terms should indeed appear in $f_2$. 

If we also included the remaining term in $f_2$, we would have $\om^{\Int} = \om^{\sp} + (s_1^{\sp})^2 + w_1 s_1^{\sp} + w_1^2 + w_2$. But when $\rho_s$ is trivial (i.e. $G_b$ acts trivially on space), this would give us $\om^{\Int} = \om^{\sp} + (s_1^{\sp})^2$, which would give an equivalence between any system that has ${\bf T}^2 = +1$ with a system that has ${\bf T}^2 = (-1)^F$. This is a contradiction. Therefore we take $f_2 = w_1 s_1^{\sp} + w_1^2 + w_2$. This implies Eq.~\eqref{eq:fCEP}.

This heuristic argument also indicates why unitary translations do not appear in the formula. This is because the cohomology of $\RR^d$ vanishes, so $[\alpha_i] \in \H^i(\RR^d,\Z_2)$ is always trivial and therefore so is $[\vec{R}^* \alpha_i]$.

\section{\texorpdfstring{Calculations when $G_f$ is abelian}{Relations between Q2, Qm and Qpsi when Gf is abelian}}\label{app:abelian}

Here we will use the notation set up in Section \ref{sec:abelian}. We assume $G_f = \Z_{2N_0}^f \times A$ where $N_0$ is even, or $G_f = U(1)^f \times A$. In both cases $A$ is abelian. ${\bf h}$ is defined as the order 2 element of $\Z_{2N_0}^f$ or $U(1)^f$. We have $e^{2\pi i Q_a} := \eta_a({\bf h},{\bf h})$ for $a=e,m,\psi$.

First we derive the equations for even $c_-$. We use Eq.~\eqref{eq:eta:U} along with the fusion rules for even $c_-$, 
\begin{equation}
    e^2 = m^2 = 1; \quad e \times m = \psi.
\end{equation}
These equations directly imply that $2 Q_e = 2 Q_m = 0 \mod 1$ and $Q_e + Q_m = Q_{\psi} \mod 1$. The main computation is now to prove that if there exists some ${\bf k} \in G_b$ with $n_1({\bf k}) = 1$, then we also have 
\begin{equation}
    Q_e = Q_m \mod 1.
\end{equation}
Equivalently, $\eta_e({\bf h},{\bf h}) = \eta_m({\bf h},{\bf h})$. 

Assuming that such a ${\bf k}$ exists, we can use Eq.~\eqref{eq:etacoc} with different permutations of the group elements ${\bf h},{\bf h},{\bf k}$ to obtain the following set of equations:
\begin{align}
    \eta_e({\bf k},{\bf h}) \eta_e({\bf kh},{\bf h}) &= \eta_m({\bf h,{\bf h}}) \\
    \eta_e({\bf h},{\bf k}) \eta_e({\bf kh},{\bf h}) &= \eta_e({\bf h,{\bf kh}}) \eta_e({\bf k},{\bf h}) \\
    \eta_e({\bf h},{\bf h}) &= \eta_e({\bf h,{\bf kh}}) \eta_e({\bf h},{\bf k}).
\end{align}
We used the facts that ${\bf hk} = {\bf kh}$ (since $G_b$ is abelian) and that $^{\bf k}e = m$. Eliminating $\eta_e({\bf kh},{\bf h}),\eta_e({\bf h,{\bf kh}})$ from this system of equations gives
\begin{equation}
    \eta_m({\bf h},{\bf h}) = \eta_e({\bf h},{\bf h}) \left(\frac{\eta_e({\bf k},{\bf h})}{\eta_e({\bf h},{\bf k})}\right)^2.
\end{equation}
The square term is trivial, because we must have $\eta_e^2 = 1$, from Eq.~\eqref{eq:eta:U}. Now substituting the remaining $\eta$ symbols with the defintion of $Q_e, Q_m$, we obtain the claimed result.

For odd $c_-$, the fusion rules are
\begin{equation}
    v^2 = \bar{v}^2 = \psi; \quad v \times \bar{v} = 1.
\end{equation}
These equations imply that $2 Q_v = 2 Q_{\bar{v}} = Q_{\psi} \mod 1$ and $Q_v + Q_{\bar{v}} = 0 \mod 1$. Now we show that we should also have $Q_v = Q_{\bar{v}}$. Arguing exactly as in the even $c_-$ case, we obtain three equations for $\eta$ and eliminate two terms to obtain the following:
\begin{equation}
    \eta_{\bar{v}}({\bf h},{\bf h}) = \eta_v({\bf h},{\bf h}) \left(\frac{\eta_v({\bf k},{\bf h})}{\eta_v({\bf h},{\bf k})}\right)^2.
\end{equation}
Now using Eq.~\eqref{eq:eta:U}, the square term becomes $\frac{\eta_{\psi}({\bf k},{\bf h})}{\eta_{\psi}({\bf h},{\bf k})} = (-1)^{\om({\bf k},{\bf h}) - \om({\bf h},{\bf k})}$. If this equals $-1$, that would imply that ${\bf h},{\bf k}$ anticommute on fermions, even though they commute on bosons. But if this were the case, $G_f$ would have to be non-abelian. By assumption, therefore, this square term must equal 1, so we recover the claimed result for odd $c_-$ as well.

\section{Calculations for charge conservation symmetry}\label{app:insulator}

In Section \ref{sec:charge} we showed that if $G_f = U(1)^f \times H$, the system cannot have unpaired MZMs; we also argued that if $G_f = \Oo(2)^f$, the system can have unpaired MZMs when $c_-$ is odd. Here we prove the second statement for a more general symmetry group in which there are group elements which act by charge conjugation. The idea is to identify the relevant properties of the group $\Oo(2) \cong U(1) \rtimes \Z_2$, which is the simplest group with charge conjugation, and embed them in a more general group through the mathematical operation of pullbacks.

Consider $G_f = U(1)^f \rtimes H$ where the symbol $\rtimes$ corresponds to (unitary) charge conjugation. The charge conjugating elements are specified by a homomorphism $\PH: H \rightarrow \Z_2$.
Let's assume that the elements that act trivially form a subgroup $H_0 \neq H$. Note that there is a map $\phi:\U(1)\rtimes H \rightarrow \U(1)\rtimes \ZZ_2 \cong \Oo(2)$ given by $\phi(z,\gbf) = (z,\PH(\gbf))$. It is straightforward to check that
\[
G_b = \U(1)\rtimes H,
\]
$\om = \phi^*\ww_2 $, and $\phi^*\ww_1 =\PH$, where $\ww_i$ are the i-th SW classes of $\Oo(2)$. Using the last relation, we obtain a chain of equalities $\Sq^1(\om) = \Sq^1(\phi^*\ww_2)=\phi^*(\Sq^1(\ww_2)) := \phi^*(\ww_1\ww_2)=\PH \cup \om$. The result $\Sq^1(\ww_2) = \ww_1 \ww_2$ can be found in Theorem 1.2 of Ref.~\cite{brown1982cohomology}. Thus, for $G_b$ as above, we always have
\begin{equation}
    \PH \cup \om = \om \cup_1 \om.
\end{equation}
This means that there is a solution to Eq.~\eqref{eq:n2unitary} whenever $n_1=c_- \PH$ for $c_-\in \ZZ$. In particular, systems with odd $c_-$ can have MZMs at defects of charge-conjugating group elements, while systems with even $c_-$ do not.

Next we show that the above are the only possible solutions. The idea is to find the most general expression for $n_1$ and check the corresponding obstruction using Eq.~\eqref{eq:n2unitary}. By looking at the spectral sequence for the short exact sequence
\[
1\rightarrow \U(1) 
\rightarrow G_b 
\rightarrow H 
\rightarrow 1
\]
with coefficients in $\ZZ_2$, we find that 
\[
\H^1(G_b,\ZZ_2)\cong \H^1(H,\ZZ_2)
\]
because $\H^1(\U(1),\ZZ_2)=0$ and there cannot be a no-trivial differential hitting or going out of $E_2^{1,0}$.

The cocycles in $\H^1(H,\ZZ_2)$ correspond to one dimensional representations. The isomorphism is given by $\rho = (-1)^{\nu}$ where $\rho$ is the irrep and $\nu\in\H^1(H,\ZZ_2)$. If $\H^1(H,\ZZ_2) = \ZZ_2^{M+1}$ then we have $M+1$ different root representations that generate all of them by tensor product. Denote them by $\rho_{0},\rho_1,\dots,\rho_M$ and set $\rho_0(\gbf)=(-1)^{\PH(\gbf)}$. As these irreps form a basis for $\H^1(G,\ZZ_2)$, there must be elements $\{\hbf_0,\hbf_1,\dots,\hbf_M\}$ such that $\rho_{I}(\hbf_J)=(-1)^{\delta_{IJ}}$ \footnote{One can prove this by contradiction. If there is no such set, then at least one of the irreps could be expressed as a linear combination of the others thus violating the assumption of them generating $\H^1(H,\ZZ_2)$.  }. 

Next, we can write the most general $n_1$ as 
\[
n_1 = \sum_{i=0}^{M}\mu_{i}\nu_{(i)},
\]
where $\rho_i = (-1)^{\nu_{(i)}}$. The $\mathcal{O}_3$ obstruction reads
\[
\mathcal{O}_3 = n_1\cup\om + c_- \Sq^1(\om) = n_1\cup\om + c_- \PH \cup \om.
\]
In order to show that only $n_1=c_- \PH$ is a solution, we calculate the following invariant (using normalized cochains):
\[
\nabla^{\pi}_2\nabla^{\hbf_i}_1\mathcal{O}_3=(\iota_{\hbf_i}\mathcal{O}_3)(\pi,\pi) = \mu_i + c_- \delta_{i,0} \mod 2.
\]
where $\pi $ is the order two element of $\U(1)$ which commutes with each $\hbf_i$. For the rhs to be trivial mod 2, it is clear that we must have $\mu_i = c_- \delta_{i,0}$, implying that $n_1=c_- \PH$.

\section{$\mathcal{O}_4$ obstruction calculations}\label{app:tHooftObstruction}

This appendix aims to explain the calculations involving the $\mathcal{O}_4$ obstruction. We start with some general considerations and then proceed to the examples.

In this appendix we restrict to the groups $G_b$ such that the Bockstein map $\beta:\H^d(G_b,\ZZ)\rightarrow \H^{d+1}(G_b,\U(1))$ is an isomorphism for $d=3,4$. This is true when $G_b$ is a compact Lie group or a wallpaper group.

\subsection{Comments about the $\mathcal{O}_4$ obstruction}\label{app:CommentstHooft}

\paragraph{Order of the obstruction:}
From the explicit formula for $\Omc_{4}$ in terms of $c_-,n_1,n_2,\omega_2$, \[
\Omc_4 = \frac{1}{8} W_4 \mod 1
\]
for some $W_4\in C^4(G_b,\ZZ)$. As $\Omc_4$ is a cocycle ($\dd \Omc_4=0$), we must have  $\dd{W_4}=0 \mod 8$. This allows us to define $\w_4\in \H^{4}(G_b,\ZZ_8)$ with $\w_4 = W_4 \mod 8$ and write $\Omc_4 = j^*_{8} \w_4$ with $j_{n}:\ZZ_n \rightarrow \U(1)$ the inclusion map $j(m)=\frac{m}{n}$. In some cases, $\w_4$ is divisible by $2$ or $4$. For instance, when $c_-$ is an even integer, $\w_4$ is always even so we can write $\Omc_{4}= j^*_4\left(\frac{\w_4}{2}\right)$. Furthermore, when $n_1=0$ and $4|c_-$, $\w_4$ is divisible by 4, so $\Omc_4=j^*_2\left(\w_4/4\right)$.

When $\Omc_4 = j_n^*\l_4$, we must have $n \Omc_4=0$. Furthermore, if $\beta$ is the Bockstein map associated to
\[
\begin{tikzcd}
  1 \arrow[r] & \ZZ \arrow[hook,r, "\iota "] & \RR \arrow[r, "\rho_{1}"] & \U(1) \ar[r] & 1
\end{tikzcd}\]
to the above sequence, then $\beta\Omc_4 \in \H^5(G_b,\ZZ)$. As $\beta$ is a group homomorphism we have $n \beta \Omc_4=\beta (n\Omc_4) =0$.

Write $\H^5(G_b,\ZZ) = \ZZ^{n_5} \oplus \bigoplus_{p \in \mathbb{P}} T_{p}$ where we decompose $\H^5(G_b,\ZZ)$ into its free part $(\ZZ^{n_5})$ and split the torsion subgroup into contributions from prime powers. Here $T_p :=\bigoplus_{m=1}^{M_p}(\ZZ_{p^{m}})^{l_m}$ where $p$ is a prime number. Because $n\beta \Omc_4=0$, $\beta \Omc_4$ can only have image in the torsion parts that satisfy $p|n$. We see this as follows. First, it cannot have image in the free part ($\ZZ^{n_5}$) because $\beta\Omc_4$ is torsion. Secondly, if $p\ndiv n$, multiplication by $n$ does not kill any non-zero element in $T_p$.\footnote{ To see this, denote by $e_p$ the exponent of $T_p$, {i.e.} the smallest integer such that every element of $T_p$ is send to the trivial element by multipliying by $p^{e_p}$. If $p\ndiv n$, then $\gcd(n,p^{e_p})=1$. This implies that we can find two integers $l,m$ such that $l n +m p^{e_p} =1$. Then take any element $\l \in T_{p}$ and write $1\cdot \l = (l n +m p^{e_p})\l = l\cdot (n\l)$. Thus, if $\l$ is a non-zero element, $n\l$ must also be non-trivial.} Thus we focus on $T_2$ because we always have $n = 2,4$ or $8$. In what follows we we denote by $e$ the smallest integer such that $2^e$ kills every element in $T_2$.

\paragraph{A detecting map:} Let $\rho_{2^{e}}:\ZZ \rightarrow\ZZ_{2^e}$ denote reduction modulo $2^e$. Then $\rho_{2^{e}}^*$ restricted to $T_2$ is injective. To see this, look at the LES induced from 
\[
\begin{tikzcd}
  1 \arrow[r] & \ZZ\arrow[r, "\q_{2^e} "] & \ZZ_{} \arrow[r, "\rho_{2^e}"] & \ZZ_{2^{e}} \ar[r] & 1.
\end{tikzcd}\]
where $\q_{n}$ denotes multiplication by $n$: 
\[
\begin{tikzcd}
  \H^5(G_b,\ZZ)\arrow[r, "\q^*_{2^e} "] &  \H^5(G_b,\ZZ) \arrow[r, "\rho^*_{2^e}"] &  \H^5(G_b,\ZZ_{2^e}).
\end{tikzcd}\]
Note that in degree 5, $\img[\q_{2^e}^*] \cong (2^e\ZZ)^{n_5}\oplus \bigoplus_{p\in \PP_{\text{odd}}} T_p \cong \ker(\rho_{2^e}^*)$ so $T_2$ is not in the kernel of $\rho_{2^e}^*$. This proves the claim.

As $\beta$ is an isomorphism and $\r_{2^e}$ is injective, $\rho_{2^e}^*\circ\beta$ is also injective. Therefore, to check if $\Omc_4$ is trivial or not it is enough to evaluate $\rho_{2^e}^*\beta \Omc_4= \rho_{2^e}^*\beta j^*_{2^k} (\omega_4/2^{3-k})$. Thus we need to study $\beta_{2^e,2^k}:=(\rho_{2^e}^*\circ \beta\circ j_{2^k}^*)$ more carefully. 

From App.~\ref{app:coho}, $\beta_{2^e,2^k}$ is the Bockstein map for the SES
\[
\begin{tikzcd}
  1 \arrow[r] & \ZZ_{2^e}\arrow[r, "\q_{2^e} "] & \ZZ_{2^{e+k}} \arrow[r, "\rho_{2^k}"] & \ZZ_{2^{k}} \ar[r] & 1.
\end{tikzcd}\]
that in the special case of $k=e=1$ it is known to reduce to $\Sq^1 = \beta_{2,2}$. $\beta_{2^e,2^k}$ is a detecting map for the obstruction whenever $\beta: \H^4(G_b,\U(1)) \rightarrow \H^5(G_b,\Z)$ is an isomorphism, as we assume throughout. Note that it is common to find the combination $\beta_{2,2}\circ \Sq^2= \Sq^3$, which automatically vanishes on elements of $\H^{d}(G_b,\ZZ_2)$ for $d<3$ for dimensional reasons.

\paragraph{Finding solutions for $\nu_3$ when $\Omc_4$ is trivial:} Using the Universal Coefficient Theorem, we obtain the following decomposition:
\[
\H^4(G_b,\ZZ_8) = \left(\H^4(G_b,\ZZ)\otimes\ZZ_8 \right)\oplus \Tor^{\ZZ}_{1}\left(\H^5(G_b,\ZZ),\ZZ_8\right).
\]
Elements of the first factor can be written as the mod $8$ reduction of an integer cochain. The second factor corresponds to elements that are not of this form. 

Given $Z\in \H^4(G_b,\ZZ)$, let $z \in \H^3(G_b,\U(1))$ be defined by $\beta z = Z$. Let $\zeta$ be an $\RR$ lift of some representative of $z$. A representative of $Z$ is $\dd{\zeta}$ because of the explicit description of $\beta$. Any other representative has the form $\dd{\zeta} + \dd B$ with $B $ a integer 3-cochain.

Then if $\Omc_4= j_8^*Z$, we have 
\[
\Omc_4 = \frac{\dd\zeta}{8}= \dd{\left(\frac{\zeta}{8}\right)} \mod 1.
\]
Thus we find that $\Omc_4$ is a coboundary and we can take $\nu_3 = \frac{\zeta}{8}$. Since $\zeta \mod 1$ represents $z$, a class in $\H^3(G_b,\U(1))$, sometimes it is possible to interpret $\nu_3$ as a `fraction' of an element of $\H^3(G_b,\U(1))$. This can be made slightly more general.\footnote{If $\Omc_4 = j_{k}^* \dd{\zeta}$ for some $\zeta \in C^3(G_b,\RR)$ such that $\zeta \mod 1$ represents a class $z \in \H^3(G_b,\U(1))$, we can write $\nu_3 = \frac{\zeta}{k} + \nu_3^{0}$ where $\nu_3^{0} \in Z^3(G_b,\U(1))$. Then $k\cdot \nu_3=  \zeta + k\cdot \nu_3^{0}$ lies in a cohomology class in $z + k\cdot \H^3(G_b,\U(1))$ such that $\beta(k\nu_3)\in \dd{\zeta} + k \cdot \H^4(G_b,\ZZ)$.}

\subsection{$G_f=\Spin(N)^f$}\label{app:SON}
For $G_b=\SO(N)$, there is a well-known relation\footnote{For instance see Theorem 
C of Ref.~\cite{Thomas1960OnCohomology} with $i=1$ and $W_1=0$ because our bundles are orientable. }:
\begin{equation}
    \mathcal{P}(\ww_2) = p_1 + 2 \cdot (\ww_4) \mod 4
\end{equation}
where $p_1 \in \H^4(\SO(N),\ZZ)$ is the first Pontryagin class and $\ww_4\in \H^4(\SO(N),\ZZ_2) $ is the fourth SW class of the vector representation. Note that $\ww_4$ is trivial in $\H^4(\SO(N),\ZZ_2)$ for $N<4$.

For $G_b=\SO(N)$, we have $\H^2(G_b,\ZZ_2)=\ZZ_2$ generated by the second SW class $\ww_2$. We can fix $n_2=0$ by the equivalence relation $n_2\sim n_2+\om $. When $2| c_-$ the $\Omc_4$ obstruction reduces to 
\begin{align}
    \dd \n_3 = \frac{c_-}{2}\frac{p_1+ 2\ww_4}{4} \mod 1.
\end{align}
\def\gb{\mathbf{g}}
Since $p_1$ is an integer class, it does not contribute to the obstruction. Equivalently there is some $\mu_3 \in C^3(G_b,\U(1))$ such that $\dd \mu_3= \frac{p_1}{4} \mod 1$, as we argued in the previous section.
We are left with 
\[
\dd(\nu_3-\frac{c_-}{2}\mu_3) = \frac{c_-}{2}\frac{\ww_4}{2} \mod 1.
\]
The RHS is non-trivial for $N\ge 5$. The easiest way to see this is by noting that if we restrict to $\SO(3)\times\SO(2)\subset \SO(N)$, we have that $\ww_4 \rightarrow \ww_2^{\SO(2)}\cup\ww_2^{\SO(3)}$ (see for instance Theorem 1.2 of Ref.~\cite{brown1982cohomology}). Then we can take $g = (\pi, R_x), h=(0,R_y)  \in \SO(2)\times\SO(3)$ and evaluate the invariant 
\begin{align}
    \mathcal{I}[\omega_4]
    &=\prod_{j=0}^{1}\frac{1}{(\iota_{h}\omega_4)(g,g^{j},g) }\\
    &= \prod_{j=0}^{1}\frac{\omega_4(h,g,g^{j},g)\omega_4(g,g^{j},h,g)}{\omega_4(g,h,g^{j},g)\omega_4(g,g^{j},g,h)}.
\end{align}
Taking $\omega_4 = \exp(i \pi \ww_2^{\SO(2)}\cup \ww_2^{\SO(3)})$ we obtain 
\begin{align}
    \mathcal{I}(\omega_4)
    &= \frac{\exp(i\pi \ww_2^{\SO(2)}(\pi,\pi)\ww_2^{\SO(3)}(R_y,R_x))}{\exp(i\pi \ww_2^{\SO(2)}(\pi,\pi)\ww_2^{\SO(3)}(R_x,R_y))}\\
    &= \exp(i\pi 1\times 1 )=-1.
\end{align}
This is a non-trivial value, thus $\omega_4$ (and hence $\Omc_4$) is a non-trivial element in cohomology. \footnote{A more abstract way to study the RHS, is by noting that $\Omc_4 = j^*_2 \ww_4$, so the obstruction is non-trivial if $\beta\Omc_4 = \delta \ww_4$ is non-trivial (recall that $\delta = \beta\circ j^*_2)$. Then we invoke Theorem 1.5 of Ref.~\cite{brown1982cohomology} to claim that $\d(\ww_4)$ is non-zero in $\H^5(\SO(N),\ZZ)$ for $N\ge 5$ and trivial for $N<5$.}

\subsection{Proof that there is a solution with $c_-=4$ for suitable choices of $G_b$}\label{app:SomeGfAdmitc4}

Consider $G_b$ to be a compact unitary Lie group  such that 
\begin{equation}\label{eq:2TorsionH5}
    \mathrm{Tor}^\ZZ_{1}\left[\H^5(G_b,\ZZ),\ZZ_{4}\right] \cong \ZZ_2^{m}
\end{equation}
for some integer $m$. This is equivalent to requiring that the exponent of $T_2$ is 1, {i.e.} $e=1$.

The $\Omc_3$ equation can be solved by setting $2|c_-$, $n_1=0$ and $n_2=0$. The $\Omc_4$ obstruction reduces to 
\[
\Omc_4 = \frac{c_-}{4} \frac{2\Pmc(\om)}{4} \mod 1.
\]
We can rewrite the above as $\Omc_4= (c_-/4)j^*_2 (\om\cup\om)$ because $\Pmc(x)=x\cup x\mod 2 $. If $8|c_-$, the obstruction obviously vanishes so we restrict to the case $c_-=4\mod 8$.

From the general discussion in App.~\ref{app:CommentstHooft}, the obstruction is non-trivial if $\beta_{2^e,2^1}(\om\cup\om)$ is non-trivial. From the condition in Eq.~\ref{eq:2TorsionH5}, we see that $e=1$ so we need to study $\beta_{2,2}\om^2$. Recall that $\beta_{2,2}=\Sq^1$. So $\beta_{2,2}\om^2 = \Sq^1(\om)\om+\om\Sq^1(\om)=2\om\Sq^1(\om) =0 $ (the  manipulations are done in $\H^5(G_b,\ZZ_2)$).

Thus we have found a sufficient condition (Eq.~\ref{eq:2TorsionH5}) for the existence of  a solution with $c_-=4$ and $n_1=n_2=0$. Now we discuss how to obtain $\nu_3$. In general, we cannot take $\nu_3 =0 \mod 1$, because even though $\om^2/2$ is trivial in $\H^4(G_b,\U(1))$, $\om^2$ is not necessarily zero in $\H^4(G_b,\ZZ_2)$. In this situation, $\om^2$ must be the mod 2 reduction of some class in $\H^4(G_b,\ZZ)$, with cochain $W_4 = \dd{\zeta}$ for $\zeta \in C^3(G_b,\RR)$. Then we can take $\nu_3 = \frac{1}{2}\zeta \mod 1$. 

As the only nontrivial data are $c_-=4+8n$ and $\nu_3$, stacking two copies of this system amounts to taking $c_-^{\text{tot}}=2c_-=8+16n$ and $\nu_3^{\text{tot}} = 2\nu_3 = \zeta$. As $\dd\zeta =W_4$, we have $\beta(\nu_3^{\text{tot}})= \dd{\zeta} = W_4 = \om^2 \mod 2$, where $\beta$ is the Bockstein map. This corresponds to a bosonic $E_8$ phase (with $c_-=8$), stacked with a bosonic SPT (with cocycle $\zeta$) and a trivial fermionic layer.

\subsubsection{Intrincically interacting example with $c_-=4$}\label{sec:D8xL}

Consider the group $G_b= (\ZZ_2\times\ZZ_2)^{\times L}$, with $L$ some positive integer. We now construct a solution with $c_-=4$, and moreover argue that for $L>3$ this solution corresponds to an intrinsically interacting invertible phase that cannot have any free fermion realization.

Let $a_i$ and $b_i$ be the projections onto the first and second $\ZZ_2$ factors, respectively, in the $i$-th $\ZZ_2\times\ZZ_2$ subgroup. Consider
\[
\om=\sum_{j=1}^{L}a_j \cup b_j.
\]

It is known that $\H(G_b,\ZZ)$ only contains factors of $\ZZ_2$ in positive degrees (we can use induction and the K\"{u}nneth formula for instance). So our criterion in Eq.~\ref{eq:2TorsionH5} is satisfied. 

To find $\nu_3$, we express $\om^2$ as an element of $\H^{4}(G_b,\ZZ_2)$:
\begin{equation}
    \begin{split}
        \om^2 &= \Sq^2(\om) \\
        &= \sum_{j=1}^{L} \Sq^{2}(a_jb_j) \\
        &= \sum_{j=1}^{L} (a_jb_j)^2 \\
        &= \sum_{j=1}^{L} a_j^2 \cup b_j^2.
    \end{split}
\end{equation}
Now we note that $a_j^2 = \beta a_j =\frac{1}{2}\dd{[a_j]_2} \mod 2$; this gives an integral lift of $\om^2$. We can then take
\begin{equation}
    2\nu_3 = \sum_{j=1}^L \frac{1}{4}[a_j]_2 \cup \dd{ [b_j]_2} \mod 1.
\end{equation}

Next, we study free fermion invertible phases with the same symmetry. Note that $G_f$ is a central product of $L$ copies of $\DD_8$, because the restriction to each $\ZZ_2\times \ZZ_2$ factor has $\om=a_j\cup b_j$. We also have $|G_f|=2^{2L+1}$. The one dimensional representations of each $\DD_8$ are representations of $G_f$ and are independent. There are $4^L$ of them but in all of them $\ZZ_2^f$ acts trivially, hence they cannot describe how fermions transform under $G_f$. However, there is one last representation, corresponding to the tensor product of all the 2d irreps of each $\DD_8$. It has dimension $2^L$. Consider the character of this representation. The only values at which it is non zero are $\chi(e)=2^L$ and $\chi(P_f)=-2^L$. Thus $\expval{\chi,\chi} = \frac{\chi(e)^2+\chi^2(P_f)}{|G_f|} = \frac{2\times 2^{2L}}{2^{2L+1}}=1$, meaning that the rep is irreducible. It is also easy to see that the irrep is real because $\sum_{g\in G_f}\chi(g^2) = \chi(e^2)+\chi(P_f^2)=2\chi(e)=2^{2L+1}=|G_f|$.

Therefore, $G_f$ has a unique fermionic irrep. It has dimension $2^{L}$, so free fermion phases constructed with this irrep have chiral central charge equal to some multiple of $2^{L-1}$. Since invertible bosonic phases as well as free fermion phases must have $8|c_-$ (when $L>3$), the $c_-=4$ solution given above describes an intrinsically interacting fermionic phase.

\subsection{Obstruction calculation for $G_f = \SU(2)^f \times H$}\label{app:su2anom}

In the main text, we saw that when $G_f = \SU(2)^f \times H$, $c_-$ must be even, and also $n_1 = 0$, i.e. the system cannot have unpaired MZMs. We also claimed that the unique choice of $n_2$ is $n_2 = 0$. To see this, note that $\H^2(\SO(3),\ZZ_2)=\ZZ_2$ is generated by $\ww_2$ and satisfies $\Sq^1(\ww_2)=\ww_3$, where $\ww_3$ is the generator of $\H^3(\SO(3),\ZZ_2)=\ZZ_2$. Thus when $G_b = SO(3) \times H$, and $dn_2 = 0$, we can always gauge fix $n_2$ to be an element of $\H^2(H,\ZZ_2)$ by using the equivalence $n_2\sim n_2+\om$. 

Then for any $c_-= 0 \mod 2$ and $n_1=0$, the $\mathcal{O}_4$ obstruction is
\begin{equation}
    \Omc_4= \frac{c_-}{2} \frac{\mathcal{P}(\om)}{4}+ \frac{n_2(\om+n_2)}{2}.
\end{equation}
As shown in Appendix~\ref{app:SON}, the first term is trivial in $\H^4(\SO(3),\U(1))$. We focus on the remaining piece that can be written as $j^*_2\left(n_2(\w_2+n_2) \right)$.
Then consider 
\begin{equation}
    \begin{split}
        \rho_2^*\beta\Omc_4 
    &= (\rho_2^*\beta j^*_2) (n_2(\w_2+n_2))\\
    &= \Sq^1 (n_2(\w_2+n_2))\\
    &= \Sq^1(n_2)\w_2 + n_2\Sq^1(\w_2) \\
    &= \Sq^1(n_2)\w_2 + n_2 \ww_3 \mod 2.
    \end{split}
\end{equation}
The two terms belong to different factors in the decomposition 
\[
\H^5(G_b,\ZZ_2) = \bigoplus_{j=0}^5 \H^{j}(\SO(3),\ZZ_2)\otimes\H^{5-j}(H,\ZZ_2).
\]
In particular, $n_2\cup \ww_3 \in \H^3(\SO(3),\ZZ_2)\otimes\H^2(H,\ZZ_2)$ represents a non-trivial element as long as $n_2 $ is non-trivial (because $\ww_3$ is the generator of $\H^3(\SO(3),\ZZ_2)$). Thus for $\Omc_4$ to vanish we must set $n_2 = 0$, as claimed.

\subsection{ Orthogonal groups \texorpdfstring{$G_f=\Oo(2n)^f,\SO(2n)^f$}{Gf=O(2n)}}\label{app:o2nf}

\def\ia{\iota}
\def\ib{\kappa}
\def\ic{\lambda}
\def\vv{\mathrm{v}}
\def\uu{\mathrm{u}}
\def\tt{\mathrm{t}}
In this appendix we discuss in more detail the calculations for the orthogonal groups. First we establish some notation. For any of the orthgonal groups, there is a two-fold cover and a quotient group, defined by
\begin{equation}\label{eq:SES:SO}
    \begin{tikzcd}
     \Spin(2n) \arrow[r,"\ia"]\arrow[bend right=20,swap]{rr}{\ic} & \SO(2n) \arrow[r,"\ib"] & \PSO(2n)
\end{tikzcd}
\end{equation}
and
\begin{equation}\label{eq:SES:O}
    \begin{tikzcd}
     \Pin^{\pm}(2n) \arrow[r,"\ia_{\pm}"]\arrow[bend right=20,swap]{rr}{\ic_{\pm}} & \Oo(2n) \arrow[r,"\ib"] & \PO(2n).
\end{tikzcd}
\end{equation}
We denote by $\tau$ the map that forgets about reflections and sends Eq.~\ref{eq:SES:O} to Eq.~\ref{eq:SES:SO}.

The $\Spin(2n)$ groups have characteristic classes $q_i,\xi,\vv_i$ called the fractional Pontryagin class, the Euler class (of the spinor irrep), and Spin-SW classes, respectively. 

For $\SO(2n)$ we have $p_i, e, \ww_i$ which are the Pontryagin classes, the Euler class (of the vector irrep), and SW classes respectively (see e.g. \cite{brown1982cohomology}). 

For $\PSO(2n)$ we define new characteristic classes below. We also discuss similar classes for the orthogonal groups. For the sake of completeness, we will calculate $\H(\PSO(2n),\ZZ)$, $\H(\PO(2n),\ZZ)$ and  $\H(\PO(2n),\ZZ^{\eps})$ in degrees below 6 by looking at spectral sequences associated to the fibrations \footnote{These fibrations are discussed, for instance, in the last paragraph in page 3 of Ref.~\cite{Gu_2019}.}
\[
\begin{tikzcd}
     \Bc\SO(2n) \arrow[r]\arrow[d] & \Bc\PSO(2n) \arrow[r]\arrow[d] & \Bc^2\ZZ_2\\
     \Bc\Oo(2n) \arrow[r] & \Bc\PO(2n) \arrow[r] & \Bc^2\ZZ_2  \\
     \Bc\Spin(4n) \arrow[r] & \Bc\PSO(4n) \arrow[r] & \Bc^2\ZZ_2^2 \\
     \Bc\Spin(4n+2) \arrow[r] & \Bc\PSO(4n+2) \arrow[r] & \Bc^2\ZZ_4
\end{tikzcd}\]
where $\Bc^2A = \mathrm{K}(A,2)$ is the second Eilenberg-MacLane space of the abelian group $A$. The first diagram comes from the definition of $\PSO(2n)$ and $\PO(2n)$ as quotients of the orthogonal groups. The other sequences come from noting that the universal cover of $\SO(2n)$ is $\Spin(2n)$, so $\Spin(2n)$ is also a universal cover of $\PSO(2n)$. There are different groups on the right hand side because the center of $\Spin(2n)$ depends on the parity of $n$.

\subsubsection{Group cohomology of $\PSO(4n+2),\PO(4n+2)$}\label{app:Orthogonal(4n+2)}
\paragraph{\underline{$\PSO(4n+2):$}} The main result is contained in Eqs.~\eqref{eq:coho_PSO4n+2-Z2} and ~\eqref{eq:coho-PSO4n+2-Z}. Assume $n>0$ to avoid the special case $\PSO(2)\cong\SO(2)$. The universal cover of $\PSO(4n+2)$ is $\Spin(4n+2)$ with a $\ZZ_4$ center. Therefore, we use the definition $\PSO(4n+2)=\Spin(4n+2)/\ZZ_4$ and consider the fibration
\begin{equation}\label{eq:FibrationPSO(odd)}
    \begin{tikzcd}
     \Bc\Spin(4n+2) \arrow[r] & \Bc\PSO(4n+2) \arrow[r] & \Bc^2\ZZ_4. 
\end{tikzcd}
\end{equation}
For $n>0$, we have  (see for instance  \cite{BENSON199513, duan2018characteristic})
\[
\Hc^d(\Bc\Spin(4n+2))=\begin{cases}
\ZZ \quad&, d=0,4 \\
* \quad&, d=1,2,3,5 \\
\ZZ^{\q(n)} \quad&, d=6 
\end{cases}
\]
with $\q(n)= 0$ if $n>2$ and $\q(n)=1$ if $n=1$. The fractional Pontryagin class $q_1$ is the generator in degree 4.\footnote{The name is so because under pullback by the quotient map $\iota: \Spin(4n+2)\to \SO(4n+2)$, we have $\iota^*p_1 =2q_1$ where $p_1$ is the first Pontryagin class of $\SO(4n+2)$.} For $n=1$, the generator of the $d=6$ group is the third Chern class of $\SU(4)\cong\Spin(6)$, which equivalently is the Euler class of the spinor irrep.

The cohomology of $\Bc^2\ZZ_4$ is (See App.~\ref{app:FactorsB2A})
\begin{equation}
    \begin{split}
        \mathrm{H}^d(\Bc^2\ZZ_4,\ZZ) 
        &=\begin{cases}
        \ZZ, \quad &d=0\\
        *,\quad &d=1,2,4 \\
        \ZZ_{4},\quad & d=3\\
        \ZZ_{8},\quad & d=5 \\
        \ZZ_{2},\quad & d=6. \\
        \end{cases}
    \end{split}
\end{equation}
It is generated by $U_3=\beta_4 u_2$, $U_5=\beta_8 \Pmc(u_2)$, $U_3^2$ in degree 3, 5 and 6 respectively. Here $u_2\in\Hc^2(\Bc^2\Z_4,\Z_4)\cong \Z_4$ is a generator.

We consider the following spectral sequence associated to the fibration in Eq.~\ref{eq:FibrationPSO(odd)}:
\begin{equation}
    \begin{split}
  E_2^{p,q}&=\Hc^p(\Bc^2\ZZ_4,\Hc^q(\Bc\Spin(4n+2),\ZZ))\\
        E_2^{p,q} &\Rightarrow \Hc^{*}(\Bc\PSO(4n+2),\ZZ). 
    \end{split}
\end{equation}

As we only care about $d<6$, we can look at the spectral sequence below the line $p+q<7$. When $n=1$, there can be a non-trivial $\dd_3:E_3^{0,6}\cong\ZZ\rightarrow E_3^{3,4}\cong\ZZ_4$. Regardless of its image, $E_4^{0,6}\cong\ZZ$. The only relevant differential on the $E_4$ page is $\dd_4:E_4^{0,4} = \ZZ \rightarrow E_4^{5,0}\cong\ZZ_8$. This means that $\dd_4q_1=k_1\beta\Pmc(u_2)$, where $k_1$ is to be determined. 
With the above information, we know that 
\[
\H^d(\PSO(4n+2),\ZZ) = \begin{cases}
\ZZ \quad &, d=0,4 \\
* \quad &, d=1,2 \\
\ZZ_4 \quad &, d=3 \\
\ZZ_{(8,k_1)} &, d=5\\
\ZZ_2\oplus\ZZ^{\q(n)} \quad &, d=6.
\end{cases}
\]

To determine $k_1$ we use previous work. Ref.~\cite{kono1975cohomology} calculated the group structure of $\Hc(\Bc\PSO(4n+2),\ZZ_2)$ and partially characterized its ring structure\footnote{Note that in Ref.~\cite{kono1975cohomology}, the authors use $\PO(4n+2)$ for the group we call $\PSO(4n+2)$. }. In particular, their \textbf{Theorem 4.9} states that (after simplification)
\begin{equation}\label{eq:coho_PSO4n+2-Z2}
    \begin{split}
   \Hc^d(\Bc\PSO(4n+2),\ZZ_2)
        &=\begin{cases}
        \ZZ_2, \quad &d=0,2,3,4,5\\
        *,\quad &d=1 \\
        \ZZ_2^{2},\quad & d=6 \\
        \ZZ_2^{1-\q(n)},\quad & d=7.
        \end{cases}
    \end{split}
\end{equation}
The generators are $\xx_2$ ( $d=2$), $\xx_3$ ($d=3$), $\xx_2^2$($d=4$), $\xx_5$ ($d=5$), $\xx_2^3, \xx_3^2$ ($d=6$), and $\xx_7$  ($d=7$) if $n>1$. Note that $\xx_2 \xx_{2k+1} =0$.\footnote{We renamed the classes $a_2,y_1,y_2,y_3$ (from the cited reference) to $\xx_2,\xx_3,\xx_5,\xx_7$.}

Using the Universal Coefficient Theorem (see \textbf{Theorem J.5} of Ref.~\cite{Manjunath2020fqh}),
\begin{equation}
    \begin{split}
        \H^4(G,\ZZ_2) \cong &\H^4(G,\ZZ)\otimes \ZZ_2 \oplus \Tor[\H^5( G,\ZZ),\ZZ_2],
    \end{split}
\end{equation}
with $G=\PSO(4n+2)$, we obtain
\begin{equation}
    \begin{split}
        \ZZ_2 \cong \ZZ_2 \oplus \Tor[\ZZ_{{(8,k_1)}},\ZZ_2]\\
        \Rightarrow \ZZ_{{(8,k_1)}} \cong \ZZ_1.
    \end{split}
\end{equation}
Therefore $k_1$ must be odd and  so $\dd_4 q_1$ is a generator of $\ZZ_8$ thus killing $E_4^{0,5}$. 

Putting everything together,
\begin{equation}\label{eq:coho-PSO4n+2-Z}
    \H^d(\PSO(4n+2),\ZZ) = \begin{cases}
\ZZ \quad &, d=0,4 \\
* \quad &, d=1,2,5 \\
\ZZ_4 \quad &, d=3 \\
\ZZ_2\oplus\ZZ^{\q(n)} \quad &, d=6.
\end{cases}
\end{equation}
with generators $U_3=\beta_4 u_2$, $r_1$ and $U_3^2$ in degrees, 3,4 and $6$. For $n=1$, the extra generator in $d=6$ is $\q$. \footnote{We use the same symbols $u_2,U_3$ because these classes are pullbacks of the respective classes in $\Bc^2\Z_4$.} We summarize the generators of $\H^d[\PSO(4n+2),M]$ below:
\begin{align*}
	\begin{array}{|c||c|c|c|c|c|c|c|}
		\hline
		M | d &0 & 1& 2 & 3 & 4 & 5 & 6 \\ \hline\hline
		\ZZ_{\,} & 1&-& - & U_3 & r_1 & - & \q, U_3^2 \\ \hline
		\ZZ_{2} & 1&  - & \xx_2& \xx_3 & \xx_2^2 & \xx_5 & \xx_2^3,\xx_3^2 \\\hline
		\ZZ_{4} & 1&  - & u_2& u_3 & u_2^2 & u_2u_3 & u_2^3,u_3^2\\ \hline
		\U(1) & * & -& \frac{1}{4}u_2& \CS_3 & - &  \CS_5, \frac{1}{2}\xx_5 & ?\\ \hline
	 \hline
	\end{array}
\end{align*}
where $\q$ and $\CS_5$ are trivial for $n>1$.

Finally, we discuss how classes in $\H(\PSO(4n+2),\Z)$ map to those in $\H(\SO(4n+2),\Z)$ and $\H(\Spin(4n+2),\Z)$. Suppose $\hat{G}$ is a compact, connected, simply-connected Lie group and $H$ is a subgroup of its center, with $G=\hat{G}/H$. Then, the image of the pullback map $\Z\cong \Hc^4(\Bc G,\ZZ) \rightarrow \Hc^4(\Bc \hat{G},\ZZ)\cong\Z$
is $n_{\CS}(\hat{G},H)\ZZ$ \cite{390077,Gawedzki_2009}. Here, $n_{\CS}(\hat{G},H)$ is the so-called 'Chern-Simons constraint'. In our case, we have $\hat{G}=\Spin(4n+2), H=\Z_4$ and $n_{\CS}(\Spin(4n+2),\ZZ_4)=8$. Now, using that $\Hc^4(\Bc \PSO(4n+2),\ZZ)$ is generated by $r_1$ and $\Hc^4(\Bc\Spin(4n+2),\ZZ)$ is generated by $q_1$, we must have $\ic^*r_1 = \pm 8 q_1$. Without loss of generality, we can fix $\ic^*r_1 = 8 q_1$.

The maps induced from those in Eq.~\eqref{eq:SES:SO} can be written as
\begin{equation}
    \begin{split}
        \ic^*: [u_2,\xx_3,r_1, \q] &\rightarrow [0,0, 8q_1,l'l''\xi] \\
    \ib^*: [u_2,\xx_3,r_1, \q] &\rightarrow [2\ww_2,\ww_3,4p_1,l'e]
    \end{split}
\end{equation}
where $l',l''$ are non-zero integers. This concludes our discussion of $\PSO(4n+2)$.

\def\TT{\mathbb{T}}
\def\xfk{\mathfrak{x}}

\def\ii{\eps}
\paragraph{\underline{$\PO(4n+2)$:}} Next we discuss the cohomology of $\PO(4n+2)$ over $\ZZ$ and $\ZZ^{\eps}$ by considering the SES
\[
\begin{tikzcd}
      \PSO(4n+2)\arrow[r] & \PO(4n+2) \arrow[r] & \ZZ_2
\end{tikzcd}\]
with a trivial extension but with $\ZZ_2$ acting on $\PSO(4n+2)$ as reflection. This induces an action $u_2\rightarrow -u_2$.

We calculate the regular and the twisted group cohomology at the same time by considering the ring $M=\ZZ[\ii]/(\ii^2-1)$ where the reflection acts on $\ii$ by negation and $\ii^2=1$ (see App.~\ref{app:coho}).

We consider the spectral sequence starting at 
\[
E_2^{p,q} = \H^p(\ZZ_2,\H^q(\PSO(4n+2),M)).
\]
In low degrees, we apply the result from the preceding section, which forces several differentials to vanish and solves extension problems. 

Note that $E_2^{*,0} = \ZZ[\xfk_1]/(2\xfk_1)$ where $\xfk_1\in \H^1(\ZZ_2,\ZZ^{\eps})\cong\Z_2$ is the generator. We must have $E_2^{*,0}\cong E_{\infty}^{*,0}$ because $\PO(4n+2)$ is an split extension of $\ZZ_2$ by $\PSO(4n+2)$. This is a general property of split extensions; see around Eq.~\ref{eq:EdgeMorphism} for an argument based on edge morphisms.

For $q\le 6$, we use Tate cohomology (\textbf{Theorem J.3} of Ref. \cite{Manjunath2020fqh}) to calculate $E_2^{*,q}$. $E_2^{*,q}$ is trivial for $q=1,2,5$. For the other values of $q$, $E_2^{*,q}$ is obtained by a cup product of $E_2^{*,0}$ with a set of generators; there are no extra relations. For $q=3$, the generators are $W_3\in \H^3(\PO(4n+2),\ZZ)$, $\Ufk_3\in \H^3(\PO(4n+2),\ZZ^{\eps})$. For $q=4$, the generators are $r_1 \in \H^4(\PO(4n+2),\ZZ)$. For $d=6$, we have two generators, $W_6,\Ufk_6$. For $n=1$, we also have $\vartheta \in \H^6(\PO(6),\ZZ^{\eps})$ ( $\vartheta$ restricts to $\theta$ in $\PSO(6)$). Thus, in low-degrees the $E_2$ page can be written in terms of  $\xfk_1,\Ufk_3,W_3,r_1,\vartheta, \Ufk_5, W_5$. We will argue below that the spectral sequence already stabilizes for $d\le 5$.

There is a generator $\ufk_2\in \H^2(\PO(4n+2),\ZZ_4^{\eps})=\Z_4\times\ZZ_2$ that restricts to $u_2$ on $\PSO(4n+2)$ and satisfies $\beta_4^{\eps} \ufk_2=\Ufk_3$.\footnote{This corresponds to the group extension $\Pin^+(4n+2)$.} We can also define $\xx_2,\xx_3\in\H(\PSO(4n+2),\ZZ_2)$ by $\xx_2=\r_2^*\ufk_2$ and $\xx_3=\r_2^*\Ufk_3$. Because they are reductions of classes in $\H^{*}(\PO(4n+2), \ZZ_4^{\eps})$, we have $\Sq^1(\xx_2)=\xx_1\xx_2$ and $\Sq^1(\xx_3)=\xx_1\xx_3$. Note that $\tau^*:[\ufk_2,\Ufk_3]\rightarrow[u_2,U_3]$.

\begin{widetext}
We summarize the generators of $\H^d(\PO(4n+2),M)$ below:
\begin{align*}
	\begin{array}{|c||c|c|c|c|c|c|}
		\hline
		M | d &0 & 1& 2 & 3 & 4 & 5 \\ \hline\hline
		\ZZ_{\,} & 1&-& \xfk_1^2 & W_3 & r_1,\xfk_1^4,\xfk_1\Ufk_3 & \xfk_1^2W_3  \\ \hline
		\Z_{\eps}^{\,} & -&\xfk_1& - & \Ufk_3,\xfk^3_1 & \xfk_1W_3 & r_1\xfk_1,\xfk_1^5,\xfk_1^2\Ufk_3  \\\hline
		\ZZ^{}_{2} & 1&\xx_1& \xx_2,\xx_1^2 &\xx_3,\xx_1^3,\xx_1\xx_2 & \xx_1\xx_3,\xx_1^4,\xx_1^2\xx_2, \xx_2^2 & ? \\
		\hline
	 \hline
	\end{array}
\end{align*}
There is no extension problem in $d\le 5$ for the following reasons. For $d\le 3$, we have explicit descriptions for these classes in terms of projective representations of $\PSO(4n+2)$. The group generated by $r_1$ cannot be extended because its a free group. As all multiplicative generators of the slots in the range of interest survive at $E_\infty$, there is no extension problem. Above we calculated the $\ZZ_2$ line by using the Universal Coefficient Theorem (see \textbf{Theorem J.5} of Ref.~\cite{Manjunath2020fqh}). 
The images of the generators under $\kappa^*, \tau^*$ (the pullback of the maps defined in Eq.~\eqref{eq:SES:O}) are
\end{widetext}
\begin{equation}
    \begin{split}
    \kappa^*: [\xfk_1,\ufk_2,\Ufk_3,r_1, \vartheta] &\rightarrow [\wfk_1,2\ww_2,\wfk_3,p_1,l'e] \\
    \tau^*: [\xfk_1,\ufk_2,\Ufk_3,r_1, \vartheta] &\rightarrow [0,u_2,U_3,r_1,\q] .
    \end{split}
\end{equation}
Above $\wfk_i \in \H^i(\Oo(4n+2),\Z_{\eps})$ are a lift of the SW class $\ww_i$. $l'$ is a non-zero integer that we could not determine.

\subsubsection{Group cohomology of $\PSO(4n),\PO(4n)$}\label{app:Orthogonal(4n)}
\paragraph{\underline{$\PSO(4n):$}} 

The main result is contained in Eqs.~\eqref{eq:coho_PSO8n+4-Z} and ~\eqref{eq:coho_PSO8n-Z}. When $n=1$, we have $\PSO(4)=\SO(3)_L\times\SO(3)_R$. Then we can use the K\"unneth theorem to calculate $\H(\PSO(4),\ZZ)$. There are 5 generators $W_{3;R}, W_{3;L}, W_5, p_{1;L}, p_{1;R}$.  Here $W_{3;X} =\beta_2\ww_{2;X}$ $(X = L,R)$, $\ww_{2;X}$ is the second SW class of $\SO(3)_X$, $W_5=\beta_2(\ww_2^L\ww_2^R)$ and $p_{1;X}$ is the first Pontryagin class of $\SO(3)_{X}$.

For $n>1$, we need to do a spectral sequence calculation using the fibration
\begin{equation}\label{eq:FibPSO(even)}
    \begin{tikzcd}
     \Bc\Spin(4n) \arrow[r] & \Bc\PSO(4n) \arrow[r] & \Bc^2\ZZ_2^2 .
\end{tikzcd}
\end{equation}
The relevant cohomology groups are (See App.~\ref{app:FactorsB2A}) 
\begin{equation}
    \begin{split} 
        \mathrm{H}^d(\Bc^2(\ZZ_2^2),\ZZ) 
        &=\begin{cases}
        \ZZ, \quad &d=0\\
        *,\quad &d=1,2,4 \\
        \ZZ_{2}^2,\quad & d=3\\
        \ZZ_{4}^2\times\ZZ_2,\quad & d=5 \\
        \ZZ_{2}^3,\quad & d=6. \\
        \end{cases}
    \end{split}
\end{equation}
If we write $\ZZ_2^2=\ZZ_2^L\times\ZZ_2^R$, then the generators are $W_{3;X}=\beta_2 \ww_{2;X}$; $W_{5;X}=\beta_4\Pmc(\ww_{2;X})$, $W_5=\beta_2(\ww_{2;R}\ww_{2;L})$; and $W_{3;X}^2,W_{3;R}W_{3;L}$ in degree 6.

For $n>1$, we have   (see for instance  \cite{BENSON199513, duan2018characteristic})
\[
\Hc^d(\Bc\Spin(4n))=\begin{cases}
\ZZ \quad&, d=0,4 \\
* \quad&, d=1,2,3,5,6 \\
\end{cases}
\]

We start by looking at the spectral sequence with $\Z_2$ and $\Z$ coefficients. As $E_2^{*,q}$ is trivial for $1\le q \le 3$, $\Hc^d(\PSO(4n),\Z) \cong \Hc^d(\Bc^2(\Z_2^2),\Z)$ and $\Hc^d(\PSO(4n),\Z_2) \cong \Hc^d(\Bc^2(\Z_2^2),\Z_2)$ for $d=0,1,2,3$. Next, we proceed to determine the first non-trivial differential $(\dd_4)$ for the spectral sequence with $\Z$-coefficients. We only need $\dd_4:E_4^{4,0}\rightarrow E_4^{0,5}$.

From the existence of the free fermion root state, we know there is a solution to the equations defining invertible fermionic phases when $G_f=\Oo(4n)^f$ with $c_-= 2n$ and $n_1 \neq 0$. By considering the assignment of quantum numbers to the fermion parity fluxes we can show that after restriction to $\SO(4n)^f$, $n_2$ must be nontrivial, i.e. $n_2 =\ww_{2;L} + l(\ww_{2;L} + \ww_{2;R})$ with $l=0,1$. The argument resembles the one given for $\Oo(4)^f$ in the main text, Sec.~\ref{sec:o4f}.
\footnote{  
Projective reps of $\PSO(4n)$ are equivalent to linear reps of its universal cover $\Spin(4n)$ which have nontrivial charge under the center $\Z_2^L\times\Z_2^{R} \subset \Spin(4n)$. Since $e,m,\psi$ could in principle all transform projectively under $\PSO(4n)$, we can define $Q_a$ as the charge mod 2 of $a$ under $\Z_2^L\times\Z_2^R$, for $a \in \{e,m,\psi\}$. From the definition of $G_f$, $Q_{\psi}=[1,1]$. Now, reflections in $\PO(4n)$ exchange the $\Z_2^L$ and $\Z_2^R$ subgroups, and also exchange $e$ and $m$, when $n_1 \ne 0$. Thus, if $Q_e=[x,y]$, then $Q_m=[y,x]$. Finally, since $Q_\psi=Q_e+Q_m$, we must have $[x,y]=[1,0]$ or $[0,1]$. This corresponds to either $n_2= \ww_{2;L}$ or $\ww_{2;R}$. }.
Using Eq.~\eqref{eq:O4simple}, this means that the class
$$\W_5=\beta_4(n\Pmc(\ww_{2;L}+\ww_{2;R})+2\ww_{2;L}\ww_{2;R}) $$
must be trivial because it is equal to the Bockstein of the obstruction $\beta\Omc_4$. For this to happen,  $\W_5$ must be a multiple of $\dd_{4}q_1$, where $q_1$ is the fractional Pontryagin class introduced previously and generates $E_{4}^{0,4}$.

Note that the most general form of $\dd_4{q_1}$ is
\begin{equation}
    \dd_4q_1 = \beta_4(k_0 2\ww_{2;L}\ww_{2;R}+k_1\Pmc(\ww_{2;L})+k_2\Pmc(\ww_{2;R})).
\end{equation}
Since $\W_5$ is a multiple of $\dd_4q_1$, we can write $\W_5 = m\dd_4 q_1$ for $m\in\Z$. Then, looking at the coefficients of $\beta_4\Pmc(\ww_{2;L}),\beta_4\Pmc(\ww_{2;R})$ and $\beta_42\ww_{2;L}\ww_{2;R}$ gives the following relations
\begin{align}
    n = mk_1 &= mk_2 \mod 4 \label{eq:A1}\\
    2(1+n) &= 2mk_0 \mod 4 \label{eq:A2}.
\end{align}
The analysis now depends on the parity of $n$. 

Let us start with $n$ odd. Here, Eq.~\ref{eq:A1} implies that $m$ and $k_{1}=k_{2} \mod 4$ are both odd integers. From Eq.~\ref{eq:A2} we obtain that $k_0$ is even. Therefore, $\dd_4{q_1}= k_1\beta_4(\Pmc(\ww_{2;L})+\Pmc(\ww_{2;R}))$. Using $\Pmc(a+b)=\Pmc(a)+2a\cup b +\Pmc(b)$, we can write 
\begin{equation}
    \dd_{4}q_1= k_1n \W_{5} = \pm \W_5.
\end{equation}
Here we used that $4\W_5=0$ as a cohomology class. As $\dd_{4}q_1$ is an order 4 element, it will kill a $\ZZ_4$ subgroup of $E_{4}^{5,0} \cong \Hc^5(\Bc^2(\Z_2^2),\Z) \cong \Z_4^2\times\Z_2$ so that $E_{\infty}^{5,0} \cong E_{5}^{5,0}\cong \Z_4\times \Z_2 $. This implies that for $n=2n'+1>1$,
\begin{equation}\label{eq:coho_PSO8n+4-Z}
    \begin{split}
   \Hc^d(\Bc\PSO(8n'+4),\ZZ)
        &=\begin{cases}
        \ZZ, \quad &d=0,4\\
        *,\quad &d=1,2 \\
        \ZZ_2^{2},\quad & d=3 \\
        \ZZ_4\times\ZZ_2,\quad & d= 5 \\
        \ZZ_2^{3},\quad & d= 6.
        \end{cases}
    \end{split}
\end{equation}
The generators of the torsion subgroups are $W_{3;X}=\beta_2 \ww_{2;X}$,  $W_{5;L}=\beta_4\Pmc(\ww_{2;L})$, $W_5=\beta_2(\ww_{2;L}\ww_{2;R})$,  $W_{3;X}^2$ and $W_{3;R}W_{3;L}$. They are obtained by pullback from $\Hc^*(\Bc^2(\Z_2^2),\Z)$. We denote the generator for $d=4$ by $r_1$. Furthermore we have the following relation $ \beta_4(\Pmc\ww_{2;L})=-\beta_4(\Pmc\ww_{2;R})$.

Doing a similar manipulation for $n$ even, we find that $\dd_4{q_1} = \pm \W_5$. But $\W_5= \beta_2((n/2)(\ww_{2;L}+\ww_{2;R})+\ww_{2;L}\ww_{2;R})$ which is an order two element. Therefore, it will kill a $\ZZ_2$ subgroup of $E_{4}^{5,0} \cong \Hc^5(\Bc^2(\Z_2^2),\Z)$ so that $E_{\infty}^{5,0} \cong E_{5}^{5,0}\cong \Z_4^2 $. This implies that for $n=2n'>0$,
\begin{equation}\label{eq:coho_PSO8n-Z}
    \begin{split}
   \Hc^d(\Bc\PSO(8n'),\ZZ)
        &=\begin{cases}
        \ZZ, \quad &d=0,4\\
        *,\quad &d=1,2 \\
        \ZZ_2^{2},\quad & d=3 \\
        \ZZ_4^2,\quad & d= 5 \\
        \ZZ_2^{3},\quad & d= 6.
        \end{cases}
    \end{split}
\end{equation}
The generators of the torsion subgroups are $W_{3;X}=\beta_2 \ww_{2;X}$,  $W_{5;L}=\beta_4\Pmc(\ww_{2;L})$, $W_{5;R}=\beta_4\Pmc(\ww_{2;R})$, and $\{W_{3;X}^2,W_{3;R}W_{3;L}\}$. They are obtained by pullback from $\Hc^*(\Bc^2(\Z_2^2),\Z)$. We denote the generator for $d=4$ by $r_1$. Furthermore we have the relation $ \beta_2(\ww_{2;L}\ww_{2;R})= n' \beta_2(\ww_{2;L}^2+\ww_{2;R}^2)$.

The relation to classes in the cohomology of $\SO(4n)$ is given by 
\[
\kappa^*: [\ww_{2;L},\ww_{2;R},W_{5;X},r_1]\rightarrow [\ww_{2},\ww_{2},\d \ww_4,l p_1]
\]
for some integer $l$. To find $l$ we again use Refs.~\cite{390077,Gawedzki_2009} and find $l = n_{\CS}(\Spin(4n),\ZZ_2^2)/2=2/(n,2)$ by noting that we need $2|2l$ and $4|2l n$.

\def\xfk{\mathfrak{x}}
\def\Xfk{\mathfrak{X}}
\paragraph{$\underline{\PO(4n):}$}
Here we follow the steps outlined for $G_b = \PO(4n+2)$ previously: we determine the group cohomology by using the above result for $\PSO(4n)$ in the SES
\[
\begin{tikzcd}
      \PSO(4n)\arrow[r] & \PO(4n) \arrow[r] & \ZZ_2 .
\end{tikzcd}\]
We will only state the results. First, we consider $n=1$. We find that the generators in low degrees are $\xfk_1,U_3, \ufk_3, U_5, \Ufk_5 $ (all of order two) and $r_1, \efk_4$ (of infinite order) such that $\xfk_1U_3=\xfk_1\Ufk_3=0$.

For $n>1$ odd, $\efk_4$ disappears but two new classes appear. One is $\Xfk_5 \in \H^5(\PO(4n),\Z^{\eps})$ (order 4), that satisfies $\t^*\Xfk_5=\beta_4\Pmc(\ww_{2;L})$. The other one is $X_5\in \H^5(\PO(4n),\Z)$ of order 2; it satisfies $\t^*X_5=2\beta_4\Pmc(\ww_{2;L})$.

For $n$ even, the generators are $\xfk_1,U_3, \Ufk_3$ (order two), $r_1$ (infinite order) and $ U_5, \Ufk_5$ (order 4). If we let $\uu_2\in \H^2(\PO(4n),\Z_2)$ be the class corresponding to the extension to $\Oo(4n)$, we have $U_3 =\beta_2 \uu_2$, $\Ufk_3=\beta_2^{\eps} \uu_2$. Furthermore, $\t^* U_5= \beta_4(\Pmc(\ww_{2;L})+\Pmc(\ww_{2;R}))$ and $\t^* \Ufk_5= \beta_4(\Pmc(\ww_{2;L})-\Pmc(\ww_{2;R}))$. There are relations $U_i\xfk_1=\Ufk_i\xfk_1=0$.

\begin{widetext}
We summarize the generators below. For $\H^d(\PO(8n+4),M), n>0$: 
\begin{align*}
	\begin{array}{|c||c|c|c|c|c|c|}
		\hline
		M | d &0 & 1& 2 & 3 & 4 & 5 \\ \hline\hline
		\ZZ_{\,} & 1&-& \xfk_1^2 & U_3 & r_1,\xfk_1^4 & U_5,X_5  \\ \hline
		\ZZ_{\eps}^{\,} & -&\xfk_1& - & \Ufk_3,\xfk^3_1 & - & r_1\xfk_1,\xfk_1^5,\Ufk_5, \Xfk_5  \\\hline
		\ZZ^{}_{2} & 1&\xx_1& \xx_1^2,\uu_2 &\xx_1^3,\uu_3 & \xx_1^4, \uu_2^2,\uu_4,\xx_4 & ? \\
		\hline
	 \hline
	\end{array}
\end{align*}
and $\H^d(\PO(8n),M) , n>0$:
\begin{align*}
	\begin{array}{|c||c|c|c|c|c|c|}
		\hline
		M | d &0 & 1& 2 & 3 & 4 & 5 \\ \hline\hline
		\ZZ_{\,} & 1&-& \xfk_1^2 & U_3 & r_1,\xfk_1^4 & U_5  \\ \hline
		\ZZ_{\eps}^{\,} & -&\xfk_1& - & \Ufk_3,\xfk^3_1 & - & r_1\xfk_1,\xfk_1^5,\Ufk_5  \\\hline 
		\ZZ^{}_{2} & 1&\xx_1& \xx_1^2,\uu_2 &\xx_1^3,\uu_3 & \xx_1^4, \uu_2^2,\uu_4 & ? \\
		\hline
	 \hline
	\end{array}
\end{align*}
All extensions are trivial, for the same reason as in the case $G_b=\PSO(4n+2)$.
\end{widetext}

\subsubsection{Constraints for $G_f=\Oo(8n)^{f}$}\label{app:ConstrainsO(8)O(16)}

Since the constraints for $G_f = \Oo(8n+4)^f$ reduce to those for $\Oo(4)^f$, as mentioned in the main text, here we only consider $G_f=\Oo(8n)^{f}$. The constraints on $c_-,n_1$ depend on the parity of $n$, and can be obtained by restriction to the subgroups $\Oo(8m)^f$ with $m=1$ or $2$. 

First, we argue that $4m|c_-$. For $m=1$, consider the $\Spin(5)^f\subset \Oo(8)^f$ subgroup; in this case we already proved the constraint $4|c_-$ (Appendix \ref{app:SON}). Similarly, for $m=2$ we consider the $\SU(8)^f\subset \Oo(16)^f$ subgroup (see Appendix \ref{app:HSUNZ2} below) to show that $8|c_-$.

Next, we constrain $n_1$ in terms of $c_-$. For any $G_f=\Oo(4n')^f$ we find the following constraints:
\begin{align}
    n_1& =k_1\xx_1 \nonumber \\
    n_2 &= k_1\ww_{2;L} + k_2 (\ww_{2;L}+\ww_{2;R})+k_3 \ww_1^2
\end{align}
with $k_1,k_2,k_3\in \ZZ_2$. When $n' = 2n$, $c_-$ constrains $k_1$ as follows:
\begin{align}\label{eq:O8nconst-c-n1}
    n \text{ odd}, c_- = 4 \mod 8 &\implies k_1 = 1; \nonumber \\
     n \text{ odd}, c_- = 0 \mod 8 &\implies k_1 = 0; \nonumber \\
      n \text{ even} \implies c_- = 0 \mod 8, & \quad k_1 \text{ is arbitrary}.
\end{align}
To see this notice that $\H^2(\PO(4n'),\ZZ_2)\cong \ZZ_2^2$ is generated by two classes, $\xx_1^2$ and $\uu_2$, such that upon restriction to $\PSO(4n')$ they satisfy $\tau^*[\xx_1^2,\uu_2] = [0,\ww_{2;L}+\ww_{2;R}]$\footnote{This is true by using the edge morphisms for the spectral sequence with coefficients in $\ZZ_2$. $\tau: \PSO(2n')\hookrightarrow \PO(2n')$ is the inclusion map.}. Now when $k_1 = 0$, we must have $n_1=0$ and $n_2$ must be a cocycle. Thus $n_2= k_2 \ww_2 +k_3 \ww_1^2$. On the other hand when $k_1 = 1$, the reflection swaps the fermion parity fluxes, and upon restriction to $G_f=\SO(4n')^f$, $n_2 =  \ww_{2;L} + k_2(\ww_{2;L}+\ww_{2;R})$ with $k_2=0,1$, as we argued while discussing $\H(\PSO(4n'),\Z)$. Combining the two possibilities for $k_1$, $\tau^* n_2= k_1\ww_{2;L} + k_2(\ww_{2;L}+\ww_{2;R})$. Using $n_2 \simeq n_2 + \om$, we can set $k_2 = 0$.

Next we consider a putative state with $2|c_-$ (this constraint comes from the $\Omc_3$ equation) and take $n_1,n_2$ as the previous paragraph with $k_2=0$. We calculate $\Omc_4$ and restrict to $\PSO(4n')$ so that there is no contribution from $k_3$. We take $n'=2n$ even. Then the obstruction reduces to 
\begin{widetext}
    \begin{equation}
    \begin{split}
        \t^*\Omc_4 
        &= \frac{c_-}{2} \frac{\Pmc(\ww_{2;L}+\ww_{2;R})}{4} +\frac{k_1\ww_{2;L}((k_1+1)\ww_{2;L}+\ww_{2;R})}{2}\\
        &=\frac{c_-}{2} \frac{\Pmc(\ww_{2;L}+\ww_{2;R})}{4} +\frac{k_1\ww_{2;L}\ww_{2;R}}{2}\\
        \beta\t^* \Omc_4 &= \frac{c_-}{2}\beta_{4}(\Pmc(\ww_{2;L}+\ww_{2;R})) + k_1 \beta_2(\ww_{2;L}\ww_{2;R})\\
        &= (\frac{c_-}{2}+nk_1)\beta_{4}(\Pmc(\ww_{2;L}+\ww_{2;R})).
        \end{split}
    \end{equation}
\end{widetext}
We used the relation $\beta_2(\ww_{2;L}\ww_{2;R})= n'\beta_4(\Pmc(\ww_{2;L}+\ww_{2;R}))$ valid for $\PSO(4n')$ derived in the previous section. 

As we saw there, $\beta_{4}(\Pmc(\ww_{2;L}+\ww_{2;R}))$ is an order $4$ class when $n'=2n$ is even. Then if the obstruction is trivial, we must have 
\begin{equation}
    \begin{split}
        \frac{c_-}{2} + 2n k_1 &= 0 \mod 4\\ 
        \Rightarrow c_- & = 4n k_1 \mod 8. 
    \end{split}
\end{equation}
Now by considering the different possible cases, we obtain the result in Eq.~\eqref{eq:O8nconst-c-n1}.

\subsection{Group cohomology of \texorpdfstring{$\SU(n)/\ZZ_2$}{PSU(n)}}\label{app:HSUNZ2}

In the main text, we sketched how to obtain constraints on $c_-$ for $G_f = \SU(4n+2)^f, \SU(8n+4)^f$. In order to obtain constraints on $G_f = \SU(8n)^f$, we perform a spectral sequence calculation below. The main results of this calculation are contained in Eqs.~\eqref{eq:QUcoho} and~\eqref{eq:SUcoho}.

Ref.~\cite{ning2021enforced} gave a physical argument to show that when $G_f=\SU(8n)^f$ for some  natural number $n$, $c_-$ must be a multiple of $8$. This may seem to be in tension with our result in App.~\ref{app:SomeGfAdmitc4} that a wide class of groups $G_f$ can have solutions with $4|c_-$. However, Eq.~\eqref{eq:SUcoho} shows that $\H^5(G_b, \Z) = \Z_4$, which is not of the form $\Z_2^m$, and hence does not satisfy our condition. Furthermore, when $c_-=4$ we are able to show that the $\mathcal{O}_4$ obstruction is the order 2 element of $\H^4(G_b, \U(1))$.

According to \textbf{Theorem 1.1} of Ref.~\cite{Gu_2019}, the cohomology $\H(\PSU(N),\ZZ) \cong\Hc(\mathrm{B}\PSU(N),\ZZ)$ in degrees lower than 6  for $N>2$ is generated by $x_3\in \H^{2}(\PSU(N),\ZZ)$, $e_{i}\in \H^{2i}(\PSU(N),\ZZ)$ for $i=2,3$, and $x_3^2\in \H^6(\PSU(N),\ZZ) $.

The generators can be identified as follows. $x_3$ is the Bockstein map of $x_2\in \H^2(\PSU(N),U(1))$ that represents the obstruction to lifting the fundamental irrep of $\SU(N)$ to a $\PSU(N)$ representation. $e_2$ and $e_3$ can be thought of as Chern classes. The class $x_3$ satisfies the relations $Nx_3=0$ and $\mathrm{gcd}(N,2)x_3^2=0$.

Our strategy is to calculate the group cohomology of the groups $\QU(N,M)$, which are defined by the SES
\[
1\rightarrow \ZZ_{M} \rightarrow \QU(N,M) \rightarrow \PSU(N) \rightarrow 1
\]
where $N=k M$ with $k$ a positive integer. The extension is by $W_3=k x_3$. We then use the relation $\QU(kM,M)=\SU(kM)/\ZZ_k$ to obtain our final result. We use two spectral sequence calculations.  

First consider
\[
E_{2}^{p,q}= \H^p(\PSU(N),\H^q(\ZZ_{M},\ZZ)) \Longrightarrow \H^{p+q}(\QU(N,M),\ZZ)
\]
for values $p+q<7$. As $\H(\ZZ_M,\ZZ)$ is only non-trivial in even degrees, the first differential is $\dd_3$ so $E_3\cong E_3$. Below we show the $E_3$ page of the spectral sequence (we denote $M_2:=\gcd(M,2)$). Let $c$ be the generator of $\H^2(\ZZ_M,\ZZ)$. Then $\dd_3 c = kx_3$, because we know that the final group has a $\ZZ_k$ classification of projective reps. For dimensional reasons, $\dd_3y_{2,2}=0$ where $y_{2,2}$ is a generator of $E_{2}^{2,2}\cong \H^{2}(\PSU(N),\H^{2}(\ZZ_M,\ZZ))\cong\ZZ_M$. 
\begin{widetext}
\begin{center}
		\begin{sseq}[grid=chess,labelstep=1,entrysize=1.5cm]{0...6}{0...5}
		
		\ssmove{0}{0}
		\ssdrop{\ZZ}
		\ssmove{1}{0}
		\ssdrop{}
		\ssmove{1}{0}
		\ssdrop{}
		\ssmove{1}{0}
		\ssdrop{\ZZ_{N}}
		\ssmove{1}{0}
		\ssdrop{\ZZ}
		\ssmove{1}{0}
		\ssdrop{}
		\ssmove{1}{0}
		\ssdrop{\ZZ\oplus \ZZ_{N_2}}
	
		\ssmove{-6}{2}
		\ssdrop{\ZZ_M}
		\ssmove{1}{0}
		\ssdrop{}
		\ssmove{1}{0}
		\ssdrop{\ZZ_{M}}
		\ssmove{1}{0}
		\ssdrop{\ZZ_{M}}
		\ssmove{1}{0}
		\ssdrop{\ZZ_{M}}
		\ssmove{1}{0}
		\ssdrop{\ZZ_{M_2}}
		\ssmove{1}{0}
		\ssdrop{\ZZ_{M}\oplus \ZZ_{M_2}}
		
		\ssmove{-6}{2}
		\ssdrop{\ZZ_M}
		\ssmove{1}{0}
		\ssdrop{}
		\ssmove{1}{0}
		\ssdrop{\ZZ_{M}}
		\ssmove{1}{0}
		\ssdrop{\ZZ_{M}}
		\ssmove{1}{0}
		\ssdrop{\ZZ_{M}}
		\ssmove{1}{0}
		\ssdrop{\ZZ_{M_2}}
		\ssmove{1}{0}
		\ssdrop{\ZZ_{M}\oplus \ZZ_{M_2}}
	\end{sseq}
\end{center}
\end{widetext}

Then we note that for $n>1$, the rows $E_2^{2n,\bullet}=E_3^{2n,\bullet}$ can be written as $p c^n + p'y_{2,2} c^{n-1}$, where $p,p'$ are elements in $\H(\PSU(N),\ZZ)$. Since $\dd_3$ vanishes in elements of $\H(\PSU(N),\ZZ)$, we know the full action of $\dd_{3}$. In particular, $\dd_{3}c^2= 2kc x_3$ and $\dd_3 c x_3= kx_3^2$. Then on the line starting at $E_3^{4,0}$, we get the following chain of maps under $\dd_3$:
\[
\begin{tikzcd}
  1 \arrow[r] & \ZZ_{M}\arrow[r, "\q_{2k} "] & \ZZ_{M} \arrow[r, "\q_{k}"] & \ZZ_{(N,2)} \ar[r] & 1.
\end{tikzcd}\]
Let's assume that $N$ is even and $2|k$. Taking the cohomology of the above gives the following groups in the $E_4$ page: 
\begin{equation}
    \begin{split}
        E_{4}^{0,4} &= \ker[\q_{2k}: \ZZ_{M}\rightarrow \ZZ_{M}], \\ 
        E_{4}^{3,4} &= \ZZ_{(M,2k)},\\
        E_{4}^{6,0} &= \ZZ_2.
    \end{split}
\end{equation}
For dimensional reasons, the spectral sequence collapses for $p+q<6$. In degrees $0,1,2,3,5$ there is no extension problem:
\begin{equation}\label{eq:QUcoho}
    \begin{aligned}
    \H^{0}(\QU(kM,M),\ZZ) &= \ZZ \\
    \H^{1}(\QU(kM,M),\ZZ) &= \ZZ_{1} \\
    \H^{2}(\QU(kM,M),\ZZ) &= \ZZ_{1} \\
    \H^{3}(\QU(kM,M),\ZZ) &= \ZZ_{k} \\
    \H^{5}(\QU(kM,M),\ZZ) &= \ZZ_{(M,2k)},
\end{aligned}
\end{equation}
while in degree 4 there is an extension problem involving $\ZZ,\ZZ_{M}, E_{4}^{0,4}$. 

Regardless, we can use the fact that for $k=2$ and $M= 4n$, $\QU(8n,4n)\cong \SU(8n)/\ZZ_2$. In this case 
\[
\H^5(\SU(8n)/\ZZ_2,\ZZ)\cong \ZZ_4.
\]
Indeed, this is not of the form $\Z_2^m$, so the criterion which guarantees a $c_-=4$ solution does not apply when $G_f = \SU(8n)^f$.

We can solve the extension problem in degree 4 by looking at the spectral sequence associated to the fibration
\[
  \mathrm{B}\SU(kM)
\rightarrow \mathrm{B}\QU(kM,M)\rightarrow \mathrm{B}^2\ZZ_{k}. 
\]

Note that $\Hc(\Bc\SU(N),\ZZ)$ is a polynomial ring in the Chern classes $c_{i}$ $i=2,\dots, N$. $c_i$ has degree $2i$ and $\Hc(\Bc^2\Z_k,\ZZ)$ can be found in App.~\ref{app:FactorsB2A}. Therefore, the $E_2$ page looks as below: 
\begin{widetext}
\begin{center}
		\begin{sseq}[grid=chess,labelstep=1,entrysize=1.5cm]{0...5}{0...5}
		
		\ssmove{0}{0}
		\ssdrop{\ZZ}
		\ssmove{1}{0}
		\ssdrop{}
		\ssmove{1}{0}
		\ssdrop{}
		\ssmove{1}{0}
		\ssdrop{\ZZ_{k}}
		\ssmove{1}{0}
		\ssdrop{}
		\ssmove{1}{0}
		\ssdrop{\ZZ_{2k}}
        
		\ssmove{-5}{4}
		\ssdrop{\ZZ}
		\ssmove{1}{0}
		\ssdrop{}
		\ssmove{1}{0}
		\ssdrop{}
		\ssmove{1}{0}
		\ssdrop{\ZZ_{k}}
		\ssmove{1}{0}
		\ssdrop{}
		\ssmove{1}{0}
		\ssdrop{\ZZ_{2k}}
	\end{sseq}
\end{center}
\end{widetext}
We only care about degrees $p+q<6$, so the only possible non-trivial differential is $\dd_{4}:E_4^{0,4}\rightarrow E_4^{5,0}$. $\dd_4$ corresponds to multiplication by some integer such that the cokernel is $\ZZ_{(M,2k)}$ because of our previous calculation. We see that there is a single $\Z$ on the $p+q=4$ line, which solves the extension problem we had before. 

Set $k=2$ and $M=4n$. The result of the cohomology calculation is
\begin{equation}\label{eq:SUcoho}
    \begin{split}
         \H^0(\SU(8n)/\ZZ_2,\ZZ)&=\ZZ\\
    \H^1(\SU(8n)/\ZZ_2,\ZZ)&=\ZZ_1\\
    \H^2(\SU(8n)/\ZZ_2,\ZZ)&=\ZZ_1\\
    \H^3(\SU(8n)/\ZZ_2,\ZZ)&=\ZZ_2\\
    \H^4(\SU(8n)/\ZZ_2,\ZZ)&=\ZZ\\
    \H^5(\SU(8n)/\ZZ_2,\ZZ)&=\ZZ_4.
    \end{split}
\end{equation}
Now if $\l_2\in \H^2(\SU(8n)/\ZZ_2,\ZZ_2)$ is a generator, then $\Pmc(\l_2)/4$ generates $\H^4(\SU(8n)/\ZZ_2,\U(1))$ (see App.~\ref{app:FactorsB2A}). Thus $\frac{1}{2}\lambda^2_2= 2\frac{\mathcal{P}(\lambda_2)}{4}$ is non-trivial as a class in $\H^4(\SU(8)/\Z_2,\U(1))$. From here we can show that the $\mathcal{O}_4$ obstruction is non-trivial for $G_f=\SU(8n)^f$ with $c_-=4\mod 8$, because in that case $\om =\lambda$ and $\Omc_4 = \l^2/2$.

\subsection{Symplectic groups $G_f=\Sp(n)^f$}\label{app:SymplecticGroup}
We consider the symplectic groups $G_f=\Sp(n)^f$, with the normalization $\Sp(1)=\SU(2)$. 

We need to calculate the group cohomology of $G_b=\PSp(n)=\Sp(n)/\ZZ_2$ using the fibration
\[
\begin{tikzcd}
     \Bc\Sp(n) \arrow{r}{\pi} & \Bc\Sp(n) \arrow[r] & \Bc^2\ZZ_2
\end{tikzcd}\]
It is known that $\Hc(\Bc\Sp(n),\ZZ)=\ZZ[x_1,\dots,x_n]$ with $\deg(x_i)=4i$. Then the only additional information we need is the differential $d_4:E_4^{0,4}\cong\ZZ\rightarrow E_4^{0,5}\cong \ZZ_4$. If $\dd_4 p_1 = k_1 \beta_4\Pmc(\ww_2)$, we find that $\H^4(\PSp(n),\ZZ)$ is generated by a class $r_1$ that satisfies $\pi^* r_1 = k_2 p_1 $ and $k_2 k_1 = 0 \mod 4$; this comes from an edge morphism argument. From Ref.~\cite{180173}, we find that one needs $4|nk_2$ or equivalently $4/\gcd(4,n)|k_2$. This then implies that $k_1= \pm \gcd(n,4) \mod 4$. We conclude that

\begin{equation}
    \begin{split}
   \Hc^d(\Bc\PSp(n),\ZZ)
        &=\begin{cases}
        \ZZ, \quad &d=0,4\\
        *,\quad &d=1,2 \\
        \ZZ_2,\quad & d=3,6 \\
        \ZZ_{(n,4)},\quad & d= 5
        \end{cases}
    \end{split}
\end{equation}
where the generators are $r_1, W_3=\d \ww_2, W_3^2, \beta_4 \Pmc(\ww_2)$ with $(n,4)\beta_4 \Pmc(\ww_2)=0$.

As a sanity check, for $n=2$ we have $\Sp(2)=\Spin(5)$ and $\PSp(2)=\SO(5)$. This matches our calculations in App.~\ref{app:SON}. Also note that $\Sp(4)^f\subset \SU(8)^f$; indeed, from the above result we see that the $\H^5(\PSU(8),\Z) \cong \Z_4$ obstruction is detected by restriction to $\PSp(4)$.

\subsection{\texorpdfstring{$G_f=\DD_8^f\circ \QQ_8^f, G_b=\ZZ_2^4$}{D8.H8}}\label{app:finite}

In App.~\ref{app:SomeGfAdmitc4} we discussed how to define invertible phases with $4|c_-$ in a large class of symmetry groups $G_f$ where $\om$ is nontrivial. Here we discuss a finite group which admits a solution only when $c_-  = 0\mod 4$. 

Consider $G_b = \prod_{i=1}^{4}\ZZ_2^{(i)}$ where each factor is generated by $x_i$. Let $a_i \in \H^1(\ZZ_2^{(i)},\ZZ_2)$ be the projection to the $i$-th component, {i.e.} $a_i(x_j) = \delta_{ij}$. Take
\[
\om = (a_1^2+a_1 a_2+a_2^2) +a_3a_4
\]
which is the second SW class of the (orientable)  representation \footnote{\begin{align*}
    \ww_{t}(\rho_1)&=\ww_t(A_{1000})\ww_t(A_{0100})\ww_t(A_{1100})\\
&= (1+a_1t)(1+a_2t)(1+(a_1+a_2)t)\\ 
&= (1+(a_1+a_2)t+a_1a_2t^2)(1+(a_1+a_2)t) \\
&= (1+2(a_1+a_2)t+((a_1+a_2)^2+a_1a_2)t^2) \mod t^3 
\end{align*}}
\begin{align}
\rho &= \rho_1 \oplus \rho_{2} \\
    \rho_{1} &= \left(A_{1000} \oplus A_{0100} \oplus A_{1100}\right) \\
    \rho_{2} &= \left(A_{0010} \oplus A_{0001} \oplus 3 A_{0011}\right)
\end{align}
where $A_{q_1\dots q_4}$ denotes the 1d rep with charge $q_i$ under $\ZZ_2^{(i)}$. 
Now as $\Sq^1(\om)=a_1a_2(a_1+a_2) + a_3a_4(a_3+a_4)  \neq n_1 \om$ for any $n_1\in \H^1(G,\ZZ_2)$ we must have $c_-  = 0\mod 2$. Also as $G_b=\ZZ^4_2$ is abelian, we must have $n_1=0$. 

Below we show that there is no solution when $c_-=2 \mod 4$. Using again the arguments from Appendix \ref{app:SON}, the $\Omc_4$ obstruction for $c_-=2 \mod 4$ reduces to checking whether 

\[
\OO_4 =\frac{1}{2} n_1 \om \pm \frac{\mathcal{P}(\om)}{4}= \frac{1}{2}(n_2\om + \ww_4) \mod 1
\]
is trivial in $\H^4(G,\U(1))$, where $\ww_4$ is the 4th SW class of the representation $\rho$. We can show by brute force expansion that
\begin{equation}
    \begin{split}
        \ww_4(\rho) &=\ww_2(\r_1)\ww_2(\r_2)+\ww_4(\r_2)\\
        &= (a_1^2+a_1 a_2+a_2^2)a_3a_4 \\
        &\,\,+a_3^4+a_4^4+a_3a_4(a_3^2+a_3a_4+a_4^2).
    \end{split}
\end{equation}
Then we consider the most general form of $n_2$:
\begin{equation}
   n_2=A+B+C+\ww_2(\rho_1), 
\end{equation}
with $A\in\H^2(\ZZ_2^{(1)}\times \ZZ_2^{(2)},\ZZ_2)$ ,
$B\in\H^1(\ZZ_2^{(1)}\times \ZZ_2^{(2)},\ZZ_2)\cup \H^1(\ZZ_2^{(3)}\times \ZZ_2^{(4)},\ZZ_2)$ ,$C\in\H^2(\ZZ_2^{(3)}\times \ZZ_2^{(4)},\ZZ_2)$. After substituting, we find that the obstruction is non-trivial if 
\[
\mathcal{I}=\Sq^1((A+B+C)\om + \ww_2(\r_1)^2+\ww_4(\r_2)) 
\]
is non-trivial in $\H^5(\ZZ_2^4,\ZZ_2)$. The term in the middle vanishes, so next we need to constrain $A,B,C$. 

By restricting to $\ZZ_2^{(3)}\times\ZZ_2^{(4)}$ we see that we must have 
\[
\Sq^1(C a_3a_4)=a_3a_4(a_3+a_4)^3
\]
which implies that $C=a_3^2+a_4^2 + C_0 a_3a_4$. Next, restricting to $\ZZ_2^{(1)}\times\ZZ_2^{(2)}$ gives
\[
\Sq^1(A (a_1^2+a_1a_2+a_2^2))=0
\]
which implies that $A=A_0 (a_1^2+a_1a_2+a_2^2)$. But as $n_2 \rightarrow n_2 + \om $ does not change the obstruction we can set $A_0=0$. Now we look at the terms with degree $4$ in $\H^\bullet(\ZZ_2^{(1)}\times \ZZ_2^{(2)})$ and degree 1 in $\H^\bullet(\ZZ_2^{(3)}\times \ZZ_2^{(4)})$. They come from
\[
\Sq^1(B (a_1^2+a_1a_2+a_2^2)). 
\]
Writing $B=a_1b_1+a_2b_2$, with $b_i\in\H^1(\ZZ_2^{(A)}\times\ZZ_2^{(4)},\ZZ)$, the above expands to (we keep only terms of the right degree)
\[
 (a_1^2b_1+a_2^2b_2)(a_1^2+a_1a_2+a_2^2)+ (a_1b_1+a_2b_2)a_1a_2(a_1+a_2).
\]
We thus see that $b_1=b_2=0$ because they are the coefficients of $a_1^4$ and $a_2^4$. Thus we must have $B=0$. 

Putting this together, we have
\begin{align}
    \mathcal{I} &= \Sq^1( C \om) + a_3a_4(a_3+a_4)^3 \\
    &= \Sq^1(C (a_1^2+a_1a_2+a_2^2)) \\
    &= C a_1a_2(a_1+a_2) + \Sq^1(C)(a_1^2+a_1a_2+a_2^2)
\end{align}
which is non-trivial in $\H^5(G_b,\ZZ_2)$ because, for example, the coefficient of $a_3^2a_1^2a_2$ is non-zero. We conclude that when $c_-=2 \mod 4$, the obstruction is nonvanishing. Thus $c_-$ must be a multiple of 4.

The $G_f$ constructed this way is the central product of $\QQ_8$ and $\DD_8$. From this observation, one can see that the only fermionic representation has (complex) dimension $4$ and is quaternionic so it has dimension $8$ as a real representation. This gives some insight into why $c_-$ must be a multiple of 4.

\subsection{Calculations for Section \ref{sec:dirprod}} \label{app:dirprod}

In Sec.~\ref{sec:dirprod} we argued that when $G_b = G_b^A \times G_b^B$ and $\om^A, \om^B$ are nontrivial, the terms $\om^A \cup_1 \om^A$ and $\om^B \cup_1 \om^B$ should be trivial in $\H^3(G_b,\Z_2)$. We now show that the $\mathcal{O}_4$ obstruction gives a stronger constraint $\beta \om^A = \beta\om^B=0$ where $\beta: \H^2(G_b,\Z_2) \rightarrow \H^3(G_b,\Z)$, {i.e.} both $\om^{A},\om^B$ must have integral lifts.

The proof is as follows. Since $\Sq^1\om^{A/B}=0$, they admit lifts to $\ZZ_4$, say $u_2^{A/B}$. The obstruction for $c_-=1$ is $\Omc_4=\Pmc(u_2^A+u_2^B)/8=\Pmc(u_2^A)/8+ \Pmc(u_2^B)/8+u_2^Au_2^B/4$. If $\beta_4 u_2^A\neq 0$, then  $u_2^Au_2^B/4$ represents an element of $\H^2(G_B,\H^2(G_A,\U(1)))$ that is not trivial because $u_2^B$ is not trivial. Therefore $\beta_4 u_2^A=0$. By repeating the argument with $A$ and $B$ exchanged, we must also have $\beta_4 u_2^B =0$. This implies that $\beta \om^A = \beta\om^B=0$, as claimed.

\subsection{Calculations for \texorpdfstring{$G_f = \Z_n \times \Z_2^{\Tt} \times \Z_2^f$}{ Gf = Zn x Z2T x Z2f}}\label{app:znz2T}
For this group, $G_b = \Z_n \times \Z_2^{\Tt}$. First note that if $n_1 = s_1$, we have $dn_2 = s_1^3$, which is obstructed. Thus a nontrivial $n_1$ is possible only if it is nonzero on the $\Z_n$ subgroup. This requires that $n$ be even. 

Let $n_1 = w_1$ where $w_1$ generates $\H^1(\Z_n, \Z_2)$. We can solve the equation $dn_2 = w_1 s_1 w_1$ only if $n$ is a multiple of 4. In this case, an explicit expression for $n_2$ is 
\begin{equation}\label{eq:f1}
    n_2 = w_1 \cup (w_1 \cup_1 s_1) + b_1 \cup s_1
\end{equation}
where $b_1$ satisfies $db_1 = w_1 \cup w_1 \mod 2$. Let ${\bf h}$ generate the $\Z_n$ subgroup. If ${\bf h}^i{\bf T}^j$ is a general group element of $G_b$, an explicit expression for $b_1$ is
\begin{equation}
    b_1({\bf h}^i{\bf T}^j) = \frac{i(i+1)}{2} \mod 2. 
\end{equation}
Note that $b_1$ cannot be defined in this way when $n = 4k+2$ for integer $k$, because then we would have $b_1({\bf h}^n) \ne b_1({\bf h}^0)$.

The most general expressions for $s_1,n_1, n_2$ are thus
\begin{align}
    s_1({\bf h}^i{\bf T}^j) &= j \mod 2 \\
    n_1({\bf h}^i{\bf T}^j) &= i \mod 2 \\
    n_2({\bf h}^i{\bf T}^j,{\bf h}^k{\bf T}^l) &= \left(i k + \frac{i(i+1)}{2}\right) l \mod 2.
\end{align}
Finally, we evaluate the obstruction invariants associated to this choice of $n_1$ and $n_2$, assuming $n$ is a multiple of 4. A general expression for the obstruction $\mathcal{O}_4$ in the case of antiunitary symmetries is given in Ref.~\cite{barkeshli2021invertible}. Specializing to $\om = 0$, the expression is
\begin{widetext}
\begin{align}\label{eq:s1anom}
    \mathcal{O}_4({\bf g},{\bf h},{\bf k},{\bf l}) &= \frac{1}{2}n_2({\bf g},{\bf h}) (n_2({\bf k},{\bf l}) + s_1({\bf k}) n_1({\bf l})) + \frac{1}{2} n_1({\bf g})n_1({\bf h})s_1({\bf h})s_1({\bf k})n_1({\bf k})n_1({\bf l}) + \frac{1}{4}n_1({\bf g})n_1({\bf h})s_1({\bf k})n_1({\bf l}) \nonumber \\ &+ \frac{1}{2} s_1({\bf g}) n_1({\bf h}) n_2({\bf k},{\bf l}) \mod 1.
\end{align}
\end{widetext}
We compute the obstruction invariants\footnote{These invariants are complete because $\H^4(\ZZ_n\times\ZZ_2^{\Tt},\U(1)^{\Tt}) = \ZZ_2^3$ generated by $\frac{1}{2}s_1^4, \frac{1}{2}\ww_2s_1^2, \frac{1}{2}\ww_2^2$ where $\ww_2$ is the generator of $\H^2(\Z_n,\ZZ_2)$. 
\begin{equation}
    \begin{array}{c||c|c|c}
         & \frac{1}{2}s_1^4& \frac{1}{2}\ww_2s_1^2& \frac{1}{2}\ww_2^2 \\\hline\hline
      \mathcal{I}_4^{(1)}   & \frac{1}{2} &0 &0  \\
      \mathcal{I}_4^{(2)} & \frac{1}{2} & \frac{1}{2} & \frac{1}{2}\\
      \mathcal{I}_4^{(3)}&  0& \frac{1}{2} & 0
    \end{array}
\end{equation}
}
\def\gT{\gbf\Tt}
\begin{align}
\mathcal{I}_4^{(1)} &:= \mathcal{O}_4({\bf T},{\bf T},{\bf T},{\bf T}) \\
    \mathcal{I}_4^{(2)} &:= \mathcal{O}_4({\bf gT},{\bf gT},{\bf gT},{\bf gT}) \\
    \mathcal{I}_4^{(3)} &:= 
    \Omc_4(\gT,\gT,\Tt,\Tt)
    +\Omc_4(\gT,\Tt,\Tt,\gT) \nonumber \\
    &+\Omc_4(\Tt,\Tt,\gT,\gT)
    -\Omc_4(\gT,\Tt,\gT,\Tt)
    \nonumber \\
    &-\Omc_4(\Tt,\gT,\Tt,\gT)+\Omc_4(\Tt,\gT,\gT,\Tt)
\end{align}
where ${\bf g}$ is the order 2 element of $\Z_n$. Note that $n_1({\bf T}) = n_1({\bf gT}) = 0$, so only the first term in the obstruction (Eq.~\ref{eq:s1anom}) can contribute. Next, note that 
\begin{align}
    n_2({\bf T},{\bf T}) = n_2({\bf T},{\bf gT}) &= 0; \nonumber \\
    n_2({\bf gT},{\bf T}) = n_2({\bf gT},{\bf gT}) &= \frac{n}{4} \mod 1. \nonumber
\end{align}
After substituting and simplifying, we find that
\begin{align}
    \mathcal{I}^{(1)} &= 0 \\
    \mathcal{I}^{(2)} &= \frac{1}{2} \frac{n^2}{16} \mod 1 \\
    \mathcal{I}^{(3)} &=  \frac{1}{2} \frac{n^2}{16} \mod 1.
\end{align}
Now observe that if $n = 8k+4$, $\mathcal{I}^{(2)} =\mathcal{I}^{(3)} = \frac{1}{2}$, so the above choice of $n_2$ is obstructed. On the other hand, if $n = 8k$, all the invariants are trivial. Therefore we indeed have a solution with $n_1 \ne 0$ when $8|n$.

\subsection{Calculations for $G_f = \Z_{2n}^{\bf T} \times \Z_2^f$}
Here $G_b = \Z_{2n}^{\bf T}$. Let ${\bf h}$ be the generator of $G_b$. We set $n_1 = s_1$, where $s_1$ generates $\H^1(G_b,\Z_2)$. Since $\om = 0$, Eq.~\eqref{eq:eta_AU} gives
\begin{equation}
    dn_2 = s_1^3
\end{equation}
and if $n$ is even, this can be solved by setting $n_2 = b_1 s_1$, where 
\begin{equation}
    b_1({\bf h}^k) = \frac{k(k+1)}{2}
\end{equation}
for $k \in \Z_{2n}$. For odd $n$, there is no solution. 

Let us compute $\mathcal{O}_4$. The required expression is identical to Eq.~\eqref{eq:s1anom}. Defining ${\bf g} = {\bf h}^n$ as the order 2 element of $G_b$, the $\H^4(\Z_{2n},U(1))$ invariant is
\begin{align}
    \mathcal{I} &:= \mathcal{O}_4({\bf h},{\bf g},{\bf g},{\bf g}) + \mathcal{O}_4({\bf g},{\bf g},{\bf h},{\bf g}) \nonumber \\ &\quad - \mathcal{O}_4({\bf g},{\bf h},{\bf g},{\bf g}) - \mathcal{O}_4({\bf g},{\bf g},{\bf g},{\bf h}) \nonumber \\
    & \quad + \mathcal{O}_4({\bf g},{\bf g},{\bf g},{\bf g}).
\end{align}
Note that $b_1({\bf h}) = 1, b_1({\bf g}) = \frac{n}{2}$, and $s_1({\bf h}) = 1, s_1({\bf g}) = 0$. Thus
\begin{align*}
    n_2({\bf h},{\bf g}) = n_2({\bf g},{\bf g}) &= 0 \\
    n_2({\bf h},{\bf h}) &= 1 \\
    n_2({\bf g},{\bf h}) &= \frac{n}{2}.
\end{align*}
With this we have, for all even $n$,
\begin{align}
    \mathcal{I} = 0.
\end{align}
Thus there is always an unobstructed state with $n_1 = s_1$. Let us check other values of $n_2$. Suppose we consider $n_2' = n_2 + \ww_2$, where $\ww_2$ generates $\H^2(\Z_{2n},\Z_2)$. For this $n_2'$ we find that $\mathcal{I} = \frac{1}{2}$, implying that this choice of $n_2'$ is obstructed. If we first gauge the fermion parity of the invertible phase, and take $n=2$, the above results are consistent with calculations on SET phases with $G_b = \Z_4^{\bf T}$ performed in Ref.~\cite{Barkeshli2020Anomaly}.

\section{Wreath products}\label{app:WreathProducts}

In order to generalize the $\Oo(4)^f$ example we consider a wreath product defined as 
\[
H \wr \ZZ_2 = (H_L\times H_R)\rtimes_{\pi} \ZZ_2,
\]
where $H_L\cong H_R$ are isomorphic groups and the generator $C$ of $\Z_2$ permutes the two subgroups. 

Corollary 3.2 of Ref.~\cite{quillen1971adams} shows that the $\ZZ_2$-cohomology is detected by restriction to $H_L\times H_R$ and $H_{\Delta} \times \Z_2$. Here $H_{\Delta}$ is the diagonal $H$ subgroup. Denote the map  $j^*: \H^\bullet(H\wr \ZZ_2 ,\ZZ_2) \rightarrow \H^\bullet(H_L\times H_R ,\ZZ_2)\times \H^\bullet(H_{\Delta}\times\ZZ_2 ,\ZZ_2) $ which sends classes to their respective restrictions. 

Then whenever we have classes such that $j^* \mu = (\mu',0)$ and $j^* \nu = (0,\nu')$ we must have $\mu \nu=0$ because the image is zero in both factors.

In particular, whenever $j^*\om = (w_2^L+w_2^R,0)$ there is a solution for the $\Omc_3$ equation for $n_1=c_1$, where $c_1$ is the projection to the $\ZZ_2$ factor.

\def\Kc{\mathrm{K}}
\section{Cohomology of $\Bc^2A \cong \Kc(A,2)$}\label{app:FactorsB2A}
Here we review some facts about the cohomology of $\Bc^2A$, or equivalently, the Eilenberg-MacLane space $\Kc(A,2)$, for $A$ a finite abelian group. The singular cohomology is
\begin{equation}
    \Hc^d(\Kc(A,2),\Z) =\begin{cases}
    \Z \quad &,d=0\\
    * \quad &,d=1,2,4\\
    A\quad &, d=3 \\
    \Gamma(A)\quad &, d=5 \\
    \Lambda(A)\quad &, d=6 \\
    \end{cases}
\end{equation}
where $\Gamma(A)$ is the universal quadratic group of $A$ and $\Lambda(A)$ is the exterior square of $A$. If $A$ decomposes into cyclic subgroups as $A=\bigoplus_{j=1}^M \Z_{n_j}$, then 
\begin{equation}
    \begin{split}
        \Gamma(A)&\cong\bigoplus_{1\le i \le M}  \Z_{(2,n_i)n_i} \oplus \bigoplus_{1\le i<j \le M} \Z_{(n_i,n_j)}\\
    \Lambda(A)&\cong\bigoplus_{1\le i \le M}  \Z_{(2,n_i)} \oplus \bigoplus_{1\le i<j \le M} \Z_{(n_i,n_j)}
    \end{split}
\end{equation}
where $(n,m)$ denotes the greatest common divisor of the two positive integers $n$ and $m$.

Given a decomposition $A=\bigoplus_{j=1}^M \Z_{n_j}$, a basis for the elements of degree $d$ is
    \begin{align*}
	\begin{array}{|c||c|c|c|c|c|c|}
		\hline
		d & \text{Basis} & \text{Range} &\text{Order} \\ \hline\hline
		3 & W_{3;i} & 1\le i \le M & n_i\\\hline
		\multirow{ 2}{*}{5} & W_{5;i} & 1\le i \le M & (2,n_i) n_i\\
		 & W_{5;ij}& 1\le i < j\le M & (n_i,n_j)\\\hline
		\multirow{ 2}{*}{6} & W_{3;i}^2& 1\le i\le M & (2,n_i) \\
		&W_{3;i}W_{3;j}&1\le i<j\le M& (n_i,n_j)\\
		\hline
	 \hline
	\end{array}
\end{align*}
We can write the basis elements in terms of Bockstein maps of "fundamental classes". For each $j$, consider the subgroup of $\Hc^2(\Kc(A,2),\Z_{n_j}) \cong\Tor[A, \Z_{n_j}]$ that survives upon restricting $A\to \Z_{n_j}$. A fundamental class is a generator of this subgroup; we denote each one by $w_{2;j}$. We then have 
\begin{equation}
    \begin{split}
        W_{3;i}&=\beta_{n_i}[w_{2;i}], \\
        W_{5;i}&=\beta_{(n_i,2)n_i}[\Pmc(w_{2;i})], \\
        W_{5;ij}&=\beta_{(n_i,n_j)}[w_{2;i}w_{2;j}]. 
    \end{split}
\end{equation}
\subsection{Justification}
The above results can be extracted from Refs.~\cite{eilenberg1954groups,kapustin2014topological, Gu_2019}. First, Eq.~6.6 of Ref.~\cite{eilenberg1954groups} tells us that 
\begin{equation}
    \Kc(\bigoplus_{j=1}^{M} \ZZ_{n_j} ,2 ) \cong \Kc( \ZZ_{n_1} ,2 ) \times \cdots \times \Kc( \ZZ_{n_M} ,2 ).
\end{equation}
Then, the calculation of $\Hc^*(\Kc(\bigoplus_{j=1}^{M} \ZZ_{n_j} ,2 ),\Z)$ reduces to the calculation of $\Hc^*(\Kc(\Z_n ,2 ),\Z)$, and then an application of the K\"unneth theorem.

Next, from section 21 and 22 of Ref.~\cite{eilenberg1954groups}, we can extract ($\Hc_{d}(\Kc(\Pi,2),\Z)$ is $ H_{d}(\Pi,2)$ in their notation): 
\begin{equation}
    \Hc_{d}(\Kc(\Z_n,2),\Z) =
    \begin{cases}
    \Z \quad &,d=0\\
    * \quad &,d=1,3\\
    \Z_{n} \quad &,d=2\\
    \Z_{(n,2)n} \quad &,d=4\\
    \Z_{(n,2)} \quad &,d=5 \\
    \text{torsion} \quad &,d=6
    \end{cases}
\end{equation}
for $n$ a positive integer. $\Z_1\cong *$ is the trivial group. Then, the UCT gives \begin{equation}
    \Hc^{d}(\Kc(\Z_n,2),\Z) =
    \begin{cases}
    \Z \quad &,d=0\\
    * \quad &,d=1,2,4\\
    \Z_{n} \quad &,d=3\\
    \Z_{(n,2)n} \quad &,d=5\\
    \Z_{(n,2)} \quad &,d=6.
    \end{cases}
\end{equation}

In order to determine the generators, we can use the fibration 
\begin{equation}
    \begin{tikzcd}
         \Kc(\Z,2) \arrow[r]& \Kc(\Z_n,2) \arrow{r}{\chi} &\Kc(\Z,3)
    \end{tikzcd}
\end{equation}
and the ring structure of $\Hc^*(\Kc(\Z,3),\Z)$. According to \textbf{Corollary 2.15} of Ref.~\cite{Gu_2019}, we have 
\begin{equation}
    \Hc^d(\Kc(\Z,3),\Z) = \begin{cases}
    \Z \quad &, d=0,3 \\
    * \quad &, d=1,2,4,5 \\
    \Z_2 \quad &, d=6
    \end{cases}
\end{equation}
where $x_1$ generates $\Hc^3(\Kc(\Z,3),\Z) \cong \Z$ and $\Hc^6(\Kc(\Z,3),\Z) \cong\Z_2$ is generated by $x_1^2 =y_{2;0}$. As $\Kc(\Z,2)\cong \mathbb{CP}^{\infty}$, its cohomology is concentrated in even degree and generated by a class $v \in \Hc^2(\Kc(\Z,2),\Z)$.

The first differential of the above spectral sequence is $\dd_3$. Because $\Hc^2(\Kc(\Z_n,2),\Z)=*$ and $\Hc^3(\Kc(\Z_n,2),\Z)\cong \Z_n$, we must have $\dd_3 v = n x_1$. Thus $\dd_3 v^2 = 2n x_1 v$ and $\dd_3 vx_1 = nx_1^2$. 

Then $E_3^{0,2}$ and $E_3^{0,4}$ are trivial. Because $2x_1^2=0$, the kernel of $\dd_3:E_3^{3,2} \to E_3^{0,6}$ is $\frac{2}{(n,2)}\Z$ and the image is $n\Z_2$. Then $E_4^{0,6}\cong \Z_{(2,n)} $ and $E_4^{2,3} \cong \Z_{(n,2)n}$. As there are no more possible differentials, the spectral series collapses for $p+q\le 6$. In particular, the generator of $\Hc^3(\Kc(\Z_n,2),\Z)$($\Hc^6(\Kc(\Z_n,2),\Z)$) is the pull-back $\chi^* x_1$ ($\chi^* x_1^2$), (here $\chi$ is the map in the fibration). We denoted $W_3:=\chi^*x_1$ above.

Then the generator of $\Hc^5(\Kc(\Z_n,2),\Z)$ is $v x_3$. We cannot use the edge morphisms to identify these classes. Instead, we use the fact that if $w_2$ generates $\Hc^2(\Kc(\Z_n,2),\Z_n)\cong\Hom(\Z_n,\Z_n) $ then $\Hc^4(\Kc(\Z_n,2),\Z_{(n,2)n})\cong \Hom(\Z_{(n,2)n},\Z_{(n,2)n})$ is generated by $\Pmc(w_2)$ (See Ref.~\cite{kapustin2014topological} for a discussion of this point). Applying the relevant Bockstein map gives the cited relations.

\bibliography{D8phase_refs}

\begin{thebibliography}{94}%
\makeatletter
\providecommand \@ifxundefined [1]{%
 \@ifx{#1\undefined}
}%
\providecommand \@ifnum [1]{%
 \ifnum #1\expandafter \@firstoftwo
 \else \expandafter \@secondoftwo
 \fi
}%
\providecommand \@ifx [1]{%
 \ifx #1\expandafter \@firstoftwo
 \else \expandafter \@secondoftwo
 \fi
}%
\providecommand \natexlab [1]{#1}%
\providecommand \enquote  [1]{``#1''}%
\providecommand \bibnamefont  [1]{#1}%
\providecommand \bibfnamefont [1]{#1}%
\providecommand \citenamefont [1]{#1}%
\providecommand \href@noop [0]{\@secondoftwo}%
\providecommand \href [0]{\begingroup \@sanitize@url \@href}%
\providecommand \@href[1]{\@@startlink{#1}\@@href}%
\providecommand \@@href[1]{\endgroup#1\@@endlink}%
\providecommand \@sanitize@url [0]{\catcode `\\12\catcode `\$12\catcode
  `\&12\catcode `\#12\catcode `\^12\catcode `\_12\catcode `\%12\relax}%
\providecommand \@@startlink[1]{}%
\providecommand \@@endlink[0]{}%
\providecommand \url  [0]{\begingroup\@sanitize@url \@url }%
\providecommand \@url [1]{\endgroup\@href {#1}{\urlprefix }}%
\providecommand \urlprefix  [0]{URL }%
\providecommand \Eprint [0]{\href }%
\providecommand \doibase [0]{https://doi.org/}%
\providecommand \selectlanguage [0]{\@gobble}%
\providecommand \bibinfo  [0]{\@secondoftwo}%
\providecommand \bibfield  [0]{\@secondoftwo}%
\providecommand \translation [1]{[#1]}%
\providecommand \BibitemOpen [0]{}%
\providecommand \bibitemStop [0]{}%
\providecommand \bibitemNoStop [0]{.\EOS\space}%
\providecommand \EOS [0]{\spacefactor3000\relax}%
\providecommand \BibitemShut  [1]{\csname bibitem#1\endcsname}%
\let\auto@bib@innerbib\@empty
\bibitem [{\citenamefont {Read}\ and\ \citenamefont {Green}(2000)}]{read2000}%
  \BibitemOpen
  \bibfield  {author} {\bibinfo {author} {\bibfnamefont {N.}~\bibnamefont
  {Read}}\ and\ \bibinfo {author} {\bibfnamefont {D.}~\bibnamefont {Green}},\
  }\href {https://doi.org/10.1103/PhysRevB.61.10267} {\bibfield  {journal}
  {\bibinfo  {journal} {Phys. Rev. B}\ }\textbf {\bibinfo {volume} {61}},\
  \bibinfo {pages} {10267} (\bibinfo {year} {2000})}\BibitemShut {NoStop}%
\bibitem [{\citenamefont {Kitaev}(2001)}]{kitaev2001}%
  \BibitemOpen
  \bibfield  {author} {\bibinfo {author} {\bibfnamefont {A.~Y.}\ \bibnamefont
  {Kitaev}},\ }\href@noop {} {\bibfield  {journal} {\bibinfo  {journal}
  {Physics-Uspekhi}\ }\textbf {\bibinfo {volume} {44}},\ \bibinfo {pages} {131}
  (\bibinfo {year} {2001})}\BibitemShut {NoStop}%
\bibitem [{\citenamefont {Nayak}\ \emph {et~al.}(2008)\citenamefont {Nayak},
  \citenamefont {Simon}, \citenamefont {Stern}, \citenamefont {Freedman},\ and\
  \citenamefont {Sarma}}]{nayak2008}%
  \BibitemOpen
  \bibfield  {author} {\bibinfo {author} {\bibfnamefont {C.}~\bibnamefont
  {Nayak}}, \bibinfo {author} {\bibfnamefont {S.~H.}\ \bibnamefont {Simon}},
  \bibinfo {author} {\bibfnamefont {A.}~\bibnamefont {Stern}}, \bibinfo
  {author} {\bibfnamefont {M.}~\bibnamefont {Freedman}},\ and\ \bibinfo
  {author} {\bibfnamefont {S.~D.}\ \bibnamefont {Sarma}},\ }\href@noop {}
  {\bibfield  {journal} {\bibinfo  {journal} {Rev. Mod. Phys.}\ }\textbf
  {\bibinfo {volume} {80}},\ \bibinfo {pages} {1083} (\bibinfo {year}
  {2008})}\BibitemShut {NoStop}%
\bibitem [{\citenamefont {Hasan}\ and\ \citenamefont {Kane}(2010)}]{hasan2010}%
  \BibitemOpen
  \bibfield  {author} {\bibinfo {author} {\bibfnamefont {M.~Z.}\ \bibnamefont
  {Hasan}}\ and\ \bibinfo {author} {\bibfnamefont {C.~L.}\ \bibnamefont
  {Kane}},\ }\href {https://doi.org/10.1103/RevModPhys.82.3045} {\bibfield
  {journal} {\bibinfo  {journal} {Rev. Mod. Phys.}\ }\textbf {\bibinfo {volume}
  {82}},\ \bibinfo {pages} {3045} (\bibinfo {year} {2010})}\BibitemShut
  {NoStop}%
\bibitem [{\citenamefont {Qi}\ and\ \citenamefont {Zhang}(2011)}]{qi2010RMP}%
  \BibitemOpen
  \bibfield  {author} {\bibinfo {author} {\bibfnamefont {X.-L.}\ \bibnamefont
  {Qi}}\ and\ \bibinfo {author} {\bibfnamefont {S.-C.}\ \bibnamefont {Zhang}},\
  }\href {https://doi.org/10.1103/RevModPhys.83.1057} {\bibfield  {journal}
  {\bibinfo  {journal} {Rev. Mod. Phys.}\ }\textbf {\bibinfo {volume} {83}},\
  \bibinfo {pages} {1057} (\bibinfo {year} {2011})}\BibitemShut {NoStop}%
\bibitem [{\citenamefont {Alicea}(2012)}]{alicea2012review}%
  \BibitemOpen
  \bibfield  {author} {\bibinfo {author} {\bibfnamefont {J.}~\bibnamefont
  {Alicea}},\ }\href@noop {} {\bibfield  {journal} {\bibinfo  {journal}
  {Reports on Progress in Physics}\ }\textbf {\bibinfo {volume} {75}},\
  \bibinfo {pages} {076501} (\bibinfo {year} {2012})}\BibitemShut {NoStop}%
\bibitem [{\citenamefont {Sarma}\ \emph {et~al.}(2015)\citenamefont {Sarma},
  \citenamefont {Freedman},\ and\ \citenamefont {Nayak}}]{Dassarma2015mzm}%
  \BibitemOpen
  \bibfield  {author} {\bibinfo {author} {\bibfnamefont {S.~D.}\ \bibnamefont
  {Sarma}}, \bibinfo {author} {\bibfnamefont {M.}~\bibnamefont {Freedman}},\
  and\ \bibinfo {author} {\bibfnamefont {C.}~\bibnamefont {Nayak}},\ }\href
  {https://doi.org/10.1038/npjqi.2015.1} {\bibfield  {journal} {\bibinfo
  {journal} {npj Quantum Information}\ }\textbf {\bibinfo {volume} {1}},\
  \bibinfo {pages} {15001} (\bibinfo {year} {2015})}\BibitemShut {NoStop}%
\bibitem [{\citenamefont {Sato}\ and\ \citenamefont
  {Ando}(2017)}]{Sato2017tsc}%
  \BibitemOpen
  \bibfield  {author} {\bibinfo {author} {\bibfnamefont {M.}~\bibnamefont
  {Sato}}\ and\ \bibinfo {author} {\bibfnamefont {Y.}~\bibnamefont {Ando}},\
  }\href {https://doi.org/10.1088/1361-6633/aa6ac7} {\bibfield  {journal}
  {\bibinfo  {journal} {Reports on Progress in Physics}\ }\textbf {\bibinfo
  {volume} {80}},\ \bibinfo {pages} {076501} (\bibinfo {year}
  {2017})}\BibitemShut {NoStop}%
\bibitem [{\citenamefont {Sau}\ and\ \citenamefont {Tewari}(2021)}]{sau2021}%
  \BibitemOpen
  \bibfield  {author} {\bibinfo {author} {\bibfnamefont {J.}~\bibnamefont
  {Sau}}\ and\ \bibinfo {author} {\bibfnamefont {S.}~\bibnamefont {Tewari}},\
  }\href@noop {} {\bibfield  {journal} {\bibinfo  {journal} {arXiv preprint
  arXiv:2105.03769}\ } (\bibinfo {year} {2021})}\BibitemShut {NoStop}%
\bibitem [{\citenamefont {Kitaev}(2009)}]{Kitaev2009periodic}%
  \BibitemOpen
  \bibfield  {author} {\bibinfo {author} {\bibfnamefont {A.}~\bibnamefont
  {Kitaev}},\ }\href {https://doi.org/10.1063/1.3149495} {\bibfield  {journal}
  {\bibinfo  {journal} {AIP Conference Proceedings}\ }\textbf {\bibinfo
  {volume} {1134}},\ \bibinfo {pages} {22} (\bibinfo {year} {2009})},\ \Eprint
  {https://arxiv.org/abs/https://aip.scitation.org/doi/pdf/10.1063/1.3149495}
  {https://aip.scitation.org/doi/pdf/10.1063/1.3149495} \BibitemShut {NoStop}%
\bibitem [{\citenamefont {Ryu}\ \emph {et~al.}(2010)\citenamefont {Ryu},
  \citenamefont {Schnyder}, \citenamefont {Furusaki},\ and\ \citenamefont
  {Ludwig}}]{ryu2010}%
  \BibitemOpen
  \bibfield  {author} {\bibinfo {author} {\bibfnamefont {S.}~\bibnamefont
  {Ryu}}, \bibinfo {author} {\bibfnamefont {A.~P.}\ \bibnamefont {Schnyder}},
  \bibinfo {author} {\bibfnamefont {A.}~\bibnamefont {Furusaki}},\ and\
  \bibinfo {author} {\bibfnamefont {A.~W.~W.}\ \bibnamefont {Ludwig}},\
  }\href@noop {} {\bibfield  {journal} {\bibinfo  {journal} {New J. Phys.}\
  }\textbf {\bibinfo {volume} {12}},\ \bibinfo {pages} {065010} (\bibinfo
  {year} {2010})}\BibitemShut {NoStop}%
\bibitem [{\citenamefont {Chen}\ \emph {et~al.}(2013)\citenamefont {Chen},
  \citenamefont {Gu}, \citenamefont {Liu},\ and\ \citenamefont
  {Wen}}]{Chen2013}%
  \BibitemOpen
  \bibfield  {author} {\bibinfo {author} {\bibfnamefont {X.}~\bibnamefont
  {Chen}}, \bibinfo {author} {\bibfnamefont {Z.-C.}\ \bibnamefont {Gu}},
  \bibinfo {author} {\bibfnamefont {Z.-X.}\ \bibnamefont {Liu}},\ and\ \bibinfo
  {author} {\bibfnamefont {X.-G.}\ \bibnamefont {Wen}},\ }\href
  {https://doi.org/10.1103/PhysRevB.87.155114} {\bibfield  {journal} {\bibinfo
  {journal} {Phys. Rev. B}\ }\textbf {\bibinfo {volume} {87}},\ \bibinfo
  {pages} {155114} (\bibinfo {year} {2013})}\BibitemShut {NoStop}%
\bibitem [{\citenamefont {Kapustin}(2014)}]{kapustin2014SPTbeyond}%
  \BibitemOpen
  \bibfield  {author} {\bibinfo {author} {\bibfnamefont {A.}~\bibnamefont
  {Kapustin}},\ }\href@noop {} {\bibfield  {journal} {\bibinfo  {journal}
  {arXiv preprint arXiv:1403.1467}\ } (\bibinfo {year} {2014})}\BibitemShut
  {NoStop}%
\bibitem [{\citenamefont {Kapustin}\ \emph {et~al.}(2015)\citenamefont
  {Kapustin}, \citenamefont {Thorngren}, \citenamefont {Turzillo},\ and\
  \citenamefont {Wang}}]{kapustin2015fSPT}%
  \BibitemOpen
  \bibfield  {author} {\bibinfo {author} {\bibfnamefont {A.}~\bibnamefont
  {Kapustin}}, \bibinfo {author} {\bibfnamefont {R.}~\bibnamefont {Thorngren}},
  \bibinfo {author} {\bibfnamefont {A.}~\bibnamefont {Turzillo}},\ and\
  \bibinfo {author} {\bibfnamefont {Z.}~\bibnamefont {Wang}},\ }\href@noop {}
  {\bibfield  {journal} {\bibinfo  {journal} {Journal of High Energy Physics}\
  }\textbf {\bibinfo {volume} {2015}},\ \bibinfo {pages} {1} (\bibinfo {year}
  {2015})}\BibitemShut {NoStop}%
\bibitem [{\citenamefont {Senthil}(2015)}]{senthil2015}%
  \BibitemOpen
  \bibfield  {author} {\bibinfo {author} {\bibfnamefont {T.}~\bibnamefont
  {Senthil}},\ }\href@noop {} {\bibfield  {journal} {\bibinfo  {journal}
  {Annual Review of Condensed Matter Physics}\ }\textbf {\bibinfo {volume}
  {6}},\ \bibinfo {pages} {299} (\bibinfo {year} {2015})}\BibitemShut {NoStop}%
\bibitem [{\citenamefont {Chiu}\ \emph {et~al.}(2016)\citenamefont {Chiu},
  \citenamefont {Teo}, \citenamefont {Schnyder},\ and\ \citenamefont
  {Ryu}}]{Chiu2016review}%
  \BibitemOpen
  \bibfield  {author} {\bibinfo {author} {\bibfnamefont {C.-K.}\ \bibnamefont
  {Chiu}}, \bibinfo {author} {\bibfnamefont {J.~C.~Y.}\ \bibnamefont {Teo}},
  \bibinfo {author} {\bibfnamefont {A.~P.}\ \bibnamefont {Schnyder}},\ and\
  \bibinfo {author} {\bibfnamefont {S.}~\bibnamefont {Ryu}},\ }\href
  {https://doi.org/10.1103/RevModPhys.88.035005} {\bibfield  {journal}
  {\bibinfo  {journal} {Rev. Mod. Phys.}\ }\textbf {\bibinfo {volume} {88}},\
  \bibinfo {pages} {035005} (\bibinfo {year} {2016})}\BibitemShut {NoStop}%
\bibitem [{\citenamefont {Freed}\ and\ \citenamefont
  {Hopkins}(2021)}]{Freed:2016rqq}%
  \BibitemOpen
  \bibfield  {author} {\bibinfo {author} {\bibfnamefont {D.~S.}\ \bibnamefont
  {Freed}}\ and\ \bibinfo {author} {\bibfnamefont {M.~J.}\ \bibnamefont
  {Hopkins}},\ }\href {https://doi.org/10.2140/gt.2021.25.1165} {\bibfield
  {journal} {\bibinfo  {journal} {Geom. Topol.}\ }\textbf {\bibinfo {volume}
  {25}},\ \bibinfo {pages} {1165} (\bibinfo {year} {2021})},\ \Eprint
  {https://arxiv.org/abs/1604.06527} {arXiv:1604.06527 [hep-th]} \BibitemShut
  {NoStop}%
\bibitem [{\citenamefont {Wang}\ and\ \citenamefont {Gu}(2020)}]{Wang2020fSPT}%
  \BibitemOpen
  \bibfield  {author} {\bibinfo {author} {\bibfnamefont {Q.-R.}\ \bibnamefont
  {Wang}}\ and\ \bibinfo {author} {\bibfnamefont {Z.-C.}\ \bibnamefont {Gu}},\
  }\href {https://doi.org/10.1103/PhysRevX.10.031055} {\bibfield  {journal}
  {\bibinfo  {journal} {Phys. Rev. X}\ }\textbf {\bibinfo {volume} {10}},\
  \bibinfo {pages} {031055} (\bibinfo {year} {2020})}\BibitemShut {NoStop}%
\bibitem [{\citenamefont {Barkeshli}\ \emph
  {et~al.}(2019{\natexlab{a}})\citenamefont {Barkeshli}, \citenamefont
  {Bonderson}, \citenamefont {Cheng},\ and\ \citenamefont
  {Wang}}]{Barkeshli2019}%
  \BibitemOpen
  \bibfield  {author} {\bibinfo {author} {\bibfnamefont {M.}~\bibnamefont
  {Barkeshli}}, \bibinfo {author} {\bibfnamefont {P.}~\bibnamefont
  {Bonderson}}, \bibinfo {author} {\bibfnamefont {M.}~\bibnamefont {Cheng}},\
  and\ \bibinfo {author} {\bibfnamefont {Z.}~\bibnamefont {Wang}},\ }\href
  {https://doi.org/10.1103/PhysRevB.100.115147} {\bibfield  {journal} {\bibinfo
   {journal} {Phys. Rev. B}\ }\textbf {\bibinfo {volume} {100}},\ \bibinfo
  {pages} {115147} (\bibinfo {year} {2019}{\natexlab{a}})}\BibitemShut
  {NoStop}%
\bibitem [{\citenamefont {Barkeshli}\ \emph {et~al.}(2021)\citenamefont
  {Barkeshli}, \citenamefont {Chen}, \citenamefont {Hsin},\ and\ \citenamefont
  {Manjunath}}]{barkeshli2021invertible}%
  \BibitemOpen
  \bibfield  {author} {\bibinfo {author} {\bibfnamefont {M.}~\bibnamefont
  {Barkeshli}}, \bibinfo {author} {\bibfnamefont {Y.-A.}\ \bibnamefont {Chen}},
  \bibinfo {author} {\bibfnamefont {P.-S.}\ \bibnamefont {Hsin}},\ and\
  \bibinfo {author} {\bibfnamefont {N.}~\bibnamefont {Manjunath}},\ }\href@noop
  {} {\bibinfo {title} {Classification of (2+1)d invertible fermionic
  topological phases with symmetry}} (\bibinfo {year} {2021}),\ \Eprint
  {https://arxiv.org/abs/2109.11039} {arXiv:2109.11039 [cond-mat.str-el]}
  \BibitemShut {NoStop}%
\bibitem [{\citenamefont {Bulmash}\ and\ \citenamefont
  {Barkeshli}(2022{\natexlab{a}})}]{bulmashSymmFrac}%
  \BibitemOpen
  \bibfield  {author} {\bibinfo {author} {\bibfnamefont {D.}~\bibnamefont
  {Bulmash}}\ and\ \bibinfo {author} {\bibfnamefont {M.}~\bibnamefont
  {Barkeshli}},\ }\href {https://doi.org/10.1103/PhysRevB.105.125114}
  {\bibfield  {journal} {\bibinfo  {journal} {Phys. Rev. B}\ }\textbf {\bibinfo
  {volume} {105}},\ \bibinfo {pages} {125114} (\bibinfo {year}
  {2022}{\natexlab{a}})}\BibitemShut {NoStop}%
\bibitem [{\citenamefont {Aasen}\ \emph {et~al.}(2021)\citenamefont {Aasen},
  \citenamefont {Bonderson},\ and\ \citenamefont
  {Knapp}}]{aasen2021characterization}%
  \BibitemOpen
  \bibfield  {author} {\bibinfo {author} {\bibfnamefont {D.}~\bibnamefont
  {Aasen}}, \bibinfo {author} {\bibfnamefont {P.}~\bibnamefont {Bonderson}},\
  and\ \bibinfo {author} {\bibfnamefont {C.}~\bibnamefont {Knapp}},\
  }\href@noop {} {\bibfield  {journal} {\bibinfo  {journal} {arXiv preprint
  arXiv:2109.10911}\ } (\bibinfo {year} {2021})}\BibitemShut {NoStop}%
\bibitem [{\citenamefont {Ivanov}(2001)}]{ivanov2001hqv}%
  \BibitemOpen
  \bibfield  {author} {\bibinfo {author} {\bibfnamefont {D.~A.}\ \bibnamefont
  {Ivanov}},\ }\href {https://doi.org/10.1103/PhysRevLett.86.268} {\bibfield
  {journal} {\bibinfo  {journal} {Phys. Rev. Lett.}\ }\textbf {\bibinfo
  {volume} {86}},\ \bibinfo {pages} {268} (\bibinfo {year} {2001})}\BibitemShut
  {NoStop}%
\bibitem [{\citenamefont {Das~Sarma}\ \emph {et~al.}(2006)\citenamefont
  {Das~Sarma}, \citenamefont {Nayak},\ and\ \citenamefont
  {Tewari}}]{dassarma2006hqv}%
  \BibitemOpen
  \bibfield  {author} {\bibinfo {author} {\bibfnamefont {S.}~\bibnamefont
  {Das~Sarma}}, \bibinfo {author} {\bibfnamefont {C.}~\bibnamefont {Nayak}},\
  and\ \bibinfo {author} {\bibfnamefont {S.}~\bibnamefont {Tewari}},\ }\href
  {https://doi.org/10.1103/PhysRevB.73.220502} {\bibfield  {journal} {\bibinfo
  {journal} {Phys. Rev. B}\ }\textbf {\bibinfo {volume} {73}},\ \bibinfo
  {pages} {220502} (\bibinfo {year} {2006})}\BibitemShut {NoStop}%
\bibitem [{\citenamefont {Asahi}\ and\ \citenamefont
  {Nagaosa}(2012)}]{Asahi2012mzm}%
  \BibitemOpen
  \bibfield  {author} {\bibinfo {author} {\bibfnamefont {D.}~\bibnamefont
  {Asahi}}\ and\ \bibinfo {author} {\bibfnamefont {N.}~\bibnamefont
  {Nagaosa}},\ }\href {https://doi.org/10.1103/PhysRevB.86.100504} {\bibfield
  {journal} {\bibinfo  {journal} {Phys. Rev. B}\ }\textbf {\bibinfo {volume}
  {86}},\ \bibinfo {pages} {100504} (\bibinfo {year} {2012})}\BibitemShut
  {NoStop}%
\bibitem [{\citenamefont {Hughes}\ \emph {et~al.}(2014)\citenamefont {Hughes},
  \citenamefont {Yao},\ and\ \citenamefont {Qi}}]{hughes2014mzm}%
  \BibitemOpen
  \bibfield  {author} {\bibinfo {author} {\bibfnamefont {T.~L.}\ \bibnamefont
  {Hughes}}, \bibinfo {author} {\bibfnamefont {H.}~\bibnamefont {Yao}},\ and\
  \bibinfo {author} {\bibfnamefont {X.-L.}\ \bibnamefont {Qi}},\ }\href
  {https://doi.org/10.1103/PhysRevB.90.235123} {\bibfield  {journal} {\bibinfo
  {journal} {Phys. Rev. B}\ }\textbf {\bibinfo {volume} {90}},\ \bibinfo
  {pages} {235123} (\bibinfo {year} {2014})}\BibitemShut {NoStop}%
\bibitem [{\citenamefont {Qian}\ \emph {et~al.}(2014)\citenamefont {Qian},
  \citenamefont {Liu}, \citenamefont {Fu},\ and\ \citenamefont
  {Li}}]{Qian2014QSH}%
  \BibitemOpen
  \bibfield  {author} {\bibinfo {author} {\bibfnamefont {X.}~\bibnamefont
  {Qian}}, \bibinfo {author} {\bibfnamefont {J.}~\bibnamefont {Liu}}, \bibinfo
  {author} {\bibfnamefont {L.}~\bibnamefont {Fu}},\ and\ \bibinfo {author}
  {\bibfnamefont {J.}~\bibnamefont {Li}},\ }\href
  {https://doi.org/10.1126/science.1256815} {\bibfield  {journal} {\bibinfo
  {journal} {Science}\ }\textbf {\bibinfo {volume} {346}},\ \bibinfo {pages}
  {1344} (\bibinfo {year} {2014})},\ \Eprint
  {https://arxiv.org/abs/https://www.science.org/doi/pdf/10.1126/science.1256815}
  {https://www.science.org/doi/pdf/10.1126/science.1256815} \BibitemShut
  {NoStop}%
\bibitem [{\citenamefont {Hsu}\ \emph {et~al.}(2020)\citenamefont {Hsu},
  \citenamefont {Cole}, \citenamefont {Zhang},\ and\ \citenamefont
  {Sau}}]{Hsu2020htsc}%
  \BibitemOpen
  \bibfield  {author} {\bibinfo {author} {\bibfnamefont {Y.-T.}\ \bibnamefont
  {Hsu}}, \bibinfo {author} {\bibfnamefont {W.~S.}\ \bibnamefont {Cole}},
  \bibinfo {author} {\bibfnamefont {R.-X.}\ \bibnamefont {Zhang}},\ and\
  \bibinfo {author} {\bibfnamefont {J.~D.}\ \bibnamefont {Sau}},\ }\href
  {https://doi.org/10.1103/PhysRevLett.125.097001} {\bibfield  {journal}
  {\bibinfo  {journal} {Phys. Rev. Lett.}\ }\textbf {\bibinfo {volume} {125}},\
  \bibinfo {pages} {097001} (\bibinfo {year} {2020})}\BibitemShut {NoStop}%
\bibitem [{\citenamefont {Geier}\ \emph {et~al.}(2018)\citenamefont {Geier},
  \citenamefont {Trifunovic}, \citenamefont {Hoskam},\ and\ \citenamefont
  {Brouwer}}]{geier2018htsc}%
  \BibitemOpen
  \bibfield  {author} {\bibinfo {author} {\bibfnamefont {M.}~\bibnamefont
  {Geier}}, \bibinfo {author} {\bibfnamefont {L.}~\bibnamefont {Trifunovic}},
  \bibinfo {author} {\bibfnamefont {M.}~\bibnamefont {Hoskam}},\ and\ \bibinfo
  {author} {\bibfnamefont {P.~W.}\ \bibnamefont {Brouwer}},\ }\href
  {https://doi.org/10.1103/PhysRevB.97.205135} {\bibfield  {journal} {\bibinfo
  {journal} {Phys. Rev. B}\ }\textbf {\bibinfo {volume} {97}},\ \bibinfo
  {pages} {205135} (\bibinfo {year} {2018})}\BibitemShut {NoStop}%
\bibitem [{\citenamefont {Khalaf}(2018)}]{khalaf2018hoti}%
  \BibitemOpen
  \bibfield  {author} {\bibinfo {author} {\bibfnamefont {E.}~\bibnamefont
  {Khalaf}},\ }\href {https://doi.org/10.1103/PhysRevB.97.205136} {\bibfield
  {journal} {\bibinfo  {journal} {Phys. Rev. B}\ }\textbf {\bibinfo {volume}
  {97}},\ \bibinfo {pages} {205136} (\bibinfo {year} {2018})}\BibitemShut
  {NoStop}%
\bibitem [{\citenamefont {Benalcazar}\ \emph {et~al.}(2014)\citenamefont
  {Benalcazar}, \citenamefont {Teo},\ and\ \citenamefont
  {Hughes}}]{Benalcazar2014}%
  \BibitemOpen
  \bibfield  {author} {\bibinfo {author} {\bibfnamefont {W.~A.}\ \bibnamefont
  {Benalcazar}}, \bibinfo {author} {\bibfnamefont {J.~C.~Y.}\ \bibnamefont
  {Teo}},\ and\ \bibinfo {author} {\bibfnamefont {T.~L.}\ \bibnamefont
  {Hughes}},\ }\href {https://doi.org/10.1103/PhysRevB.89.224503} {\bibfield
  {journal} {\bibinfo  {journal} {Phys. Rev. B}\ }\textbf {\bibinfo {volume}
  {89}},\ \bibinfo {pages} {224503} (\bibinfo {year} {2014})}\BibitemShut
  {NoStop}%
\bibitem [{\citenamefont {Wang}\ \emph
  {et~al.}(2018{\natexlab{a}})\citenamefont {Wang}, \citenamefont {Lin},\ and\
  \citenamefont {Hughes}}]{wang2018htsc}%
  \BibitemOpen
  \bibfield  {author} {\bibinfo {author} {\bibfnamefont {Y.}~\bibnamefont
  {Wang}}, \bibinfo {author} {\bibfnamefont {M.}~\bibnamefont {Lin}},\ and\
  \bibinfo {author} {\bibfnamefont {T.~L.}\ \bibnamefont {Hughes}},\ }\href
  {https://doi.org/10.1103/PhysRevB.98.165144} {\bibfield  {journal} {\bibinfo
  {journal} {Phys. Rev. B}\ }\textbf {\bibinfo {volume} {98}},\ \bibinfo
  {pages} {165144} (\bibinfo {year} {2018}{\natexlab{a}})}\BibitemShut
  {NoStop}%
\bibitem [{\citenamefont {Wang}\ \emph
  {et~al.}(2018{\natexlab{b}})\citenamefont {Wang}, \citenamefont {Liu},
  \citenamefont {Lu},\ and\ \citenamefont {Zhang}}]{wang2018maj}%
  \BibitemOpen
  \bibfield  {author} {\bibinfo {author} {\bibfnamefont {Q.}~\bibnamefont
  {Wang}}, \bibinfo {author} {\bibfnamefont {C.-C.}\ \bibnamefont {Liu}},
  \bibinfo {author} {\bibfnamefont {Y.-M.}\ \bibnamefont {Lu}},\ and\ \bibinfo
  {author} {\bibfnamefont {F.}~\bibnamefont {Zhang}},\ }\href
  {https://doi.org/10.1103/PhysRevLett.121.186801} {\bibfield  {journal}
  {\bibinfo  {journal} {Phys. Rev. Lett.}\ }\textbf {\bibinfo {volume} {121}},\
  \bibinfo {pages} {186801} (\bibinfo {year} {2018}{\natexlab{b}})}\BibitemShut
  {NoStop}%
\bibitem [{\citenamefont {Yan}\ \emph {et~al.}(2018)\citenamefont {Yan},
  \citenamefont {Song},\ and\ \citenamefont {Wang}}]{yan2018maj}%
  \BibitemOpen
  \bibfield  {author} {\bibinfo {author} {\bibfnamefont {Z.}~\bibnamefont
  {Yan}}, \bibinfo {author} {\bibfnamefont {F.}~\bibnamefont {Song}},\ and\
  \bibinfo {author} {\bibfnamefont {Z.}~\bibnamefont {Wang}},\ }\href
  {https://doi.org/10.1103/PhysRevLett.121.096803} {\bibfield  {journal}
  {\bibinfo  {journal} {Phys. Rev. Lett.}\ }\textbf {\bibinfo {volume} {121}},\
  \bibinfo {pages} {096803} (\bibinfo {year} {2018})}\BibitemShut {NoStop}%
\bibitem [{\citenamefont {Liu}\ \emph {et~al.}(2018)\citenamefont {Liu},
  \citenamefont {He},\ and\ \citenamefont {Nori}}]{liu2018maj}%
  \BibitemOpen
  \bibfield  {author} {\bibinfo {author} {\bibfnamefont {T.}~\bibnamefont
  {Liu}}, \bibinfo {author} {\bibfnamefont {J.~J.}\ \bibnamefont {He}},\ and\
  \bibinfo {author} {\bibfnamefont {F.}~\bibnamefont {Nori}},\ }\href
  {https://doi.org/10.1103/PhysRevB.98.245413} {\bibfield  {journal} {\bibinfo
  {journal} {Phys. Rev. B}\ }\textbf {\bibinfo {volume} {98}},\ \bibinfo
  {pages} {245413} (\bibinfo {year} {2018})}\BibitemShut {NoStop}%
\bibitem [{\citenamefont {Ueno}\ \emph {et~al.}(2013)\citenamefont {Ueno},
  \citenamefont {Yamakage}, \citenamefont {Tanaka},\ and\ \citenamefont
  {Sato}}]{ueno2013sr}%
  \BibitemOpen
  \bibfield  {author} {\bibinfo {author} {\bibfnamefont {Y.}~\bibnamefont
  {Ueno}}, \bibinfo {author} {\bibfnamefont {A.}~\bibnamefont {Yamakage}},
  \bibinfo {author} {\bibfnamefont {Y.}~\bibnamefont {Tanaka}},\ and\ \bibinfo
  {author} {\bibfnamefont {M.}~\bibnamefont {Sato}},\ }\href
  {https://doi.org/10.1103/PhysRevLett.111.087002} {\bibfield  {journal}
  {\bibinfo  {journal} {Phys. Rev. Lett.}\ }\textbf {\bibinfo {volume} {111}},\
  \bibinfo {pages} {087002} (\bibinfo {year} {2013})}\BibitemShut {NoStop}%
\bibitem [{\citenamefont {Ning}\ \emph {et~al.}(2021)\citenamefont {Ning},
  \citenamefont {Qi}, \citenamefont {Gu},\ and\ \citenamefont
  {Wang}}]{ning2021enforced}%
  \BibitemOpen
  \bibfield  {author} {\bibinfo {author} {\bibfnamefont {S.-Q.}\ \bibnamefont
  {Ning}}, \bibinfo {author} {\bibfnamefont {Y.}~\bibnamefont {Qi}}, \bibinfo
  {author} {\bibfnamefont {Z.-C.}\ \bibnamefont {Gu}},\ and\ \bibinfo {author}
  {\bibfnamefont {C.}~\bibnamefont {Wang}},\ }\href@noop {} {\bibfield
  {journal} {\bibinfo  {journal} {arXiv preprint arXiv:2109.15307}\ } (\bibinfo
  {year} {2021})}\BibitemShut {NoStop}%
\bibitem [{\citenamefont {Pustogow}(2019)}]{pustogow2019sr}%
  \BibitemOpen
  \bibfield  {author} {\bibinfo {author} {\bibfnamefont {A.~t.}\ \bibnamefont
  {Pustogow}},\ }\href@noop {} {\bibfield  {journal} {\bibinfo  {journal}
  {Nature}\ }\textbf {\bibinfo {volume} {574}},\ \bibinfo {pages} {72}
  (\bibinfo {year} {2019})}\BibitemShut {NoStop}%
\bibitem [{\citenamefont {Ishida}\ \emph {et~al.}(2020)\citenamefont {Ishida},
  \citenamefont {Manago}, \citenamefont {Kinjo},\ and\ \citenamefont
  {Maeno}}]{Ishida2020sr}%
  \BibitemOpen
  \bibfield  {author} {\bibinfo {author} {\bibfnamefont {K.}~\bibnamefont
  {Ishida}}, \bibinfo {author} {\bibfnamefont {M.}~\bibnamefont {Manago}},
  \bibinfo {author} {\bibfnamefont {K.}~\bibnamefont {Kinjo}},\ and\ \bibinfo
  {author} {\bibfnamefont {Y.}~\bibnamefont {Maeno}},\ }\href
  {https://doi.org/10.7566/JPSJ.89.034712} {\bibfield  {journal} {\bibinfo
  {journal} {Journal of the Physical Society of Japan}\ }\textbf {\bibinfo
  {volume} {89}},\ \bibinfo {pages} {034712} (\bibinfo {year} {2020})},\
  \Eprint {https://arxiv.org/abs/https://doi.org/10.7566/JPSJ.89.034712}
  {https://doi.org/10.7566/JPSJ.89.034712} \BibitemShut {NoStop}%
\bibitem [{\citenamefont {Ghosh}(2021)}]{ghosh2021sr}%
  \BibitemOpen
  \bibfield  {author} {\bibinfo {author} {\bibfnamefont {S.~t.}\ \bibnamefont
  {Ghosh}},\ }\href@noop {} {\bibfield  {journal} {\bibinfo  {journal} {Nat
  Phys.}\ }\textbf {\bibinfo {volume} {17}},\ \bibinfo {pages} {199} (\bibinfo
  {year} {2021})}\BibitemShut {NoStop}%
\bibitem [{\citenamefont {Wang}(2016)}]{wang2016tsc}%
  \BibitemOpen
  \bibfield  {author} {\bibinfo {author} {\bibfnamefont {C.}~\bibnamefont
  {Wang}},\ }\href {https://doi.org/10.1103/PhysRevB.94.085130} {\bibfield
  {journal} {\bibinfo  {journal} {Phys. Rev. B}\ }\textbf {\bibinfo {volume}
  {94}},\ \bibinfo {pages} {085130} (\bibinfo {year} {2016})}\BibitemShut
  {NoStop}%
\bibitem [{\citenamefont {Guo}\ \emph {et~al.}(2020)\citenamefont {Guo},
  \citenamefont {Ohmori}, \citenamefont {Putrov}, \citenamefont {Wan},\ and\
  \citenamefont {Wang}}]{guo2020fermionic}%
  \BibitemOpen
  \bibfield  {author} {\bibinfo {author} {\bibfnamefont {M.}~\bibnamefont
  {Guo}}, \bibinfo {author} {\bibfnamefont {K.}~\bibnamefont {Ohmori}},
  \bibinfo {author} {\bibfnamefont {P.}~\bibnamefont {Putrov}}, \bibinfo
  {author} {\bibfnamefont {Z.}~\bibnamefont {Wan}},\ and\ \bibinfo {author}
  {\bibfnamefont {J.}~\bibnamefont {Wang}},\ }\href@noop {} {\bibfield
  {journal} {\bibinfo  {journal} {Communications in Mathematical Physics}\
  }\textbf {\bibinfo {volume} {376}},\ \bibinfo {pages} {1073} (\bibinfo {year}
  {2020})}\BibitemShut {NoStop}%
\bibitem [{\citenamefont {Sullivan}\ and\ \citenamefont
  {Cheng}(2020)}]{sullivan2020interacting}%
  \BibitemOpen
  \bibfield  {author} {\bibinfo {author} {\bibfnamefont {J.}~\bibnamefont
  {Sullivan}}\ and\ \bibinfo {author} {\bibfnamefont {M.}~\bibnamefont
  {Cheng}},\ }\href@noop {} {\bibfield  {journal} {\bibinfo  {journal} {SciPost
  Physics}\ }\textbf {\bibinfo {volume} {9}},\ \bibinfo {pages} {016} (\bibinfo
  {year} {2020})}\BibitemShut {NoStop}%
\bibitem [{\citenamefont {Thorngren}\ and\ \citenamefont
  {Else}(2018)}]{Thorngren2018}%
  \BibitemOpen
  \bibfield  {author} {\bibinfo {author} {\bibfnamefont {R.}~\bibnamefont
  {Thorngren}}\ and\ \bibinfo {author} {\bibfnamefont {D.~V.}\ \bibnamefont
  {Else}},\ }\href {https://doi.org/10.1103/PhysRevX.8.011040} {\bibfield
  {journal} {\bibinfo  {journal} {Phys. Rev. X}\ }\textbf {\bibinfo {volume}
  {8}},\ \bibinfo {pages} {011040} (\bibinfo {year} {2018})}\BibitemShut
  {NoStop}%
\bibitem [{\citenamefont {Else}\ and\ \citenamefont
  {Thorngren}(2019{\natexlab{a}})}]{else2019crystalline}%
  \BibitemOpen
  \bibfield  {author} {\bibinfo {author} {\bibfnamefont {D.~V.}\ \bibnamefont
  {Else}}\ and\ \bibinfo {author} {\bibfnamefont {R.}~\bibnamefont
  {Thorngren}},\ }\href@noop {} {\bibfield  {journal} {\bibinfo  {journal}
  {Physical Review B}\ }\textbf {\bibinfo {volume} {99}},\ \bibinfo {pages}
  {115116} (\bibinfo {year} {2019}{\natexlab{a}})}\BibitemShut {NoStop}%
\bibitem [{\citenamefont {Debray}(2021)}]{debray2021invertible}%
  \BibitemOpen
  \bibfield  {author} {\bibinfo {author} {\bibfnamefont {A.}~\bibnamefont
  {Debray}},\ }\href@noop {} {\bibinfo {title} {Invertible phases for mixed
  spatial symmetries and the fermionic crystalline equivalence principle}}
  (\bibinfo {year} {2021}),\ \Eprint {https://arxiv.org/abs/2102.02941}
  {arXiv:2102.02941 [math-ph]} \BibitemShut {NoStop}%
\bibitem [{\citenamefont {Song}\ \emph {et~al.}(2020)\citenamefont {Song},
  \citenamefont {Fang},\ and\ \citenamefont {Qi}}]{song2020}%
  \BibitemOpen
  \bibfield  {author} {\bibinfo {author} {\bibfnamefont {Z.}~\bibnamefont
  {Song}}, \bibinfo {author} {\bibfnamefont {C.}~\bibnamefont {Fang}},\ and\
  \bibinfo {author} {\bibfnamefont {Y.}~\bibnamefont {Qi}},\ }\bibfield
  {journal} {\bibinfo  {journal} {Nature Communications}\ }\textbf {\bibinfo
  {volume} {11}},\ \href {https://doi.org/10.1038/s41467-020-17685-5}
  {10.1038/s41467-020-17685-5} (\bibinfo {year} {2020})\BibitemShut {NoStop}%
\bibitem [{\citenamefont {Zhang}\ \emph {et~al.}(2020)\citenamefont {Zhang},
  \citenamefont {Wang}, \citenamefont {Yang}, \citenamefont {Qi},\ and\
  \citenamefont {Gu}}]{zhang2020const}%
  \BibitemOpen
  \bibfield  {author} {\bibinfo {author} {\bibfnamefont {J.-H.}\ \bibnamefont
  {Zhang}}, \bibinfo {author} {\bibfnamefont {Q.-R.}\ \bibnamefont {Wang}},
  \bibinfo {author} {\bibfnamefont {S.}~\bibnamefont {Yang}}, \bibinfo {author}
  {\bibfnamefont {Y.}~\bibnamefont {Qi}},\ and\ \bibinfo {author}
  {\bibfnamefont {Z.-C.}\ \bibnamefont {Gu}},\ }\href@noop {} {\bibfield
  {journal} {\bibinfo  {journal} {Physical Review B}\ }\textbf {\bibinfo
  {volume} {101}},\ \bibinfo {pages} {100501} (\bibinfo {year}
  {2020})}\BibitemShut {NoStop}%
\bibitem [{\citenamefont {Zhang}\ \emph {et~al.}(2022)\citenamefont {Zhang},
  \citenamefont {Yang}, \citenamefont {Qi},\ and\ \citenamefont
  {Gu}}]{zhang2020realspace}%
  \BibitemOpen
  \bibfield  {author} {\bibinfo {author} {\bibfnamefont {J.-H.}\ \bibnamefont
  {Zhang}}, \bibinfo {author} {\bibfnamefont {S.}~\bibnamefont {Yang}},
  \bibinfo {author} {\bibfnamefont {Y.}~\bibnamefont {Qi}},\ and\ \bibinfo
  {author} {\bibfnamefont {Z.-C.}\ \bibnamefont {Gu}},\ }\href@noop {}
  {\bibfield  {journal} {\bibinfo  {journal} {Physical Review Research}\
  }\textbf {\bibinfo {volume} {4}},\ \bibinfo {pages} {033081} (\bibinfo {year}
  {2022})}\BibitemShut {NoStop}%
\bibitem [{\citenamefont {Zhang}\ \emph {et~al.}(2019)\citenamefont {Zhang},
  \citenamefont {Cole}, \citenamefont {Wu},\ and\ \citenamefont
  {Das~Sarma}}]{Zhang2019FeTSC}%
  \BibitemOpen
  \bibfield  {author} {\bibinfo {author} {\bibfnamefont {R.-X.}\ \bibnamefont
  {Zhang}}, \bibinfo {author} {\bibfnamefont {W.~S.}\ \bibnamefont {Cole}},
  \bibinfo {author} {\bibfnamefont {X.}~\bibnamefont {Wu}},\ and\ \bibinfo
  {author} {\bibfnamefont {S.}~\bibnamefont {Das~Sarma}},\ }\href
  {https://doi.org/10.1103/PhysRevLett.123.167001} {\bibfield  {journal}
  {\bibinfo  {journal} {Phys. Rev. Lett.}\ }\textbf {\bibinfo {volume} {123}},\
  \bibinfo {pages} {167001} (\bibinfo {year} {2019})}\BibitemShut {NoStop}%
\bibitem [{\citenamefont {Shaffer}\ \emph {et~al.}(2021)\citenamefont
  {Shaffer}, \citenamefont {Wang},\ and\ \citenamefont
  {Santos}}]{shaffer2021hofsc}%
  \BibitemOpen
  \bibfield  {author} {\bibinfo {author} {\bibfnamefont {D.}~\bibnamefont
  {Shaffer}}, \bibinfo {author} {\bibfnamefont {J.}~\bibnamefont {Wang}},\ and\
  \bibinfo {author} {\bibfnamefont {L.~H.}\ \bibnamefont {Santos}},\ }\href
  {https://doi.org/10.1103/PhysRevB.104.184501} {\bibfield  {journal} {\bibinfo
   {journal} {Phys. Rev. B}\ }\textbf {\bibinfo {volume} {104}},\ \bibinfo
  {pages} {184501} (\bibinfo {year} {2021})}\BibitemShut {NoStop}%
\bibitem [{\citenamefont {Shiozaki}\ and\ \citenamefont
  {Sato}(2014)}]{Shiozaki2014TopologyCTSc}%
  \BibitemOpen
  \bibfield  {author} {\bibinfo {author} {\bibfnamefont {K.}~\bibnamefont
  {Shiozaki}}\ and\ \bibinfo {author} {\bibfnamefont {M.}~\bibnamefont
  {Sato}},\ }\href {https://doi.org/10.1103/PhysRevB.90.165114} {\bibfield
  {journal} {\bibinfo  {journal} {Phys. Rev. B}\ }\textbf {\bibinfo {volume}
  {90}},\ \bibinfo {pages} {165114} (\bibinfo {year} {2014})}\BibitemShut
  {NoStop}%
\bibitem [{\citenamefont {Barkeshli}\ and\ \citenamefont
  {Cheng}(2020)}]{Barkeshli2020Anomaly}%
  \BibitemOpen
  \bibfield  {author} {\bibinfo {author} {\bibfnamefont {M.}~\bibnamefont
  {Barkeshli}}\ and\ \bibinfo {author} {\bibfnamefont {M.}~\bibnamefont
  {Cheng}},\ }\bibfield  {journal} {\bibinfo  {journal} {SciPost Physics}\
  }\textbf {\bibinfo {volume} {8}},\ \href
  {https://doi.org/10.21468/scipostphys.8.2.028} {10.21468/scipostphys.8.2.028}
  (\bibinfo {year} {2020})\BibitemShut {NoStop}%
\bibitem [{\citenamefont {Barkeshli}\ \emph
  {et~al.}(2019{\natexlab{b}})\citenamefont {Barkeshli}, \citenamefont
  {Bonderson}, \citenamefont {Cheng}, \citenamefont {Jian},\ and\ \citenamefont
  {Walker}}]{barkeshli2019tr}%
  \BibitemOpen
  \bibfield  {author} {\bibinfo {author} {\bibfnamefont {M.}~\bibnamefont
  {Barkeshli}}, \bibinfo {author} {\bibfnamefont {P.}~\bibnamefont
  {Bonderson}}, \bibinfo {author} {\bibfnamefont {M.}~\bibnamefont {Cheng}},
  \bibinfo {author} {\bibfnamefont {C.-M.}\ \bibnamefont {Jian}},\ and\
  \bibinfo {author} {\bibfnamefont {K.}~\bibnamefont {Walker}},\ }\bibfield
  {journal} {\bibinfo  {journal} {Communications in Mathematical Physics}\
  }\href {https://doi.org/10.1007/s00220-019-03475-8}
  {10.1007/s00220-019-03475-8} (\bibinfo {year} {2019}{\natexlab{b}}),\ \Eprint
  {https://arxiv.org/abs/arXiv:1612.07792} {arXiv:1612.07792} \BibitemShut
  {NoStop}%
\bibitem [{\citenamefont {Metlitski}\ \emph {et~al.}(2014)\citenamefont
  {Metlitski}, \citenamefont {Fidkowski}, \citenamefont {Chen},\ and\
  \citenamefont {Vishwanath}}]{metlitski2014}%
  \BibitemOpen
  \bibfield  {author} {\bibinfo {author} {\bibfnamefont {M.~A.}\ \bibnamefont
  {Metlitski}}, \bibinfo {author} {\bibfnamefont {L.}~\bibnamefont
  {Fidkowski}}, \bibinfo {author} {\bibfnamefont {X.}~\bibnamefont {Chen}},\
  and\ \bibinfo {author} {\bibfnamefont {A.}~\bibnamefont {Vishwanath}},\
  }\href@noop {} {\bibfield  {journal} {\bibinfo  {journal} {arXiv preprint
  arXiv:1406.3032}\ } (\bibinfo {year} {2014})}\BibitemShut {NoStop}%
\bibitem [{\citenamefont {Tata}\ \emph {et~al.}(2021)\citenamefont {Tata},
  \citenamefont {Kobayashi}, \citenamefont {Bulmash},\ and\ \citenamefont
  {Barkeshli}}]{tata2021anomalies}%
  \BibitemOpen
  \bibfield  {author} {\bibinfo {author} {\bibfnamefont {S.}~\bibnamefont
  {Tata}}, \bibinfo {author} {\bibfnamefont {R.}~\bibnamefont {Kobayashi}},
  \bibinfo {author} {\bibfnamefont {D.}~\bibnamefont {Bulmash}},\ and\ \bibinfo
  {author} {\bibfnamefont {M.}~\bibnamefont {Barkeshli}},\ }\href@noop {}
  {\bibfield  {journal} {\bibinfo  {journal} {arXiv preprint arXiv:2104.14567}\
  } (\bibinfo {year} {2021})}\BibitemShut {NoStop}%
\bibitem [{\citenamefont {Qi}\ \emph {et~al.}(2009)\citenamefont {Qi},
  \citenamefont {Hughes}, \citenamefont {Raghu},\ and\ \citenamefont
  {Zhang}}]{Qi2009TRS-TSC}%
  \BibitemOpen
  \bibfield  {author} {\bibinfo {author} {\bibfnamefont {X.-L.}\ \bibnamefont
  {Qi}}, \bibinfo {author} {\bibfnamefont {T.~L.}\ \bibnamefont {Hughes}},
  \bibinfo {author} {\bibfnamefont {S.}~\bibnamefont {Raghu}},\ and\ \bibinfo
  {author} {\bibfnamefont {S.-C.}\ \bibnamefont {Zhang}},\ }\href
  {https://doi.org/10.1103/PhysRevLett.102.187001} {\bibfield  {journal}
  {\bibinfo  {journal} {Phys. Rev. Lett.}\ }\textbf {\bibinfo {volume} {102}},\
  \bibinfo {pages} {187001} (\bibinfo {year} {2009})}\BibitemShut {NoStop}%
\bibitem [{\citenamefont {Zhang}\ \emph {et~al.}(2013)\citenamefont {Zhang},
  \citenamefont {Kane},\ and\ \citenamefont {Mele}}]{Zhang2013MKP}%
  \BibitemOpen
  \bibfield  {author} {\bibinfo {author} {\bibfnamefont {F.}~\bibnamefont
  {Zhang}}, \bibinfo {author} {\bibfnamefont {C.~L.}\ \bibnamefont {Kane}},\
  and\ \bibinfo {author} {\bibfnamefont {E.~J.}\ \bibnamefont {Mele}},\ }\href
  {https://doi.org/10.1103/PhysRevLett.111.056402} {\bibfield  {journal}
  {\bibinfo  {journal} {Phys. Rev. Lett.}\ }\textbf {\bibinfo {volume} {111}},\
  \bibinfo {pages} {056402} (\bibinfo {year} {2013})}\BibitemShut {NoStop}%
\bibitem [{\citenamefont {Else}\ and\ \citenamefont
  {Thorngren}(2019{\natexlab{b}})}]{Else2019}%
  \BibitemOpen
  \bibfield  {author} {\bibinfo {author} {\bibfnamefont {D.~V.}\ \bibnamefont
  {Else}}\ and\ \bibinfo {author} {\bibfnamefont {R.}~\bibnamefont
  {Thorngren}},\ }\href {https://doi.org/10.1103/PhysRevB.99.115116} {\bibfield
   {journal} {\bibinfo  {journal} {Phys. Rev. B}\ }\textbf {\bibinfo {volume}
  {99}},\ \bibinfo {pages} {115116} (\bibinfo {year}
  {2019}{\natexlab{b}})}\BibitemShut {NoStop}%
\bibitem [{\citenamefont {Yao}\ and\ \citenamefont {Ryu}(2013)}]{YaoTScwithRT}%
  \BibitemOpen
  \bibfield  {author} {\bibinfo {author} {\bibfnamefont {H.}~\bibnamefont
  {Yao}}\ and\ \bibinfo {author} {\bibfnamefont {S.}~\bibnamefont {Ryu}},\
  }\href {https://doi.org/10.1103/PhysRevB.88.064507} {\bibfield  {journal}
  {\bibinfo  {journal} {Phys. Rev. B}\ }\textbf {\bibinfo {volume} {88}},\
  \bibinfo {pages} {064507} (\bibinfo {year} {2013})}\BibitemShut {NoStop}%
\bibitem [{\citenamefont {Song}\ \emph {et~al.}(2017)\citenamefont {Song},
  \citenamefont {Huang}, \citenamefont {Fu},\ and\ \citenamefont
  {Hermele}}]{song2017}%
  \BibitemOpen
  \bibfield  {author} {\bibinfo {author} {\bibfnamefont {H.}~\bibnamefont
  {Song}}, \bibinfo {author} {\bibfnamefont {S.-J.}\ \bibnamefont {Huang}},
  \bibinfo {author} {\bibfnamefont {L.}~\bibnamefont {Fu}},\ and\ \bibinfo
  {author} {\bibfnamefont {M.}~\bibnamefont {Hermele}},\ }\href
  {https://doi.org/10.1103/PhysRevX.7.011020} {\bibfield  {journal} {\bibinfo
  {journal} {Phys. Rev. X}\ }\textbf {\bibinfo {volume} {7}},\ \bibinfo {pages}
  {011020} (\bibinfo {year} {2017})}\BibitemShut {NoStop}%
\bibitem [{\citenamefont {Song}\ \emph {et~al.}(2019)\citenamefont {Song},
  \citenamefont {Huang}, \citenamefont {Qi}, \citenamefont {Fang},\ and\
  \citenamefont {Hermele}}]{song2019topcryst}%
  \BibitemOpen
  \bibfield  {author} {\bibinfo {author} {\bibfnamefont {Z.}~\bibnamefont
  {Song}}, \bibinfo {author} {\bibfnamefont {S.-J.}\ \bibnamefont {Huang}},
  \bibinfo {author} {\bibfnamefont {Y.}~\bibnamefont {Qi}}, \bibinfo {author}
  {\bibfnamefont {C.}~\bibnamefont {Fang}},\ and\ \bibinfo {author}
  {\bibfnamefont {M.}~\bibnamefont {Hermele}},\ }\href
  {https://doi.org/10.1126/sciadv.aax2007} {\bibfield  {journal} {\bibinfo
  {journal} {Science Advances}\ }\textbf {\bibinfo {volume} {5}},\ \bibinfo
  {pages} {eaax2007} (\bibinfo {year} {2019})},\ \Eprint
  {https://arxiv.org/abs/https://www.science.org/doi/pdf/10.1126/sciadv.aax2007}
  {https://www.science.org/doi/pdf/10.1126/sciadv.aax2007} \BibitemShut
  {NoStop}%
\bibitem [{\citenamefont {Turzillo}\ and\ \citenamefont
  {You}(2019)}]{turzillo2019fmps}%
  \BibitemOpen
  \bibfield  {author} {\bibinfo {author} {\bibfnamefont {A.}~\bibnamefont
  {Turzillo}}\ and\ \bibinfo {author} {\bibfnamefont {M.}~\bibnamefont {You}},\
  }\href {https://doi.org/10.1103/PhysRevB.99.035103} {\bibfield  {journal}
  {\bibinfo  {journal} {Phys. Rev. B}\ }\textbf {\bibinfo {volume} {99}},\
  \bibinfo {pages} {035103} (\bibinfo {year} {2019})}\BibitemShut {NoStop}%
\bibitem [{\citenamefont {Bourne}\ and\ \citenamefont
  {Ogata}(2021)}]{bourne2021}%
  \BibitemOpen
  \bibfield  {author} {\bibinfo {author} {\bibfnamefont {C.}~\bibnamefont
  {Bourne}}\ and\ \bibinfo {author} {\bibfnamefont {Y.}~\bibnamefont {Ogata}},\
  }\bibfield  {journal} {\bibinfo  {journal} {Forum of Mathematics, Sigma}\
  }\textbf {\bibinfo {volume} {9}},\ \href
  {https://doi.org/10.1017/fms.2021.19} {10.1017/fms.2021.19} (\bibinfo {year}
  {2021})\BibitemShut {NoStop}%
\bibitem [{\citenamefont {Aksoy}\ and\ \citenamefont
  {Mudry}(2022)}]{aksoy2022}%
  \BibitemOpen
  \bibfield  {author} {\bibinfo {author} {\bibfnamefont {O.~M.}\ \bibnamefont
  {Aksoy}}\ and\ \bibinfo {author} {\bibfnamefont {C.}~\bibnamefont {Mudry}}\
  }(\bibinfo {year} {2022})\BibitemShut {NoStop}%
\bibitem [{\citenamefont {Thomas}(1960{\natexlab{a}})}]{thomas1960cohomology}%
  \BibitemOpen
  \bibfield  {author} {\bibinfo {author} {\bibfnamefont {E.}~\bibnamefont
  {Thomas}},\ }\href@noop {} {\bibfield  {journal} {\bibinfo  {journal}
  {Transactions of the American Mathematical Society}\ }\textbf {\bibinfo
  {volume} {96}},\ \bibinfo {pages} {67} (\bibinfo {year}
  {1960}{\natexlab{a}})}\BibitemShut {NoStop}%
\bibitem [{\citenamefont {Brown~Jr}(1982)}]{brown1982cohomology}%
  \BibitemOpen
  \bibfield  {author} {\bibinfo {author} {\bibfnamefont {E.~H.}\ \bibnamefont
  {Brown~Jr}},\ }\href@noop {} {\bibfield  {journal} {\bibinfo  {journal}
  {Proceedings of the American mathematical society}\ ,\ \bibinfo {pages}
  {283}} (\bibinfo {year} {1982})}\BibitemShut {NoStop}%
\bibitem [{\citenamefont {Chen}\ \emph {et~al.}(2019)\citenamefont {Chen},
  \citenamefont {Kapustin}, \citenamefont {Turzillo},\ and\ \citenamefont
  {You}}]{Chen2019freeinteracting}%
  \BibitemOpen
  \bibfield  {author} {\bibinfo {author} {\bibfnamefont {Y.-A.}\ \bibnamefont
  {Chen}}, \bibinfo {author} {\bibfnamefont {A.}~\bibnamefont {Kapustin}},
  \bibinfo {author} {\bibfnamefont {A.}~\bibnamefont {Turzillo}},\ and\
  \bibinfo {author} {\bibfnamefont {M.}~\bibnamefont {You}},\ }\href
  {https://doi.org/10.1103/PhysRevB.100.195128} {\bibfield  {journal} {\bibinfo
   {journal} {Phys. Rev. B}\ }\textbf {\bibinfo {volume} {100}},\ \bibinfo
  {pages} {195128} (\bibinfo {year} {2019})}\BibitemShut {NoStop}%
\bibitem [{\citenamefont {Kitaev}(2006)}]{Kitaev2006}%
  \BibitemOpen
  \bibfield  {author} {\bibinfo {author} {\bibfnamefont {A.}~\bibnamefont
  {Kitaev}},\ }\href
  {https://doi.org/https://doi.org/10.1016/j.aop.2005.10.005} {\bibfield
  {journal} {\bibinfo  {journal} {Annals of Physics}\ }\textbf {\bibinfo
  {volume} {321}},\ \bibinfo {pages} {2} (\bibinfo {year} {2006})},\ \bibinfo
  {note} {january Special Issue}\BibitemShut {NoStop}%
\bibitem [{\citenamefont {Sato}(1999)}]{Sato1999}%
  \BibitemOpen
  \bibfield  {author} {\bibinfo {author} {\bibfnamefont {H.}~\bibnamefont
  {Sato}},\ }\href@noop {} {\emph {\bibinfo {title} {Algebraic Topology: An
  Intuitive Approach}}}\ (\bibinfo  {publisher} {Amer Mathematical Society},\
  \bibinfo {year} {1999})\BibitemShut {NoStop}%
\bibitem [{\citenamefont {Gu}\ and\ \citenamefont
  {Levin}(2014)}]{Gu2014interaction}%
  \BibitemOpen
  \bibfield  {author} {\bibinfo {author} {\bibfnamefont {Z.-C.}\ \bibnamefont
  {Gu}}\ and\ \bibinfo {author} {\bibfnamefont {M.}~\bibnamefont {Levin}},\
  }\href {https://doi.org/10.1103/PhysRevB.89.201113} {\bibfield  {journal}
  {\bibinfo  {journal} {Phys. Rev. B}\ }\textbf {\bibinfo {volume} {89}},\
  \bibinfo {pages} {201113} (\bibinfo {year} {2014})}\BibitemShut {NoStop}%
\bibitem [{\citenamefont {Grigoletto}\ and\ \citenamefont
  {Putrov}(2021)}]{grigoletto2021spin}%
  \BibitemOpen
  \bibfield  {author} {\bibinfo {author} {\bibfnamefont {A.}~\bibnamefont
  {Grigoletto}}\ and\ \bibinfo {author} {\bibfnamefont {P.}~\bibnamefont
  {Putrov}},\ }\href@noop {} {\bibfield  {journal} {\bibinfo  {journal} {arXiv
  preprint arXiv:2106.16247}\ } (\bibinfo {year} {2021})}\BibitemShut {NoStop}%
\bibitem [{\citenamefont {Rao}\ and\ \citenamefont
  {Sodemann}(2021)}]{rao2021theory}%
  \BibitemOpen
  \bibfield  {author} {\bibinfo {author} {\bibfnamefont {P.}~\bibnamefont
  {Rao}}\ and\ \bibinfo {author} {\bibfnamefont {I.}~\bibnamefont {Sodemann}},\
  }\href@noop {} {\bibfield  {journal} {\bibinfo  {journal} {Physical Review
  Research}\ }\textbf {\bibinfo {volume} {3}},\ \bibinfo {pages} {023120}
  (\bibinfo {year} {2021})}\BibitemShut {NoStop}%
\bibitem [{\citenamefont {Raghu}\ \emph {et~al.}(2010)\citenamefont {Raghu},
  \citenamefont {Kapitulnik},\ and\ \citenamefont
  {Kivelson}}]{raghu2010tsc-sr}%
  \BibitemOpen
  \bibfield  {author} {\bibinfo {author} {\bibfnamefont {S.}~\bibnamefont
  {Raghu}}, \bibinfo {author} {\bibfnamefont {A.}~\bibnamefont {Kapitulnik}},\
  and\ \bibinfo {author} {\bibfnamefont {S.~A.}\ \bibnamefont {Kivelson}},\
  }\href {https://doi.org/10.1103/PhysRevLett.105.136401} {\bibfield  {journal}
  {\bibinfo  {journal} {Phys. Rev. Lett.}\ }\textbf {\bibinfo {volume} {105}},\
  \bibinfo {pages} {136401} (\bibinfo {year} {2010})}\BibitemShut {NoStop}%
\bibitem [{\citenamefont {Skurativska}\ \emph {et~al.}(2020)\citenamefont
  {Skurativska}, \citenamefont {Neupert},\ and\ \citenamefont
  {Fischer}}]{Skurativska2020}%
  \BibitemOpen
  \bibfield  {author} {\bibinfo {author} {\bibfnamefont {A.}~\bibnamefont
  {Skurativska}}, \bibinfo {author} {\bibfnamefont {T.}~\bibnamefont
  {Neupert}},\ and\ \bibinfo {author} {\bibfnamefont {M.~H.}\ \bibnamefont
  {Fischer}},\ }\href@noop {} {\bibfield  {journal} {\bibinfo  {journal}
  {Physical Review Research}\ }\textbf {\bibinfo {volume} {2}},\ \bibinfo
  {pages} {013064} (\bibinfo {year} {2020})}\BibitemShut {NoStop}%
\bibitem [{\citenamefont {Fidkowski}\ and\ \citenamefont
  {Kitaev}(2011)}]{fidkowski2011}%
  \BibitemOpen
  \bibfield  {author} {\bibinfo {author} {\bibfnamefont {L.}~\bibnamefont
  {Fidkowski}}\ and\ \bibinfo {author} {\bibfnamefont {A.}~\bibnamefont
  {Kitaev}},\ }\href {https://doi.org/10.1103/PhysRevB.83.075103} {\bibfield
  {journal} {\bibinfo  {journal} {Phys. Rev. B}\ }\textbf {\bibinfo {volume}
  {83}},\ \bibinfo {pages} {075103} (\bibinfo {year} {2011})}\BibitemShut
  {NoStop}%
\bibitem [{\citenamefont {Tantivasadakarn}(2017)}]{Tantiv2017}%
  \BibitemOpen
  \bibfield  {author} {\bibinfo {author} {\bibfnamefont {N.}~\bibnamefont
  {Tantivasadakarn}},\ }\href {https://doi.org/10.1103/PhysRevB.96.195101}
  {\bibfield  {journal} {\bibinfo  {journal} {Phys. Rev. B}\ }\textbf {\bibinfo
  {volume} {96}},\ \bibinfo {pages} {195101} (\bibinfo {year}
  {2017})}\BibitemShut {NoStop}%
\bibitem [{\citenamefont {Bulmash}\ and\ \citenamefont
  {Barkeshli}(2022{\natexlab{b}})}]{bulmashAnomaly}%
  \BibitemOpen
  \bibfield  {author} {\bibinfo {author} {\bibfnamefont {D.}~\bibnamefont
  {Bulmash}}\ and\ \bibinfo {author} {\bibfnamefont {M.}~\bibnamefont
  {Barkeshli}},\ }\href@noop {} {\bibfield  {journal} {\bibinfo  {journal}
  {Physical Review B}\ }\textbf {\bibinfo {volume} {105}},\ \bibinfo {pages}
  {155126} (\bibinfo {year} {2022}{\natexlab{b}})}\BibitemShut {NoStop}%
\bibitem [{\citenamefont {Manjunath}\ and\ \citenamefont
  {Barkeshli}(2020)}]{Manjunath2020fqh}%
  \BibitemOpen
  \bibfield  {author} {\bibinfo {author} {\bibfnamefont {N.}~\bibnamefont
  {Manjunath}}\ and\ \bibinfo {author} {\bibfnamefont {M.}~\bibnamefont
  {Barkeshli}},\ }\href@noop {} {\bibfield  {journal} {\bibinfo  {journal}
  {arXiv preprint arXiv:2012.11603}\ } (\bibinfo {year} {2020})}\BibitemShut
  {NoStop}%
\bibitem [{\citenamefont {RAMOS}(2017)}]{ramos2017spectral}%
  \BibitemOpen
  \bibfield  {author} {\bibinfo {author} {\bibfnamefont {A.~D.}\ \bibnamefont
  {RAMOS}},\ }\href@noop {} {\bibfield  {journal} {\bibinfo  {journal} {arXiv
  preprint arXiv:1702.00666}\ } (\bibinfo {year} {2017})}\BibitemShut {NoStop}%
\bibitem [{\citenamefont {Brown}(2012)}]{brown2012cohomology}%
  \BibitemOpen
  \bibfield  {author} {\bibinfo {author} {\bibfnamefont {K.~S.}\ \bibnamefont
  {Brown}},\ }\href@noop {} {\emph {\bibinfo {title} {Cohomology of groups}}},\
  Vol.~\bibinfo {volume} {87}\ (\bibinfo  {publisher} {Springer Science \&
  Business Media},\ \bibinfo {year} {2012})\BibitemShut {NoStop}%
\bibitem [{\citenamefont {(https://mathoverflow.net/users/360/tyler
  lawson)}()}]{316315}%
  \BibitemOpen
  \bibfield  {author} {\bibinfo {author} {\bibfnamefont {T.~L.}\ \bibnamefont
  {(https://mathoverflow.net/users/360/tyler lawson)}},\ }\href
  {https://mathoverflow.net/q/316315} {\bibinfo {title} {Convergence of the
  lyndon-hochschild-serre spectral sequence as an algebra}},\ \bibinfo
  {howpublished} {MathOverflow},\ \bibinfo {note}
  {uRL:https://mathoverflow.net/q/316315 (version: 2018-11-27)},\ \Eprint
  {https://arxiv.org/abs/https://mathoverflow.net/q/316315}
  {https://mathoverflow.net/q/316315} \BibitemShut {NoStop}%
\bibitem [{\citenamefont {Greenblatt}(2006)}]{greenblatt2006homology}%
  \BibitemOpen
  \bibfield  {author} {\bibinfo {author} {\bibfnamefont {R.}~\bibnamefont
  {Greenblatt}},\ }\href@noop {} {\bibfield  {journal} {\bibinfo  {journal}
  {Homology, Homotopy and Applications}\ }\textbf {\bibinfo {volume} {8}},\
  \bibinfo {pages} {91} (\bibinfo {year} {2006})}\BibitemShut {NoStop}%
\bibitem [{\citenamefont
  {Thomas}(1960{\natexlab{b}})}]{Thomas1960OnCohomology}%
  \BibitemOpen
  \bibfield  {author} {\bibinfo {author} {\bibfnamefont {E.}~\bibnamefont
  {Thomas}},\ }\href {http://www.jstor.org/stable/1993484} {\bibfield
  {journal} {\bibinfo  {journal} {Transactions of the American Mathematical
  Society}\ }\textbf {\bibinfo {volume} {96}},\ \bibinfo {pages} {67} (\bibinfo
  {year} {1960}{\natexlab{b}})}\BibitemShut {NoStop}%
\bibitem [{\citenamefont {Gu}(2019)}]{Gu_2019}%
  \BibitemOpen
  \bibfield  {author} {\bibinfo {author} {\bibfnamefont {X.}~\bibnamefont
  {Gu}},\ }\href {https://doi.org/10.1142/s1793525320500211} {\bibfield
  {journal} {\bibinfo  {journal} {Journal of Topology and Analysis}\ }\textbf
  {\bibinfo {volume} {13}},\ \bibinfo {pages} {535} (\bibinfo {year}
  {2019})}\BibitemShut {NoStop}%
\bibitem [{\citenamefont {Benson}\ and\ \citenamefont
  {Wood}(1995)}]{BENSON199513}%
  \BibitemOpen
  \bibfield  {author} {\bibinfo {author} {\bibfnamefont {D.}~\bibnamefont
  {Benson}}\ and\ \bibinfo {author} {\bibfnamefont {J.~A.}\ \bibnamefont
  {Wood}},\ }\href
  {https://doi.org/https://doi.org/10.1016/0040-9383(94)E0019-G} {\bibfield
  {journal} {\bibinfo  {journal} {Topology}\ }\textbf {\bibinfo {volume}
  {34}},\ \bibinfo {pages} {13} (\bibinfo {year} {1995})}\BibitemShut {NoStop}%
\bibitem [{\citenamefont {Duan}(2018)}]{duan2018characteristic}%
  \BibitemOpen
  \bibfield  {author} {\bibinfo {author} {\bibfnamefont {H.}~\bibnamefont
  {Duan}},\ }\href@noop {} {\bibfield  {journal} {\bibinfo  {journal} {arXiv
  preprint arXiv:1810.03799}\ } (\bibinfo {year} {2018})}\BibitemShut {NoStop}%
\bibitem [{\citenamefont {Kono}\ and\ \citenamefont
  {Mimura}(1975)}]{kono1975cohomology}%
  \BibitemOpen
  \bibfield  {author} {\bibinfo {author} {\bibfnamefont {A.}~\bibnamefont
  {Kono}}\ and\ \bibinfo {author} {\bibfnamefont {M.}~\bibnamefont {Mimura}},\
  }\href@noop {} {\bibfield  {journal} {\bibinfo  {journal} {Publications of
  the Research Institute for Mathematical Sciences}\ }\textbf {\bibinfo
  {volume} {10}},\ \bibinfo {pages} {691} (\bibinfo {year} {1975})}\BibitemShut
  {NoStop}%
\bibitem [{\citenamefont {(https://mathoverflow.net/users/3473/konrad
  waldorf)}()}]{390077}%
  \BibitemOpen
  \bibfield  {author} {\bibinfo {author} {\bibfnamefont {K.~W.}\ \bibnamefont
  {(https://mathoverflow.net/users/3473/konrad waldorf)}},\ }\href
  {https://mathoverflow.net/q/390077} {\bibinfo {title} {Low dimensional
  integral cohomology of $bpso(4n)$}},\ \bibinfo {howpublished}
  {MathOverflow},\ \bibinfo {note} {uRL:https://mathoverflow.net/q/390077
  (version: 2021-04-14)},\ \Eprint
  {https://arxiv.org/abs/https://mathoverflow.net/q/390077}
  {https://mathoverflow.net/q/390077} \BibitemShut {NoStop}%
\bibitem [{\citenamefont {Gawedzki}\ and\ \citenamefont
  {Waldorf}(2009)}]{Gawedzki_2009}%
  \BibitemOpen
  \bibfield  {author} {\bibinfo {author} {\bibfnamefont {K.}~\bibnamefont
  {Gawedzki}}\ and\ \bibinfo {author} {\bibfnamefont {K.}~\bibnamefont
  {Waldorf}},\ }\href {https://doi.org/10.1088/1126-6708/2009/09/073}
  {\bibfield  {journal} {\bibinfo  {journal} {Journal of High Energy Physics}\
  }\textbf {\bibinfo {volume} {2009}},\ \bibinfo {pages} {073} (\bibinfo {year}
  {2009})}\BibitemShut {NoStop}%
\bibitem [{\citenamefont {Henriques}()}]{180173}%
  \BibitemOpen
  \bibfield  {author} {\bibinfo {author} {\bibfnamefont {A.}~\bibnamefont
  {Henriques}},\ }\href {https://mathoverflow.net/q/180173} {\bibinfo {title}
  {H4(bg,z) torsion free for g a connected lie group}},\ \bibinfo
  {howpublished} {MathOverflow},\ \bibinfo {note}
  {uRL:https://mathoverflow.net/q/180173 (version: 2018-02-17)},\ \Eprint
  {https://arxiv.org/abs/https://mathoverflow.net/q/180173}
  {https://mathoverflow.net/q/180173} \BibitemShut {NoStop}%
\bibitem [{\citenamefont {Quillen}(1971)}]{quillen1971adams}%
  \BibitemOpen
  \bibfield  {author} {\bibinfo {author} {\bibfnamefont {D.}~\bibnamefont
  {Quillen}},\ }\href@noop {} {\bibfield  {journal} {\bibinfo  {journal}
  {Topology}\ }\textbf {\bibinfo {volume} {10}},\ \bibinfo {pages} {67}
  (\bibinfo {year} {1971})}\BibitemShut {NoStop}%
\bibitem [{\citenamefont {Eilenberg}\ and\ \citenamefont
  {MacLane}(1954)}]{eilenberg1954groups}%
  \BibitemOpen
  \bibfield  {author} {\bibinfo {author} {\bibfnamefont {S.}~\bibnamefont
  {Eilenberg}}\ and\ \bibinfo {author} {\bibfnamefont {S.}~\bibnamefont
  {MacLane}},\ }\href@noop {} {\bibfield  {journal} {\bibinfo  {journal}
  {Annals of Mathematics}\ ,\ \bibinfo {pages} {49}} (\bibinfo {year}
  {1954})}\BibitemShut {NoStop}%
\bibitem [{\citenamefont {Kapustin}\ and\ \citenamefont
  {Thorngren}(2014)}]{kapustin2014topological}%
  \BibitemOpen
  \bibfield  {author} {\bibinfo {author} {\bibfnamefont {A.}~\bibnamefont
  {Kapustin}}\ and\ \bibinfo {author} {\bibfnamefont {R.}~\bibnamefont
  {Thorngren}},\ }\href@noop {} {\bibfield  {journal} {\bibinfo  {journal}
  {Advances in Theoretical and Mathematical Physics}\ }\textbf {\bibinfo
  {volume} {18}},\ \bibinfo {pages} {1233} (\bibinfo {year}
  {2014})}\BibitemShut {NoStop}%
\end{thebibliography}%

\end{document}